\documentclass[a4paper,10pt]{report}
\usepackage{amsmath}
\usepackage{amssymb}
\usepackage{textcomp}
\usepackage{bm}
\usepackage{float}
\usepackage{epsfig}
\usepackage{graphics}
\usepackage{a4wide}
\usepackage[nottoc]{tocbibind}

\newcommand{\hc}{\dagger}
\newcommand{\ket}[1]{|#1\rangle}

\newcommand{\vev}[1]{\langle #1\rangle}          
\newcommand{\Lag}{\mathcal{L}}              
\newcommand{\ten}[1]{\,\cdot 10^{#1}}       
\newcommand{\units}[1]{\,\rm{#1}}
\newcommand{\mP}{m_{\rm P}}
\newcommand{\msoft}{m_{\rm soft}}
\newcommand{\const}{{\rm const.}}
\newcommand{\du}[2]{_{#1}^{\phantom{#1}#2}}
\newcommand{\spl}{\:{\scriptstyle +}\,}
\newcommand{\smi}{\,{\scriptstyle -}\,}

\newcommand{\pb}{\bar{\phi}}
\newcommand{\Hb}{\bar{H}}
\newcommand{\si}{\sigma}
\newcommand{\ka}{\kappa}
\newcommand{\la}{\lambda}
\newcommand{\ga}{\gamma}
\newcommand{\eff}{\text{eff}}
\newcommand{\pder}[2]{\frac{\partial #1}{\partial #2}}
\newcommand{\pd}{\partial}
\newcommand{\PR}{P_{\mathcal R}^{1/2}}
\newcommand{\ns}{n_{\rm s}}
\newcommand{\eeq}{\end{equation}}
\newcommand{\beq}{\begin{equation}}
\newcommand{\ba}{\begin{array}}
\newcommand{\ea}{\end{array}}
\newcommand{\bea}{\begin{eqnarray}}
\newcommand{\eea}{\end{eqnarray}}
\newcommand{\1}{\mathit{1}}
\newcommand{\m}{\mathit{m}}
\newcommand{\ti}{\mathit{t_1}}
\newcommand{\tii}{\mathit{t_2}}
\DeclareMathOperator{\tr}{Tr}
\DeclareMathOperator{\Real}{Re}

\DeclareMathOperator{\diag}{diag}
\renewcommand{\det}{\operatorname{Det}}

\newlength{\figwidth}
\begin{document}

\begin{titlepage}
\begin{center}
\vspace*{2cm}
{\Huge\bf Phenomenology and Cosmology of\\[5pt]
Supersymmetric Grand Unified\\[5pt] Theories}
\par
\vspace{3cm} {\LARGE\bf Achilleas Vamvasakis}
\par
\vspace{2cm} {\LARGE A thesis submitted for the degree of
Doctor of Philosophy.\\[5pt]
Supervised by Prof.\@ George Lazarides.}
\par
\vspace{2cm}
{\LARGE\it Physics Division, School of Technology,\\[5pt]
Aristotle University of Thessaloniki,\\[5pt]
Thessaloniki 54124, Greece.}
\par
\vfill {\LARGE\bf June 2009}
\end{center}
\end{titlepage}

\newpage
\thispagestyle{empty}
\mbox{}
\setcounter{page}{0}

\tableofcontents


\chapter*{Acknowledgements}
\addcontentsline{toc}{chapter}{Acknowledgements}

I am grateful to the man who has guided me from my
first steps through physics and has stood by my
side ever since. I avail myself of this opportunity
to express my gratitude to Prof.\@ George Lazarides,
not only for his priceless contribution to this
thesis, but also for every aspect of my personality
that he has affected as my teacher. Many thanks to
Prof.\@ Nikos Ganoulis for his continuous support and
assistance regarding technical matters. I would also
like to thank the Autonomous University of Madrid,
and especially Profs.\@ Luis Ibanez and Juan
Garcia-Bellido, for their hospitality during my
three month visit in Madrid. I owe special thanks
to Prof.\@ Juan Garcia-Bellido for our cooperation
and useful discussions during that time. This
work was partially supported by the Greek Ministry
of Education and Religion and the EPEAK program
``Pythagoras''

\newpage
\thispagestyle{empty}
\mbox{}

\chapter{Preface}
\label{sec:PREFACE}

Until relatively recently, only before about
thirty years ago, a physicist would divide
contemporary knowledge in physics into two
major, largely independent sectors, cosmology
and particle physics. This division would reflect
the two greatest scientific achievements of
twentieth century theoretical physics, general
relativity and quantum mechanics. And although
we have not yet concluded, despite our great
effort, in a unique verifiable theory, which
would engage the above two in a single
theoretical framework, interestingly enough,
in recent years, particle physics and cosmology
are found to be intimately connected in such
an extent that we often include both of them
under the general title ``High Energy Physics''
(HEP). Physicists in our days can hardly
concoct particle physics without worrying about
the cosmological consequences of their models
and vice versa. This has been so mainly due to
``inflation'', a model of cosmological evolution
that has brought a change of paradigm in
cosmology since 1981, when Alan Guth first
published his idea.

This thesis belongs to this category of modern
physics, this hybrid sector that incorporates
research on the smallest and the largest, the
particles and the universe. More specifically,
we explore here the phenomenological and
cosmological consequences of a specific model
belonging to a class of models called
``supersymmetric''. Supersymmetry (SUSY) is
basically a symmetry, as its name suggests.
But it is more than that. It is a new way of
making physics, it is an ordering system within
physics, it is a theory by itself. We will have
more to say about supersymmetry in
Chap.~\ref{sec:MSSM}. To summarize, in this
thesis we work with supersymmetry, a subcategory
of particle physics, and with inflation, a
subcategory of cosmology. The classification is
rough since, as we explained, the boundary
between particle physics and cosmology is in our
days obscure. We have chosen a specific
supersymmetric model, constructed some years
ago to cope with a known particle physics
problem, and explored its surprisingly rich
inflationary cosmology.

In Chap.~\ref{sec:MSSM} we review some of the
salient features of SUSY. We show how one can
complement the Standard Model (SM) of particle
physics to become compatible with SUSY and we
deduce the minimal particle spectrum as it is
expected to be if SUSY is realized in nature.
Finally, we mention the case that SUSY becomes
local, known as ``supergravity'' (SUGRA), and show
its effect on the scalar potential of the theory,
which is important for cosmology. This chapter
is not intended to be a comprehensible
introduction to SUSY but rather a ``notation and
conventions'' chapter, included to make the reader
familiar with the specific notation used in this
thesis.

In Chap.~\ref{sec:HYBRID} we give a short
introduction to hybrid inflation, the kind of
inflation most usually found in SUSY models.
We begin by introducing the original model and
then we go on to consider its natural realization
within supersymmetric theories. Then, we describe
in some detail two variants of SUSY hybrid
inflation, namely ``smooth'' and ``shifted'' hybrid
inflation. These models were introduced to solve
the problem of monopole overproduction after the
end of inflation in SUSY Grand Unified Theories
(GUTs) that predict their existence. These
variants of hybrid inflation also appear
naturally in the cosmology of the model
we investigate in this thesis.

Next, in Chap.~\ref{sec:QUASI}, the specific
model that will be the object of our survey, is
introduced. We display the symmetries and describe
the particle spectrum of the model. We exhibit the
new lagrangian terms and give briefly the reasoning
for their introduction. Finally, we shortly
summarize the main property of the model,
i.e.~Yukawa quasi-unification. As the reader may
notice, cosmology and inflation are not mentioned
in this chapter. This reflects the fact that this
model was first introduced to deal with a problem
completely irrelevant to cosmology and it was
only later realized that this model also contains
interesting inflationary phenomenology.

In the following chapters we start unfolding the
rich ``inflationary variety'' of the model.
In Chap.~\ref{sec:NSHIFT} we describe the first
cosmological scenario, named ``new shifted'' hybrid
inflation. The name suggests its resemblance with
standard shifted hybrid inflation, discussed in
Sec.~\ref{sec:HYBRIDshifted}. Its novelty consists
of the fact that, in contrast to the standard
scenario, it is realized only by renormalizable
interactions in the lagrangian.
Of course, the mere fact that this inflationary
scenario naturally arises from a specific viable
particle physics model has its own value. We first
present the scenario in global SUSY and summarize
the calculation of radiative corrections, which
are important for driving inflation. Then, we
consider the changes that are brought to the
model by making SUSY local, i.e. by including
SUGRA corrections. In particular, it is shown
that the scenario remains viable and that
inclusion of SUGRA does not ruin inflation.

An other, qualitatively different, inflationary
scenario, contained in the same particle physics
model for a wide range of the parameter space, is
presented in Chap.~\ref{sec:SSHIFT}. It is called
``semi-shifted'' hybrid inflation and, as its name
suggests, it also bears similarities to shifted
hybrid inflation. However, in this case, the
GUT gauge group is not completely broken to
the SM gauge group during inflation but it
carries an extra unbroken $\rm U(1)$ symmetry,
which breaks immediately after inflation leading
to the formation of cosmic strings. These strings
can then contribute a small amount to the
primordial curvature perturbation, giving thus a
different cosmological situation. We first
introduce the model in global SUSY and give some
details of the calculation of the radiative
corrections, which, as usual, are important for
driving inflation. Then, we account for SUGRA
corrections in the model and present the
necessary framework for dealing with cosmic
strings. Finally, we display our numerical
results and show how this model can become
compatible with gauge coupling constant
unification.

In Chap.~\ref{sec:NSMOOTH} we introduce the
``new smooth'' hybrid inflation model. Its name,
as compared to ``new shifted'' hybrid inflation,
suggests the resemblance of this model with
smooth hybrid inflation, discussed in
Sec.~\ref{sec:HYBRIDsmooth}. Again, one
interesting feature of the new model, in
contrast to the old one, is its realization
with the sole use of renormalizable
interactions. However, this is not its only
advantageous point. It turns out that the
model provides us with one extra degree of
freedom in fitting the cosmological data.
Thus, in the case of global SUSY, we can
achieve spectral indices much lower than the
ones expected from the old model. Nevertheless,
when we go on to consider the effect of SUGRA
corrections, we will see that this degree of
freedom is not enough, by itself, to account
for a spectral index compatible with recent
data. Thus, in Sec.~\ref{sec:NSMOOTHsugra},
we explain how we can work this problem out,
by including non-minimal terms in the K\"ahler
potential.

An other feature of the ``new smooth'' model that
can be exploited in our favor is the fact that
the ``new smooth'' inflationary phase follows
continuously from the standard hybrid phase,
also contained in the same particle physics
model under consideration. This guides us to
the construction of a two-stage inflation model,
which incorporates the two aforementioned
consecutive stages and leads to increased
freedom in fitting the cosmological data. This
scenario, called ``standard-smooth'' hybrid
inflation, is described in Chap.~\ref{sec:STSMOOTH}.
We first present, as usual, the model in the
global SUSY case. Then, we introduce SUGRA
corrections and show that the predicted
parameters can easily become compatible with
the data, even in the case of minimal SUGRA. In
the last section of this chapter, we again
give a brief account of gauge coupling
constant unification.

Finally, in Chap.~\ref{sec:CONC}, we summarize
our conclusions from the survey of this model,
which, if nothing else, it demonstrates the
usefulness of relating cosmology with particle
physics, as well as the wealth of new
cosmological models that can emerge even
from a single particle physics model.

\chapter{Supersymmetry and the Minimal
Supersymmetric Standard Model}
\label{sec:MSSM}

\section{Introduction}
\label{sec:MSSMintro}

The Standard Model (SM) of high energy physics,
incorporating the Glashow-Weinberg-Salam model
of electroweak interactions and the theory of
Quantum Chromodynamics into a single theory, based
on the semi-simple gauge group
$G_{\rm SM}=\rm SU(3)_c\times SU(2)_L\times U(1)_Y$,
has been proven a very successful framework for
the description of particle physics experiments
at energies up to a few hundreds GeV.
Nevertheless, a number of theoretical reasons
indicate that the SM may not be a correct theory
at higher energies.

First of all, it is not a complete theory since
it does not include gravity and any attempt so
far towards this direction has failed. But,
regardless of that purely theoretical reason,
there are also hitches of a more ``quantitative''
nature. In particular, the so called ``hierarchy
problem'', establishes a huge discrepancy between
the experimentally anticipated order of magnitude
for the Higgs boson mass and its theoretical
prediction. We know from the experiment that if
the Higgs boson is responsible for the masses of
all the particles in the SM, then its vacuum
expectation value (VEV) is of the order
$\vev{H}\sim170\units{GeV}$. Any reasonable
symmetry breaking theory should contain a Higgs
field with a mass between 100 and 1000 GeV if it
is to predict the correct VEV for it without
significant fine tuning. The problem is that such
a low mass parameter is susceptible to large
radiative corrections, which render it of the
order of the unification scale of the theory.
Since we are not willing to abandon the idea
that the Higgs field is the origin of the masses
of all the massive particles (this is the reason
it was introduced in the first place), the only
way for such a low mass parameter to be stable
under radiative corrections is through some kind
of special symmetry, able to fix the positive
and negative parts of the radiative corrections
to be exactly equal, so that they can cancel
each other.

Physicists have come up with such a symmetry,
called supersymmetry (SUSY), although in its
early stages its development was not driven by
the need for a solution to the hierarchy problem.
Regarding this, it is quite impressive that a
theory introduced for some theoretical reasons
was later proven to also provide a solution to
the hierarchy problem. Since radiative
corrections include contributions from both
boson and fermion loops, it is obvious that
this symmetry should relate bosons and fermions
in some profound way. Thus, the underlying
transformation should transform bosons into
fermions and vice versa. The operator $Q$ that
generates these transformations will act on
bosonic and fermionic states, turning them into
fermionic and bosonic respectively, according to
the scheme
\beq\label{eq:SUSYoperator}
Q\ket{{\rm Boson}}=\ket{{\rm Fermion}},
\qquad
Q\ket{{\rm Fermion}}=\ket{{\rm Boson}}.
\eeq
It is apparent from Eq.~\eqref{eq:SUSYoperator}
that $Q$ is a complex spinor operator that
carries spin angular momentum $1/2$. Thus, its
hermitian conjugate, $Q^\dag$, is also of the
same nature. It can be proven
\cite{ColemanMandula} that, for realistic
theories with chiral fermions, the generators
$Q$ and $Q^\hc$ must satisfy an algebra of
commutation and anticommutation relations, called
the SUSY algebra, of the form
\begin{align}
& \{Q_\alpha,Q_{\dot{\beta}}^\hc\}=
2\si^\mu_{\alpha\dot{\alpha}}P_\mu,
\label{eq:algebra1}\\
& \{Q_\alpha,Q_\beta\}=
\{Q_{\dot{\alpha}}^\hc,Q_{\dot{\beta}}^\hc\}=0,
\label{eq:algebra2}\\
& [P^\mu,Q_\alpha]=[P^\mu,Q_{\dot{\alpha}}^\hc]=0,
\label{eq:algebra3}
\end{align}
where $P^\mu$ is the generator of space-time
translations, $\si^\mu$ represents the Pauli
sigma matrices and $\alpha$, $\dot{\alpha}$,
$\beta$, $\dot{\beta}$ are two component spinor
indices (see e.g.~\cite{SUSYprimer}).
The relations in
Eqs.~\eqref{eq:algebra1}-\eqref{eq:algebra3}
can be used to show that each supermultiplet
(i.e.~each irreducible representation of the SUSY
algebra) must contain an equal number of
fermionic and bosonic degrees of freedom.
From Eq.~\eqref{eq:algebra3} it follows that the
operator $P^\mu P_\mu$ commutes with $Q$ and
$Q^\hc$, which implies that particles in the same
supermultiplet must have equal masses. The SUSY
generators $Q$ and $Q^\hc$ also commute with the
generators of gauge transformations. Therefore,
particles in the same supermultiplet must also
reside in the same representation of the gauge
group. Since we do not know of any such particles
having equal masses, SUSY must be broken.

The solution to the hierarchy problem by
supersymmetry is achieved by the cancellation
of the quadratic divergencies,
$\Lambda_{\rm UV}^2$, coming from the radiative
corrections and this requires that the associated
dimensionless couplings of fermion and boson
superpartners, say schematically $\la_S$ and
$\la_f$, are related, for example by a relation
such as $\la_S=|\la_f|^2$
(see e.g.~\cite{SUSYprimer}). In fact, unbroken
SUSY guarantees that the quadratic divergencies
in scalar squared masses vanish to all orders in
perturbation theory. But what about broken SUSY?
Broken SUSY should still provide a solution to
the hierarchy problem, which means that the
relations between dimensionless couplings must
be maintained. This leads to a very important
consequence about broken SUSY, i.e.~SUSY must be
``softly'' broken. This means that the effective
lagrangian can be written in the form
\beq
\Lag=\Lag_{\rm SUSY}+\Lag_{\rm soft},
\eeq
where $\Lag_{\rm SUSY}$ preserves supersymmetry
and $\Lag_{\rm soft}$ violates supersymmetry but
contains only mass terms and couplings with
positive mass dimension (i.e.~it does not contain
Yukawa couplings). It turns out that there are
plenty natural theoretical models for SUSY
breaking with this property.

If the largest mass scale in the lagrangian
$\Lag_{\rm soft}$ is denoted by $\msoft$, then
the SUSY breaking corrections to the Higgs
$\rm mass^2$, $\Delta m_H^2$, should vanish in
the limit $\msoft\to0$, so they can not be
proportional to $\Lambda_{\rm UV}^2$.
Furthermore, the corrections cannot go like
$\Delta m_H^2\sim\msoft\Lambda_{\rm UV}$, because
the loop integrals always diverge either
quadratically or logarithmically and never
linearly with $\Lambda_{\rm UV}$. So, by
dimensional analysis, they must be of the form
\beq\label{eq:HiggsCorrection}
\Delta m_H^2=\msoft^2\left[\frac{\la}{16\pi^2}
\ln(\Lambda_{\rm UV}/\msoft)+\dots\right],
\eeq
where the dots represent either terms that are
independent of $\Lambda_{\rm UV}$ or higher
order terms. Eq.~\eqref{eq:HiggsCorrection} shows
that the superpartner masses cannot be too big if
we want to cure the hierarchy problem with broken
SUSY and no fine tuning. This is the reason why
we expect the consequences of supersymmetry to
arise not very much higher than about
$1\units{TeV}$.

In the rest of this section we will briefly state
a few things about the notation and the
conventions used in this thesis. We will use two
component Weyl spinor notation instead of four
component Dirac spinors, as well because the
description of SUSY in this context is much
simpler, as because in SUSY models the minimal
building blocks of matter are supermultiplets
containing a single two component Weyl fermion.

A four component Dirac fermion $\Psi_D$ with mass
$M$ is described by the lagrangian
\beq\label{eq:LDirac}
\Lag_{\rm Dirac}=-i\bar{\Psi}_D
\ga^\mu\pd_\mu\Psi_D
-M\bar{\Psi}_D\Psi_D,
\eeq
where the space-time metric is
$\eta_{\mu\nu}=\diag(-1,1,1,1)$ and we use the
following representation for the gamma and sigma
matrices
\begin{gather}
\ga_\mu=\left(
\ba{cc}0 & \si_\mu \\ \bar{\si}_\mu & 0 \ea\right),
\qquad \ga_5=\left(
\ba{cc} 1 & 0 \\0 & -1 \ea\right),\\[3pt]
\si_0=\bar{\si}_0=\left(
\ba{cc} 1 & 0 \\0 & 1 \ea\right),\qquad
\si_1=-\bar{\si}_1=\left(
\ba{cc} 0 & 1 \\1 & 0 \ea\right),\\[3pt]
\si_2=-\bar{\si}_2=\left(
\ba{cc} 0 & -i \\i & 0 \ea\right),\qquad
\si_3=-\bar{\si}_3=\left(
\ba{cc} 1 & 0 \\0 & -1 \ea\right).
\end{gather}
In this bases, a four component Dirac spinor is
written in terms of two component, complex,
anticommuting Weyl spinors $\xi_\alpha$ and
$\chi^{\hc\dot{\alpha}}$
($\alpha,\dot{\alpha}=1,2$), as
\beq
\Psi_D=\left(\ba{c}\xi_\alpha \\[2pt]
\chi^{\hc\dot{\alpha}}\ea\right),\qquad
\bar{\Psi}_D=\left(\ba{cc}\chi^\alpha &
\xi^{\hc}_{\dot{\alpha}}\ea\right).
\eeq
Here the notation is somewhat misleading since
the indices $\alpha$ and $\dot{\alpha}$ do not
appear on the left hand side. What this notation
really means is that the undotted indices are
used for the first two components of a Dirac
spinor while the dotted are used for the last
two ones. Note also that the location of an
index (up or down) is important. The spinor
indices are raised and lowered using the
antisymmetric symbol $\epsilon^{\alpha\beta}$,
with components $\epsilon^{12}=-\epsilon^{21}=
-\epsilon_{12}=\epsilon_{21}=1$ and
$\epsilon^{11}=\epsilon^{22}=\epsilon_{11}=
\epsilon_{22}=0$. The spinor $\xi$ represents a
left handed Weyl spinor while $\chi^\hc$ is a
right handed one. The notation implies that the
hermitian conjugate of a left handed Weyl spinor
is a right handed Weyl spinor and vice versa,
$(\psi_\alpha)^\hc=\psi^\hc_{\dot{\alpha}}$.

While $\xi$ and $\chi$ are anticommuting objects,
the relation $\xi\chi=\chi\xi$ holds because of
the abbreviation used here, which is $\xi\chi=
\xi^\alpha\chi_\alpha=\xi^\alpha
\epsilon_{\alpha\beta}\chi^\beta$. Likewise,
$\xi^\hc\chi^\hc=\xi^\hc_{\dot{\alpha}}
\chi^{\hc\dot{\alpha}}=(\xi\chi)^*$, the complex
conjugate of $\xi\chi$. Similarly,
$\xi^\hc_{\dot{\alpha}}
(\bar{\si}^\mu)^{\dot{\alpha}\alpha}\chi_\alpha
\equiv\xi^\hc\bar{\si}^\mu\chi=
-\chi\si^\mu\xi^\hc=(\chi^\hc\bar{\si}^\mu\xi)^*
=-(\xi\si^\mu\chi^\hc)^*$. With these conventions,
the Dirac lagrangian in Eq.~\eqref{eq:LDirac} can
be written as
\beq
\Lag_{\rm Dirac}=
-i\xi^\hc\bar{\si}^\mu\pd_\mu\xi
-i\chi^\hc\bar{\si}^\mu\pd_\mu\chi
-M(\xi\chi+\xi^\hc\chi^\hc),
\eeq
where we have dropped a total derivative piece,
$i\pd_\mu(\chi^\hc\bar{\si}^\mu\chi)$, which
does not affect the action.

More generally any theory involving spin-$1/2$
fermions can always be written in terms of a
collection of left handed Weyl spinors $\psi_i$
with the kinetic part of the lagrangian being
\beq
\Lag_{\rm kin}=-i\psi^{i\hc}\bar{\si}^\mu
\pd_\mu\psi_i.
\eeq
For a Dirac fermion there is a different $\psi_i$
for its left handed part and for the hermitian
conjugate of its right handed part. According to
these general rules, we give names to the left
handed spinors of the SM particles as follows
\bea
Q_i&=&(u,d),\; (c,s),\; (t,b),
\label{eq:SMfields1}\\
\bar{u}_i&=&\bar{u},\; \bar{c},\; \bar{t},
\label{eq:SMfields2}\\
\bar{d}_i&=&\bar{d},\; \bar{s},\; \bar{b},
\label{eq:SMfields3}\\
L_i&=&(\nu_e,e),\;(\nu_\mu,\mu),\;(\nu_\tau,\tau),
\label{eq:SMfields4}\\
\bar{e}_i&=&\bar{e},\;\bar{\mu},\;\bar{\tau},
\label{eq:SMfields5}
\eea
where $i$ is the family index and $Q_i$ and $L_i$
are doublets under the weak isospin symmetry
$\rm SU(2)_L$ of $G_{\rm SM}$. The bars on the
fields are part of their names and they do not
indicate any kind of conjugation. In particular
they are introduced to declare that the barred
fields come from the conjugates of the right
handed parts of the corresponding Dirac spinors.
For example the Dirac spinor for the electron is
written in terms of its two component left handed
spinors as
\beq
\Psi_e=\left(\ba{c} e \\ \bar{e}^\hc\ea\right)
=\left(\ba{c} e_L \\ e_R\ea\right).
\eeq
Note that the neutrinos are not part of a Dirac
spinor in the SM. Of course, it is common in
Grand Unified Theory (GUT) models to complete
the picture with right handed parts for the
neutrinos, which usually acquire large masses of
the order of the GUT scale and explain the small
left handed neutrino masses through the see-saw
mechanism. Suppressing all color and weak isospin
indices, the purely kinetic part of the SM
fermion lagrangian is
\beq
\Lag^{\rm SM}_{\rm kin}=
-iQ^{i\hc}\bar{\si}^\mu\pd_\mu Q_i
-i\bar{u}^{i\hc}\bar{\si}^\mu\pd_\mu\bar{u}_i
-i\bar{d}^{\,i\hc}\bar{\si}^\mu\pd_\mu\bar{d}_i
-iL^{i\hc}\bar{\si}^\mu\pd_\mu L_i
-i\bar{e}^{i\hc}\bar{\si}^\mu\pd_\mu\bar{e}_i.
\eeq

\section{Supersymmetric lagrangians}
\label{sec:MSSMlag}

Now, let us turn to the construction of
supersymmetric lagrangians using the notation
introduced in the previous section. The aim of
the description here is to make the reader
familiar with the notation and the style of
this thesis and not to present a pedagogical
introduction to supersymmetry. We begin by
considering the simplest example of a
supersymmetric theory in four dimensions.

The simplest possibility is a supermultiplet
containing a single Weyl fermion (with two
degrees of freedom) and two real scalars (each
with one degree of freedom). Furthermore, one
can assemble the two real scalars into a complex
scalar field. This set of a two component Weyl
fermion and a complex scalar field is called a
``chiral'' or ``scalar'' supermultiplet. The next
simplest possibility for a supermultiplet
contains a spin-$1$ vector boson, which is
massless before the gauge symmetry breaking, so
it has two bosonic degrees of freedom. Its
superpartner is therefore a spin-$1/2$ Weyl
fermion with two fermionic degrees of freedom.
This collection of fields is called a ``gauge''
or ``vector'' supermultiplet. Gauge bosons
transform as the adjoint representation of the
gauge group, so their fermionic partners, called
``gauginos'', must also transform according to
the same representation. The adjoint
representation of a gauge group is always its
own conjugate, so these fermions have left and
right handed components with the same
transformation properties. This is in contrast
with the fermions of the Standard Model which
have left and right handed components belonging
to different representations of the gauge group.
Thus, in any supersymmetric extension of the SM
all the known fermions must be included in chiral
supermultiplets while all the known vector bosons
will necessarily belong to gauge supermultiplets.

The simplest SUSY theory consists of a free
chiral supermultiplet, containing a left handed
Weyl fermion $\psi$ and a complex scalar $\phi$,
with only kinetic terms, the simplified
Wess-Zumino model \cite{WessZumino}
\beq
\Lag=-\pd^\mu\phi^*\pd_\mu\phi
-i\psi^\hc\bar{\si}^\mu\pd_\mu\psi.
\eeq
It can be proven (see e.g.~\cite{SUSYprimer})
that the action, $S=\int d^4x\;\Lag$, of this
simple model is invariant under the SUSY
transformations when the model is on-shell,
i.e.~when the equation of motion
$\bar{\si}^\mu\pd_\mu\psi=0$, following
from the action, is employed. Still, we would
like supersymmetry to hold even off-shell.This
can be fixed by a trick. We invent a new complex
scalar field $F$ with no kinetic terms. Such a
field is called auxiliary and it is not a real
degree of freedom but only an object used to
render the action invariant off-shell. The
lagrangian density for $F$ is just
$\Lag_{\rm aux}=F^*F$. The dimensions of $F$ are
$\rm mass^2$ unlike ordinary scalar fields. The
equation of motion for $F$ in the non-interacting
theory is trivial, $F=0$, but it becomes
non-trivial in the interacting case. In general,
if we have a collection of chiral supermultiplets
labelled by the index $i$, the free part of the
lagrangian is written as
\beq
\Lag_{\rm chiral}^{\rm (free)}=
-\pd^\mu\phi^{i*}\pd_\mu\phi_i
-i\psi^{i\hc}\bar{\si}^\mu\pd_\mu\psi_i
+F^{i*}F_i.
\eeq

The next thing is to do is to write down the
lagrangian for a gauge supermultiplet. Consider
a massless gauge boson $A_\mu^a$ and a
corresponding set of Weyl fermion gauginos
$\la^a_\alpha$, where the index $a$ runs over the
adjoint representation of the gauge group. The
on-shell degrees of freedom for $A_\mu^a$ and
$\la^a_\alpha$ amount to two bosonic and two
fermionic degrees of freedom for each $a$, as
required by SUSY. However, the off-shell degrees
of freedom do not much and we need to invent an
auxiliary field, called $D^a$, in order to make
the lagrangian supersymmetric. This field will
also transform as the adjoint representation of
the gauge group. Like the auxiliary field $F$,
it has dimensions of $\rm mass^2$ and thus no
kinetic terms. Without any further justification
we write down the lagrangian for a gauge
supermultiplet,
\beq\label{eq:Lgauge}
\Lag_{\rm gauge}^{\rm (free)}=
-\frac{1}{2}F_{\mu\nu}^aF^{\mu\nu a}
-i\la^{a\hc}\bar{\si}^\mu D_\mu\la^a
+\frac{1}{2}D^aD^a,
\eeq
where
\beq
F_{\mu\nu}^a=\pd_\mu A_\nu^a-\pd_\nu
A_\mu^a-gf^{abc}A_\mu^b A_\nu^c
\eeq
is the usual Yang-Mills field strength and
\beq
D_\mu\la^a=\pd_\mu\la^a-gf^{abc}A_\mu^b\la^c
\eeq
is the covariant derivative for the gaugino field.
It can be proven (see e.g.~\cite{SUSYprimer}) that
this lagrangian is invariant under both gauge and
SUSY transformations. Again, the equations of
motion for the auxiliary fields are $D^a=0$, but
this is no longer true in the interacting theory.

Up to now we have only dealt with free theories,
not of much practical importance. We now turn to
the description of interacting theories in the
context of SUSY. Starting with the case of
chiral supermultiplets, we will argue that the
most general set of renormalizable interactions
for these fields can be written as
\beq\label{eq:Lint}
\Lag_{\rm chiral}^{\rm (int)}=
-\frac{1}{2}W^{ij}\psi_i\psi_j+W^iF_i+c.c.,
\eeq
where $W^{ij}$ and $W^i$ are some functions of
the scalar fields with dimensions of $\rm mass$
and $\rm mass^2$ respectively. It follows from
Eq.~\eqref{eq:Lint} that, if the lagrangian is
renormalizable, by dimensional analysis
$W^{ij}$ and $W^i$ can not contain fermion or
auxiliary fields. Furthermore, $W^i$ will be a
quadratic polynomial and $W^{ij}$ linear in the
fields $\phi_i$ and $\phi^{i*}$. Also, the
lagrangian can not contain terms that are
functions of the scalar fields only, because it
can be shown that these terms do not respect
SUSY. So, Eq.~\eqref{eq:Lint} is indeed the
most general possibility. One can show, by
requiring that the lagrangian is invariant
under SUSY transformations, that
$\pd W^{ij}/\pd\phi^{k*}=0$, i.e.~the
function $W^{ij}$ is analytic in the fields
$\phi_k$. In addition, from Eq.~\eqref{eq:Lint}
$W^{ij}$ can be taken to be symmetric under
interchange of the indices $i,j$, so it can be
written as
\beq
W^{ij}=M^{ij}+y^{ij\,k}\phi_k,
\eeq
where $M^{ij}$ represents a symmetric fermion
mass matrix and $y^{ij\,k}$ the Yukawa couplings
of a scalar with two fermions, totally symmetric
under interchange of its indices. It is
convenient to write
\beq\label{eq:SuperpotDef}
W^{ij}=\frac{\delta^2}
{\delta\phi_i\delta\phi_j}\,W
\eeq
introducing the ``superpotential'' $W$, given by
\beq\label{eq:SuperpotExpr}
W=\frac{1}{2}M^{ij}\phi_i\phi_j
+\frac{1}{6}\,y^{ij\,k}\phi_i\phi_j\phi_k.
\eeq
$W$ is by no means a scalar potential in the
ordinary sense, since it is not even real, but
it is an analytic function of the complex
variables $\phi_i$. Continuing to pursue the
implications of the requirement that the
lagrangian should respect SUSY and taking into
account Eq.~\eqref{eq:SuperpotDef}, one can
prove that the function $W^i$ should be
\beq
W^i=\frac{\delta W}{\delta\phi_i}=
M^{ij}\phi_j+\frac{1}{2}y^{ij\,k}\phi_j\phi_k.
\eeq
Now it is clear why we have used for the two
functions $W^{ij}$ and $W^i$ the same symbol.

To summarize the results in the case of chiral
supermultiplets, we have found that the most
general interactions can be determined simply by
a single analytic function of the complex scalar
fields, the superpotential $W$. The auxiliary
field $F_i$ can be eliminated from the final
form of the lagrangian using the classical
equations of motion. From the lagrangian
$\Lag_{\rm chiral}^{\rm (free)}+
\Lag_{\rm chiral}^{\rm (int)}$, one finds
$F_i=-W_i^*$ and $F^{i*}=-W^i$ and thus, the
lagrangian for the chiral supermultiplets takes
the form
\beq\label{eq:Lchiral}
\Lag_{\rm chiral}=
-\pd^\mu\phi^{i*}\pd_\mu\phi_i
-i\psi^{i\hc}\bar{\si}^\mu\pd_\mu\psi_i
-\frac{1}{2}\left(W^{ij}\psi_i\psi_j+h.c.\right)
-W^iW_i^*.
\eeq
It is clear from that equation that the scalar
potential of the theory is completely determined
by the superpotential and it reads
\bea
V(\phi,\phi^*)=W^iW_i^*
&=&M^{ji}M_{i\,l}^*\phi_j\phi^{\,l*}
+\frac{1}{2}M^{ji}y_{i\,lm}^*
\phi_j\phi^{\,l*}\phi^{m*}\nonumber\\
& &+\frac{1}{2}M_{li}^*\,y^{ij\,k}
\phi^{\,l*}\phi_j\phi_k
+\frac{1}{4}y^{j\,ki}y_{i\,lm}^*
\phi_j\phi_k\phi^{\,l*}\phi^{m*}
\eea
Of course, this is only the case when SUSY is
unbroken. In broken SUSY there are also other
terms in the scalar potential, which are
responsible for SUSY breaking and which are not
so strictly determined. The only requirement for
these terms is that they should generate soft
SUSY breaking, a restriction discussed earlier
in Sec.~\ref{sec:MSSMintro}. The supersymmetric
scalar potential is automatically bounded from
below. In fact, since it is a sum of squares of
absolute values, it is always non-negative.

Finally let us consider supersymmetric gauge
interactions. Suppose that the chiral
supermultiplets transform under the gauge
group in a representation
$(T^a)_i^{\phantom{i}j}$ satisfying
$[T^a,T^b]=if^{abc}T^c$, where $f^{abc}$ are
the structure constants. Following well known
techniques, to get a gauge invariant lagrangian
we need to turn the ordinary derivatives into
covariant derivatives, as
\beq
\pd_\mu\phi_i\to
D_\mu\phi_i+igA_\mu^a(T^a\phi)_i,\quad
\pd_\mu\psi_i\to
D_\mu\psi_i+igA_\mu^a(T^a\psi)_i.
\eeq
Yet, this is not the end of the story since we
have to consider whether there are any other
interactions allowed by gauge invariance
involving the gaugino and $D^a$ fields.
Indeed, there are three such renormalizable
terms, which read
\beq
(\phi^*T^a\psi)\la^a,\qquad
\la^{a\hc}(\psi^\hc T^a\phi)\qquad
{\rm and}\qquad(\phi^*T^a\phi)D^a.
\eeq
One can add these terms with arbitrary
dimensionless coupling constants and demand that
the whole lagrangian be real and invariant under
SUSY transformations, up to a total divergence.
This fixes the coefficients of these extra terms
and the total lagrangian becomes
\beq\label{eq:Ltotal}
\Lag=\Lag_{\rm chiral}
+\Lag_{\rm gauge}^{\rm (free)}
-\sqrt{2}\,g\left[(\phi^*T^a\psi)\la^a
+\la^{a\hc}(\psi^\hc T^a\phi)\right]
+g(\phi^*T^a\phi)D^a,
\eeq
where $\Lag_{\rm chiral}$ is the lagrangian given
in Eq.~\eqref{eq:Lchiral} but with ordinary
derivatives replaced by gauge covariant
derivatives and $\Lag_{\rm gauge}^{\rm (free)}$
is the lagrangian given in Eq.~\eqref{eq:Lgauge}.
From this lagrangian one can find the equations
of motion for the $D^a$ fields, which are
\beq
D^a=-g(\phi^*T^a\phi).
\eeq
Replacing $D^a$ in Eq.~\eqref{eq:Ltotal}, one
finds that the complete scalar potential is
given by
\beq
V(\phi,\phi^*)=F^{i*}F_i+\frac{1}{2}D^aD^a=
W^iW_i^*+\frac{1}{2}\sum_a g_a^2(\phi^*T^a\phi)^2.
\eeq
The two types of terms in this expression are
called ``F-terms'' and ``D-terms''
respectively. The index $a$ in the gauge coupling
constant $g_a$ is introduced to include the case
that the gauge group has several distinct factors
with different gauge couplings, as is the case
with $G_{\rm SM}$. Note that the final scalar
potential is always non-negative in a
supersymmetric theory, since it is a sum of
absolute squares. A very interesting and
unique feature of unbroken SUSY is that the
scalar potential is completely determined by
the superpotential and the gauge interactions
in the theory.

\section{The MSSM superpotential}
\label{sec:MSSMsuperpot}

Given the supermultiplet content of the theory,
the form of the superpotential is restricted by
gauge invariance. In any given theory, only a
subset of the couplings $M^{ij}$ and $y^{ij\,k}$
in Eq.~\eqref{eq:SuperpotExpr} is allowed to be
non-zero. In this section we will roughly describe
the simplest possible supersymmetric extension of
the SM, called Minimal Supersymmetric Standard
Model (MSSM).

The field content of the theory is the one given
in Eqs.~\eqref{eq:SMfields1}-\eqref{eq:SMfields5},
supplemented with their corresponding scalar
superpartners and the usual gauge bosons of the
SM along with their gaugino superpartners. In
addition, one has to include a Higgs field,
responsible for the breaking of the electroweak
symmetry, accompanied by its fermionic
superpartner, called the ``higgsino''. In
general, the nomenclature for a fermionic
superpartner of a SM bosonic field is to append
the ending \mbox{``-ino''} to its name. On the
other hand, the names for the scalar
superpartners of the SM fermions are constructed
by prepending an ``s-'' to their names. For
example, the superpartners of the quarks and
leptons are generically called squarks and
sleptons. In this thesis we will denote the
scalar component of a chiral supermultiplet with
the same symbol as the one used for the
supermultiplet itself. The fermionic component of
the supermultiplet, which is a two component Weyl
spinor, will be denoted by $\psi_x$, where $x$ is
the symbol used for the supermultiplet (and its
scalar component). For example, the electron,
which is the superpartner of the selectron $e$,
is denoted by $\psi_e$. For a gauge
supermultiplet, the gaugino corresponding to a
gauge boson $A_\mu^a$ will be denoted by $\la^a$,
with $T^a$ being the generator to which the gauge
boson corresponds. In some circumstances, as
e.g.~in Table~\ref{tab:MSSMparticles}, the
gaugino will be simply denoted by $\la_X$, if $X$
is the symbol used for the gauge boson.

Before writing down the full particle content of
the MSSM, let us first say a few words about the
Higgs boson. In the SM there was a need for only
one such field with a scalar potential,
responsible for giving mass to all the massive
particles of the theory. Because of the structure
of the superpotential, which is an analytic
function of the scalar fields, in the MSSM one
needs at least two scalar fields, with different
transformation properties under $\rm U(1)_Y$ of
$G_{\rm SM}$, to give masses to the up and down
type quarks. In particular, as one can see from
Table~\ref{tab:MSSMparticles}, one needs an
$\rm SU(2)_L$ doublet with $Y=1/2$, denoted by
$H_u$, to give mass to up type quarks and one
with $Y=-1/2$, denoted by $H_d$, to give mass
to down type quarks. This minimal choice for
the Higgs bosons completes the particle spectrum
of the MSSM, which is given collectively in
Table~\ref{tab:MSSMparticles} along with the
transformation properties of the supermultiplets
under $G_{\rm SM}$.

\begin{table}[tbp]
\centering
\caption{The MSSM particle content}
\label{tab:MSSMparticles}
\vspace{3mm}
\begin{tabular}[c]{|c|c|c|c|c|}
\hline
\multicolumn{2}{|c|}{Names} & SM particles &
superpartners & $G_{\rm SM}$\\
\hline \hline
&&&&\\[-3mm]
quarks, squarks & $Q$ &
$(\psi_u\;\;\psi_d\,)$ & $(u\;\;d\,)$ &
$(\bm{3},\bm{2},\frac{1}{6})$ \\[1mm]
($\times 3$ families) & $\bar{u}$ &
$\psi_{\bar{u}}$ & $\bar{u}$ &
$(\bar{\bm{3}},\bm{1},-\frac{2}{3})$ \\[1mm]
& $\bar{d}$ & $\psi_{\bar{d}}$ & $\bar{d}$ &
$(\bar{\bm{3}},\bm{1},\frac{1}{3})$\\[1mm]
\hline
&&&&\\[-3mm]
leptons, sleptons & $L$ &
$(\psi_\nu\;\;\psi_e)$ & $(\nu\;\;e)$ &
$(\bm{1},\bm{2},-\frac{1}{2})$ \\[1mm]
($\times 3$ families) & $\bar{e}$ &
$\psi_{\bar{e}}$ & $\bar{e}$ &
$(\bm{1},\bm{1},1)$ \\[1mm]
\hline
&&&&\\[-3mm]
Higgs, higgsinos & $H_u$ &
$(\psi_{H_u^+}\;\;\psi_{H_u^0})$ &
$(H_u^+\;\;H_u^0)$ &
$(\bm{1},\bm{2},+\frac{1}{2})$ \\[1mm]
& $H_d$ & $(\psi_{H_d^0}\;\;\psi_{H_d^-})$ &
$(H_d^0\;\;H_d^-)$ &
$(\bm{1},\bm{2},-\frac{1}{2})$ \\[1mm]
\hline
&&&&\\[-3mm]
gluon, gluino & $g$ & $g$ & $\la_g$ &
$(\bm{8},\bm{1},0)$ \\[1mm]
W bosons, winos & $W$ & $W^{\pm},W^0$ &
$\la_{W^\pm},\la_{W^0}$ &
$(\bm{1},\bm{3},0)$ \\[1mm]
B boson, bino & $B$ & $B^0$ & $\la_{B^0}$ &
$(\bm{1},\bm{1},0)$ \\[1mm]
\hline
\end{tabular}
\end{table}

Now that we have the full particle content of the
theory, we are ready to write down the correct
superpotential. Except for SUSY and gauge
invariance, we will postulate that the
superpotential respects an extra discrete
symmetry, known as ``$R$-parity'' or ``matter
parity''. Define the operator
\beq
P_M=(-1)^{3(B-L)},
\eeq
where $B$ and $L$ are the baryon and lepton
number operators respectively. $R$-parity
conservation consists of the principle that a
term in the superpotential is allowed only if
the product of $P_M$ for all of the fields in
it is $+1$. This symmetry prevents terms in the
superpotential, like $LL\bar{e}$, that respect
gauge invariance but violate baryon and lepton
number and lead to fast proton decay, a process
strictly constrained by experiment. The
superpotential for the MSSM, containing all the
possible renormalizable terms that respect gauge
invariance and $R$-parity, is
\beq\label{eq:WMSSM}
W_{\rm MSSM}=\bar{u}\,\bm{y}_u Q H_u
-\bar{d}\,\bm{y}_d\,Q H_d
-\bar{e}\,\bm{y}_e L H_d+\mu H_u H_d,
\eeq
where all the gauge and family indices are
suppressed and the dimensionless parameters
$\bm{y}_u$, $\bm{y}_d$ and $\bm{y}_e$ are
$3\times3$ matrices in family space. The
``$\mu$-term'', as it is traditionally called,
is the supersymmetric version of the Higgs boson
mass and it can be written out analytically as
$\mu(H_u)_\alpha(H_d)_\beta
\epsilon^{\alpha\beta}$. The minus signs in
Eq.~\eqref{eq:WMSSM} are chosen by convention so
that when the matrices $\bm{y}_u$, $\bm{y}_d$ and
$\bm{y}_e$ are diagonal, the terms that will
give masses to the quarks and leptons after
electroweak symmetry breaking appear in the
superpotential with positive sign. For example,
if we use the approximation
$\bm{y}_u=\diag(0,0,y_t)$,
$\bm{y}_d=\diag(0,0,y_b)$,
$\bm{y}_e=\diag(0,0,y_\tau)$,
the superpotential reads
\bea
W_{\rm MSSM}&=&y_t\bar{t}tH_u^0+y_b\bar{b}bH_d^0
+y_\tau\bar{\tau}\tau H_d^0-y_t\bar{t}bH_u^+
-y_b\bar{b}tH_d^- -y_\tau\bar{\tau}\nu_\tau H_d^-
\nonumber\\ & &+\mu(H_u^+H_d^- -H_u^0H_d^0).
\eea

From the $\mu$-term of the superpotential, one
can derive the mass terms of the Higgs scalar
potential,
\beq\label{eq:VHiggs}
V_{\rm Higgs}\supset |\mu|^2\left(|H_u^0|^2+
|H_d^0|^2+|H_u^+|^2+|H_d^-|^2\right).
\eeq
Since Eq.~\eqref{eq:VHiggs} is positive definite,
it is clear that we cannot understand electroweak
symmetry breaking without including supersymmetry
breaking soft terms, which can give negative
$\rm mass^2$ terms. Thus, electroweak breaking is
closely related with SUSY breaking. This leads to
the infamous ``$\mu$-problem''. As we have
already pointed out in the introduction, we
expect that $\mu$ should be roughly of order
$10^2$ - $10^3\units{GeV}$, in order to allow a
Higgs VEV of order $170\units{GeV}$ without too
much fine tuning between $|\mu|^2$ and the
negative $\rm mass^2$ terms coming from soft SUSY
breaking terms. The problem is that, although, in
contrast with the SM, $\mu$ will now be stable
under radiative corrections if SUSY is softly
broken, we do not have an explanation of why it
should be so small in the first place compared
to e.g.~the Planck mass $M_{\rm P}$. In
particular, the fact that it is roughly of the
order $\msoft$ (see Sec.~\ref{sec:MSSMintro})
suggests that the $\mu$-term is probably not an
independent parameter of the theory but is
intimately connected with SUSY breaking. Several
different solutions to this problem have been
proposed. They all assume the parameter $\mu$
to be absent at tree level, usually by invoking
some additional symmetry of the superpotential,
as for example a $Z_2$ symmetry. The $\mu$-term
is assumed to be dynamically generated at some
stage of the history of the early universe by
the VEV of some field. In this way, the value of
the effective parameter $\mu$ need no longer be
conceptually distinct from the mechanism of SUSY
breaking. However, from the point of view of the
MSSM, one can treat $\mu$ as an independent
parameter.

\section{Supergravity}
\label{sec:MSSMsugra}

Most symmetries in particle physics are realized
as local symmetries, i.e.~the parameters of a
transformation are functions of the space-time
point $x_\mu$. In particular, because the
SUSY algebra contains the generator of
translations $P_\mu$, we should consider
translations that vary from point to point in
space-time. Thus, we expect local SUSY to be a
theory of general coordinate transformations
of space-time, i.e.~a theory of gravity.
Therefore, the theory of local SUSY is referred
to as supergravity (SUGRA) (for an introduction
see e.g.~\cite{SUGRAintro}). In SUGRA, the spin-2
graviton has a spin-$3/2$ fermion superpartner,
called the ``gravitino''. As long as SUSY is
unbroken, the graviton and the gravitino are both
massless. Once SUSY is spontaneously broken, the
gravitino acquires a mass, which is traditionally
denoted by $m_{3/2}$.

SUGRA is a non-renormalizable theory and so it
can only be thought of as a low energy
approximation of some more complete theory,
e.g.~some string theory. For most practical
purposes, the non-renormalizable
terms can be neglected from the lagrangian,
because they are suppressed by powers of $E/\mP$,
with $\mP$ being the reduced Planck mass and $E$
represents the energy scales accessible to
experiment. However, there are several reasons
why one might be interested in non-renormalizable
contributions to the lagrangian. For example,
some very rare processes, like proton decay, can
only be described by an effective lagrangian with
non-renormalizable terms, since we know that the
proton does not decay through renormalizable
interactions. But, most importantly, the study
of the early universe and cosmology are fields
that SUGRA is expected to have significant
consequences, as they refer to an era of high
energy processes. Thus, in this section we will
very roughly sketch some aspects of SUGRA and
in particular its effect on the scalar potential.

Let us consider a supersymmetric theory
containing some chiral and gauge supermultiplets.
If one attempts to make SUSY local, it turns out
that the part of the lagrangian containing terms
up to two space-time derivatives is completely
determined by specifying three independent
functions of the scalar fields treated as complex
variables. These are the superpotential
$W(\phi_i)$, the ``K\"ahler potential''
$K(\phi_i,\phi^{i*})$ and the ``gauge kinetic
function'' $f_{\alpha\beta}(\phi_i)$. $W$ has
dimensions $\rm mass^3$, $K$ has dimensions
$\rm mass^2$ and $f_{\alpha\beta}$ is
dimensionless. Unlike the superpotential, $K$ is
real and analytic in the scalar fields. The
K\"ahler potential does not appear in the
renormalizable lagrangian for global SUSY because
at tree level there is only one possibility for
it, namely $K=\phi^{i*}\phi_i$. On the other
hand, the gauge kinetic function is, like the
superpotential, an analytic function of the
scalar fields. The indices $\alpha$ and $\beta$
run over the adjoint representation of the gauge
group and $f_{\alpha\beta}$ is symmetric under
interchange of its two indices. In global SUSY,
$f_{\alpha\beta}$ is independent of the fields
and it equals the identity matrix divided by the
square of the gauge coupling constant,
$f_{\alpha\beta}=\delta_{\alpha\beta}/g^2$.

The whole lagrangian with up to two derivatives
can now be written down in terms of these
functions. To proceed, let us define one
extra function, the ``K\"ahler function''
\beq
G=K/\mP^2+\ln(W/\mP^3)+\ln(W^*/\mP^3).
\eeq
From $G$ one can construct its derivatives with
respect to the scalar fields and their complex
conjugates, using the convention that a raised
(lowered) index $i$ corresponds to derivation
with respect to $\phi_i$ ($\phi^{i*}$),
e.g.~$G^i=\pd G/\pd\phi_i$,
$G_i=\pd G/\pd\phi^{i*}$ and
$G_i^j=\pd G/\pd\phi^{i*}\pd\phi_j$. Note that
$G_i^j$, called the ``K\"ahler metric'', depends
only on $K$, since $G_i^j=K_i^j/\mP^2$. The
inverse of this matrix is denoted by
$(G^{-1})_i^j$, so that $(G^{-1})_i^k G_k^j=
\delta_i^j$. Similarly, the inverse of the matrix
$K_i^j$ is denoted by $(K^{-1})_i^j$. In terms of
these objects, the generalization of the F-term
contribution to the scalar potential in SUGRA is
given by (see e.g.~\cite{SUSYprimer})
\beq\label{eq:VSUGRAG}
V_{\rm SUGRA}^F=\mP^4\;e^G
\left[(G^{-1})_i^j\;G^iG_j-3\right].
\eeq
Written in terms of the superpotential and the
K\"ahler potential, this equation takes the form
\beq\label{eq:VSUGRAF}
V_{\rm SUGRA}^F=e^{K/\mP^2}\left[(K^{-1})_i^j\;
F^{i*}F_j-3|W|^2/\mP^2\right],
\eeq
with
\beq\label{eq:Fterms}
F^{i*}=-(W^i+WK^i/\mP^2)\quad {\rm and}\quad
F_j=-(W_j^*+W^*K_j/\mP^2).
\eeq
Now, if one assumes a ``minimal'' K\"ahler
potential $K=\phi^{i*}\phi_i$, then
$K_i^j=(K^{-1})_i^j=\delta_i^j$ and
Eqs.~\eqref{eq:VSUGRAF}-\eqref{eq:Fterms},
expanded to lowest order in $1/\mP$, become
$V_{\rm SUGRA}^F=F^{i*}F_i$, $F^{i*}=-W^i$
and $F_i=-W_i$, i.e.~they take on their form
in the global SUSY case. The D-term
contribution to the scalar potential is given by
\beq
V_{\rm SUGRA}^D=\frac{1}{2}
\Real f_{\alpha\beta}^{-1}\;\hat{D}^\alpha
\hat{D}^\beta,\quad{\rm with}\quad\hat{D}^\alpha=
-K^i(T^\alpha)_i^{\phantom{i}j}\phi_j,
\eeq
where $\Real f_{\alpha\beta}^{-1}$ is the inverse
of the real part of the gauge kinetic function,
viewed as a matrix. In the case of minimal
K\"ahler potential and $f_{\alpha\beta}=
\delta_{\alpha\beta}/g^2$, this just reproduces
the result for the global SUSY case,
i.e.~$V_{\rm SUGRA}^D=1/2\,D^\alpha D^\alpha$.
There are also many contributions to the
lagrangian other than the scalar potential,
which depend on the three functions $W$, $K$
and $f_{\alpha\beta}$, which we will not deal
with here. The only thing that we will mention,
in order to have a complete picture of the
dynamics of a scalar field during inflation, is
the form of the kinetic terms, which become
\beq
\Lag_{\rm SUGRA}^{\rm (kin)}=-K_i^j\;
\pd^\mu\phi^{i*}\,\pd_\mu\phi_j.
\eeq
It should be noted that, unlike the
case of global SUSY, the scalar potential in
SUGRA is not necessarily non-negative, because
of the $-3$ term in Eq.~\eqref{eq:VSUGRAG}.

\chapter{Hybrid inflation and extensions}
\label{sec:HYBRID}

\section{Introduction}
\label{sec:HYBRIDintro}

The next thing one needs to consider in order to
proceed to the study of the early universe is,
of course, cosmology. Early cosmology, soon
after Einstein published his General Theory of
Relativity, was haunted by the idea of a static
universe, without any expansion or contraction.
In contrast to this idea, Einstein's equations
kept predicting an expanding universe, which led
him to introduce the famous cosmological
constant in order to make them compatible with
contemporary belief. But, when in 1929 Edwin
Hubble formulated his law of expansion of the
universe, after nearly a decade of observations,
the picture changed radically, causing Einstein
to make the legendary statement that the work
on the cosmological constant was his greatest
blunder. Hubble expansion, along with the later
discovery of the cosmic microwave background
radiation (CMBR) in 1964, established the hot
big band (HBB) model as the standard cosmological
model for the years to come. Today, the HBB model
has been replaced by the theory of inflation
which, although it constitutes a change of
paradigm in cosmology, has kept many of the
salient features of its predecessor.

Central and firm feature in all theories of
modern cosmology has been the cosmological or
Copernican principle, postulating that the
universe is pretty much the same everywhere. The
strongest evidence for this principle, that holds
only on the largest scales of observation of the
universe, is the observed isotropy of the CMBR.
Mathematically, this principle is expressed by
the notions of homogeneity and isotropy.
Homogeneity is the statement that the metric is
the same throughout the space. Isotropy applies
at some specific point in space and states that
space looks the same along all directions of
observation through this point. Note that, if
space is isotropic around one point and also
homogeneous, then it will be isotropic around
every point. Applying these notions to the metric
one ends up with only one possible form for it,
the Robertson-Walker metric (see
e.g.~\cite{GenRelatIntro})
\beq\label{eq:RWmetric}
ds^2=-dt^2+a^2(t)\left[\frac{dr^2}{1-kr^2}+
r^2(d\theta^2+\sin^2\theta\,d\phi^2)\right],
\eeq
where $r$, $\phi$ and $\theta$ are ``comoving''
coordinates, which remain fixed for objects that
have no other motion other than the general
expansion of the universe. The parameter $k$ is
the ``scalar curvature'' of the 3-space and
$k=0$, $>0$ or $<0$ corresponds to flat, closed
or open universe. The dimensionless parameter
$a(t)$ is called the ``scale factor'' of the
universe and describes the cosmological
expansion.

Up to now, we have not made use of Einstein's
equations,
\beq\label{eq:Einstein}
G_{\mu\nu}=8\pi G T_{\mu\nu},
\eeq
where $G_{\mu\nu}=R_{\mu\nu}-1/2 Rg_{\mu\nu}$ is
the Einstein tensor, $R_{\mu\nu}$ and $R$ are the
Ricci tensor and scalar, $T_{\mu\nu}$ is the
energy-momentum tensor and $G=M_{\rm P}^{-2}$ is
Newton's constant. For a homogeneous and
isotropic universe, the energy-momentum tensor
takes the diagonal form $T\du{\mu}{\nu}=
T_{\mu\si}g^{\si\nu}=\diag(-\rho,p,p,p)$, where
$\rho$ is the energy density of the universe and
$p$ the pressure. With this form for the
energy-momentum tensor, Einstein's equations give
\begin{gather}
\frac{\ddot{a}}{a}=-\frac{4\pi G}{3}\,(\rho+3p),
\label{eq:Friedmann1}\\
\left(\frac{\dot{a}}{a}\right)^2=
\frac{8\pi G}{3}\,\rho-\frac{k}{a^2},
\label{eq:Friedmann2}
\end{gather}
where a dot represents a derivative with respect
to the cosmic time $t$. Together, these two
equations are known as the Friedmann equations
and metrics of the form of
Eq.~\eqref{eq:RWmetric} which obey these
equations define Friedmann-Robertson-Walker
(FRW) universes. From the Friedmann equations,
or directly from conservation of energy and
momentum $T_{\mu\phantom{\nu};\nu}^{\phantom{\mu}
\nu}=0$, one obtains the continuity equation
\beq\label{eq:continuity}
\dot{\rho}=-3H(t)(\rho+p),
\eeq
where the Hubble parameter $H(t)\equiv\dot{a}/a$
characterizes the rate of expansion of the
universe. The value of the Hubble parameter
at the present epoch is the Hubble constant,
$H_0$. Another useful quantity is the density
parameter, $\Omega=\rho/\rho_c=8\pi G\rho/3H^2$,
where the critical density, $\rho_c=3H^2/8\pi G$,
is the energy density corresponding to a flat
universe. Thus, $\Omega=1$, $>1$ or $<1$
corresponds to flat, closed or open universe.

It is possible to solve the Friedmann equations
in a number of simple cases. To do that, one
needs an extra condition, known as the equation
of state, which is a relation between $\rho$ and
$p$ that depends on the form of the energy that
the universe contains. Most of the perfect fluids
relevant to cosmology obey the simple equation
of state $\rho+p=\ga\rho$, where $\ga$ is a
constant independent of time. With this
assumption, Eq.~\eqref{eq:continuity} becomes
$\rho\propto a^{-3\ga}$ and substituting in
Eq.~\eqref{eq:Friedmann2}, in the case of a flat
universe ($k=0$), we obtain
\beq
a(t)=a_0(t/t_0)^{2/3\ga},
\eeq
where $a_0$ is the scale factor at a cosmic time
$t=t_0$. It is common to take $t_0$ to represent
the present time and define $a_0=1$. For a
universe dominated by pressureless matter we
have $p=0$ and thus $\ga=1$. In the case of a
radiation dominated universe, $p=\rho/3$ and
$\ga=4/3$. So, for a matter dominated universe
we get the expansion law $a(t)=(t/t_0)^{2/3}$,
while for a radiation dominated universe we
get $a(t)=(t/t_0)^{1/2}$. Both of these solutions
(and many others) predict that at time $t\to 0$,
$a(t)\to 0$ and the energy density of the
universe becomes infinite. This particular time
instance $t=0$, had been considered by many
physicists, although outside the validity of
any real knowledge regarding that instance, to
represent the creation of everything, including
space-time itself, a process called the Big Bang.
These solutions, along with their observational
confirmation by Hubble, signify the beginning of
the HBB era.

The HBB cosmological model achieved many great
successes in the explanation of observations,
such as the Hubble expansion, the existence of
the CMBR and the abundances of light elements,
which were formed during primordial
nucleosynthesis. But it also came up against a
number of shortcomings, such as the horizon and
flatness problems and the magnetic monopole
problem, when it is combined with GUTs that
predict their existence. The horizon problem is
the difficulty in explaining the isotropy of the
CMBR, since it seems to come from regions of the
sky that have never communicated causally in the
past. The flatness problem consists of the fact
that the energy density of the observable
universe is at present very close to its critical
energy density, so that, at the beginning of its
evolution, $\Omega$ should have been inexplicably
close to 1. Also, combined with GUTs that predict
the existence of heavy magnetic monopoles, the
HBB model leads to a cosmological catastrophe
due to the overproduction of these monopoles.
Finally, even if one takes the isotropy of the
CMBR for granted, there is no explanation of
the observed temperature fluctuations in it,
or of the origin of the small density
perturbations required for structure formation.

Inflation came as a solution to these problems
\cite{GuthInf}. The main idea underlying all
versions of the inflationary universe scenario
is that, in the very early stages of its
evolution, the universe fell in a metastable,
vacuum-like state with high energy density
(for an introduction see e.g.~\cite{LindeInf,
LazaridesReview}). In such a state $\rho=\const$
and Eq.~\eqref{eq:continuity} gives $p=-\rho$.
Then, from Eq.~\eqref{eq:Friedmann2} with $k=0$,
which corresponds to a flat universe as
observations suggest, one gets
\beq
a(t)\propto e^{Ht},\quad
H=\sqrt{\frac{8\pi G}{3}\,\rho},
\eeq
i.e.~the universe experiences an exponential
expansion. Inflation ends when the universe
leaves this metastable state, either by
tunnelling out of it, or by slow rolling towards
a critical point, depending on the model. It is
assumed that after inflation, during which the
universe has cooled down, reheating occurs and
the universe continues its evolution according
to the HBB model.

To briefly describe the salient properties of
inflation, consider a real scalar field $\phi$
whose evolution is driven by the lagrangian
density
\beq
\Lag=\frac{1}{2}\,\pd_\mu\phi\,
\pd^\mu\phi-V(\phi),
\eeq
where $V(\phi)$ is the potential energy density,
which we assume to be quite flat near some point
$\phi=\phi_0$. The energy-momentum tensor is
found to be
\beq
T\du{\mu}{\nu}=-\pd_\mu\phi\,\pd^\nu\phi
+\delta\du{\mu}{\nu}\left(\frac{1}{2}\,
\pd_\la\phi\,\pd^\la\phi-V(\phi)\right).
\eeq
Now, if we assume that there is a large region
in space where the field $\phi$ is essentially
homogeneous with a value near $\phi=\phi_0$,
which changes very slowly with time due to the
flatness of the potential, then the
energy-momentum tensor takes the form
$T\du{\mu}{\nu}\simeq-V_0\delta\du{\mu}{\nu}$,
where $V_0=V(\phi_0)$. This means that $\rho
\simeq-p\simeq V_0$ and the conditions for
an exponential expansion of the scale factor
are fulfilled. The equation of motion of the
homogeneous field $\phi$, derived from the
lagrangian, reads
\beq\label{eq:evolution}
\ddot{\phi}+3H\dot{\phi}+V'(\phi)=0,
\eeq
where a dot represents the derivative $d/dt$,
while a prime represents $d/d\phi$. Inflation
is by definition the situation where the
``kinetic'' term $\ddot{\phi}$ is subdominant
to the ``friction'' term $3H\dot{\phi}$ and
Eq.~\eqref{eq:evolution} reduces to the
inflationary equation
\beq
3H\dot{\phi}=-V'(\phi).
\eeq
The conditions for the validity of the
inflationary equation can be summarized in the
form of restrictions imposed to the two slow
roll parameters, $\eta$ and $\epsilon$, as
(see e.g.~\cite{LazaridesReview})
\beq
|\eta|\equiv\mP^2\left|\frac{V''(\phi)}
{V(\phi)}\right|\leq 1\quad,\quad
\epsilon\equiv\frac{\mP^2}{2}\left(
\frac{V'(\phi)}{V(\phi)}\right)^2\leq 1.
\eeq
The end of the slow roll occurs when either of
these inequalities is saturated. The exponential
expansion of the universe during inflation can
be measured by the number of the ``e-foldings'',
defined as the logarithmic growth of the scale
factor between an initial time $t_i$ and a final
time $t_f$,
\beq\label{eq:efoldings}
N\equiv\ln\frac{a(t_f)}{a(t_i)}=\int_{t_i}^{t_f}
H\,dt=\frac{1}{\mP^2}\int_{\phi_f}^{\phi_i}
\frac{V(\phi)}{V'(\phi)}\,d\phi.
\eeq

The great success of the inflationary
cosmological model is that for $N\gtrsim 55$ all
three problems of the HBB model mentioned above
can be simultaneously solved (see
e.g.~\cite{LazaridesReview}). Furthermore,
inflation can explain the origin of the density
perturbations required for structure formation
in the universe. To understand this, one should
note that an exponentially expanding space,
called ``de Sitter'' space, can be considered as
a black hole turned inside out, i.e.~a black hole
that surrounds the space from all sides. Then,
exactly as in a black hole, there are quantum
fluctuations governed by the equivalent Hawking
temperature $T_H=H/2\pi$. Skipping all the
details, the power spectrum of the primordial
curvature perturbation at a scale $k_0$ can be
approximated by (see e.g.~\cite{InfReview})
\beq\label{eq:perturbations}
\PR\simeq\frac{1}{2\pi\sqrt{3}}\,
\frac{V^{3/2}(\phi_Q)}{\mP^3V'(\phi_Q)},
\eeq
where $\phi_Q$ is the value of the inflaton
field when the scale $k_0$ crossed outside the
inflationary horizon. This curvature perturbation
is not the same for all length scales. The
running of $\PR$ with respect to $k$ is governed
by a power law, with $k$ raised to an exponent
called the ``spectral index'' $\ns$. In addition,
the perturbations may also have a significant
``tensor'' component, measured by the ``tensor
to scalar ratio'' $r$. The spectral index, the
tensor to scalar ratio and the running of the
spectral index $d\ns/d\ln k$, in the slow roll
approximation, are given by (see
e.g.~\cite{InfReview})
\beq\label{eq:nsrdns}
\ns\simeq 1+2\eta-6\epsilon,\quad
r\simeq 16\epsilon,\quad\frac{d\ns}{d\ln k}
\simeq16\epsilon\eta-24\epsilon^2-2\xi^2,
\eeq
where
\beq
\xi^2=\mP^4\frac{V'V'''}{V^2}
\eeq
is the third slow roll parameter. By measuring
various properties of the CMBR one can extract
experimental values for $\PR$, $\ns$,
$d\ns/d\ln k$ and $r$ and compare the
corresponding theoretical values with them. In
order for a specific model to be realistic, it
should predict values for the above parameters
that lie within the experimental limits of the
observed values.

Inflation is the most successful cosmological
model so far and it is certainly the most
promising, not only because of its success in
explaining the shortcomings of the HBB model,
but also because it has provided us with an
unprecedented way to link cosmology and particle
physics and with the ability to confront particle
physics theories with cosmological observations.
Yet, we are still very far from deciding which is
the right model that describes best the
realization of inflation in the universe. Many
models have been proposed, each with different
appealing characteristics. Among them, hybrid
inflation is one of the most prevalent, often
making its appearance spontaneously in particle
physics models. In the sections that follow, we
will briefly describe the original model and some
of its most successful extensions within the
context of SUSY.

\section{The standard non-supersymmetric version}
\label{sec:HYBRIDstandard}

Standard hybrid inflation was initially proposed
by Linde \cite{hybrid} in an attempt to construct
new inflationary models by making the hybrids of
some known ones, such as ``chaotic'' and ``new''
inflation. Hybrid inflation helped to solve some
of the problems of the old models and turned out
to be a very fruitful arena for inflationary
model building. The idea is to use two real
scalar fields $\chi$ and $\si$, of which $\chi$
provides the vacuum energy density that drives
inflation and $\si$ is the slowly varying
inflaton field. Inflation ends by a rapid rolling
of the field $\chi$, called ``waterfall'',
triggered by the slow rolling of the field $\si$,
when the latter reaches a critical value $\si_c$.

The scalar potential of the original model is
of the form (for a review see
e.g.~\cite{LazaridesReview})
\beq\label{eq:VLinde}
V(\chi,\si)=\ka^2\left(M^2-\frac{\chi^2}{4}\right)^2
+\frac{\la^2}{4}\chi^2\si^2+\frac{m^2}{2}\,\si^2,
\eeq
where $\ka$, $\la$ are dimensionless parameters
and $M$, $m$ are mass parameters. This potential
has two degenerate global minima at $\vev{\si}=0$,
$\vev{\chi}=\pm 2M$. In the limit $m\to0$, $V$
possesses a flat direction at $\chi=0$ with
$V(0,\si)=\ka^2M^4$. The mass squared of the
field $\chi$ along this direction is
$m_\chi^2=-\ka^2M^2+\la^2\si^2/2$ and it follows
that the critical value of $\si$ at which the
flat direction becomes unstable and the waterfall
occurs is $\si_c=\sqrt{2}\ka M/\la$. For
$|\si|>\si_c$ and $m=0$ we obtain a flat valley
of minima, while setting $m\neq 0$ this valley
acquires a non zero slope that can drive the
inflaton field $\si$ toward its critical value.
On this flat valley, with potential energy
density $V=\ka^2M^4$, the system can inflate
while $\si$ is slowly rolling towards the
critical point.

The $\epsilon$ and $\eta$ criteria imply that
the mass parameter $m$ should be $m/M<\ka M/\mP$,
where $\mP\simeq2.43\ten{18}\units{GeV}$ is the
reduced Planck mass, for the slow roll to occur
on the inflationary path. If this is satisfied,
inflation continues until $\si$ reaches $\si_c$,
where it terminates by a waterfall, i.e.~a sudden
entrance into an oscillatory phase about a global
minimum. Since the system can fall into either of
the two minima with equal probability,
topological defects (monopoles, cosmic strings or
domain walls) are copiously produced if they are
predicted by the particular particle physics
model employed. So, if the underlying GUT gauge
symmetry breaking (by the field $\chi$) leads to
the existence of monopoles or domain walls, we
encounter a cosmological catastrophe, while if
it leads to the existence of cosmic strings, then
their contribution to the CMBR power spectrum
should comply with observational bounds
\cite{semishifted}. The curvature perturbation
can be easily estimated in this model, using
Eq.~\eqref{eq:perturbations}, to be
\beq
\PR\simeq\frac{1}{\sqrt{6}\pi}\frac{\la M}
{|\eta|\mP}e^{-|\eta|N_Q},
\eeq
where $|\eta|\simeq m^2\mP^2/\ka^2M^4<1$ is the
$\eta$ parameter during inflation and $N_Q$ is
the number of e-foldings from the time when the
pivot scale $k_0$ crossed outside the
inflationary horizon until the end of inflation.
For example, if we set M equal to the SUSY GUT
scale $M_{\rm GUT}\simeq2.86\ten{16}\units{GeV}$,
$N_Q=55$ and $|\eta|=0.1$, the three-year WMAP
\cite{WMAP3} result $\PR\simeq4.85\ten{-5}$ can
be reproduced with $\la\simeq0.78$. From the
constraint $|\eta|=0.1$ and assuming $\ka=\la$,
we obtain $m=8.28\ten{13}\units{GeV}$.

\section{The supersymmetric version}
\label{sec:HYBRIDsusy}

Hybrid inflation appears ``naturally'' in
supersymmetric theories. To see this, consider
the simple model given by the superpotential
\beq\label{eq:Wstandard}
W=\ka S(-M^2+\pb\phi),
\eeq
where S is a gauge singlet and $\phi$, $\pb$ are
two fields belonging to non-trivial conjugate
representations of the GUT gauge group G and
whose VEVs break this group down to a group
$G'$ containing $G_{\rm SM}$. The parameters
$\ka$ and M can be made real and positive by
field redefinitions. The scalar potential
derived from this superpotential reads
\beq\label{eq:Vstandard}
V(S,\phi,\pb)=\ka^2|M^2-\pb\phi|^2+\ka^2|S|^2
(|\phi|^2+|\pb|^2)+{\rm D-terms},
\eeq
where now the symbols $\phi$ and $\pb$ are
used for the SM singlet components of the
corresponding multiplets. The D-terms vanish for
$|\phi|=|\pb|$, which can be expressed as
$\pb^*=e^{i\vartheta}\phi$. The SUSY vacua lie at
the direction $\vartheta=0$, with $S=0$,
$|\phi|=M$ and $\pb=\phi^*$. The superpotential
possesses a $\rm U(1)_R$ R-symmetry, under which
$\pb\phi\to\pb\phi$, $S\to e^{ia}S$, $W\to
e^{ia}W$. Actually, $W$ in
Eq.~\eqref{eq:Wstandard} is the most general
renormalizable superpotential allowed by $G$ and
$\rm U(1)_R$. If we stick to the direction
$\vartheta=0$ containing the SUSY vacua and bring
$S$, $\phi$ and $\pb$ to the real axis by $G$ and
$\rm U(1)_R$ transformations, we can write $\phi=
\pb\equiv\chi/2$ and $S\equiv\si/\sqrt{2}$, where
$\chi$ and $\si$ are real scalars with normalized
kinetic terms, and the scalar potential takes the
form
\beq
V=\ka^2\left(M^2-\frac{\chi^2}{4}\right)^2
+\frac{\ka^2}{4}\chi^2\si^2.
\eeq
Comparing this with Eq.~\eqref{eq:VLinde}, we
see that the scalar potential obtained from
this simple supersymmetric model is the same
as Linde's potential if we set $\ka=\la$ and
take $m=0$.

Instead of the mass term, the slope along the
inflationary path, which corresponds to $\phi=
\pb=0$ and $|S|>S_c\equiv M$, is generated in
this model by the radiative corrections to the
potential. SUSY breaking by the potential energy
density $\ka^2M^4$ along this valley causes a
mass splitting in the supermultiplets $\phi$,
$\pb$. The scalar mass terms in the lagrangian,
calculated from the potential in
Eq.~\eqref{eq:Vstandard}, are
\beq
V\supset \ka^2|S|^2(|\phi|^2+|\pb|^2)
-\ka^2M^2(\pb\phi+c.c.).
\eeq
Transforming to the fields $\phi^{\pm}=
(\phi\pm\pb^*)/\sqrt{2}$, one obtains the mass
squared matrix
\beq
M_{\pm}^2=\ka^2\left(\ba{cc}
|S|^2-M^2 & 0\\[3pt] 0 & |S|^2+M^2
\ea\right).
\eeq
So, we have obtained two complex scalars with
masses squared $\ka^2(|S|^2\pm M^2)$. In the
fermionic sector, one can use
Eq.~\eqref{eq:Lchiral} to calculate the masses
directly from the superpotential in
Eq.~\eqref{eq:Wstandard}. We obtain two Weyl
fermions, both with mass squared $\ka^2|S|^2$.
This mass splitting leads to the existence of
one-loop radiative corrections to the potential
on the inflationary valley, which can be
calculated from the Coleman-Weinberg formula
\cite{ColemanWeinberg}
\beq\label{eq:ColemanWeinberg}
\Delta V=\frac{1}{64\pi^2}\,\sum_i(-1)^{F_i}
M_i^4\ln\frac{M_i^2}{\Lambda^2},
\eeq
where the sum extends over all helicity states
$i$, $F_i$ and $M_i^2$ are the fermion number
and mass squared of the $i$th state and $\Lambda$
is a renormalization mass scale. The calculation
gives
\beq\label{eq:SUSYdV}
\Delta V=\ka^2M^4\,\frac{\ka^2N}{32\pi^2}\left(
2\ln\frac{\ka^2|S|^2}{\Lambda^2}+(z+1)^2
\ln(1+z^{-1})+(z-1)^2\ln(1-z^{-1})\right),
\eeq
where $z=|S|^2/M^2$ and $N$ is the dimensionality
of the representations to which $\phi$ and $\pb$
belong. It is crucial to note that the slope
generated from this radiative correction is
$\Lambda$-independent.

The $\epsilon$ and $\eta$ parameters are
calculated from Eq.~\eqref{eq:SUSYdV} to be
\begin{gather}
\epsilon\simeq\left(\frac{\ka^2N}{16\pi^2}
\right)^2\frac{\mP^2}{M^2}\;z\left[(z+1)
\ln(1+z^{-1})+(z-1)\ln(1-z^{-1})\right],\\[3pt]
\eta\simeq\frac{\ka^2 N}{16\pi^2}
\frac{\mP^2}{M^2}\left[(3z+1)\ln(1+z^{-1})+
(3z-1)\ln(1-z^{-1})\right].
\end{gather}
Note that $\eta\to-\infty$ as $z\to1$. However,
for most relevant values of the parameters, the
slow roll conditions are violated only very close
to the critical point at $z=1$ and we can assume
that for all practical purposes inflation ends at
$|S|=S_c$. The curvature perturbation power
spectrum amplitude is given in this model by
\begin{gather}
\PR\simeq\frac{8\pi}{\sqrt{3}\ka N}
\frac{M^3}{\mP^3}\frac{M}{\si_Q}\Pi(z_Q)^{-1},
\label{eq:SUSYpert}\\[3pt]
\Pi(z)=(z+1)\ln(1+z^{-1})+(z-1)\ln(1-z^{-1}),
\end{gather}
where $z_Q=\si_Q^2/2M^2$ and $\si_Q$ is the
value of $\si$ when the present horizon scale
crossed outside the inflationary horizon. The
above equations are rather complicated but they
can be simplified by a trick. The number of
e-foldings of the present horizon scale during
inflation is
\beq
N_Q\simeq\frac{1}{\mP^2}\int_{\si_c}^{\si_Q}
\frac{16\pi^2}{\ka^2N}\,\frac{M^2}{\si}
\,\Pi(\si^2/2M^2)^{-1}\,d\si=
\frac{8\pi^2}{\ka^2N}\,\frac{M^2}{\mP^2}
\int_1^{z_Q}\frac{dz}{z}\Pi(z)^{-1}.
\eeq
Multiplying Eq.~\eqref{eq:SUSYpert} with
$(N_Q/N_Q)^{1/2}$ and setting $x_Q=z_Q^{1/2}$,
we obtain
\beq\label{eq:SUSYpert2}
\PR\simeq\sqrt{\frac{4N_Q}{3N}}\;
\frac{M^2}{\mP^2}\;x_Q^{-1}y_Q^{-1}\Pi(z_Q)^{-1},
\qquad y_Q^2=\int_1^{z_Q}\frac{dz}{z}\Pi(z)^{-1}.
\eeq
Now, for $x_Q\to\infty$ we have that
$y_Q\to x_Q$ and $x_Q y_Q\Pi(z_Q)\to1$,
so, assuming that $x_Q$ is large enough,
Eq.~\eqref{eq:SUSYpert2} becomes
\beq
\PR\simeq\sqrt{\frac{4N_Q}{3N}}\;
\frac{M^2}{\mP^2}.
\eeq
If we take $N_Q=55$ and $N=8$ for an order of
magnitude estimate, the WMAP3 \cite{WMAP3}
result, $\PR\simeq4.85\ten{-5}$, can be
reproduced with $M\simeq9.8\ten{15}\units{GeV}$,
a value that is somewhat lower than the SUSY GUT
scale $M_{\rm GUT}=2.86\ten{16}\units{GeV}$, but
quite close to it. Detailed investigation (see
e.g.~\cite{HybridExample}) shows that the
spectral index lies in the range
$\ns\simeq0.98-0.985$, values that are outside
the 1-$\si$ range of the WMAP3 \cite{WMAP3}
result, $\ns=0.958\pm0.016$, although within
the 2-$\si$ range.

Since we are dealing with a supersymmetric theory
it would be wise to consider making SUSY local.
It is known \cite{HybridSUGRA} that SUGRA
corrections to the scalar potential in general
tend to spoil slow roll inflation, due to the
infamous $\eta$-problem. To see this, take the
general form of the scalar potential in SUGRA,
given in Sec.~\ref{sec:MSSMsugra}, which we
repeat here for convenience. If we assume that
the D-term is flat along the inflationary
trajectory, only the F-term scalar potential is
relevant,
\beq\label{eq:VSUGRAF2}
V_{\rm SUGRA}^F=e^{K/\mP^2}\left[(K^{-1})_i^j\;
F^{i*}F_j-3|W|^2/\mP^2\right].
\eeq
Now, if the K\"ahler potential is expanded as
$K=|S|^2+|\phi|^2+|\pb|^2+k_S\,|S|^4/4\mP^2+
\cdots$, then the term $|S|^2$ in the exponential
of Eq.~\eqref{eq:VSUGRAF2}, could generate a mass
term on the inflationary path for the field $S$
of the form $(\ka^2M^4/\mP^2)|S|^2\sim
H^2|S|^2$. This leads directly to an extra term
in the $\eta$ parameter of the order 1 and the
slow roll is ruined. However, interestingly
enough, this does not happen in the specific
model under consideration and in many other
supersymmetric hybrid inflation models. The
reason for this is that this mass term is
cancelled in the potential. The linearity of $W$
in $S$, guaranteed to all orders by $\rm U(1)_R$,
is crucial for this cancellation. The $|S|^4$
term in $K$ also generates a mass term for $S$
through the factor $(\pd^2K/\pd S\pd S^*)^{-1}=
1-k_S\,|S|^2/\mP^2+\cdots$, which is not
cancelled. In order to avoid ruining inflation,
one has then to assume that $|k_S|$ is small
enough ($\lesssim 10^{-2}$). Actually, it
has been shown \cite{SUSYnonminimal} that the
existence of a large enough positive $k_S$ can
help reducing the spectral index, which in the
case of a minimal K\"ahler potential turns out
to exceed its value in the global SUSY case, to
make it lie within the observational bounds. All
higher order terms in $K$ give suppressed
contributions on the inflationary path, since
$|S|\ll\mP$.

\section{Smooth hybrid inflation}
\label{sec:HYBRIDsmooth}

In trying to apply SUSY hybrid inflation to
higher GUT gauge groups which predict the
existence of monopoles, we encounter a
cosmological catastrophe. Inflation is
terminated abruptly as the system reaches
the critical point on the inflationary path
and is followed by the waterfall regime, during
which the scalar fields $\phi$, $\pb$ develop
their VEVs and the spontaneous breaking of the
GUT gauge symmetry takes place. The fields
$\phi$, $\pb$ can end up at any point of the
vacuum manifold with equal probability and
thus monopoles are copiously produced through
the Kibble mechanism \cite{Kibble}. One of
the simplest GUTs predicting monopoles is the
Pati-Salam (PS) model \cite{PatiSalam} with
gauge group $G_{\rm PS}=\rm SU(4)_c\times
SU(2)_L\times SU(2)_R$. Solutions to the
monopole problem have been proposed
\cite{smooth,shifted} within the SUSY PS
model, that lead to extensions of standard
hybrid inflation.

In the PS model, the spontaneous breaking of
$G_{\rm PS}$ to $G_{\rm SM}$ is achieved via the
VEVs of a conjugate pair of Higgs fields
\bea
H^c\:({\bf \bar{4},1,2})&=&\left(\ba{cccc}
u_H^c & u_H^c & u_H^c & \nu_H^c\\
d_H^c & d_H^c & d_H^c & e_H^c\ea\right),\\[3pt]
\Hb^c\:({\bf 4,1,2})&=&\left(\ba{cccc}
\bar{u}_H^c & \bar{u}_H^c &
\bar{u}_H^c & \bar{\nu}_H^c\\
\bar{d}_H^c & \bar{d}_H^c &
\bar{d}_H^c & \bar{e}_H^c\ea\right),
\eea
in the $\nu_H^c$, $\bar{\nu}_H^c$ directions.
For simplicity, we will adopt our standard
convention to denote the SM singlet component
of a gauge multiplet with the same symbol as
the multiplet itself, hopping that the meaning
of the symbol is clear from the context. The
relevant part of the superpotential, including
the leading non-renormalizable term, is
\beq\label{eq:Wextend}
W=\ka S(-M^2+H^c\Hb^c)+\beta S
\frac{(H^c\Hb^c)^2}{M_S^2},
\eeq
where $M_S\sim 5\ten{17}\units{GeV}$ is a
superheavy string scale \cite{shifted}. Note
that the existence of the non-renormalizable
coupling is an automatic consequence of the
first two couplings, which constitute the
standard superpotential for SUSY hybrid
inflation. Indeed, the operator $H^c\Hb^c$
is neutral under all symmetries of the
superpotential and thus the above coupling,
which is crucial for the specific inflationary
scheme, cannot be forbidden. In fact, all higher
order couplings of the form $S(H^c\Hb^c)^n/
M_S^{2(n-1)}$ with $n\geq3$ are also allowed.
They are, however, subdominant to the term with
$n=2$ in the relevant region of the field space,
even if their coefficients are of order one.

If we impose an extra $Z_2$ symmetry
\cite{smooth} in the superpotential, under which
$H^c\to-H^c$, the hole structure of the model
remains unchanged except that now only even
powers of the combination $H^c\Hb^c$ are allowed.
If in Eq.~\eqref{eq:Wextend} we absorb the
parameters $\ka$ and $\beta$ in $M$ and $M_S$,
the new superpotential is written as
\beq\label{eq:Wsmooth}
W=S\left[-\mu^2+\frac{(H^c\Hb^c)^2}
{M_S^2}\right],
\eeq
where $\mu$ and $M_S$ are taken to be real and
positive by field redefinitions. The scalar
potential derived from $W$ is
\beq
V=\left|\mu^2-\frac{(H^c\Hb^c)^2}{M_S^2}\right|^2
+\frac{4|S|^2|H^c|^2|\Hb^c|^2}{M_S^4}\;
(|H^c|^2+|\Hb^c|^2).
\eeq
To go on, D-flatness implies that $\Hb^{c*}=
e^{i\vartheta}H^c$ and we can restrict ourselves
to the direction with $\vartheta=0$, which
contains the SUSY vacua. Then, after rotating the
fields $S$, $H^c$ and $\Hb^c$ to the real axis by
gauge and $\rm U(1)_R$ transformations, we can
set $H^c=\Hb^c=\chi/2$ and $S=\si/\sqrt{2}$ and
the scalar potential takes the simple form
\beq
V=\left(\mu^2-\frac{\chi^4}{16M_S^2}\right)^2
+\frac{\si^2\chi^6}{16M_S^4}.
\eeq

The emerging picture is completely different.
The flat direction at $\chi=0$ is now a valley of
local maxima for all values of $\si$ and two new
symmetric valleys of minima appear \cite{smooth} at
\beq
\chi=\pm\sqrt{6}\,\si\left(-1+\sqrt{1+
\frac{4\mu^2M_S^2}{9\si^4}}\;\;\right)^{1/2}.
\eeq
They contain the SUSY vacua, which lie at $\chi=
\pm 2\sqrt{\mu M_S}$, $\si=0$. These valleys are
not classically flat. In fact, they possess a
slope already at the classical level, which can
drive the inflaton towards the vacua. Thus, there
is no need of radiative corrections in this case.
For large enough values of $\si$, the value of
$\chi^2$ and the potential along the inflationary
path can be expanded as
\beq\label{eq:SmoothPot}
\chi^2\simeq\frac{2\mu^2M_S^2}{3\si^2},\quad
V\simeq\mu^4\left(1-\frac{\mu^2M_S^2}{27\si^4}
\right),\quad{\rm for}\quad\si\gg
\sqrt{2\mu M_S/3}.
\eeq
The system follows, from the beginning, a
particular inflationary path and ends up at a
particular point of the vacuum manifold, thus
not producing any monopoles after inflation.
The end of inflation is not abrupt in this case,
since the inflationary path is stable with
respect to $\chi$ for all values of $\si$.

The value $\si_f$ at which inflation is
terminated smoothly is found from the $\epsilon$
and $\eta$ criteria. The $\epsilon$ and $\eta$
parameters are given by
\beq
\epsilon\simeq\frac{8\mu^4M_S^4\,\mP^2}
{729\,\si^{10}},\qquad
\eta\simeq-\frac{20\mu^2M_S^2\,\mP^2}{27\,\si^6}
\eeq
and the $\eta$ criterion, which is more stringent
than the $\epsilon$ one in this case, gives
\beq
\si_f\simeq \left(\frac{20}{27}\right)^{1/6}
\,(\mu M_S\,\mP)^{1/3},
\eeq
a value that is within the range of approximation
of Eq.~\eqref{eq:SmoothPot}. The number of
e-foldings suffered by our present horizon scale
is found to be
\beq\label{eq:smoothNQ}
N_Q\simeq\frac{9}{8\mu^2M_S^2\,\mP^2}\;
(\si_Q^6-\si_f^6)=\frac{9\,\si_Q^6}
{8\mu^2M_S^2\,\mP^2}-\frac{5}{6}.
\eeq
The power spectrum of the primordial curvature
perturbation is calculated from
Eq.~\eqref{eq:perturbations} to be
\beq
\PR\simeq\frac{27}{8\pi\sqrt{3}}\;
\frac{\si_Q^5}{M_S^2\mP^3}.
\eeq
Finally, the tensor to scalar ratio is
negligible while the spectral index of density
perturbations is found, with the aid of
Eq.~\eqref{eq:smoothNQ}, to be
\beq
\ns\simeq1+2\eta\simeq1-\frac{5/3}{N_Q+5/6}.
\eeq
As a numerical example, we can take the common
VEV of $H^c$ and $\Hb^c$, $\sqrt{\mu M_S}$, to
be equal to the SUSY GUT scale,
$M_{\rm GUT}=2.86\ten{16}\units{GeV}$.
Then, Eq.~\eqref{eq:smoothNQ} for $N_Q=55$ gives
$\si_Q\simeq2.41\ten{17}\units{GeV}$ and the
WMAP3 \cite{WMAP3} normalization,
$\PR\simeq4.85\ten{-5}$, can be satisfied with
$M_S\simeq8.48\ten{17}\units{GeV}$,
$\mu\simeq9.65\ten{14}\units{GeV}$, values that
are quite natural. The spectral index turns out
to be $\ns\simeq0.97$, which is closer to the
WMAP3 result, $\ns=0.958\pm0.016$, than the $\ns$
predicted by standard SUSY hybrid inflation. As
in the case of standard SUSY hybrid inflation,
minimal SUGRA corrections do not ruin inflation
but tend to increase the value of the spectral
index above unity \cite{SenoguzShafi}. One may
then use a non-minimal K\"ahler potential
\cite{smoothnonminimal} in order to achieve
a spectral index compatible with WMAP3 (see also
Sec.~\ref{sec:NSMOOTHsugra}).

\section{Shifted hybrid inflation}
\label{sec:HYBRIDshifted}

A different scenario emerges \cite{shifted} if
one keeps all the terms in
Eq.~\eqref{eq:Wextend}, which reads
\beq\label{eq:Wshifted}
W=\ka S(-M^2+H^c\Hb^c)-\beta S
\frac{(H^c\Hb^c)^2}{M_S^2}.
\eeq
Here we have set $\beta\to-\beta$, which is
appropriate for this model. Note that $\beta$ can
in general be complex, but we take it to be real
and positive for simplicity. The parameters
$\ka$, $M$ and $M_S$ can be made real and
positive by field redefinitions. The scalar
potential derived from $W$ is
\beq
V=\left|\ka(-M^2+H^c\Hb^c)-\beta
\frac{(H^c\Hb^c)^2}{M_S^2}\right|^2
+\ka^2|S|^2\left|1-\frac{2\beta}{\ka}
\frac{H^c\Hb^c}{M_S^2}\right|^2
(|H^c|^2+|\Hb^c|^2).
\eeq
Once again, D-flatness implies $\Hb^{c*}=
e^{i\vartheta}H^c$ and we restrict ourselves to
the direction with $\vartheta=0$ which contains
the SUSY vacua (see below). Defining the
dimensionless variables $w=|S|/M$ and
$y=|H^c|/M$, we obtain
\beq\label{eq:Vshifted}
V=V_0\Big[(1-y^2+\xi y^4)^2+2w^2y^2
(1-2\xi y^2)^2\Big],
\eeq
were we have set $V_0\equiv\ka^2M^4$ and
$\xi\equiv\beta M^2/\ka M_S^2$. This potential
is a simple extension of the standard potential
for SUSY hybrid inflation (which corresponds to
$\xi=0$) and appears in a wide class of models
incorporating the leading non-renormalizable
correction to the standard superpotential.

For constant $w$ (or $|S|$), the potential in
Eq.~\eqref{eq:Vshifted} has extrema at
\beq
y_1=0,\quad y_2=\frac{1}{\sqrt{2\xi}}, \quad
y_{3\pm}=\frac{1}{\sqrt{2\xi}}\,\sqrt{1-6\xi w^2
\pm\sqrt{(1-6\xi w^2)^2-4\xi(1-w^2)}}.
\eeq
Note that the first two extrema ($y_1$ and $y_2$)
are $S$-independent and thus correspond to
classically flat directions, the trivial one
at $y_1=0$ with $V_1=V_0$ and the ``shifted''
one at $y_2=1/\sqrt{2\xi}$ with
$V_2=V_0(1/4\xi-1)^2$, which can be used as an
inflationary path. The trivial trajectory is a
valley of minima for $w>1$, while the shifted
one for $w>w_0\equiv(1/8\xi-1/2)^{1/2}$, which
is the critical point. We take $\xi<1/4$ so
that $w_0>0$ and the shifted path is destabilized
(in the chosen direction $\Hb^{c*}=H^c$) before
$w$ reaches zero. The extrema at $y_{3\pm}$,
which are $S$-dependent and non-flat, do not
exist for all values of $w$ and $\xi$. These two
extrema, at $w=0$, become the SUSY vacua. The
vacuum where the system most probably ends up
after inflation (see below) corresponds to
$y_{3-}|_{w=0}$ and thus, the common VEV of
$H^c$ and $\Hb^c$ is given by
\beq\label{eq:shiftedVEV}
\frac{|H^c|^2}{M^2}=\frac{1}{2\xi}\,
(1-\sqrt{1-4\xi}\,).
\eeq

We will now discuss the structure of V and the
inflationary history for $1/6<\xi<1/4$. For
fixed $w>1$, there exist two local minima at
$y_1=0$ and $y_2=1/\sqrt{2\xi}$, which has lower
potential energy density, and a local maximum
at $y_{3+}$ between the minima. As $w$ becomes
smaller than unity, the extremum at $y_1$ turns
into a local maximum, while the extremum at
$y_{3+}$ disappears. The system then can fall
into the shifted path, in case it had started at
$y_1=0$. As we further decrease $w$ below
$(2-\sqrt{36\xi-5}\,)^{1/2}/\sqrt{18\xi}$, a pair
of new extrema, a local minimum at $y_{3-}$ and a
local maximum at $y_{3+}$, are created between
$y_1$ and $y_2$. As $w$ crosses $w_0$, the local
maximum at $y_{3+}$ crosses $y_2$ becoming a
local minimum. At the same time, the local
minimum at $y_2$ turns into a local maximum and
inflation ends with the system falling into the
local minimum at $y_{3-}$ which, at $w=0$,
becomes the SUSY vacuum. After inflation, the
system could fall into the minimum at $y_{3+}$
instead of the one at $y_{3-}$. However, it is
most probable that the system will end up at
$y_{3-}$, since in the last e-folding or so the
barrier separating  the minima at $y_{3-}$ and
$y_2$ is considerably reduced and the decay of
the ``false vacuum'' at $y_2$ to the minimum at
$y_{3-}$ can be completed before the $y_{3+}$
minimum even appears. This transition is further
accelerated by the inflationary density
perturbations. We see that, in this scenario,
inflation takes place on the shifted path, where
$G_{\rm PS}$ is already broken to $G_{\rm SM}$
and thus no monopoles are produced at the
waterfall.

If we evaluate the mass spectrum on the shifted
path \cite{shifted}, we find that the only mass
splitting in supermultiplets occurs in the
$\nu_H^c$, $\bar{\nu}_H^c$ sector. Specifically,
we obtain one Majorana fermion with
${\rm mass}^2$ equal to $4\ka^2|S|^2$ and two
normalized real scalars with $m_\pm^2=4\ka^2|S|^2
\mp2\ka^2m^2$, where $m=M(1/4\xi-1)^{1/2}$. The
radiative corrections on the shifted path can
then be constructed using
Eq.~\eqref{eq:ColemanWeinberg} and one finds that
the effective potential on the inflationary path
is given by
\beq
V_\eff=V_0\left\{1+\frac{\ka^2}{16\pi^2}\left[
2\ln\frac{2\ka^2\si^2}{\Lambda^2}+(z+1)^2\ln
(1+z^{-1})+(z-1)^2\ln(1-z^{-1})\right]\right\},
\eeq
with $V_0=\ka^2m^4$. Here $z=\si^2/m^2$ and
$\si=\sqrt{2}\,S$ is the real normalized inflaton
field. Then, as in the case of standard SUSY
hybrid inflation, the power spectrum of the
primordial curvature perturbation can be
approximated, for $x_Q\equiv|\si_Q|/m\gg1$, by
\beq
\PR\simeq\sqrt{\frac{2N_Q}{3}}\,
\frac{m^2}{\mP^2}.
\eeq
If we take as a numerical example $\xi=1/5$
and $N_Q=55$, the WMAP3 \cite{WMAP3} value,
$\PR\simeq4.85\ten{-5}$, can be met with
$m\simeq6.89\ten{15}\units{GeV}$,
$M\simeq1.38\ten{16}\units{GeV}$ and the
common VEV of $H^c$ and $\Hb^c$ in the SUSY
vacuum $|H^c|\simeq1.62\ten{16}\units{GeV}$.
The spectral index, $\ns$, depends strongly
on the parameter $\ka$ \cite{shifted} and
can take values between about $0.9$ and $1$.
Finally, again as in standard SUSY hybrid
inflation, minimal SUGRA corrections do not
ruin the inflationary scenario, although they
tend to increase the predicted value of the
spectral index.

\chapter{The extended SUSY Pati-Salam model
with Yukawa quasi-unification}
\label{sec:QUASI}

In the previous chapter we described hybrid
inflation and its extensions in the context
of the SUSY Pati-Salam model, smooth and
shifted hybrid inflation, which solve the
monopole problem by introducing the leading
non-renormalizable term in the superpotential.
However, these variants of hybrid inflation
can also arise without the need of this term.
In this chapter we will briefly describe the
extended supersymmetric Pati-Salam model with
Yukawa quasi-unification \cite{quasi}, a model
that was introduced to cope with a problem
completely irrelevant to inflation. In the
following chapters, we will describe what is
intended to be the main matter of this thesis,
the rich cosmology that this model can exhibit.

\section{Introduction}
\label{sec:QUASIintro}

The most restrictive version of the MSSM with
gauge coupling unification, radiative electroweak
breaking and universal boundary conditions from
gravity mediated soft SUSY breaking, known as
constrained MSSM (CMSSM) \cite{Cmssm}, can be
made even more predictive if we impose Yukawa
unification (YU), i.e.~assume that the three
third generation Yukawa coupling constants unify
at the SUSY GUT scale, $M_{\rm GUT}$. The
requirement of YU can be achieved by embedding
the MSSM in a SUSY GUT with a gauge group
containing $\rm SU(4)_c$ and $\rm SU(2)_R$.
Indeed, assuming that the electroweak Higgs
superfields $H_u$, $H_d$ and the third family
right handed quark superfields $\bar{t}$,
$\bar{b}$ form $\rm SU(2)_R$ doublets, we
obtain \cite{pana} the ``asymptotic'' Yukawa
coupling relation $h_t=h_b$ and hence large
$\tan\beta\sim m_t/m_b$. Moreover, if the
third generation quark and lepton $\rm SU(2)_L$
doublets [singlets] $Q_3$ and $L_3$
[$\bar{b}$ and $\bar{\tau}$] form a
$\rm SU(4)_c$ $\bm{4}$-plet [$\bar{\bm{4}}$-plet]
and the Higgs doublet $H_d$ which couples to them
is a $\rm SU(4)_c$ singlet, we get $h_b=h_{\tau}$
and the ``asymptotic'' relation $m_{b}=m_{\tau}$
follows. The simplest GUT gauge group which
contains both $\rm SU(4)_c$ and $\rm SU(2)_R$ is
the Pati-Salam group $G_{\rm PS}$ and the model
we will describe is based on this group.

However, applying YU in the context of the CMSSM
and given the experimental values of the
top-quark and tau-lepton masses (which naturally
restrict $\tan\beta\simeq50$), the resulting
value of the $b$-quark mass turns out to be
unacceptable. This is due to the fact that, in
the large $\tan\beta$ regime, the tree level
$b$-quark mass receives sizeable SUSY corrections
\cite{copw,pierce,susy,king} (about 20$\%$),
which have the sign of $\mu$ (with the standard
sign convention \cite{SugraColl}) and drive, for
$\mu>[<]~0$, the corrected $b$-quark mass at
$M_Z$, $m_b(M_Z)$, well above [somewhat below]
its $95\%$ confidence level (c.l.) experimental
range:
\beq\label{eq:mbrg}
2.684~\units{GeV}\lesssim m_b(M_Z)\lesssim
3.092~\units{GeV},\quad{\rm with}\quad
\alpha_s(M_Z)=0.1185.
\eeq
This is derived by appropriately evolving
\cite{quasi} the corresponding range of
$m_b(m_b)$ in the $\overline{MS}$ scheme
(i.e.~$3.95-4.55~\units{GeV}$) up to $M_{Z}$ in
accordance with \cite{baermb}. We see that, for
both signs of $\mu$, YU leads to an unacceptable
$b$-quark mass with the $\mu<0$ case being less
disfavored.

A way out of this $m_b$ problem can be found
\cite{quasi} without abandoning the CMSSM
(in contrast to the usual strategy
\cite{king,raby,baery,nath}) or YU altogether.
Instead, we can rather modestly correct YU by
including an extra $\rm SU(4)_c$ non-singlet
Higgs superfield with Yukawa couplings to the
quarks and leptons. The Higgs $\rm SU(2)_L$
doublets contained in this superfield can
naturally develop \cite{wetterich} subdominant
VEVs and mix with the main electroweak doublets,
which are assumed to be $\rm SU(4)_c$ singlets
and form a $\rm SU(2)_R$ doublet. This mixing
can, in general, violate $\rm SU(2)_R$.
Consequently, the resulting electroweak Higgs
doublets $H_u$, $H_d$ do not form a
$\rm SU(2)_R$ doublet and also break
$\rm SU(4)_c$. The required deviation from YU
is expected to be more pronounced for $\mu>0$.
Despite this, we will describe here this case,
since the $\mu<0$ case has been excluded
\cite{cd2} by combining the WMAP restrictions
\cite{WMAP1} on the cold dark matter (CDM) in
the universe with the experimental results
\cite{cleo} on the inclusive branching ratio
${\rm BR}(b\rightarrow s\gamma)$.

\section{The SUSY GUT model}
\label{sec:QUASImodel}

We take the SUSY GUT model of shifted hybrid
inflation \cite{shifted} (see also
Secs.~\ref{sec:HYBRIDsmooth},
\ref{sec:HYBRIDshifted}) as our starting point.
It is based on $G_{\rm PS}$, which is the
simplest gauge group that can lead to YU. The
representations under $G_{\rm PS}$ and the
global charges of the various matter and Higgs
superfields contained in this model are presented
in Table~\ref{tab:content}, which also contains
the extra Higgs superfields required for
accommodating an adequate violation of YU
(see below). The matter superfields are $F_i$
and $F^c_i$ ($i=1,2,3$ \ is the family index),
while the electroweak Higgs doublets belong to
the superfield $h$. The particle content of
these superfields in terms of SM fields is
\bea
F_i\:({\bf 4,2,1})&=&\left(\ba{cccc}
u_i & u_i & u_i & \nu_i\\
d_i & d_i & d_i & e_i\ea\right),\\[3pt]
F_i^c\:({\bf \bar{4},1,2})&=&\left(\ba{cccc}
\bar{u}_i & \bar{u}_i &
\bar{u}_i & \bar{\nu}_i\\
\bar{d}_i & \bar{d}_i &
\bar{d}_i & \bar{e}_i\ea\right),\\[3pt]
h\:({\bf 1,2,2})&=&\left(\ba{cc}
h_2^+ & h_1^0\\ h_2^0 & h_1^-\ea\right),
\eea
so, all the requirements for exact YU are
fulfilled. The breaking of $G_{\rm PS}$ down to
$G_{\rm SM}$ is achieved by the superheavy VEVs
($\sim M_{\rm GUT}$) of the right handed neutrino
type components of a conjugate pair of Higgs
superfields $H^c$, $\Hb^c$, written as
\bea
H^c\:({\bf \bar{4},1,2})&=&\left(\ba{cccc}
u_H^c & u_H^c & u_H^c & \nu_H^c\\
d_H^c & d_H^c & d_H^c & e_H^c\ea\right),\\[3pt]
\Hb^c\:({\bf 4,1,2})&=&\left(\ba{cccc}
\bar{u}_H^c & \bar{u}_H^c &
\bar{u}_H^c & \bar{\nu}_H^c\\
\bar{d}_H^c & \bar{d}_H^c &
\bar{d}_H^c & \bar{e}_H^c\ea\right).
\eea
The model also contains a gauge singlet $S$
which triggers the breaking of $G_{\rm PS}$, a
$\rm SU(4)_c$ ${\bf 6}$-plet $G$ which gives
\cite{leontaris} masses to the right handed down
quark type components of $H^c$, $\Hb^c$ and a
pair of gauge singlets $N$, $\bar{N}$ for
solving \cite{rsym} the $\mu$ problem of the
MSSM via a Peccei-Quinn (PQ) symmetry. In
addition to $G_{\rm PS}$, the model possesses
two global $\rm U(1)$ symmetries, namely a R
and a PQ symmetry, as well as a discrete
$Z_2^{\rm mp}$ symmetry (``matter parity'').
A moderate violation of exact YU can be
naturally accommodated \cite{quasi} in this
model by adding a new Higgs superfield $h'$
with Yukawa couplings $FF^ch'$. Actually,
$({\bf 15,2,2})$ is the only representation,
besides $({\bf 1,2,2})$, which possesses such
couplings to the fermions. In order to give
superheavy masses to the color non-singlet
components of $h'$, one needs to include one
more Higgs superfield $\bar{h}'$ with the
superpotential coupling $\bar{h}'h'$, whose
coefficient is of the order of $M_{\rm GUT}$.

\begin{table}[!t]
\centering
\caption{Superfield Content of the Model}
\label{tab:content}
\begin{tabular}{@{}ccccc@{}}
\hline \hline & & & &\\[-1.5ex]
Superfields & Representations &
\multicolumn{3}{c}{Global Symmetries}\\[1ex]
\multicolumn{1}{c}{} & under $G_{\rm PS}$ &
$R$ & \ \ $PQ$ & $Z^{\rm mp}_2$\\[1ex]
\hline & & & &\\[-1.5ex]
\multicolumn{5}{c}{Matter Fields}\\[1ex]
\hline & & & &\\[-1.5ex]
$F_i$ & $({\bf 4, 2, 1})$ & $1/2$ & $-1$ & $1$\\[1ex]
$F^c_i$ & $({\bf \bar{4}, 1, 2})$ & $1/2$ & $0$ & $-1$\\[1ex]
\hline & & & &\\[-1.5ex]
\multicolumn{5}{c}{Higgs Fields}\\[1ex]
\hline & & & &\\[-1.5ex]
$h$ & $({\bf 1, 2, 2})$ & $0$ & $1$ & $0$\\[1ex]
$H^c$ & $({\bf \bar{4}, 1, 2})$ & $0$ & $0$ & $0$ \\[1ex]
$\bar{H}^c$ & $({\bf 4, 1, 2})$ & $0$ & $0$ & $0$\\[1ex]
$S$ & $({\bf 1, 1, 1})$ & $1$ & $0$ & $0$\\[1ex]
$G$ & $({\bf 6, 1, 1})$ & $1$ & $0$ & $0$\\[1ex]
$N$ & $({\bf 1, 1, 1})$ & $1/2$ & $-1$ & $0$\\[1ex]
$\bar{N}$ & $({\bf 1, 1, 1})$ & $0$ & $1$ & $0$\\[1ex]
\hline & & & &\\[-1.5ex]
\multicolumn{5}{c}{Extra Higgs Fields}\\[1ex]
\hline & & & &\\[-1.5ex]
$h'$ & $({\bf 15, 2, 2})$ & $0$ & $1$ &$0$\\[1ex]
$\bar{h}'$ & $({\bf 15, 2, 2})$ & $1$ & $-1$ & $0$\\[1ex]
$\phi$ & $({\bf 15, 1, 3})$ & $0$ & $0$ &$0$\\[1ex]
$\bar{\phi}$ & $({\bf 15, 1, 3})$ & $1$ & $0$ &$0$\\[1ex]
\hline\hline
\end{tabular}
\end{table}

After the breaking of $G_{\rm PS}$ to
$G_{\rm SM}$, the two color singlet $\rm SU(2)_L$
doublets $h'_1$, $h'_2$ contained in $h'$ can mix
with the corresponding doublets $h_1$, $h_2$ in
$h$. This mainly happens due to the terms
$\bar{h}'h'$ and $H^c\Hb^c\bar{h}'h$.
Actually, since
\bea
& & H^c\Hb^c=({\bf\bar 4,1,2})({\bf 4,1,2})=
({\bf 15, 1, 1+3})+\cdots,\\
& &\bar{h}'h=({\bf 15,2,2})({\bf 1,2,2})=
({\bf 15, 1, 1+3})+\cdots,
\eea
there are two independent couplings of the type
$H^c\bar{H}^c\bar{h}^{\prime}h$ (both suppressed
by the string scale $M_S\sim5\ten{17}\units{GeV}$,
being non-renormalizable). One of them is between
the $\rm SU(2)_R$ singlets in $H^c\Hb^c$ and
$\bar{h}'h$, and the other between the
$\rm SU(2)_R$ triplets in these combinations. So,
we obtain two bilinear terms $\bar{h}'_1h_1$ and
$\bar{h}'_2h_2$ with different coefficients,
which are suppressed by $M_{\rm GUT}/M_S$. These
terms together with the terms $\bar{h}'_1h'_1$
and $\bar{h}'_2h'_2$ from $\bar{h}'h'$, which
have equal coefficients, generate different
mixings between $h_1$, $h'_1$ and $h_2$, $h'_2$.
Consequently, the resulting electroweak doublets
$H_u$, $H_d$ contain $\rm SU(4)_c$-violating
components suppressed by $M_{\rm GUT}/M_S$ and
fail to form a $\rm SU(2)_R$ doublet by an
equally suppressed amount. So, YU is moderately
violated. Unfortunately, this violation is not
adequate for correcting the $b$-quark mass
within the CMSSM for $\mu>0$.

In order to allow for a more sizable violation
of YU, the model is further extend by including
the superfield $\phi$ with the coupling
$\phi\bar{h}'h$. To give superheavy masses to
the color non-singlets in $\phi$, one needs to
introduce one more superfield, $\pb$, with the
coupling $\pb\phi$, whose coefficient is of order
$M_{\rm GUT}$. The terms $\pb\phi$ and
$\pb H^c\Hb^c$ imply that, after the breaking of
$G_{\rm PS}$ to $G_{\rm SM}$, $\phi$ acquires a
superheavy VEV of order $M_{\rm GUT}$. The
coupling $\phi\bar{h}'h$ then generates
$\rm SU(2)_R$-violating unsuppressed bilinear
terms between the doublets in $\bar{h}'$ and $h$.
These terms can certainly overshadow the
corresponding ones from the non-renormalizable
term $H^c\Hb^c\bar{h}'h$. The resulting
$\rm SU(2)_R$-violating mixing of the doublets
in $h$ and $h'$ is then unsuppressed and we can
obtain stronger violation of YU.

\section{The Yukawa quasi-unification condition}
\label{sec:QUASIcondition}

To further analyze the mixing of the doublets in
$h$ and $h'$, observe that the part of the
superpotential corresponding to the symbolic
couplings $\bar{h}'h'$, $\phi\bar{h}'h$ is
properly written as
\beq\label{eq:expmix}
m\,\tr\{\bar{h}'\epsilon \tilde{h}'\epsilon\}+
p\,\tr\{\bar{h}'\epsilon\phi\tilde{h}\epsilon\},
\eeq
where $\epsilon$ is the antisymmetric $2\times2$
matrix with $\epsilon_{12}=+1$, $\tr$ denotes
trace taken with respect to the $\rm SU(4)_c$
and $\rm SU(2)_L$ indices and a tilde denotes
the transpose of a matrix. After the breaking
of $G_{\rm PS}$ to $G_{\rm SM}$, $\phi$ acquires
a VEV $\vev{\phi}\sim M_{\rm GUT}$. If we
substitute $\phi$ by its VEV in the above
couplings, we obtain
\bea
& & \tr\{\bar{h}'\epsilon\tilde{h}'\epsilon\}=
\tilde{\bar{h}}'_1\epsilon h'_2+\tilde{h}'_1
\epsilon\bar{h}'_2+\cdots,  \label{eq:mass}\\
& & \tr\{\bar{h}'\epsilon\phi\tilde{h}\epsilon\}=
\frac{\vev{\phi}}{\sqrt{2}}\,\tr\{
\bar{h}'\epsilon\si_3\tilde{h}\epsilon\}=
\frac{\vev{\phi}}{\sqrt{2}}\,(\tilde{\bar{h}}'_1
\epsilon h_2-\tilde{h}_1\epsilon\bar{h}'_2),
\label{eq:triplet}
\eea
where the ellipsis in Eq.~\eqref{eq:mass}
contains the colored components of $\bar{h}'$,
$h'$ and $\si_3=\diag(1,-1)$. Inserting
Eqs.~\eqref{eq:mass} and \eqref{eq:triplet}
into Eq.~\eqref{eq:expmix}, we obtain
\beq\label{eq:superheavy}
m\,\tilde{\bar{h}}'_1\epsilon(h'_2-\alpha_1h_2)+
m\,(\tilde{h}'_1+\alpha_1\tilde{h}_1)\epsilon
\bar{h}'_2,\quad{\rm with}\quad
\alpha_1=-p\,\vev{\phi}/\sqrt{2}m.\quad
\eeq
So, we get two pairs of superheavy doublets with
mass $m$. They are predominantly given by
\beq\label{eq:superdoublets}
\bar{h}'_1~,~\frac{h'_2-\alpha_1h_2}
{\sqrt{1+|\alpha_1|^2}}\quad{\rm and}\quad
\frac{h'_1+\alpha_1h_1}{\sqrt{1+|\alpha_1|^2}}
~,~\bar{h}'_2.
\eeq
The orthogonal combinations of $h_1$, $h'_1$ and
$h_2$, $h'_2$ constitute the electroweak doublets
\beq\label{eq:ew}
H_d=\frac{h_1-\alpha_1^*h'_1}
{\sqrt{1+|\alpha_1|^2}}\quad{\rm and}\quad
H_u=\frac{h_2+\alpha_1^*h'_2}
{\sqrt{1+|\alpha_1|^2}}.
\eeq
The superheavy doublets in
Eq.~\eqref{eq:superdoublets} must have vanishing
VEVs, which readily implies that $\vev{h'_1}=
-\alpha_1\vev{h_1}$, $\vev{h'_2}=\alpha_1
\vev{h_2}$. Eq.~\eqref{eq:ew} then gives
\beq
\vev{H_d}=\sqrt{1+|\alpha_1|^2}\,\vev{h_1}
\quad{\rm and}\quad
\vev{H_u}=\sqrt{1+|\alpha_1|^2}\,\vev{h_2}.
\eeq
From the third generation Yukawa couplings
$y_{33}F_3hF_3^c$, $2y'_{33}F_3h'F_3^c$,
we obtain
\bea
& & m_t=|y_{33}\vev{h_2}+y'_{33}\vev{h'_2}|=
\left|\frac{1+\rho\alpha_1/\sqrt{3}}
{\sqrt{1+|\alpha_2|^2}}\,y_{33}\vev{H_u}\right|,
\label{eq:top}\\[3pt]
& & m_b=\left|\frac{1-\rho\alpha_1/\sqrt{3}}
{\sqrt{1+|\alpha_1|^2}}\,y_{33}\vev{H_d}\right|,
\quad m_\tau=\left|\frac{1+\sqrt{3}\rho\alpha_1}
{\sqrt{1+|\alpha_1|^2}}\,y_{33}\vev{H_d}\right|.
\label{eq:bottomtau}
\eea
where $\rho=y'_{33}/y_{33}$. From
Eqs.~\eqref{eq:top} and \eqref{eq:bottomtau},
we see that YU is now replaced by the Yukawa
quasi-unification condition (YQUC),
\beq\label{eq:minimal}
h_t:h_b:h_\tau=(1+c):(1-c):(1+3c),\quad
{\rm with}\quad 0<c=\rho\alpha_1/\sqrt{3}<1.
\eeq
For simplicity, we restricted ourselves to real
values of $c$ only, which lie between zero and
unity.

It turns out \cite{quasi} that this YQUC can
allow for an acceptable $b$-quark mass within
the CMSSM with $\mu>0$ and universal boundary
conditions. Furthermore, there exists a wide
and natural range of parameters consistent with
cosmological and phenomenological requirements.
In particular, the model was successfully
confronted with data from CDM considerations,
the branching ratio $b\rightarrow s\gamma$,
the muon anomalous magnetic moment and the
Higgs boson masses. Interestingly enough, apart
from its success in the $b$-quark mass problem,
the model also revealed a quite rich cosmological
phenomenology, incorporating and expanding all
the extensions of supersymmetric hybrid inflation
mentioned in Chap.~\ref{sec:HYBRID}. The detailed
study of the cosmology of the model is the subject
of the remaining chapters of this thesis.

\chapter{New shifted hybrid inflation}
\label{sec:NSHIFT}

\section{Introduction}
\label{sec:NSHIFTintro}

In Chap.~\ref{sec:HYBRID} we saw that the
monopole problem of hybrid inflation in SUSY
GUTs and in particular in the SUSY Pati-Salam
model with gauge group $G_{\rm PS}=\rm SU(4)_c
\times SU(2)_L\times SU(2)_R$, can be solved by
taking into account the leading
non-renormalizable term in the superpotential
(see also \cite{LazaridesReview} for a review).
It was argued that this term cannot be excluded
by any symmetry and can be comparable to the
trilinear term of the standard superpotential.
The coexistence of both these terms (see
Sec.~\ref{sec:HYBRIDshifted} and
Ref.~\cite{shifted}) leads to the appearance of
a ``shifted'' classically flat valley of local
minima where the GUT gauge symmetry is broken.
This valley acquires a slope at the one-loop
level and can be used  as an alternative
inflationary path. In this scenario, which is
known as shifted hybrid inflation, there is no
formation of topological defects at the end of
inflation and hence the potential monopole
problem is avoided. This is crucial for the
compatibility of the SUSY PS model with hybrid
inflation since this model predicts the
existence of magnetic monopoles.

It would be desirable to solve the magnetic
monopole problem of hybrid inflation in SUSY
GUTs with the GUT gauge group broken directly
to $G_{\rm SM}$ (the monopole problem could also
be solved by employing \cite{twostep} an
intermediate symmetry breaking scale or by
other mechanisms, e.g.~\cite{fate}), without
relying on the presence of non-renormalizable
superpotential terms. In this chapter, we show
how a new version of shifted hybrid inflation
\cite{newshifted} can take place in the extended
SUSY PS model described in Chap.~\ref{sec:QUASI},
without invoking any non-renormalizable
superpotential terms. This feature is caused
by the inclusion of the conjugate pair of
superfields $\phi$ and $\pb$, which belong to
the representation $({\bf 15, 1, 3})$ of
$G_{\rm PS}$ (see Sec.~\ref{sec:QUASImodel}).
These fields lead to three new renormalizable
terms in the part of the superpotential which
is relevant for inflation, which is
\beq\label{eq:Wnshift}
W=\ka S(H^c\Hb^c-M^2)-\beta S\phi^2+
m\phi\pb+\la\pb H^c\Hb^c,
\end{equation}
where $M$ and $m$ are superheavy masses of the
order of $M_{\rm GUT}$ and $\ka$, $\beta$ and
$\la$ are dimensionless coupling constants.
These parameters are normalized so that they
correspond to the couplings between the SM
singlet components of the superfields. We can
take $M,~m,~\ka,~\la>0$ by field redefinitions.
For simplicity, we also take $\beta>0$, although
it can be generally complex.

\section{New shifted hybrid inflation in
global SUSY} \label{sec:NSHIFTsusy}

The scalar potential obtained from $W$ is given by
\bea\label{Vnshift}
V&=&|\kappa(H^c\Hb^c-M^2)-\beta\phi^2|^2+
|-2\beta S\phi+m\pb|^2+|m\phi+\la H^c\Hb^c|^2
\nonumber \\
& &+|\ka S+\la\pb|^2\left(|H^c|^2+|\Hb^c|^2
\right)+{\rm D-terms},
\eea
where the complex scalar fields which belong to
the SM singlet components of the superfields are
denoted by the same symbols as the corresponding
superfields. As usual, the vanishing of the
D-terms yields $\Hb^{c*}=e^{i\vartheta}H^c$
($H^c$, $\Hb^c$ lie in the $\nu^c_H$,
$\bar{\nu}^c_H$ direction). We restrict ourselves
to the direction with $\vartheta=0$ which contains
the ``new shifted'' inflationary path and the SUSY
vacua (see below). Performing an appropriate
global transformation, we can bring the complex
scalar field $S$ to the positive real axis. Also,
by a gauge transformation, the fields $H^c$,
$\Hb^c$ can be made positive.

From the potential in Eq.~\eqref{Vnshift}, we
find that the SUSY vacuum lies at
\beq\label{VACnshift}
\frac{H^c\Hb^c}{M^2}\equiv\left(\frac{v_0}
{M}\right)^2=\frac{1}{2\xi}\left(1-\sqrt{1-4\xi}
\,\right),\quad S=0,\quad \frac{\phi}{M}=
-\frac{\ka^{\frac{1}{2}}\xi^{\frac{1}{2}}}
{\beta^{\frac{1}{2}}}\left(\frac{v_0}{M}
\right)^2,\quad\pb=0,
\eeq
where $\xi=\beta\la^2M^2/\ka m^2<1/4$. Here, we
chose the vacuum with the smallest $v_0~(>0)$
for the same reasons as in simple shifted hybrid
inflation (see Sec.~\ref{sec:HYBRIDshifted}).
The derivatives of the potential with respect
to the scalar fields considered as complex
variables, are
\begin{flalign}
&\pder{V}{S^*}=(-2\beta S\phi+m\pb)(-2\beta\phi^*)
+\ka(\ka S+\la\pb)(|H^c|^2+|\Hb^c|^2),&
\label{eq:dVdSnsh}\\[3pt]
&\pder{V}{\pb^*}=(-2\beta S\phi+m\pb)m
+\la(\ka S+\la\pb)(|H^c|^2+|\Hb^c|^2),&
\label{eq:dVdpbnsh}\\[3pt]
&\pder{V}{\phi^*}=\Big[\ka(H^c\Hb^c-M^2)
-\beta\phi^2\Big](-2\beta\phi^*)
+(m\phi+\la H^c\Hb^c)m
+(-2\beta S\phi+m\pb)(-2\beta S^*),&
\label{eq:dVdphinsh}\\[3pt]
&\pder{V}{H^{c*}}=\Big[\ka(H^c\Hb^c-M^2)
-\beta\phi^2\Big]\ka \Hb^{c*}+(m\phi+\la H^c\Hb^c)
\la\Hb^{c*}+|\ka S+\la\pb|^2H^c,&
\label{eq:dVdHnsh}\\[3pt]
&\pder{V}{\Hb^{c*}}=\Big[\ka(H^c\Hb^c-M^2)
-\beta\phi^2\Big]\ka H^{c*}+(m\phi+\la H^c\Hb^c)
\la H^{c*}+|\ka S+\la\pb|^2\Hb^c.&
\label{eq:dVdHbnsh}
\end{flalign}
From these partial derivatives one can see that
the potential possesses in general three flat
directions. The trivial one is at $H^c=\Hb^c=
\phi=\pb=0$ with $V=\ka^2M^4$. The second is
defined from the equations
\beq
-2\beta S\phi+m\pb=0,\quad H^c=\Hb^c=0,
\eeq
which come from setting the partial derivatives
of the potential with respect to $S^*$ and
$\pb^*$, Eqs.~\eqref{eq:dVdSnsh} and
\eqref{eq:dVdpbnsh}, equal to zero. We will
deal with this case in Chap.~\ref{sec:SSHIFT}.
The third one is defined from
\beq
-2\beta S\phi+m\pb=0,\quad \ka S+\la\pb=0,\quad
H^c,\Hb^c\neq0,
\eeq
which is the other case that one obtains
from setting Eqs.~\eqref{eq:dVdSnsh}
and \eqref{eq:dVdpbnsh} equal to zero.
The VEVs of the fields along this direction are
\beq\label{eq:PATHnshift}
\frac{H^c\Hb^c}{M^2}\equiv\left(\frac{v}{M}
\right)^2=\frac{2\ka^2(1+1/4\xi)+\la^2/\xi}
{2(\kappa^2+\lambda^2)},\quad S>0,\quad
\frac{\phi}{M}=-\frac{\ka^{\frac{1}{2}}}
{2\beta^{\frac{1}{2}}\xi^{\frac{1}{2}}},
\quad\pb=-\frac{\ka}{\la}S,
\eeq
with
\beq
\frac{V_0}{M^4}=\frac{\ka^2\la^2}{\ka^2+
\la^2}\left(\frac{1}{4\xi}-1\right)^2,
\eeq
This is a flat direction with the properties of
the shifted path described in
Sec.~\ref{sec:HYBRIDshifted}, along which
$G_{\rm PS}$ is broken to $G_{\rm SM}$ since
$H^c,\Hb^c\neq0$, which can be used as an
inflationary path.

\section{One-loop radiative corrections}
\label{sec:NSHIFTcorr}

As in the case of simple shifted hybrid
inflation, which is based on non-renormalizable
superpotential terms, the constant classical
energy density on this ``new shifted'' path
breaks SUSY and implies the existence of
one-loop radiative corrections which lift the
classical flatness of this path, yielding the
necessary inclination for driving the inflaton
towards the SUSY vacuum. The one-loop radiative
correction to the potential along this path is
calculated by using the Coleman-Weinberg formula
given in Eq.~\eqref{eq:ColemanWeinberg} and
repeated here for convenience,
\beq\label{eq:nshiftCW}
\Delta V=\frac{1}{64\pi^2}\,\sum_i(-1)^{F_i}
M_i^4\ln\frac{M_i^2}{\Lambda^2}.
\eeq
In order to use this formula for creating a
logarithmic slope to the potential, one has
first to derive the mass spectrum of the model
on the new shifted inflationary path.

As mentioned, during inflation, $H^c$, $\Hb^c$
acquire constant values in the $\nu_H^c$,
$\bar{\nu}_H^c$ directions which are equal to
$v~(>0)$ and break $G_{\rm PS}$ to $G_{\rm SM}$.
We can then write $\nu_H^c=v+\delta\nu_H^c$,
$\bar{\nu}_H^c=v+\delta\bar{\nu}_H^c$, where
$\delta\nu_H^c$, $\delta\bar{\nu}_H^c$ are
complex scalar fields. The (complex) deviations
of the fields $S$, $\phi$, $\pb$ from their
values at a point on the new shifted path
(corresponding to $S>0$) are similarly denoted
as $\delta S$, $\delta\phi$, $\delta\pb$. We
define the complex scalar fields
\bea
\theta &=&\frac{\delta\nu_H^c+\delta\bar{\nu}^c_H}
{\sqrt{2}},\qquad\eta=\frac{\delta\nu_H^c-
\delta\bar{\nu}_H^c}{\sqrt{2}},\\
\zeta &=&\frac{\ka\delta S+\la\delta\pb}
{(\ka^2+\la^2)^{\frac{1}{2}}},\qquad
\varepsilon=\frac{\la\delta S-\ka\delta\pb}
{(\ka^2+\la^2)^{\frac{1}{2}}}\cdot
\eea
We find that $\eta$ and $\varepsilon$ do not
acquire any masses from the scalar potential in
Eq.~\eqref{Vnshift}. Actually, $\varepsilon$ (and
its SUSY partner) remains massless even after
including the gauge interactions (see below). It
corresponds to the complex inflaton field
$\Sigma=(\la S-\ka\pb)/(\ka^2+\la^2)^{1/2}$,
which on the new shifted path takes the form
$\Sigma=(\ka^2+\la^2)^{1/2}S/\la$. So, in this
case, the real normalized inflaton field is
$\si=2^{1/2}(\ka^2+\la^2)^{1/2}S/\la$.

Contrary to $\eta$ and $\varepsilon$, the
complex scalars $\theta$, $\delta\phi$ and
$\zeta$ acquire masses from the potential in
Eq.~\eqref{Vnshift}. Expanding these scalars
in real and imaginary parts, $\chi=(\chi_1+
i\chi_2)/\sqrt{2}$ ($\chi=\theta,\delta\phi,
\zeta$), we find that the mass squared matrices
$M_+^2$ and $M_-^2$ of $\theta_1$, $\delta
\phi_1$, $\zeta_1$ and $\theta_2$, $\delta
\phi_2$, $\zeta_2$ are given by
\beq\label{eq:nshiftM+-}
M_{\pm}^2=M^2\left(\ba{ccc}
a^2 & ab & 0 \\ab & b^2+c^2\pm f^2 & -cb\\
0 & -cb & a^2 + b^2\ea\right),
\eeq
where $a^2=2\ka^2(1/4\xi+1)+\la^2/\xi$, $b^2=
\beta(\ka^2+\la^2)/\ka\xi$, $c^2=2\beta^2\la^2
\si^2/M^2(\ka^2+\la^2)$, $f^2=2\ka\beta\la^2
(1/4\xi-1)/(\ka^2+\la^2)$ ($a$, $b$, $c$, $f>0$).

One can show that, for $\si\to\infty$
($c\to\infty$), all the eigenvalues of these two
mass squared matrices are positive. So, for large
values of $\si$, the new shifted path is a valley
of local minima. As $\si$ decreases, one
eigenvalue may become negative destabilizing the
trajectory. From continuity, no eigenvalue can
become negative without passing from zero. So, the
critical point on the new shifted trajectory is
encountered when one of the determinants of the
matrices in Eq.~\eqref{eq:nshiftM+-}, which are
$\det\{M_\pm^2\}=M^6a^2[a^2c^2\pm f^2(a^2+b^2)]$,
vanishes. We see that $\det\{M_+^2\}$ is always
positive, while $\det\{M_-^2\}$ vanishes at
$c^2=f^2(1+b^2/a^2)$, which corresponds to the
critical point of the new shifted path, given by
\beq\label{eq:scnshift}
\left(\frac{\si_c}{M}\right)^2=\frac{\ka}{\beta}
\left(\frac{1}{4\xi}-1\right)\frac{2\ka^2\left(1+
\frac{\ka+2\beta}{4\ka\xi}\right)+\frac{\la^2
(\ka+\beta)}{\ka\xi}}{2\ka^2\left(1+\frac{1}{4\xi}
\right)+\frac{\la^2}{\xi}}\cdot
\eeq

The superpotential in Eq.~\eqref{eq:Wnshift} gives
rise to mass terms between the fermionic partners
of $\theta$, $\delta\phi$ and $\zeta$. The square
of the corresponding mass matrix is found to be
\beq\label{eq:nshiftM0}
M_{0}^2=M^2\left(\ba{ccc}
a^2 & ab & 0 \\ab & b^2+c^2 & -cb\\
0 & -cb & a^2+b^2\ea\right).
\eeq

To complete the spectrum in the SM singlet
sector, which consists of the superfields
$\nu_H^c$, $\bar{\nu}_H^c$, $S$, $\phi$ and
$\pb$ (SM singlet directions), we must consider
the following D-terms in the scalar potential:
\beq\label{eq:Dtermsnshift}
\frac{1}{2}g^2\sum_{a}(H^{c*}\,T^aH^c+
\Hb^{c*}\,T^a\Hb^c)^2,
\eeq
where $g$ is the $G_{\rm PS}$ gauge coupling
constant and the sum extends over all the
generators $T^a$ of $G_{\rm PS}$. The part of
this sum over the generators
$T^{15}=(1/\sqrt{24})~\diag(1,1,1,-3)$ of
$\rm SU(4)_c$ and $T^3=(1/2)~\diag(1, -1)$ of
$\rm SU(2)_R$ gives rise to a mass term for the
normalized real scalar field $\eta_1$ with
$m^2=5g^2v^2/2$. The field $\eta_2$, however, is
left massless by the D-terms and is absorbed by
the gauge boson $A^{\perp}=-(3/5)^{1/2}A^{15}+
(2/5)^{1/2}A^{3}$ ($A^{15}$, $A^{3}$ are the
gauge bosons corresponding to $T^{15}$, $T^{3}$)
which becomes massive with $m^2=5g^2v^2/2$.

Contributions to the fermion masses also arise
from the Lagrangian terms
(see Eq.~\eqref{eq:Ltotal})
\beq\label{eq:fermionsnshift}
-\sqrt{2}g\sum_{a}\la^a(H^{c*}\,T^a\psi_{H^c}
+\Hb^{c*}\,T^a\psi_{\Hb^c})+{\rm h.c.},
\eeq
where $\la^a$ is the gaugino corresponding to
$T^a$ and $\psi_{H^c}$, $\psi_{\Hb^c}$ represent
the chiral fermions in the superfields $H^c$,
$\Hb^c$. Concentrating again on $T^{15}$ and
$T^3$, we obtain a Dirac mass term between the
chiral fermion in the $\eta$ direction and
$-i\la^{\perp}$ (with $\la^{\perp}$ being the
SUSY partner of $A^{\perp}$) with
$m^2=5g^2v^2/2$. The SM singlet components
of $\phi$ and $\pb$ do not contribute to
bosonic and fermionic couplings analogous to
the ones in Eqs.~\eqref{eq:Dtermsnshift} and
\eqref{eq:fermionsnshift} since they commute
with $T^{15}$ and $T^{3}$.

This completes the analysis of the SM singlet
sector of the model. In summary, we found two
groups of three real scalars with mass squared
matrices $M_\pm^2$ and three two component
fermions with mass matrix squared $M_0^2$. Also,
one Dirac fermion (with four components), one
gauge boson and one real scalar, all of them
having the same mass squared $m^2=5g^2v^2/2$
and thus not contributing to the one-loop
radiative correction. From
Eq.~\eqref{eq:nshiftCW}, we find that the
contribution of the SM singlet sector to the
radiative correction along the new shifted
path is given by
\beq\label{eq:ssect}
\Delta V=\frac{1}{64\pi^2}\tr\left\{M_+^4
\ln\frac{M_+^2}{\Lambda^2}+M_-^4\ln\frac{M_-^2}
{\Lambda^2}-2M_0^4\ln\frac{M_0^2}{\Lambda^2}
\right\}.
\eeq
One can show that, in this sector, $\tr\{M^2\}=0$
and $\tr\{M^4\}=2M^4f^4$, which is
$\si$-independent and thus the generated slope
on the inflationary path is $\Lambda$-independent.

We now turn to the $u^c$ and $\bar{u}^c$ type
fields which are color antitriplets with charge
$-2/3$ and color triplets with charge $2/3$
respectively. Such fields exist in $H^c$,
$\Hb^c$, $\phi$ and $\pb$ and we denote
them by $u^c_H$, $\bar{u}^c_H$, $u^c_{\phi}$,
$\bar{u}^c_{\phi}$, $u^c_{\pb}$ and
$\bar{u}^c_{\pb}$. The relevant expansion
of $\phi$ is
\beq\label{eq:phiu}
\phi=\left[\frac{1}{\sqrt{12}}\left(\ba{cc}
\bm{1}_3&0\\0&-3\ea\right),\frac{1}{\sqrt{2}}
\left(\ba{cc}1&0\\0&-1\ea\right)\right]\phi+
\left(\ba{cc}0&\bm{0}_3\\1&0\ea\right)u^c_{\phi}+
\left(\ba{cc}0&1\\\bm{0}_3&0\ea\right)
\bar{u}^c_{\phi}+\cdots,
\eeq
where the SM singlet in $\phi$ (denoted by the
same symbol) is also shown with the first
(second) matrix in the brackets belonging to
the algebra of $\rm SU(4)_c$ ($\rm SU(2)_R$).
Here, $\bm{1}_3$ and $\bm{0}_3$ denote the
$3\times3$ unit and zero matrices respectively.
The fields $u^c_{\phi}$, $\bar{u}^c_{\phi}$
are $\rm SU(2)_R$ singlets, so only their
$\rm SU(4)_c$ structure is shown and summation
over their $\rm SU(3)_c$ indices is implied in
the ellipsis. The field $\pb$ can be similarly
expanded.

In the bosonic $u^c$, $\bar{u}^c$ type sector,
we find that the mass squared matrices
$M_{u\pm}^2$  of the complex scalars
$u_{\chi\pm}^c=(u^c_\chi\pm\bar{u}^{c*}_\chi)/
\sqrt{2}$ \ ($\chi=H,\phi,\pb$), are given by
\beq\label{eq:nshiftMu+}
M_{u+}^2=M^2\left(\ba{ccc}
\frac{4\ka^2c^2}{9\beta^2}\spl
\frac{2\la^2a^2\beta}{3\ka\xi b^2}\spl
\frac{2\ka f^2}{3\beta}&
\smi\frac{\sqrt{2}\la^2\beta a}{\sqrt{3}\ka\xi b}&
\smi\frac{2\sqrt{2}\ka^{\frac{1}{2}}\la ca}{3\sqrt{3}
\beta^{\frac{1}{2}}\xi^{\frac{1}{2}}b}\\[6pt]
\smi\frac{\sqrt{2}\la^2\beta a}{\sqrt{3}\ka\xi b}&
\frac{\beta\la^2}{\ka\xi}\spl
{\scriptstyle c^2}\smi{\scriptstyle f^2}&
\smi\frac{\la\beta^{\frac{1}{2}}c}
{\ka^{\frac{1}{2}}\xi^{\frac{1}{2}}}\\[6pt]
\smi\frac{2\sqrt{2}\ka^{\frac{1}{2}}\la ca}
{3\sqrt{3}\beta^{\frac{1}{2}}\xi^{\frac{1}{2}}b}&
\smi\frac{\la\beta^{\frac{1}{2}}c}
{\ka^{\frac{1}{2}}\xi^{\frac{1}{2}}}&
\frac{\beta \la^2}{\ka\xi}\spl
\frac{2\la^2a^2\beta}{3\kappa\xi b^2}
\ea\right)
\eeq
and
\beq\label{eq:nshiftMu-}
M_{u-}^2= M^2\left(\ba{ccc}
\frac{4\ka^2c^2}{9\beta^2}\spl\frac{2\la^2a^2\beta}
{3\ka\xi b^2}\smi\frac{2\ka f^2}{3\beta}\spl
\frac{g^2a^2\beta}{2\ka\xi b^2}&
\frac{\smi\sqrt{2}\la^2\beta a}{\sqrt{3}\ka\xi b}\spl
\frac{g^2a}{\sqrt{6}\xi b}&
\frac{\smi2\sqrt{2}\ka^{\frac{1}{2}}\la ca}
{3\sqrt{3}\beta^{\frac{1}{2}}\xi^{\frac{1}{2}}b}\spl
\frac{g^2\ka^{\frac{1}{2}}ac}{\sqrt{6}
\xi^{\frac{1}{2}}\beta^{\frac{1}{2}}\la b}\\[6pt]
\frac{\smi\sqrt{2}\la^2\beta a}{\sqrt{3}\ka\xi b}\spl
\frac{g^2a}{\sqrt{6}\xi b}&
\frac{\beta\la^2}{\ka\xi}\spl
{\scriptstyle c^2}\spl{\scriptstyle f^2}\spl
\frac{g^2\ka}{3\beta\xi}&
\frac{\smi\la\beta^{\frac{1}{2}}c}
{\ka^{\frac{1}{2}}\xi^{\frac{1}{2}}}\spl
\frac{g^2c\ka^{\frac{3}{2}}}
{3\xi^{\frac{1}{2}}\beta^{\frac{3}{2}}\la}\\[6pt]
\frac{\smi2\sqrt{2}\ka^{\frac{1}{2}}\la ca}
{3\sqrt{3}\beta^{\frac{1}{2}}\xi^{\frac{1}{2}}b}\spl
\frac{g^2\ka^{\frac{1}{2}}ac}{\sqrt{6}
\xi^{\frac{1}{2}}\beta^{\frac{1}{2}}\la b}&
\frac{\smi\la\beta^{\frac{1}{2}}c}
{\ka^{\frac{1}{2}}\xi^{\frac{1}{2}}}\spl
\frac{g^2c\ka^{\frac{3}{2}}}
{3\xi^{\frac{1}{2}}\beta^{\frac{3}{2}}\la}&
\frac{\beta\la^2}{\ka\xi}\spl\frac{2\la^2a^2\beta}
{3\ka\xi b^2}\spl\frac{g^2\ka^2c^2}{3\la^2\beta^2}
\ea\right).
\eeq
The mass squared matrix $M_{u+}^2$ has one zero
eigenvalue corresponding to the Goldstone boson
which is absorbed by the superhiggs mechanism.
This is easily checked by showing that
$\det\{M_{u+}^2\}=0$. However, it does no harm to
keep this Goldstone mode since it has vanishing
contribution to the radiative corrections in
Eq.~\eqref{eq:nshiftCW} anyway.

In the $u^c$, $\bar{u}^c$ type sector, we obtain
four Dirac fermions (per color)
$\psi^D_{u^c_\chi}=\psi_{u^c_\chi}+
\psi^c_{\bar{u}^c_\chi}$, with $\chi=H,\phi,\pb$
and $-i\la^D=-i(\la^{+}+\la^{-c})$. Here,
$\la^\pm=(\la^1\pm i\la^2)/\sqrt{2}$,
where $\la^1$ ($\la^2$) is the gaugino
color triplet corresponding to the $\rm SU(4)_c$
generators with $1/2$ ($-i/2$) in the $i4$ and
$1/2$ ($i/2$) in the $4i$ entry ($i=1,2,3$). The
fermionic mass matrix is
\beq\label{eq:nshiftMpsiu}
M_{\psi_u}=M\left(\ba{cccc}
\frac{2\ka c}{3\beta}&0&\smi\frac{\sqrt{2}
\beta^{\frac{1}{2}}\la a}{\sqrt{3}
\ka^{\frac{1}{2}}\xi^{\frac{1}{2}}b}&
\frac{ga\beta^{\frac{1}{2}}}{\sqrt{2}
\ka^{\frac{1}{2}}\xi^{\frac{1}{2}}b}\\[6pt]
0&\smi c&\frac{\beta^{\frac{1}{2}}\la}
{\ka^{\frac{1}{2}}\xi^{\frac{1}{2}}}&
\frac{g \ka^{\frac{1}{2}}}{\sqrt{3}
\beta^{\frac{1}{2}}\xi^{\frac{1}{2}}}\\[6pt]
\smi\frac{\sqrt{2}\beta^{\frac{1}{2}}\la a}
{\sqrt{3}\ka^{\frac{1}{2}}\xi^{\frac{1}{2}}b}
&\frac{\beta^{\frac{1}{2}}\la}
{\ka^{\frac{1}{2}}\xi^{\frac{1}{2}}}&0&
\frac{g\ka c}{\sqrt{3}\la\beta}\\[6pt]
\frac{ga\beta^{\frac{1}{2}}}{\sqrt{2}
\ka^{\frac{1}{2}}\xi^{\frac{1}{2}}b}&
\frac{g\ka^{\frac{1}{2}}}{\sqrt{3}
\beta^{\frac{1}{2}}\xi^{\frac{1}{2}}}&
\frac{g\ka c}{\sqrt{3}\la\beta}&0\ea\right).
\eeq
To complete this sector, we must also include the
gauge bosons $A^\pm$ which are associated with
$\la^\pm$. They acquire a mass squared $M_g^2
=g^2M^2(a^2\beta/2\ka\xi b^2+\ka/3\beta\xi+
\ka^2c^2/3\beta^2\la^2)$.

The overall contribution of the $u^c$,
$\bar{u}^c$ type sector to $\Delta V$ in
Eq.~\eqref{eq:nshiftCW} is
\beq\label{eq:usect}
\Delta V=\frac{3}{32\pi^2}\tr\left\{M_{u+}^4
\ln\frac{M_{u+}^2}{\Lambda^2}+M_{u-}^4
\ln\frac{M_{u-}^2}{\Lambda^2}-2M_{\psi_u}^4
\ln\frac{M_{\psi_u}^2}{\Lambda^2}+3M_g^4
\ln\frac{M_g^2}{\Lambda^2}\right\}.
\eeq
In this sector, $\tr\{M^2\}=0$ and $\tr\{M^4\}
=12M^4f^4(1+4\ka^2/9\beta^2-2g^2\ka^2/3\beta^2
\la^2)$. So, the contribution of this sector to
the slope of the new shifted path is also
$\Lambda$-independent.

We will now discuss the contribution from the
$e^c$, $\bar{e}^c$ type sector consisting of
color singlets with charge $1$, $-1$. Such
fields exist in $H^c$, $\Hb^c$, $\phi$, $\pb$
and we denote them by $e^c_H$, $\bar{e}^c_H$,
$e^c_{\phi}$, $\bar{e}^c_{\phi}$, $e^c_{\pb}$,
$\bar{e}^c_{\pb}$. The field $\phi$ can be
expanded in $e^c_{\phi}$, $\bar{e}^c_{\phi}$
as follows:
\beq\label{eq:phie}
\phi=\left[\frac{1}{\sqrt{12}}\left(\ba{cc}
\bm{1}_3&0\\0&-3\ea\right),\left(\ba{cc}
0&1\\0&0\ea\right)e^c_{\phi}+\left(\ba{cc}
0&0\\1&0\ea\right)\bar{e}^c_{\phi}\right]+\cdots,
\eeq
with the same notation as in Eq.~\eqref{eq:phiu}.
A similar expansion holds for $\pb$. The analysis
in this sector is similar to the one in the $u^c$,
$\bar{u}^c$ type sector and we only summarize the
results.

In the bosonic sector, we obtain two groups, each
consisting of three complex scalars with mass
squared matrices
\beq\label{eq:nshiftMe+}
M_{e+}^2=M^2\left(\ba{ccc}
\frac{\ka^2c^2}{\beta^2}\spl\frac{\la^2a^2\beta}
{\ka\xi b^2}\spl\frac{\ka f^2}{\beta}&
\frac{\la^2\beta a}{\ka\xi b}&
\frac{\ka^{\frac{1}{2}}\la ca}
{\beta^{\frac{1}{2}}\xi^{\frac{1}{2}}b}\\[6pt]
\frac{\la^2\beta a}{\ka\xi b}&
\frac{\beta\la^2}{\ka\xi}\spl
{\scriptstyle c^2}\smi{\scriptstyle f^2}&
\smi\frac{\la \beta^{\frac{1}{2}}c}
{\ka^{\frac{1}{2}}\xi^{\frac{1}{2}}}\\[6pt]
\frac{\ka^{\frac{1}{2}}\la ca}
{\beta^{\frac{1}{2}}\xi^{\frac{1}{2}}b}&
\smi\frac{\la\beta^{\frac{1}{2}}c}
{\ka^{\frac{1}{2}}\xi^{\frac{1}{2}}}&
\frac{\beta\la^2}{\ka\xi}\spl
\frac{\la^2a^2\beta}{\ka\xi b^2}\ea\right)
\eeq
and
\beq\label{eq:nshiftMe-}
M_{e-}^2=M^2\left(\ba{ccc}
\frac{\ka^2c^2}{\beta^2}\spl
\frac{\la^2a^2\beta}{\ka\xi b^2}\smi
\frac{\ka f^2}{\beta}\spl
\frac{g^2a^2\beta}{2\ka\xi b^2}&
\frac{\la^2\beta a}{\ka\xi b}\smi
\frac{g^2a}{2\xi b}&
\frac{\ka^{\frac{1}{2}}\la ca}
{\beta^{\frac{1}{2}}\xi^{\frac{1}{2}}b}\smi
\frac{g^2\ka^{\frac{1}{2}}ac}{2\xi^{\frac{1}{2}}
\beta^{\frac{1}{2}}\la b}\\[6pt]
\frac{\la^2\beta a}{\ka\xi b}\smi
\frac{g^2a}{2\xi b}&
\frac{\beta\la^2}{\ka\xi}\spl
{\scriptstyle c^2}\spl{\scriptstyle f^2}
\spl\frac{g^2\ka}{2\xi\beta}&
\frac{\smi\la\beta^{\frac{1}{2}}c}
{\ka^{\frac{1}{2}}\xi^{\frac{1}{2}}}\spl
\frac{g^2\ka^{\frac{3}{2}}c}{2\xi^{\frac{1}{2}}
\beta^{\frac{3}{2}}\la}\\[6pt]
\frac{\ka^{\frac{1}{2}}\la ca}
{\beta^{\frac{1}{2}} \xi^{\frac{1}{2}} b}\smi
\frac{g^2\ka^{\frac{1}{2}}ac}
{2\xi^{\frac{1}{2}}\beta^{\frac{1}{2}}\la b}&
\frac{\smi\la\beta^{\frac{1}{2}}c}
{\ka^{\frac{1}{2}}\xi^{\frac{1}{2}}}\spl
\frac{g^2\ka^{\frac{3}{2}}c}
{2\xi^{\frac{1}{2}}\beta^{\frac{3}{2}}\la}&
\frac{\beta\la^2}{\ka\xi}\spl\frac{\la^2a^2\beta}
{\ka\xi b^2}\spl\frac{g^2\ka^2c^2}{2\la^2\beta^2}
\ea\right).
\eeq
The matrix $M_{e+}^2$, similarly to $M_{u+}^2$ in
the $u^c$, $\bar{u}^c$ type sector, has one zero
eigenvalue corresponding to the Goldstone mode
absorbed by the superhiggs mechanism.

In the fermion sector, we obtain four Dirac
fermions with mass matrix given by
\beq\label{eq:nshiftMpsie}
M_{\psi_e}=M\left(\ba{cccc}
\frac{\ka c}{\beta}&0&
\frac{\beta^{\frac{1}{2}}\la a}
{\ka^{\frac{1}{2}}\xi^{\frac{1}{2}}b}&
\frac{ga\beta^{\frac{1}{2}}}{\sqrt{2}
\ka^{\frac{1}{2}}\xi^{\frac{1}{2}}b}\\[6pt]
0&\smi c&\frac{\beta^{\frac{1}{2}}\la}
{\ka^{\frac{1}{2}}\xi^{\frac{1}{2}}}&
\smi\frac{g\ka^{\frac{1}{2}}}{\sqrt{2}
\beta^{\frac{1}{2}}\xi^{\frac{1}{2}}}\\[6pt]
\frac{\beta^{\frac{1}{2}}\la a}
{\ka^{\frac{1}{2}}\xi^{\frac{1}{2}}b}&
\frac{\beta^{\frac{1}{2}}\la}
{\ka^{\frac{1}{2}}\xi^{\frac{1}{2}}}&0&
\smi\frac{g\ka c}{\sqrt{2}\la\beta}\\[6pt]
\frac{ga\beta^{\frac{1}{2}}}{\sqrt{2}
\ka^{\frac{1}{2}}\xi^{\frac{1}{2}}b}&
\smi\frac{g\ka^{\frac{1}{2}}}{\sqrt{2}
\beta^{\frac{1}{2}}\xi^{\frac{1}{2}}}&
\smi\frac{g\ka c}{\sqrt{2}\la\beta}&0\ea\right).
\eeq
Finally, we also obtain in this sector one complex
gauge boson with mass squared given by
$\hat{M}_{g}^2=g^2M^2(a^2\beta/2\ka\xi b^2+
\ka/2\beta\xi+\ka^2c^2/2\beta^2\la^2)$.

The contribution of the $e^c$, $\bar{e}^c$ type
sector to $\Delta V$ is
\beq\label{eq:esect}
\Delta V=\frac{1}{32\pi^2}\tr\left\{M_{e+}^4
\ln\frac{M_{e+}^2}{\Lambda^2}+M_{e-}^4
\ln\frac{M_{e-}^2}{\Lambda^2}- 2M_{\psi_e}^4
\ln\frac{M_{\psi_e}^2}{\Lambda^2}+3\hat{M}_{g}^4
\ln\frac{\hat{M}_{g}^2}{\Lambda^2}\right\}.
\eeq
One can show that $\tr\{M^2\}=0$ and $\tr\{M^4\}
=4M^4f^4(1+\ka^2/\beta^2-g^2\ka^2/\beta^2\la^2)$
in this sector and thus its contribution to the
inflationary slope is again $\Lambda$-independent.

We next consider the $d^c$ and $\bar{d}^c$ type
sector consisting of color antitriplets with
charge $1/3$ and color triplets with charge
$-1/3$. We have the fields $d^c_H$, $\bar{d}^c_H$,
$d^c_{\phi}$, $\bar{d}^c_{\phi}$, $d^c_{\pb}$,
$\bar{d}^c_{\pb}$, coming from $H^c$, $\Hb^c$,
$\phi$ and $\pb$. Note that $\phi$ can be
expanded as
\beq\label{eq:phid}
\phi=\left[\left(\ba{cc}0&\bm{0}_3\\1&0\ea\right),
\left(\ba{cc}0&1\\0&0\ea\right)\right]d^c_{\phi}
+\left[\left(\ba{cc}0&1\\\bm{0}_3&0\ea\right),
\left(\ba{cc}0&0\\1&0\ea\right)\right]
\bar{d}^c_{\phi}+\cdots,
\eeq
with the notation of Eq.~\eqref{eq:phiu}. The
field $\pb$ is similarly expanded. The model
also contains \cite{quasi} a $\rm SU(4)_c$
$\bm{6}$-plet superfield $G=(\bm{6},\bm{1},
\bm{1})$ with the superpotential couplings
$xGH^cH^c$, $yG\Hb^c\Hb^c$, in order to give
\cite{leontaris} superheavy masses to $d^c_H$
and $\bar{d}^c_H$. The field $G$ splits under
$G_{\rm SM}$ into the fields $g^c=(\bar{\bm{3}},
\bm{1},1/3)$ and $\bar{g}^c=(\bm{3},\bm{1},-1/3)$.

The mass terms of the complex scalars $d^c_H$,
$\bar{d}^c_H$, $d^c_{\phi}$, $\bar{d}^c_{\phi}$,
$d^c_{\pb}$, $\bar{d}^c_{\pb}$, $g^c$ and
$\bar{g}^c$ are
\bea
\Lag_m(d)&=&M^2\Bigg\{\left[\frac{\ka^2c^2}
{9\beta^2}+\frac{2a^2\beta}{\ka\xi b^2}
\Big(\frac{2\la^2}{3}+x^2\Big)\right]
|d^c_H|^2+\left[\frac{\ka^2c^2}{9\beta^2}+
\frac{2a^2\beta}{\ka\xi b^2}\Big(\frac{2\la^2}
{3}+y^2\Big)\right]|\bar{d}^c_H|^2\nonumber \\
&+&\Big(\frac{\beta\la^2}{\ka\xi}+c^2\Big)
(|d^c_{\phi}|^2+|\bar{d}^c_{\phi}|^2)+
\Big(\frac{\beta\la^2}{\ka\xi}+
\frac{4\la^2a^2\beta}{3\ka\xi b^2}\Big)
(|d^c_{\pb}|^2+|\bar{d}^c_{\pb}|^2)\nonumber \\
&+&\frac{2a^2\beta^2}{\ka\xi b^2}
(y^2|g^c|^2+x^2|\bar{g}^c|^2)+\bigg[\frac{\ka f^2}
{3\beta}d^c_H\bar{d}^c_H-\frac{2\la^2\beta a}
{\sqrt{3}\ka\xi b}(d^c_Hd^{c*}_{\phi}+
\bar{d}^c_H\bar{d}^{c*}_{\phi})\nonumber \\
&-&\frac{2\ka^{\frac{1}{2}}\la ca}{3\sqrt{3}
\beta^{\frac{1}{2}}\xi^{\frac{1}{2}}b}
(d^c_Hd^{c*}_{\pb}+\bar{d}^c_H\bar{d}^{c*}_{\pb})
-\frac{\sqrt{2}\ka^{\frac{1}{2}}ca}
{3\beta^{\frac{1}{2}}\xi^{\frac{1}{2}}b}
(yd^c_Hg^{c*}+x\bar{d}^c_H\bar{g}^{c*})
-f^2d^c_{\phi}\bar{d}^c_{\phi}\nonumber \\
&-&\frac{\la\beta^{\frac{1}{2}}c}
{\ka^{\frac{1}{2}}\xi^{\frac{1}{2}}}
(d^c_{\phi}d^{c*}_{\pb}+
\bar{d}^c_{\phi}\bar{d}^{c*}_{\pb})+
\frac{2\sqrt{2}\la a^2\beta}{\sqrt{3}\ka\xi b^2}
(y d^c_{\pb}g^{c*}+x\bar{d}^c_{\pb}\bar{g}^{c*})+
{\rm h.c.}\bigg]\Bigg\}.
\eea
From these mass terms one can construct the
$8\times 8$ mass squared matrix $M_d^2$ of
the complex scalar fields $d^c_H$,
$\bar{d}^{c*}_H$, $d^c_{\phi}$,
$\bar{d}^{c*}_{\phi}$, $d^c_{\pb}$,
$\bar{d}^{c*}_{\pb}$, $g^c$, $\bar{g}^{c*}$.

In the fermion sector, we obtain four Dirac
fermions per color with mass matrix
\beq\label{eq:nshiftMpsid}
M_{\psi_d}=M\left(\ba{cccc}
\frac{\ka c}{3\beta}&0&
\smi\frac{2\beta^{\frac{1}{2}}\la a}
{\sqrt{3}\ka^{\frac{1}{2}}\xi^{\frac{1}{2}}b}&
\smi\frac{\sqrt{2}\beta^{\frac{1}{2}}ax}
{\ka^{\frac{1}{2}}\xi^{\frac{1}{2}}b}\\[6pt]
0&\smi c&\frac{\beta^{\frac{1}{2}}\la}
{\ka^{\frac{1}{2}}\xi^{\frac{1}{2}}}&0\\[6pt]
\smi\frac{2\beta^{\frac{1}{2}}\la a}{\sqrt{3}
\ka^{\frac{1}{2}}\xi^{\frac{1}{2}}b}&
\frac{\beta^{\frac{1}{2}}\la}{\ka^{\frac{1}{2}}
\xi^{\frac{1}{2}}}&0&0\\[6pt]
\smi\frac{\sqrt{2}\beta^{\frac{1}{2}}ay}
{\ka^{\frac{1}{2}}\xi^{\frac{1}{2}}b}&0&0&0
\ea\right).
\eeq

Note that there are no D-terms, gauge bosons or
gauginos in this sector. The contribution of the
$d^c$, $\bar{d}^c$ type sector to $\Delta V$ is
given by
\beq\label{eq:dsect}
\Delta V=\frac{3}{32\pi^2}\tr\left\{M_{d}^4
\ln\frac{M_{d}^2}{\Lambda^2}-
2(M_{\psi_d}M_{\psi_d}^\hc)^2\ln\frac{M_{\psi_d}
M_{\psi_d}^\hc}{\Lambda^2}\right\}.
\eeq
We find that $\tr\{M^2\}=0$ and $\tr\{M^4\}=
12M^4f^4(1+\ka^2/9\beta^2)$ in this sector,
implying that its contribution to the inflationary
slope is again $\Lambda$-independent.

Finally, we consider the $q^c$ and $\bar{q}^c$
type superfields which are color antitriplets
with charge $-5/3$ and color triplets with charge
$5/3$. They exist in $\phi$, $\pb$ and we call
them $q^c_{\phi}$, $\bar{q}^c_{\phi}$,
$q^c_{\pb}$, $\bar{q}^c_{\pb}$. The relevant
expansion of $\phi$ is
\beq\label{eq:phiq}
\phi=\left[\left(\ba{cc}0&\bm{0}_3\\1&0\ea\right),
\left(\ba{cc}0&0\\1&0\ea\right)\right]q^c_{\phi}+
\left[\left(\ba{cc}0&1\\\bm{0}_3&0\ea\right),
\left(\ba{cc}0&1\\0&0\ea\right)\right]
\bar{q}^c_{\phi}+\cdots,
\eeq
with the notation of Eq.~\eqref{eq:phiu}. A similar
expansion holds for $\pb$.

One finds that the mass squared matrices in the
$q^c$, $\bar{q}^c$ type bosonic sector are given by
\beq\label{eq:nshiftMq+-}
M_{q\pm}^2=M^2\left(\ba{cc}
\frac{\beta\la^2}{\ka\xi}\spl
\scriptstyle{c^2\,\mp\,f^2}&
\smi\frac{\la\beta^{\frac{1}{2}}c}
{\ka^{\frac{1}{2}}\xi^{\frac{1}{2}}}\\[6pt]
\smi\frac{\la\beta^{\frac{1}{2}}c}
{\ka^{\frac{1}{2}}\xi^{\frac{1}{2}}}&
\frac{\beta\la^2}{\ka\xi}\ea\right).
\eeq

The fermion mass matrix in this sector is given by
\beq\label{eq:nshiftMpsiq}
M_{\psi_q}=M\left(\ba{cc}
\smi c&\frac{\beta^{\frac{1}{2}}\la}
{\ka^{\frac{1}{2}}\xi^{\frac{1}{2}}}\\[6pt]
\frac{\beta^{\frac{1}{2}}\la}
{\ka^{\frac{1}{2}}\xi^{\frac{1}{2}}}&0
\ea\right).
\eeq

Furthermore, in $\phi$, $\pb$, there exist color
octet, $\rm SU(2)_R$ triplet superfields:
$\phi^0_8$, $\phi^\pm_8$, $\pb^0_8$, $\pb^\pm_8$
with charge $0$, $1$, $-1$ as indicated. The
relevant expansion of $\phi$ is
\beq\label{eq:phi8}
\phi=\left[\left(\ba{cc}
T_8&0\\0&0\ea\right)\;,\;
\frac{1}{\sqrt{2}}\left(\ba{cc}
1&0\\0&-1\ea\right)\phi_8^0
+\left(\ba{cc}0&1\\0&0\ea\right)\phi_8^+
+\left(\ba{cc}0&0\\1&0\ea\right)\phi_8^-
\right]+\dots,
\eeq
where $T_8$ represents the eight $\rm SU(3)_c$
generators appropriately normalized. A similar
expansion holds for $\pb$. It turns out that
the mass squared matrices in this sector are
the same as the ones in the $q^c$, $\bar{q}^c$
sector given in Eqs.~\eqref{eq:nshiftMq+-} and
\eqref{eq:nshiftMpsiq}.

The combined contribution from the $q^c$,
$\bar{q}^c$ type and color octet fields to
$\Delta V$ is
\beq\label{eq:qsect}
\Delta V=\frac{15}{32\pi^2}\tr\left\{M_{q+}^4
\ln\frac{M_{q+}^2}{\Lambda^2}+M_{q-}^4\ln
\frac{M_{q-}^2}{\Lambda^2}-2M_{\psi_q}^4\ln
\frac{M_{\psi_q}^2}{\Lambda^2}\right\}.
\eeq
Of course, $\tr\{M^2\}$ is vanishing in this
combined sector too and $\tr\{M^4\}=60M^4f^4$,
so that we again have a $\Lambda$-independent
contribution to the inflationary slope.

The final overall $\Delta V$ is found by adding
the contributions from the SM singlet sector in
Eq.~\eqref{eq:ssect}, the $u^c$, $\bar{u}^c$ type
sector in Eq.~\eqref{eq:usect}, the $e^c$,
$\bar{e}^c$ type sector in Eq.~\eqref{eq:esect},
the $d^c$, $\bar{d}^c$ type sector in
Eq.~\eqref{eq:dsect} and the combined $q^c$,
$\bar{q}^c$ type and color octet sector in
Eq.~\eqref{eq:qsect}. These one-loop
radiative corrections are added to $V_0$
yielding the effective potential $V(\si)$
along the new shifted inflationary trajectory.
They generate a slope on this trajectory which
is necessary for driving the system towards
the vacuum. The overall $\tr\{M^4\}=2M^4f^4
(45+16\ka^2/3\beta^2-6g^2\ka^2/\beta^2\la^2)$.
This implies that the overall slope is
$\Lambda$-independent. This is in fact a
crucial property of the model since otherwise
observable quantities like the power spectrum
amplitude $\PR$ of the primordial curvature
perturbation would depend on the scale $\Lambda$
which remains undetermined.

\begin{figure}[tp]
\centering
\includegraphics[width=\figwidth]{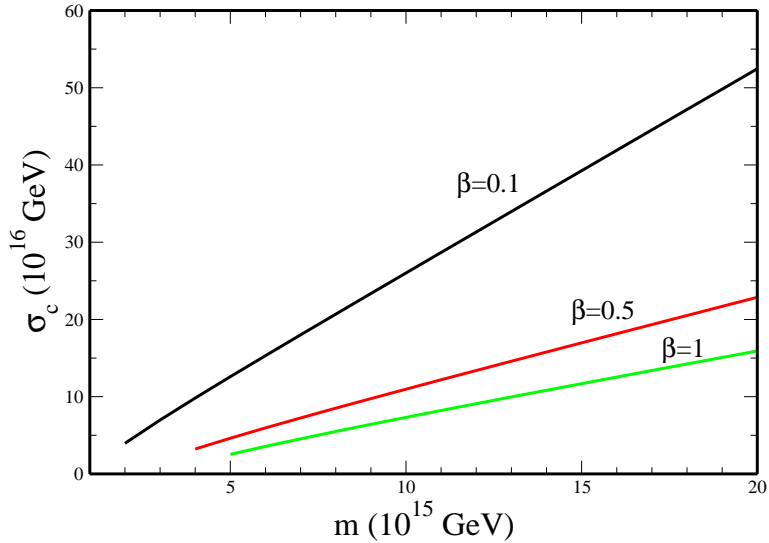}
\medskip
\caption{The critical value $\si_c$ of the
inflaton field as a function of the mass
parameter $m$ for $\ka=\la=3\ten{-3}$ and
$\beta=0.1$, $0.5$ and $1$, as indicated
on the curves.}
\label{fig:sgc}
\end{figure}

As can be easily seen from the relevant
expressions above, the effective potential
$V(\si)$ depends on the following parameters:
$M$, $m$, $\ka$, $\beta$, $\la$ and $g$.
We fix the gauge coupling constant at $M_{\rm
GUT}$ to the value $g=0.7$, which leads to the
correct values of the SM gauge coupling
constants at $M_Z$. We also assume
\cite{newshifted} that the VEV
$v_0=\vev{H^c}=\vev{\Hb^c}$ at the SUSY
vacuum is equal to the SUSY GUT scale
$M_{\rm GUT}\simeq2.86\ten{16}\units{GeV}$.
This allows us to determine the mass scale
$M$ in terms of the parameters $m$, $\ka$,
$\beta$ and $\la$. However, one finds
\cite{newshifted} that the requirement that
$M$ be real restricts the possible values of
these parameters. For instance, $\la\lesssim5
\ten{-3}$ for $m\simeq10^{16}\units{GeV}$,
$\ka\simeq10^{-3}$ and $\beta\simeq1$. In
Fig.~\ref{fig:sgc}, we present the critical
value $\si_c$ of the inflaton field, defined in
Eq.~\eqref{eq:scnshift}, as a function of the
mass scale $m$, for $\ka=\la=3\ten{-3}$ and
$\beta=0.1$, $0.5$ and $1$. As can be seen
from this figure, the smallest values of
$\si_c$ correspond to $\beta=1$. However,
in this case, the mass scale $m$ has to be
$\gtrsim5\ten{15}\units{GeV}$ to avoid
complex values of $M$. The value of the
inflaton field $\si_f$ at which inflation
terminates cannot be smaller than its critical
value $\si_c$ where the new shifted path
becomes unstable anyway. Thus, in order to
reduce the effect of SUGRA corrections which
could spoil \cite{HybridSUGRA} the flatness of
the inflationary path, one would be tempted to
choose values for the parameters which minimize
$\si_c$. A possible set of such values
\cite{newshifted} is $m=5\ten{15}\units{GeV}$,
$\ka=\la=3\ten{-3}$ and $\beta=1$, which yield
$\si_c\simeq4\ten{16}\units{GeV}$. However, in
this case, the condition $|\eta|=1$ implies that
inflation ends at $\si_f\simeq1.5\ten{18}
\units{GeV}$, which is quite large. Moreover,
it turns out that $\si_Q\simeq1.6\ten{19}
\units{GeV}$, which is much bigger than $\mP$
and, thus, this case is unacceptable.

A better set of values \cite{newshifted} is
$m=2\ten{15}\units{GeV}$, $\ka=\la=5\ten{-3}$
and $\beta=0.1$, which also yield $\si_c\simeq
4\ten{16}\units{GeV}$. In this case, $\si_f
\simeq1.7\ten{17}\units{GeV}$ and $\si_Q\simeq
1.6\ten{18}\units{GeV}$, which are much smaller
but still close to $\mP$. Values of $\beta$
smaller than $0.1$ (with suitable values of the
other parameters) give also very similar results.
Actually, it turns out \cite{newshifted} that a
general feature of the new shifted hybrid
inflationary model is that the relevant part of
inflation occurs at large values of $\si$, which
are close to $\mP$. Consequently, one is obliged
to consider the SUGRA corrections to the scalar
potential and invoke \cite{newshifted} some
mechanism to ensure that the new shifted
inflationary path remains flat. We will
address this issue in the next section.

We will now shortly discuss the constraints
imposed on the parameter space by the
measurements on the power spectrum amplitude
$\PR$ of the primordial curvature perturbation.
For a fixed $\PR$ we can determine one of the
free parameters (say $\beta$) in terms of the
others ($m$, $\ka$ and $\la$). For instance,
the COBE \cite{cobe} constraint $\PR\simeq
5.11\ten{-5}$, corresponds to $\beta=0.1$ if
$m=4.35\ten{15}\units{GeV}$ and $\ka=\la=
3\ten{-2}$. In this case,
$\si_c\simeq3.55\ten{16}\units{GeV}$,
$\si_f\simeq1.7\ten{17}\units{GeV}$ and
$\si_Q\simeq1.6\ten{18}\units{GeV}$. Also,
$M\simeq2.66\ten{16}\units{GeV}$,
$N_Q\simeq57.7$ and $\ns\simeq0.98$. We see
that the constraint on the power spectrum
amplitude can be easily satisfied with natural
values of the parameters. Moreover, superheavy
SM non-singlets with masses $\ll M_{\rm GUT}$,
which could disturb the unification of the SM
gauge couplings, are not encountered.

\section{Supergravity corrections}
\label{sec:NSHIFTsugra}

As we emphasized, new shifted hybrid inflation
occurs at values of $\si$ which are quite close
to the reduced Planck scale. Thus, one cannot
ignore the SUGRA corrections to the scalar
potential. The F-term scalar potential in SUGRA
is given by Eq.~\eqref{eq:VSUGRAF}, which is
rewritten here for convenience
\beq\label{eq:nshiftVSUGRAF}
V=e^{K/\mP^2}\left[(K^{-1})_i^j\;
F^{i*}F_j-3|W|^2/\mP^2\right],
\eeq
with $F^{i*}=-(W^i+WK^i/\mP^2)$, $F_j=-(W_j^*+
W^*K_j/\mP^2)$. $(K^{-1})_i^j$ is the inverse of
the K\"{a}hler metric $K_i^j$ and a raised
(lowered) index $i$ corresponds to derivation
with respect to $\phi_i$ ($\phi^{i*}$).

Consider a (complex) inflaton $\Sigma$
corresponding to a flat direction of global SUSY
with $W_{i\Sigma}=0$. We assume that the potential
on this path depends only on $|\Sigma|$, which
holds in this model due to a global symmetry.
From Eq.~\eqref{eq:nshiftVSUGRAF}, we find that
the SUGRA corrections lift the flatness of the
$\Sigma$ direction by generating a mass squared
for $\Sigma$ (see e.g.~\cite{lyth})
\beq\label{eq:mSigma}
m_\Sigma^2=\frac{V_0}{\mP^2}-
\frac{|W_\Sigma|^2}{\mP^2}+\sum_{i,j}
W^{i*}(K^{-1})_{i\Sigma\Sigma^*}^jW_j+\cdots,
\eeq
where the right hand side (RHS) is evaluated
on the flat direction with the explicitly
displayed terms taken at $\Sigma=0$. The
ellipsis represents higher order terms which
are suppressed by powers of $|\Sigma|/\mP$.
The slow roll parameter $\eta$ then becomes
\beq\label{eq:etasugra}
\eta=1-\frac{|W_\Sigma|^2}{V_0}+
\frac{\mP^2}{V_0}\sum_{i,j}W^{i*}
(K^{-1})_{i\Sigma\Sigma^*}^jW_j+\cdots,
\eeq
which, in general, could be of order unity and
thus invalidate \cite{HybridSUGRA} inflation.
This is the well known $\eta$ problem of
inflation in local SUSY. Several proposals have
been made in the literature to overcome this
difficulty (for a review see e.g.~\cite{lyth}).

In standard and shifted hybrid inflation, there
is an automatic mutual cancellation between the
first two terms in the RHS of
Eq.~\eqref{eq:etasugra}. This is due to the fact
that $W_n=0$ on the inflationary path for all
field directions $n$ which are perpendicular to
this path, which implies that $|W_\Sigma|^2=V_0$
on the path. This is an important feature of
these models since, in general, the sum of the
first two terms in the RHS of
Eq.~\eqref{eq:etasugra} is positive and of order
unity, thereby ruining inflation. It is easily
checked that these properties persist in this
inflationary model too. In particular, the
superpotential on the new shifted inflationary
path takes the form $W=V_0^{1/2}\Sigma$.

In all these hybrid inflation models, the only
non-zero contribution from the sum which appears
in the RHS of Eq.~\eqref{eq:etasugra} originates
from the term with $i=j=\Sigma$ (recall that
$W_n=0$ on the path). This contribution is equal
to the dimensionless coefficient of the quartic
term $|\Sigma|^4/4\mP^2$ in the K\"{a}hler
potential. For inflation to remain intact, we
need to assume that this coefficient is somewhat
small. The remaining terms give negligible
contributions to $\eta$ provided that
$|\Sigma|\ll\mP$. The latter is true for
standard and shifted hybrid inflation. So, we
see that, in these models, a mild tuning of
just one parameter is adequate for protecting
inflation from SUGRA corrections.

In the present model, however, inflation takes
place at values of $|\Sigma|$ close to $\mP$.
So, the terms in the ellipsis in the RHS of
Eq.~\eqref{eq:etasugra} cannot be ignored and
may easily invalidate inflation. Thus, one needs
to invoke \cite{newshifted} here a mechanism
which can ensure that the SUGRA corrections do
not lift the flatness of the inflationary path to
all orders. A suitable scheme has been suggested
in \cite{panagiotak}. It has been argued that
special forms of the K\"{a}hler potential can
lead to the cancellation of the SUGRA corrections
which spoil slow roll inflation to all orders. In
particular, a specific form of $K(\Sigma)$ (used
in no-scale SUGRA models) was employed and a gauge
singlet field $Z$ with a similar $K(Z)$ was
introduced. It was pointed out that, by assuming
a superheavy VEV for the $Z$ field through D-terms,
an exact cancellation of the inflaton mass on the
inflationary trajectory can be achieved.

\par
The mechanism of Ref.~\cite{panagiotak} can be
readily incorporated \cite{newshifted} in the
new shifted hybrid inflation model we have been
discussing, to ensure that the SUGRA corrections
do not lift the flatness of the inflationary path.
The only alteration caused to the lagrangian along
this path is that the kinetic term of $\si$ is now
non-minimal. This affects the equation of motion
of $\si$ and, consequently, the slow roll
conditions, $\PR$ and $N_Q$. The form of the
K\"{a}hler potential for $\Sigma$ used
in~\cite{panagiotak} is
\beq\label{eq:nshiftkaehler}
K(|\Sigma|^2)=-N\mP^2\ln
\left(1-\frac{|\Sigma|^2}{N\mP^2}\right),
\eeq
where $N=1$ or $2$. Here we take $N=2$. In this
case, the kinetic term of the real normalized
inflaton field $\si$ (recall that $|\Sigma|=
\si/\sqrt{2}$) is $(1/2)(\pd^2K/\pd\Sigma\pd
\Sigma^*)\dot{\si}^2$, where the overdot denotes
derivation with respect to the cosmic time $t$
and $\pd^2K/\pd\Sigma\pd\Sigma^*=
(1-\si^2/2N\mP^2)^{-2}$. Thus, the lagrangian on
the new shifted path is given by
\beq\label{eq:nshiftlag}
L=\int_{-\infty}^{\infty}dt\int d^3x\,a^3(t)
\left[\frac{1}{2}\,\dot{\si}^2\left(1-
\frac{\si^2}{2 N\mP^2}\right)^{-2}-V(\si)\right],
\eeq
where $a(t)$ is the scale factor of the universe.

The evolution equation of $\si$ is found by
varying this lagrangian with respect to $\si$
\beq\label{eq:motion}
\left[\ddot{\si}+3H\dot{\si}+\dot{\si}^2
\left(1-\frac{\si^2}{2N\mP^2}\right)^{-1}
\frac{\si}{N\mP^2}\right]\left(1-\frac{\si^2}
{2N\mP^2}\right)^{-2}+V'(\si)=0,
\eeq
where $H$ is the Hubble parameter. During
inflation, the ``friction'' term $3H\dot{\si}$
dominates over the other two terms in the
brackets in Eq.~\eqref{eq:motion}. Thus, this
equation reduces to the ``modified''
inflationary equation
\beq\label{eq:infeq}
\dot{\si}=-\frac{V'(\si)}{3H}\left(1-
\frac{\si^2}{2N\mP^2}\right)^2.
\eeq
Note that, for $\si\ll\sqrt{2N}\mP$, this
equation reduces to the standard inflationary
equation.

To derive the slow roll conditions, we evaluate
the sum of the first and the third term in the
brackets in Eq.~\eqref{eq:motion} by using
Eq.~\eqref{eq:infeq}:
\bea
\ddot{\si}+\dot{\si}^2\left(1-\frac{\si^2}
{2N\mP^2}\right)^{-1}\frac{\si}{N\mP^2}=
\frac{V'(\si)}{3H^2}H'(\si)\dot{\si}
\left(1-\frac{\si^2}{2N\mP^2}\right)^{2}
\nonumber\\
\hspace{1cm}-\frac{V''(\si)}{3H}\dot{\si}
\left(1-\frac{\si^2}{2N\mP^2}\right)^{2}+
\frac{V'(\si)}{3H}\dot{\si}
\left(1-\frac{\si^2}{2N\mP^2}\right)
\frac{\si}{N\mP^2}\cdot
\label{eq:sigmaddot}
\eea
Comparing the first two terms in the RHS of
Eq.~\eqref{eq:sigmaddot} with $H\dot{\si}$,
we obtain
\beq\label{eq:nshiftslow}
\epsilon\simeq\frac{\mP^2}{2}\left(\frac{V'(\si)}
{V_0}\right)^2\left(1-\frac{\si^2}{2N\mP^2}
\right)^{2}\leq 1,\quad |\eta|\simeq\mP^2\left|
\frac{V''(\si)}{V_0}\right|\left(1-\frac{\si^2}
{2N\mP^2}\right)^{2}\leq1.
\eeq
The third term in the RHS of
Eq.~\eqref{eq:sigmaddot}, compared to
$H\dot{\si}$, yields $\sqrt{2}\si\epsilon^{1/2}
/N\mP\leq1$, which is automatically satisfied
provided that $\epsilon\leq1$ and $\si\leq N\mP/
\sqrt{2}$. The latter is true for the values of
$\si$ which are relevant here. We see that the
slow roll parameters $\epsilon$ and $\eta$ now
carry an extra factor $(1-\si^2/2N\mP^2)^2\leq1$.
This leads, in general, to smaller $\si_f$'s.
However, in the present case $\si_f\ll
\sqrt{2N}\mP$ (for $N=2$) and, thus, this factor
is practically equal to unity. Consequently, its
influence on $\si_f$ is negligible.

The formulas for $N_Q$ and $\PR$ are now also
modified due to the presence of the extra factor
$(1-\si^2/2N\mP^2)^2$ in Eq.~\eqref{eq:infeq}. In
particular, a factor $(1-\si^2/2N\mP^2)^{-2}$
must be included in the integrand in the RHS of
Eq.~\eqref{eq:efoldings} and a factor $(1-\si_Q^2
/2N\mP^2)^{-4}$ in the RHS of
Eq.~\eqref{eq:perturbations}. One finds
\cite{newshifted} that, for the $\si$'s under
consideration, these modifications have only a
small influence on $\si_Q$ if one uses the same
input values for the free parameters as in the
global SUSY case. On the contrary, $\PR$
increases considerably. However, we can easily
readjust the parameters so that the observational
requirements on the power spectrum are again met.
For instance, $\PR\simeq5.11\ten{-5}$ is now
obtained with $m=3.8\ten{15}\units{GeV}$, keeping
$\ka=\la=3\ten{-2}$ and $\beta=0.1$ as in global
SUSY. In this case, $\si_c\simeq2.7\ten{16}
\units{GeV}$, $\si_f\simeq1.8\ten{17}\units{GeV}$
and $\si_Q\simeq1.6\ten{18}\units{GeV}$. Also,
$M\simeq2.6\ten{16}\units{GeV}$, $N_Q\simeq57.5$
and $n\simeq0.99$.

\newpage
\thispagestyle{empty}
\mbox{}

\chapter{Semi-shifted hybrid inflation
with $\rm B-L$ cosmic strings}
\label{sec:SSHIFT}

\section{Introduction}
\label{sec:SSHIFTintro}

As we have seen, one of the most promising models
for inflation is, undoubtedly, hybrid inflation,
which is naturally realized within SUSY GUT
models. In the standard realization of SUSY
hybrid inflation, the spontaneous breaking of
the GUT gauge symmetry takes place at the end
of inflation and, thus, superheavy magnetic
monopoles \cite{monopole} are copiously produced
if they are predicted by this symmetry breaking.
In this case, a cosmological catastrophe is
encountered. In order to avoid this disaster, one
can employ the smooth or shifted variants of SUSY
hybrid inflation (see Chap.~\ref{sec:HYBRID}). In
these inflationary scenarios, which, in their
original realization, are based on
non-renormalizable superpotential terms, the GUT
gauge symmetry is broken to the SM gauge group
already during inflation and, thus, no magnetic
monopoles are produced at the termination of
inflation. A new version of the shifted
inflationary scheme can be implemented, as we saw
in Chap.~\ref{sec:NSHIFT}, with only renormalizable
superpotential terms, within the extended SUSY PS
model introduced in Chap.~\ref{sec:QUASI}.

Fitting the three-year data of the Wilkinson
microwave anisotropy probe (WMAP) satellite with
the standard power-law cosmological model with
cold dark matter and a cosmological constant
($\Lambda$CDM), one obtains \cite{WMAP3} values
of the spectral index $\ns$ which are clearly
lower than unity. However, in supergravity with
canonical K\"{a}hler potential, the above hybrid
inflation models yield \cite{SenoguzShafi}
$\ns$'s which are very close to unity or even
larger than it, although their running is
negligible. This discrepancy may be resolved
\cite{SUSYnonminimal,smoothnonminimal,hilltop}
by including non-minimal terms in the K\"{a}hler
potential. Alternatively, if we wish to stick to
minimal SUGRA, we can reduce \cite{mhin} the
spectral index predicted by the hybrid
inflationary models by restricting the number
of e-foldings suffered by our present horizon
scale during the hybrid inflation which generates
the observed curvature perturbations. The
additional number of e-foldings required for
solving the horizon and flatness problems of
standard hot big bang cosmology can be provided
by a subsequent second stage of inflation. In
Chap.~\ref{sec:STSMOOTH}, we will show that the
same extended SUSY PS model can lead to a
two-stage inflationary scenario yielding
acceptable $\ns$'s in minimal SUGRA. The first
stage of inflation, during which the cosmological
scales exit the horizon, is of the standard
hybrid type, while the second stage, which
provides the additional e-foldings, is of the
smooth hybrid type.

In this chapter, we consider an alternative
inflationary scenario \cite{semishifted} which
incorporates cosmic strings \cite{string} (for
a textbook presentation or a review, see
e.g.~\cite{vilenkin}) and can also be naturally
realized within the same extended SUSY PS model
with only renormalizable superpotential terms.
As shown in Chap.~\ref{sec:NSHIFT}, this model
possesses a shifted classically flat direction
along which $\rm U(1)_{B-L}$ is unbroken. In order
to distinguish it from the new shifted flat
direction on which $G_{\rm PS}$ is broken to
$G_{\rm SM}$, we call this flat direction
``semi-shifted''. This direction acquires, as
usual, a logarithmic slope from one-loop
radiative corrections which are due to the SUSY
breaking caused by the non-zero potential energy
density on it. So, it can perfectly well be used
as an inflationary path along which semi-shifted
hybrid inflation takes place. When the system
crosses the critical point at which this path is
destabilized, a waterfall regime occurs during
which the $\rm U(1)_{B-L}$ gauge symmetry breaks
spontaneously and local cosmic strings are
produced. The resulting string network can
then contribute to the primordial curvature
perturbations.

It has been argued \cite{battye} that in the
presence of a small contribution to the curvature
perturbation from cosmic strings, the current
cosmic microwave background data can allow values
of the spectral index that are larger than the
ones obtained in the absence of strings.
Therefore, we may hope that the semi-shifted
hybrid inflationary scenario, which does involve
cosmic strings, can be made compatible with the
CMBR data, even without the use of non-minimal
terms in the K\"{a}hler potential or a subsequent
complementary stage of inflation. Recently, a fit
to the CMBR data and the luminous red galaxy data
in the Sloan digital sky survey (SDSS) \cite{sdss}
on large length scales outside the non-linear
regime was performed \cite{bevis1} by
using field-theory simulations \cite{bevis2} of
a dynamical network of local cosmic strings. It
demonstrated that the Harrison-Zeldovich (HZ)
model (i.e. with $\ns=1$) with a fractional
contribution $f_{10}\simeq0.10$ from cosmic
strings to the temperature power spectrum at
multipole $\ell=10$, is even moderately favored
over the standard power-law model without
strings. For the power-law $\Lambda$CDM
cosmological model with cosmic strings this fit
yields $\ns=0.94-1.06$ and $f_{10}=0.02-0.18$ at
$95\%$ confidence level (c.l.). As we will see,
under these circumstances the semi-shifted
hybrid inflation model in minimal SUGRA can
easily be compatible with the data. Note that
there is obviously no formation of PS magnetic
monopoles at the end of the semi-shifted hybrid
inflation and, thus, the corresponding
cosmological catastrophe is avoided.

\section{Semi-shifted hybrid inflation
in global SUSY} \label{sec:SSHIFTsusy}

We consider the extended SUSY PS model described
in Chap.~\ref{sec:QUASI}, which can lead to a
moderate violation of the asymptotic Yukawa
unification so that, for $\mu>0$, an acceptable
$b$-quark mass is obtained even with universal
boundary conditions. The breaking of $G_{\rm PS}$
to $G_{\rm SM}$ is achieved by the superheavy
VEVs of the right handed neutrino type components
of a conjugate pair of Higgs superfields $H^c$
and $\Hb^c$ belonging to the $({\bf \bar{4},1,2})$
and $({\bf 4,1,2})$ representations of
$G_{\rm PS}$ respectively. The model also
contains a gauge singlet $S$ and a conjugate
pair of superfields $\phi$, $\pb$ belonging to
the $({\bf 15,1,3})$ representation of
$G_{\rm PS}$. The superfield $\phi$ acquires a
VEV which breaks $G_{\rm PS}$ to $G_{\rm SM}
\times{\rm U(1)_{B-L}}$. In addition to
$G_{\rm PS}$, the model possesses a $Z_2$
matter parity symmetry and two global
$\rm U(1)$ symmetries, namely a Peccei-Quinn
and a R symmetry. Such continuous global
symmetries can effectively arise \cite{laz1}
from the rich discrete symmetry groups
encountered in many compactified string
theories (see e.g.~\cite{laz2}). As we have
seen in Chap.~\ref{sec:NSHIFT}, this model can
lead to new shifted hybrid inflation based
solely on renormalizable interactions.

The superpotential terms which are relevant for
inflation are given in Eq.~\eqref{eq:Wnshift}.
These terms, with a different (more convenient
for our purposes here) choice of basic parameters
and their phases, can be rewritten as
\beq\label{eq:Wsshift}
W=\ka S(M^2-\phi^2)-\ga S H^c\Hb^c+m\phi\pb
-\la\pb H^c\Hb^c,
\eeq
where $M$, $m$ are superheavy masses of the
order of the SUSY GUT scale $M_{\rm GUT}\simeq
2.86\ten{16}\units{GeV}$ and $\ka$, $\ga$, $\la$
are dimensionless coupling constants. These
parameters are normalized so that they
correspond to the couplings between the SM
singlet components of the superfields. In a
general superpotential of the type in
Eq.~\eqref{eq:Wsshift}, $M$, $m$ and any two
of the three dimensionless parameters $\ka$,
$\ga$, $\la$ can always be made real and positive
by appropriately redefining the phases of the
superfields. The third dimensionless parameter,
however, remains generally complex. For
definiteness, we will choose here this parameter
to be real and positive too. One can show that
the superpotential in Eq.~\eqref{eq:Wnshift} with
the particular choice of the phases of its
parameters considered there, can become equivalent
to the superpotential in Eq.~\eqref{eq:Wsshift}
provided that two of its real and positive
parameters are rotated to the negative real axis.
Actually, the form of the superpotential in
Eq.~\eqref{eq:Wsshift} can be derived from the
one in Eq.~\eqref{eq:Wnshift} by the replacement:
$\ka\to-\ga$, $\la\to-\la$, $\beta\to\ka$,
$M^2\to(\ka/\ga)M^2$. The F-term scalar potential
obtained from the superpotential $W$ in
Eq.~\eqref{eq:Wsshift} is given by
\beq\label{eq:Vsshift}
V=|\ka\,(M^2-\phi^2)-\ga H^c\Hb^c|^2
+|m\pb-2\ka S\phi|^2+|m\phi-\la H^c\Hb^c|^2
+|\gamma S+\la\pb\,|^2
\left(|H^c|^2+|\Hb^c|^2\right),
\eeq
where the complex scalar fields which belong to
the SM singlet components of the superfields are
denoted by the same symbol. We will ignore
throughout the soft SUSY breaking terms
\cite{sstad} in the scalar potential since their
effect on inflationary dynamics is negligible in
our case as in the case of the conventional
realization of shifted hybrid inflation.

From Eq.~\eqref{eq:Vsshift} and the vanishing
of the D-terms (which implies that $\Hb^{c*}=
e^{i\theta}H^c$), we find \cite{newsmooth} that
there exist two distinct continua of SUSY vacua:
\bea
\phi=\phi_{+},\quad\Hb^{c*}=H^c,\quad |H^c|=
\sqrt{\frac{m\phi_{+}}{\la}} \quad (\theta=0),
\label{eq:sshvac+}\\
\phi=\phi_{-},\quad\Hb^{c*}=-H^c,\quad |H^c|=
\sqrt{\frac{-m\phi_{-}}{\la}} \quad (\theta=\pi),
\label{eq:sshvac-}
\eea
with $\pb=S=0$, where
\beq\label{eq:SUSYvacua1}
\phi_{\pm}\equiv\frac{\ga m}{2\ka\la}\left(
-1\pm\sqrt{1+\frac{4\ka^2\la^2M^2}{\ga^2m^2}}
\,\right).
\eeq
It has been shown in Chap.~\ref{sec:NSHIFT} (see
also \cite{newsmooth}) that the potential in
Eq.~\eqref{eq:Vsshift} generally possesses three
flat directions. The first one is the usual
trivial flat direction at $\phi=\pb=H^c=\Hb^c=0$
with $V=V_{\rm tr}\equiv\ka^2M^4$. The second
one, which appears at
\begin{gather}
\phi=-\frac{\ga m}{2\ka\la},\quad
\pb=-\frac{\ga}{\la}\,S,\quad H^c\Hb^c=
\frac{\ka\ga(M^2-\phi^2)+\la m\phi}{\ga^2+\la^2},\\
V=V_{\rm nsh}\equiv\frac{\ka^2\la^2}{\ga^2+\la^2}
\left(M^2+\frac{\ga^2m^2}{4\ka^2\la^2}\right)^2,
\end{gather}
exists only for $\ga\neq 0$ and is the
trajectory for the new shifted hybrid inflation.
Along this direction, $G_{\rm PS}$ is broken to
$G_{\rm SM}$. The third flat direction exists
only if $\tilde{M}^2\equiv M^2-m^2/2\ka^2>0$
and lies at
\beq\label{eq:semishift}
\phi=\pm\,\tilde{M},\quad
\pb=\frac{2\ka\phi}{m}\,S,\quad
H^c=\Hb^c=0.
\eeq
It is a ``semi-shifted'' flat direction
(in the sense that, although the field $\phi$ is
shifted from zero, the fields $H^c$, $\Hb^c$
remain zero on it) with
\beq
V=V_{\rm ssh}\equiv\ka^2(M^4-\tilde{M}^4).
\eeq
Along this direction $G_{\rm PS}$ is broken to
$G_{\rm SM}\times{\rm U(1)_{B-L}}$.

In our subsequent discussion, we will concentrate
on the case where $\tilde{M}^2>0$. It is
interesting to note that, in this case, the
trivial flat direction is unstable
\cite{newsmooth} as it is a path of saddle points
of the potential. Moreover, for $\tilde{M}^2>0$,
we always have $V_{\rm ssh}<V_{\rm nsh}$. It is,
thus, more likely that the system will eventually
settle down on the semi-shifted rather than the
new shifted flat direction. Semi-shifted hybrid
inflation can then take place as the system
slowly rolls down the semi-shifted path driven by
its logarithmic slope provided by one-loop
radiative corrections, which are due to the SUSY
breaking by the non-vanishing potential energy
density on this path. As the system crosses the
critical point of the semi-shifted path, the
$\rm U(1)_{B-L}$ gauge symmetry breaks,
generating a network of local cosmic strings,
which contribute a small amount to the CMBR
temperature power spectrum. As mentioned, for
models with local cosmic strings, it has been
shown in~\cite{bevis1} that, at $95\%$ c.l.,
$\ns=0.94-1.06$ and $f_{10}=0.02-0.18$.

\section{One-loop radiative corrections}
\label{sec:SSHIFTcorr}

\par
The one-loop radiative correction to the
potential on the semi-shifted path is
calculated by the Coleman-Weinberg formula:
\beq\label{eq:sshiftCW}
\Delta V=
\frac{1}{64\pi^2}\,\sum_i(-1)^{F_i}M_i^4
\ln\frac{M_i^2}{\Lambda^2},
\eeq
where the sum extends over all helicity states
$i$, $F_i$ and $M_i^2$ are the fermion number and
mass squared of the $i$th state and $\Lambda$ is
a renormalization mass scale. In order to use this
formula for creating a logarithmic slope in the
inflationary potential, one has first to derive
the mass spectrum of the model on the
semi-shifted path.

As mentioned, during semi-shifted hybrid
inflation, the SM singlet components of $\phi$,
$\pb$ acquire non-vanishing values and break
$G_{\rm PS}$ to $G_{\rm SM}\times
{\rm U(1)_{B-L}}$. The value of the complex
scalar field $S$ at a point of the semi-shifted
path is taken real by an appropriate R
transformation. For simplicity, we use the same
symbol $S$ for this real value of the field as
for the complex field in general since the
distinction will be obvious from the context.
The deviation of the complex scalar field $S$
from its (real) value at a point of the
inflationary path is denoted by $\delta S$. We
can further write $\phi=v+\delta\phi$,
$\pb=\bar{v}+\delta\pb$ with $v=\pm\tilde{M}$,
$\bar{v}=(2\ka v/m)\,S$ and $\delta\phi$,
$\delta\pb$ being complex scalar fields. We can
then define the canonically normalized complex
scalar fields
\beq
\zeta=\frac{2\ka v\,\delta S-m\delta\pb}
{(m^2+4\ka^2v^2)^{1/2}},\quad
\epsilon=\frac{m\delta S+2\ka v\,\delta\pb}
{(m^2+4\ka^2v^2)^{1/2}}.
\eeq
We find that $\epsilon$ remains massless on the
semi-shifted path. So, it corresponds to the
complex scalar inflaton field
$\Sigma=(mS+2\ka v\pb)/(m^2+4\ka^2v^2)^{1/2}$,
which during inflation takes the form
$\Sigma=(1+4\ka^2v^2/m^2)^{1/2}S$. Consequently,
in our case, the real canonically normalized
inflaton is
\beq\label{eq:sshiftinflaton}
\sigma=2^{1/2}(1+4\ka^2v^2/m^2)^{1/2}S,
\eeq
where $S$ is obviously rotated to be real.

Expanding the complex scalars $\zeta$,
$\delta\phi$, $H^c$ and $\Hb^c$ in real and
imaginary parts according to the prescription
$\chi=(\chi_1+i\chi_2)/\sqrt{2}$, we find that
the mass-squared matrices $M_{-}^2$ of
$\zeta_1$, $\delta\phi_1$, $M_{+}^2$ of
$\zeta_2$, $\delta\phi_2$, $M_1^2$ of $H^c_1$,
$\Hb^c_1$ and $M_2^2$ of $H^c_2$, $\Hb^c_2$
are given by
\begin{gather}
M_{\pm}^2=m^2\left(\ba{cc}
1+a^2 & s(1+a^2)^{1/2} \\
s(1+a^2)^{1/2} & 1+a^2+s^2\pm 1
\ea\right),\\[6pt]
M_{1,2}^2=m^2\left(\ba{cc}
s^2b^2 & \mp b \\
\mp b & s^2b^2
\ea\right),
\end{gather}
where $a=2\ka v/m$, $b=(\ga+\la a)/2\ka$ and
$s=2\ka S/m$. Note that the eigenvalues of the
matrices $M_{\pm}^2$ are always positive. Though,
this is not the case with $M_{1,2}^2$.
Specifically, one of the two eigenvalues of each
of these matrices is always positive while the
other one becomes negative for
$|s|<s_c\equiv 1/\sqrt{|b|}$ (we assume that
$b\neq 0$). This defines the critical point on
the semi-shifted path at which this path is
destabilized (see below).

The superpotential in Eq.~\eqref{eq:Wsshift}
gives rise to mass terms between the fermionic
partners of $\zeta$, $\delta\phi$ and $H^c$,
$\Hb^c$ (the fermionic partner of $\epsilon$
remains massless). The squares of the
corresponding mass matrices are found to be
\begin{gather}
M_0^2=m^2\left(\ba{cc}
1+a^2 & s(1+a^2)^{1/2} \\
s(1+a^2)^{1/2} & 1+a^2+s^2
\ea\right),\\[6pt]
\bar{M}_0^2=m^2\left(\ba{cc}
s^2b^2 & 0 \\ 0 & s^2b^2
\ea\right).
\end{gather}

This completes the analysis of the SM singlet
sector of the model. In summary, we found four
groups of two real scalars with mass-squared
matrices $M_{+}^2$, $M_{-}^2$, $M_{1}^2$ and
$M_{2}^2$ and two groups of two
Weyl fermions with mass matrices squared
$M_0^2$ and $\bar{M}_0^2$. The contribution
of the SM singlet sector to the radiative
corrections to the potential along the
semi-shifted path is given by
\beq\label{eq:sshift-ssect}
\Delta V=\frac{1}{64\pi^2}\,
\tr\Bigg\{M_{+}^4\ln\frac{M_{+}^2}{\Lambda^2}
+M_{-}^4\ln\frac{M_{-}^2}{\Lambda^2}
-2M_0^4\ln\frac{M_0^2}{\Lambda^2}
+M_{1}^4\ln\frac{M_{1}^2}{\Lambda^2}
+M_{2}^4\ln\frac{M_{2}^2}{\Lambda^2}
-2\bar{M}_0^4\ln\frac{\bar{M}_0^2}
{\Lambda^2}\Bigg\}.
\eeq

We now turn to the $u^c$, $\bar{u}^c$ type fields
which are color antitriplets with charge $-2/3$
and color triplets with charge $2/3$
respectively. Such fields exist in $H^c$,
$\Hb^c$, $\phi$ and $\pb$ and we shall denote
them by $u_H^c$, $\bar{u}_H^c$, $u_\phi^c$,
$\bar{u}_\phi^c$, $u_{\pb}^c$ and
$\bar{u}_{\pb}^c$. The relevant expansion of
$\phi$ is given in Eq.~\eqref{eq:phiu}.

In the bosonic $u^c$, $\bar{u}^c$ type sector,
we find that the mass-squared matrices
$M_{u\pm}^2$ of the complex scalar fields
$u_{\chi\pm}^c=(u_{\chi}^c\pm\bar{u}_{\chi}^{c*})
/\sqrt{2}$, for $\chi=H,\phi,\pb$, are
\begin{gather}
\label{eq:sshiftMu+}
M_{u+}^2=m^2\left(\ba{ccc}
c^2s^2-c & 0 & 0\\
0 & s^2 & -s\\
0 & -s & 1\ea\right),\\[6pt]
\label{eq:sshiftMu-}
M_{u-}^2=m^2\left(\ba{ccc}
c^2s^2+c & 0 & 0\\ [3pt]
0 & 2+s^2+\rho_g^2 & -s(1-\rho_g^2)\\ [3pt]
0 & -s(1-\rho_g^2) & 1+\rho_g^2s^2\ea\right),
\end{gather}
where $c=(\ga-\la a/3)/2\ka$ and
$\rho_g^2=g^2a^2/3\ka^2$ with $g$ being the
$G_{\rm PS}$ gauge coupling constant. Note that
$\rho_g^2$ parameterizes contributions arising
from the D-terms of the scalar potential and
$M_{u+}^2$ has one zero eigenvalue corresponding
to the Goldstone boson which is absorbed by the
superhiggs mechanism. Furthermore, one of the
eigenvalues $m^2(c^2s^2\mp c)$ of the matrices in
Eqs.~\eqref{eq:sshiftMu+} and \eqref{eq:sshiftMu-}
(depending on the sign of $c$) becomes negative
as soon as $s$ crosses below the point $s_c^{(1)}
\equiv 1/\sqrt{|c|}$ on the semi-shifted path.
So, if $s_c^{(1)}$ was larger than the critical
value $s_c$, the system would be destabilized
first in one of the directions $u_{H\pm}^c$. In
this case, a $\rm SU(3)_c$-breaking VEV would
develop. To avoid this, we should demand that
$s_c^{(1)}$ is located lower than the critical
point $s_c$, so that, after the end of inflation,
the correct symmetry breaking is obtained. This
gives the condition $|b|<|c|$, which we will
consider later.

In the fermionic $u^c$, $\bar{u}^c$ type sector,
we obtain four Dirac fermions (per color):
$\psi_{u_H^c}^D=\psi_{u_H^c}+\psi^c_{\bar{u}_H^c}$,
$\psi_{u_{\phi}^c}^D=\psi_{u_{\phi}^c}+
\psi^c_{\bar{u}_{\phi}^c}$, $\psi_{u_{\pb}^c}^D=
\psi_{u_{\pb}^c}+\psi^c_{\bar{u}_{\pb}^c}$ and
$-i\la^D=-i(\la^+ + \la^{-c})$. Here, $\psi_\chi$
is the fermionic partner of the complex scalar
field $\chi$ and
$\la^{\pm}=(\la^1\pm i\la^2)/\sqrt{2}$, where
$\la^1$ ($\la^2$) is the gaugino color triplet
corresponding to the ${\rm SU}(4)_c$ generators
with $1/2$ ($-i/2$) in the $i4$ and $1/2$ ($i/2$)
in the $4i$ entry ($i=1,2,3$). The fermionic
mass matrix is
\beq
M_{\psi_u}=m\left(\ba{cccc}
-cs & 0 & 0 & 0\\
0 & -s & 1 & -\rho_g\\
0 & 1 & 0 & -\rho_g s\\
0 & -\rho_g & -\rho_g s & 0
\ea\right).
\eeq
To complete this sector, we must also include the
gauge bosons $A^{1,2}$ which are associated with
$\la^{1,2}$. They acquire a mass squared
$M_g^2=m^2\rho_g^2(1+s^2)$.

The overall contribution of the $u^c$,
$\bar{u}^c$ type sector to $\Delta V$ in
Eq.~\eqref{eq:sshiftCW} is
\beq\label{eq:sshift-usect}
\Delta V=\frac{3}{32\pi^2}\tr\Bigg\{
M_{u+}^4\ln\frac{M_{u+}^2}{\Lambda^2}
+M_{u-}^4\ln\frac{M_{u-}^2}{\Lambda^2}
-2M_{\psi_u}^4\ln\frac{M_{\psi_u}^2}{\Lambda^2}
+3M_g^4\ln\frac{M_g^2}{\Lambda^2}\Bigg\}.
\eeq

We will now discuss the contribution from the
$e^c$, $\bar{e}^c$ type sector consisting of
color singlets with charge $1$, $-1$. Such
fields exist in $H^c$, $\Hb^c$, $\phi$ and
$\pb$ and we shall denote them by $e_H^c$,
$\bar{e}_H^c$, $e_\phi^c$, $\bar{e}_\phi^c$,
$e_{\pb}^c$ and $\bar{e}_{\pb}^c$. The
relevant expansion of $\phi$ is given in
Eq.~\eqref{eq:phie}. It turns out that the
mass terms in this sector are exactly the
same as in the $u^c$, $\bar{u}^c$ type sector
with $\la/3$ replaced by $\la$ and $2g^2/3$
by $g^2$. So, we will only summarize the results.

In the bosonic $e^c$, $\bar{e}^c$ type sector,
the mass-squared matrices $M_{e\pm}^2$ of the
complex scalars $e_{\chi\pm}^c=(e_{\chi}^c\pm
\bar{e}_{\chi}^{c*})/\sqrt{2}$, for
$\chi=H,\phi,\pb$, are
\begin{gather}
\label{eq:sshiftMe+}
M_{e+}^2=m^2\left(\ba{ccc}
d^2s^2-d & 0 & 0\\
0 & s^2 & -s\\
0 & -s & 1\ea\right),\\[6pt]
\label{eq:sshiftMe-}
M_{e-}^2=m^2\left(\ba{ccc}
d^2s^2+d & 0 & 0\\ [3pt]
0 & 2+s^2+\tau_g^2 & -s(1-\tau_g^2)\\ [3pt]
0 & -s(1-\tau_g^2) & 1+\tau_g^2s^2
\ea\right),
\end{gather}
where $d=(\ga-\la a)/2\ka$ and $\tau_g=
\sqrt{3/2}\,\rho_g$. Note that, again, $M_{e+}^2$
has one zero eigenvalue corresponding to the
Goldstone boson which is absorbed by the
superhiggs mechanism. Furthermore, one of the
eigenvalues $m^2(d^2s^2\mp d)$ of the matrices in
Eqs.~\eqref{eq:sshiftMe+} and \eqref{eq:sshiftMe-}
(depending on the sign of $d$) becomes negative as
$s$ crosses below $s_c^{(2)}\equiv 1/\sqrt{|d|}$
on the semi-shifted path. Therefore, we must
impose the constraint $s_c^{(2)}<s_c\Rightarrow
|b|<|d|$ for the same reason explained above.

In the fermionic $e^c$, $\bar{e}^c$ type sector,
we obtain four Dirac fermions with mass matrix
\beq
M_{\psi_e}=m\left(\ba{cccc}
-ds & 0 & 0 & 0\\
0 & -s & 1 & -\tau_g\\
0 & 1 & 0 & -\tau_g s\\
0 & -\tau_g & -\tau_g s & 0
\ea\right).
\eeq
Finally, we again obtain two gauge bosons with
mass squared $\hat{M}_g^2=m^2\tau_g^2(1+s^2)$.

The overall contribution of the $e^c$,
$\bar{e}^c$ type sector to $\Delta V$ in
Eq.~\eqref{eq:sshiftCW} is
\beq\label{eq:sshift-esect}
\Delta V=\frac{1}{32\pi^2}\tr\Bigg\{
M_{e+}^4\ln\frac{M_{e+}^2}{\Lambda^2}
+M_{e-}^4\ln\frac{M_{e-}^2}{\Lambda^2}
-2M_{\psi_e}^4\ln\frac{M_{\psi_e}^2}{\Lambda^2}
+3\hat{M}_g^4\ln\frac{\hat{M}_g^2}{\Lambda^2}
\Bigg\}.
\eeq

Let us now consider the $d^c$, $\bar{d}^c$ type
sector consisting of color antitriplets with
charge $1/3$ and color triplets with charge
$-1/3$. Such fields exist in $H^c$, $\Hb^c$,
$\phi$ and $\pb$ and we denote them by $d_H^c$,
$\bar{d}_H^c$, $d_\phi^c$, $\bar{d}_\phi^c$,
$d_{\pb}^c$ and $\bar{d}_{\pb}^c$. The field
$\phi$ can be expanded in terms of these fields
as in Eq.~\eqref{eq:phid}.

In the bosonic $d^c$, $\bar{d}^c$ type sector,
the mass-squared matrices $M_{d\pm}^2$ of the
complex scalars $d_{\chi\pm}^c=(d_{\chi}^c\pm
\bar{d}_{\chi}^{c*})/\sqrt{2}$, for
$\chi=H,\phi,\pb$, are
\beq
M_{d\pm}^2=m^2\left(\ba{ccc}
e^2s^2\mp e & 0 & 0\\
0 & 1+s^2\mp 1 & -s\\
0 & -s & 1\ea\right),
\eeq
where $e=(\ga+\la a/3)/2\ka$. Note that,
again, one of the eigenvalues $m^2(e^2s^2\mp e)$
of these matrices (depending on the sign of $e$)
becomes negative as $s$ crosses below
$s_c^{(3)}\equiv 1/\sqrt{|e|}$ on the
semi-shifted path and we, thus, have to impose
the constraint $s_c^{(3)}<s_c\Rightarrow|b|<|e|$,
so that the correct symmetry breaking pattern
occurs at the end of inflation.

In the fermionic $d^c$, $\bar{d}^c$ type sector,
we obtain three Dirac fermions (per color) with
mass matrix
\beq
M_{\psi_d}=m\left(\ba{ccc}
-es & 0 & 0\\0 & -s & 1\\0 & 1 & 0\ea\right).
\eeq
Note that there are no D-terms, gauge bosons, or
gauginos in this sector.

The contribution of this sector to $\Delta V$
in Eq.~\eqref{eq:sshiftCW} is
\beq\label{eq:sshift-dsect}
\Delta V=\frac{3}{32\pi^2}\tr\Bigg\{
M_{d+}^4\ln\frac{M_{d+}^2}{\Lambda^2}
+M_{d-}^4\ln\frac{M_{d-}^2}{\Lambda^2}
-2M_{\psi_d}^4\ln\frac{M_{\psi_d}^2}
{\Lambda^2}\Bigg\}.
\eeq

Next, we consider the $q^c$, $\bar{q}^c$
type fields which are color antitriplets with
charge $-5/3$ and color triplets with charge
$5/3$. They exist in $\phi$, $\pb$ and we call
them $q_\phi^c$, $\bar{q}_\phi^c$, $q_{\pb}^c$,
$\bar{q}_{\pb}^c$. The relevant expansion of
$\phi$ can be found in Eq.~\eqref{eq:phiq}.

In the bosonic $q^c$, $\bar{q}^c$ type sector,
the mass-squared matrices $M_{q\pm}^2$ of the
complex scalars $q_{\chi\pm}^c=(q_{\chi}^c
\pm\bar{q}_{\chi}^{c*})/\sqrt{2}$, for
$\chi=\phi,\pb$, are
\beq
M_{q\pm}^2=m^2\left(\ba{cc}
1+s^2\mp 1 & -s\\-s & 1\ea\right).
\eeq

In the fermionic $q^c$, $\bar{q}^c$ type sector,
we obtain two Dirac fermions (per color) with
mass matrix
\beq
M_{\psi_q}=m\left(\ba{cc}
-s & 1\\1 & 0\ea\right).
\eeq
There are no D-terms, gauge bosons, or
gauginos in this sector as well.

The contribution of this sector to $\Delta V$
in Eq.~\eqref{eq:sshiftCW} is
\beq\label{eq:sshift-qsect}
\Delta V=\frac{3}{32\pi^2}\tr\Bigg\{
M_{q+}^4\ln\frac{M_{q+}^2}{\Lambda^2}
+M_{q-}^4\ln\frac{M_{q-}^2}{\Lambda^2}
-2M_{\psi_q}^4\ln\frac{M_{\psi_q}^2}
{\Lambda^2}\Bigg\}.
\eeq

Finally, in $\phi$, $\pb$ there exist color
octet, $\rm SU(2)_R$ triplet superfields:
$\phi_8^0$, $\phi_8^{\pm}$, $\pb_8^0$,
$\pb_8^{\pm}$, with charge $0$, $1$, $-1$ as
indicated. The relevant expansion of $\phi$
is given in Eq.~\eqref{eq:phi8}.

In the bosonic sector, we obtain two groups of
24 complex scalars, which can be combined in
pairs of two with mass-squared matrix
\beq
M_{\phi_8\pm}^2=m^2\left(\ba{cc}
1+s^2\mp 1 & -s\\-s & 1\ea\right).
\eeq

In the fermionic sector, we find 48 Weyl fermions
which can be combined in pairs of two with
mass matrix
\beq
M_{\psi_{\phi_8}}=m\left(\ba{cc}
-s & 1\\1 & 0\ea\right).
\eeq

The contribution of this sector to $\Delta V$
in Eq.~\eqref{eq:sshiftCW} is
\beq\label{eq:sshift-phi8sect}
\Delta V=\frac{12}{32\pi^2}\tr\Bigg\{
M_{\phi_8+}^4\ln\frac{M_{\phi_8+}^2}{\Lambda^2}
+M_{\phi_8-}^4\ln\frac{M_{\phi_8-}^2}{\Lambda^2}
-2M_{\psi_{\phi_8}}^4\ln\frac{M_{\psi_{\phi_8}}^2}
{\Lambda^2}\Bigg\}.
\eeq

The final overall $\Delta V$ is found by adding
the contributions from the SM singlet sector in
Eq.~\eqref{eq:sshift-ssect}, the $u^c$, $\bar{u}^c$
type sector in Eq.~\eqref{eq:sshift-usect},
the $e^c$, $\bar{e}^c$ type sector in
Eq.~\eqref{eq:sshift-esect}, the $d^c$, $\bar{d}^c$
type sector in Eq.~\eqref{eq:sshift-dsect},
the $q^c$, $\bar{q}^c$ type sector in
Eq.~\eqref{eq:sshift-qsect} and the color octet
sector in Eq.~\eqref{eq:sshift-phi8sect}. These
one-loop radiative corrections are added to the
tree-level potential $V_{\rm ssh}$ yielding the
effective potential along the semi-shifted
inflationary path in global SUSY. They generate
a slope on this path which is necessary for
driving the system towards the vacuum. The
overall $\sum_i(-1)^{F_i}M_i^4$ is
$\si$-independent, which implies that the
overall slope of the effective potential is
$\Lambda$-independent. This is a crucial
property of the model since otherwise
observable quantities like the power spectrum
$\PR$ of the primordial curvature perturbation
or the spectral index would depend on the scale
$\Lambda$, which remains undetermined.

Let us now discuss the constraints
$0<|b|<|c|,|d|,|e|$ derived in the course of
the calculation of the mass spectrum on the
semi-shifted path. It is easy to show that
these constraints require that $v$ be in
one of the ranges
\beq\label{eq:sshiftconstr}
0>v>-\frac{\ga m}{2\ka\la}\quad\text{or}\quad
-\frac{\ga m}{2\ka\la}>v>-\frac{3\ga m}{4\ka\la}.
\eeq
These two ranges of $v$ lead, respectively, to
the two different sets of SUSY vacua of
Eqs.~\eqref{eq:sshvac+} and \eqref{eq:sshvac-}.
To see this, let us replace all the fields in
the scalar potential of Eq.~\eqref{eq:Vsshift}
except $H^c$, $\bar{H}^c$ by their values on the
semi-shifted path. Taking into account that
$\bar{H}^{c*}=e^{i\theta}H^c$ from the vanishing
of the D-terms we are left with only two
free degrees of freedom, namely $|H^c|$ and
$\theta$, and the potential becomes
\beq
V=V_{\rm ssh}+2m^2b^2\left(s^2-
\frac{\cos\theta}{b}\right)|H^c|^2
+(\ga^2+\la^2)|H^c|^4.
\eeq
It is obvious from this equation that, if $b>0$,
which is the case in the first range for $v$ in
Eq.~\eqref{eq:sshiftconstr}, the system will
get destabilized towards the direction with
$\cos\theta=1$ leading to the SUSY vacua
in Eq.~\eqref{eq:sshvac+}, while, if $b<0$, which
holds in the second range for $v$ in
Eq.~\eqref{eq:sshiftconstr}, the system will be
led to the SUSY vacua in Eq.~\eqref{eq:sshvac-}.

\section{Supergravity corrections}
\label{sec:SSHIFTsugra}

We now turn to the discussion of the SUGRA
corrections to the inflationary potential. The
F-term scalar potential in SUGRA is given by
(see Eq.~\eqref{eq:VSUGRAF})
\beq\label{eq:sshiftVSUGRAF}
V=e^{K/\mP^2}\left[(K^{-1})_i^j\;
F^{i*}F_j-3|W|^2/\mP^2\right],
\eeq
with $F^{i*}=-(W^i+WK^i/\mP^2)$, $F_j=-(W_j^*+
W^*K_j/\mP^2)$ and a raised (lowered) index $i$
corresponds to derivation with respect to
$\chi_i$ ($\chi^{i*}$). We will consider SUGRA
with minimal K\"{a}hler potential and show that
the results of the fit in Ref.~\cite{bevis1}
can be naturally met.

The minimal K\"{a}hler potential in the model
under consideration has the form
\beq\label{eq:sshMinKahler}
K^{\rm min}=|S|^2+|\phi|^2+|\pb|^2+|H^c|^2
+|\Hb^c|^2
\eeq
and the corresponding F-term scalar potential is
\beq\label{eq:sshVSUGRAmin}
V^{\rm min}=e^{K^{\rm min}/\mP^2}\;\left[
\sum_{\chi}\left|W_\chi+\frac{W\chi^*}{\mP^2}
\right|^2-3\,\frac{|W|^2}{\mP^2}\right],
\eeq
where $\chi$ stands for any of the five complex
scalar fields appearing in
Eq.~\eqref{eq:sshMinKahler}. It is quite
easily verified that, on the semi-shifted
direction, this scalar potential expanded up to
fourth order in $|S|$ takes the form (the SUGRA
corrections to the location of the semi-shifted
path are not taken into account since they are
small)
\beq
V^{\rm min}_{\rm ssh}\simeq V_{\rm ssh}\,
e^{\tilde{M}^2/\mP^2}\left[
1+\frac{1}{2}\frac{\tilde{M}^2}{\mP^2}
\frac{\si^2}{\mP^2}+\frac{1}{8}\left(
1+\frac{2\tilde{M}^2}{\mP^2}\right)
\frac{\si^4}{\mP^4}\right],
\eeq
where $V_{\rm ssh}$ is the constant classical
energy density on the semi-shifted path in the
global SUSY case and $\si$ is the canonically
normalized inflaton field defined in
Eq.~\eqref{eq:sshiftinflaton}. Thus, after
including the SUGRA corrections with minimal
K\"{a}hler potential, the effective potential
during semi-shifted hybrid inflation becomes
\beq\label{eq:sshiftVtotsugra}
V^{\rm mSUGRA}_{\rm ssh}\simeq
V^{\rm min}_{\rm ssh}+\Delta V
\eeq
with $\Delta V$ representing the overall
one-loop radiative correction calculated in
Sec.~\ref{sec:SSHIFTcorr}.

\section{Inflationary observables}
\label{sec:SSHIFTobserv}

\par
The slow-roll parameters $\epsilon$, $\eta$ and
the parameter $\xi^2$, which enters the running
of the spectral index, are given by (see
Chap.~\ref{sec:HYBRID} or Ref.~\cite{InfReview})
\beq
\epsilon \equiv \frac{\mP^2}{2}\,
\left(\frac{V'(\si)}{V(\si)}\right)^2,\quad
\eta \equiv \mP^2\,\left(
\frac{V''(\si)}{V(\si)}\right),\quad
\xi^2 \equiv \mP^4
\left(\frac{V'(\si)V'''
(\si)}{V^2(\si)}\right),
\eeq
where a prime denotes derivation with respect to
the real canonically normalized inflaton field
$\si$ defined in Eq.~\eqref{eq:sshiftinflaton}.
Here and in the subsequent formulas in
Eqs.~\ref{eq:sshiftNQ} and \ref{eq:sshiftPert},
$V$ is the effective potential
$V^{\rm mSUGRA}_{\rm ssh}$ defined in
Eq.~\ref{eq:sshiftVtotsugra}. Inflation ends at
$\si_f=\max\{\si_\eta,\si_c\}$, where
$\si_\eta>0$ denotes the value of the inflaton
field when $\eta=-1$ and $\si_c>0$ is the
critical value of $\si$ on the semi-shifted
inflationary path corresponding to $s_c$.

The number of e-foldings from the time when the
pivot scale $k_0=0.002\units{Mpc^{-1}}$ crosses
outside the inflationary horizon until the end of
inflation is (see Eq.~\eqref{eq:efoldings})
\beq\label{eq:sshiftNQ}
N_Q\simeq\frac{1}{\mP^2}\,
\int_{\si_f}^{\si_Q}
\frac{V(\si)}{V'(\si)}\,d\si,
\eeq
where $\si_Q$ is the value of the inflaton field
at horizon crossing of the scale $k_0$. The
inflation power spectrum $\PR$ of the primordial
curvature perturbation at the pivot scale $k_0$
is given by (see Eq.~\eqref{eq:perturbations})
\beq\label{eq:sshiftPert}
\PR\simeq\frac{1}{2\pi\sqrt{3}}\,
\frac{V^{3/2}(\si_Q)}{\mP^3V'(\si_Q)}.
\eeq
The spectral index $\ns$, the tensor to scalar
ratio $r$ and the running of the spectral
index $d\ns/d\ln k$ are written as (see
Eq.~\eqref{eq:nsrdns})
\beq
\ns\simeq 1+2\eta-6\epsilon,\quad
r\simeq\,16\epsilon,\quad
\frac{d\ns}{d\ln k}\simeq16\epsilon\eta
-24\epsilon^2-2\xi^2,
\eeq
where $\epsilon$, $\eta$ and $\xi^2$ are evaluated
at $\si=\si_Q$. The number of e-foldings $N_Q$
required for solving the horizon and flatness
problems of standard HBB cosmology is
approximately given by (see
e.g.~\cite{LazaridesReview})
\beq\label{eq:sshNQvsVinf}
N_Q\simeq53.76\,+\frac{2}{3}\,\ln\left(\frac{v_0}
{10^{15}\units{GeV}}\right)+\frac{1}{3}\,\ln
\left(\frac{T_{\rm r}}{10^9\units{GeV}}\right),
\eeq
where $v_0=V_{\rm ssh}^{1/4}$ is the inflationary
scale and $T_{\rm r}$ is the reheat temperature
that is expected not to exceed about
$10^9\units{GeV}$, which is the well-known
gravitino bound \cite{gravitino}. In the
following, we take $T_{\rm r}$ to saturate the
gravitino bound, i.e. $T_{\rm r}=10^9\units{GeV}$.

\section{String power spectrum}
\label{sec:SSHIFTstring}

As mentioned before, the spontaneous breaking of
the $\rm U(1)_{B-L}$ gauge symmetry at the end
of the semi-shifted hybrid inflation leads to the
formation of local cosmic strings. These strings
can contribute a small amount to the CMBR power
spectrum. Their contribution is parameterized
\cite{bevis1} to a very good approximation by the
dimensionless string tension $G\mu_{\rm s}$,
where $G$ is the Newton's gravitational constant
and $\mu_{\rm s}$ is the string tension, i.e.
the energy per unit length of the string. In
Refs.~\cite{bevis1,bevis2} local strings were
considered within the Abelian Higgs model in the
Bogomolnyi limit, i.e. with equal scalar and
vector particle masses. If this was the case in
our model, the string tension would be given by
\beq
\label{eq:stringtension}
\mu_{\rm s}=4\pi|\vev{H^c}|^2,
\eeq
where $\vev{H^c}$ is the VEV of $H^c$ in the
relevant SUSY vacuum and is responsible for the
spontaneous breaking of the $\rm U(1)_{B-L}$
gauge symmetry. However, as it turns out, the
scalar to vector mass ratio in this model is
somewhat smaller than unity. This is, though,
not expected \cite{private} to make any
appreciable qualitative difference. Also, the
strings in our model do not coincide with the
strings in the simple Abelian Higgs model due
to the presence of the field $\phi$, which
enters the string solution. We do not anticipate,
however, that this will alter the picture in any
essential way. Moreover, as one can show by using
the results of Ref.~\cite{superconduct}, charged
fermionic transverse zero energy modes do not
exist in the presence of our strings, which,
thus, do not exhibit fermionic superconductivity.
Therefore, we will apply the results of
Refs.~\cite{bevis1,bevis2} in this model and
adopt the formula in Eq.~\eqref{eq:stringtension}
for the string tension. This is certainly an
approximation, but we believe that it is adequate
for our purposes here. In \cite{bevis1}, it was
found that the best-fit value of the string
tension required to normalize the WMAP
temperature power spectrum at multipole
$\ell=10$ is
\beq
\label{eq:Gmu}
G\mu_{\rm s}=2.04\ten{-6}.
\eeq
This corresponds to $f_{10}=1$, which is, of
course, unrealistically large. The actual value
of $f_{10}$ is proportional to the actual value
of $(G\mu_{\rm s})^2$. So, for any given value of
$f_{10}$, we can calculate $\mu_{\rm s}$ using
its normalization in Eq.~\eqref{eq:Gmu}. From
Eq.~\eqref{eq:stringtension}, we can then
determine $|\vev{H^c}|$.

\section{Numerical results}
\label{sec:SSHIFTnumeric}

\begin{figure}[tp]
\centering
\includegraphics[width=\figwidth]{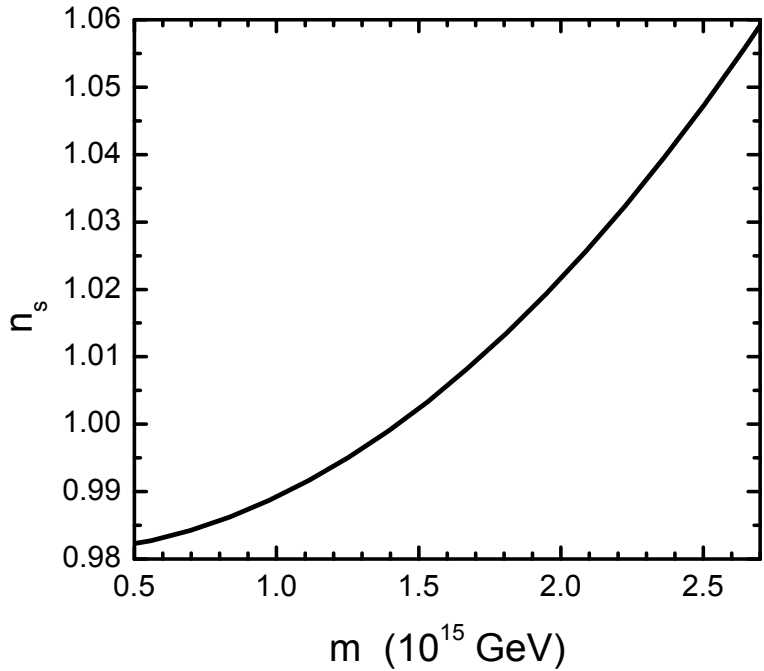}
\caption{Spectral index in semi-shifted hybrid
inflation as a function of the mass parameter $m$
in minimal SUGRA for $v=-\ga m/4\ka\la$,
$\ga/2\la=1$ and $f_{10}=0.10$.}
\label{fig:sshiftns}
\end{figure}

We choose the value $v$ of the field $\phi$ on
the semi-shifted path to lie in the first range
for $v$ in Eq.~\eqref{eq:sshiftconstr}. In
particular, we take it to be in the middle of
this range, i.e.
\beq
v=-\frac{\ga m}{4\ka\la}.
\eeq
This means, as we explained, that the universe
will end up in the vacuum of
Eq.~\eqref{eq:sshvac+}. Similar results can be
obtained if one chooses the value of $v$ to be
in the second range of
Eq.~\eqref{eq:sshiftconstr}. In order to fully
determine the five parameters of the model,
we need to make another four choices. One of
them is taken to be the ratio $\ga/2\la=1$.
Later we will comment on the dependence of the
results on variations of this ratio, which is
anyway weak. Secondly, we require the
inflationary power spectrum amplitude of the
primordial curvature perturbation at the pivot
scale $k_0$ to have its central value
in the fit of Ref.~\cite{bevis1}:
\beq
\PR\simeq 4.47\ten{-5}.
\label{eq:sshiftPRvalue}
\eeq
Further, we take, as an example, $f_{10}$ to be
equal to 0.10, its central value \cite{bevis1}.
This determines $|\vev{H^c}|$ as discussed in
Sec.~\ref{sec:SSHIFTstring}. Finally, we
calculate \cite{semishifted} numerically the
spectral index for various values of the mass
parameter $m$. The results are presented in
Fig.~\ref{fig:sshiftns} where $m$ is restricted
to be below $2.7\ten{15}\units{GeV}$, so that
the spectral index remains within its $95\%$
c.l. range.

For $m$ varying in the interval
$(0.5-2.7)\ten{15}\units{GeV}$, which is
depicted in Fig.~\ref{fig:sshiftns}, the ranges
of the various parameters of the model are
\cite{semishifted}:
$M\simeq(0.6-3.5)\ten{15}\units{GeV}$,
$\ga\simeq 0.029-0.914$,
$\la\simeq 0.0145-0.457$,
$\ka\simeq 0.73-0.67$,
$\si_Q\simeq(0.4-3.3)\ten{17}\units{GeV}$,
$\si_f\simeq(1.8-5.3)\ten{16}\units{GeV}$,
$N_Q\simeq 53.2-54.4$,
$d\ns/d\ln k\simeq-(0.1-3.1)\ten{-6}$,
$r\simeq(0.001-4.5)\ten{-5}$ and
the ratio $\si_f/\si_c\simeq 2.6-7.7$. As one
observes, we easily achieve spectral indices that
are compatible with the fit of
Ref.~\cite{bevis1}. In particular, the best-fit
value of the spectral index $\ns$ ($=1.00$)
is achieved for $m\simeq 1.40\ten{15}\units{GeV}$.
However, indices lower than about 0.98 are not
obtainable. Actually, as we lower $m$, the SUGRA
corrections become less and less important and the
spectral index decreases, tending to its value
($\approx 0.98$) in global SUSY. In all cases,
both the running of the spectral index and the
tensor-to-scalar ratio are negligibly small.

Note that our results turn out to be quite
sensitive to small changes of $\la$ (and thus
$\ga$). This is due to the fact that the
radiative correction to the inflationary
potential contains logarithms with large positive
as well as logarithms with large negative
inclination with respect to $\si$. If no
cancellation is assumed between these two
competing trends, one ends up with either a
rather fast rolling of the inflaton (dominance of
logarithms with large positive inclination) or a
negative inclination of the effective potential
for large values of $\si$ (dominance of
logarithms with large negative inclination). In
the latter case, after the inclusion of minimal
SUGRA corrections, which lift the potential for
$\si\gtrsim\mP$, a local minimum and maximum will
be generated on the inflationary path. This leads
\cite{SUSYnonminimal,smoothnonminimal} to
complications and should, therefore, be avoided.
It turns out that a cancellation to the third
significant digit between the positive and
negative contributions to the derivative of the
effective potential is needed in order to avoid
these complications and ensure that the slow-roll
conditions for the inflaton are fulfilled. This
can be achieved by a mild tuning of the parameter
$\la$ to the third significant digit. So, the
model entails a moderate tuning in one of its
parameters in order to be cosmologically viable.
Note, however, that this tuning needs only to be
performed between the various contributions to
the radiative correction and it is not spoiled
by minimal SUGRA corrections. We should also
mention that, in this model, $\si_f$ turns out
to be much larger than $\si_c$ and inflation
terminates well before the system reaches the
critical point of the semi-shifted path. This
is again due to the presence in the
inflationary potential of logarithms with large
inclination. Finally, we find \cite{semishifted}
that reducing the ratio $\ga/2\la$ generally
leads to a slight increase of the spectral
index. Though, this dependence is rather weak
and that is why we have chosen to constrain
this ratio to a constant value (instead of
setting e.g. the ratio $\ka/\la={\rm const.}$).

We observe \cite{semishifted} numerically that,
varying $f_{10}$ within its $95\%$ c.l. range
$0.02-0.18$, the value of $\ns$ changes only
in the third decimal place. So, the curve in
Fig.~\ref{fig:sshiftns}, is practically
independent of $f_{10}$. We should, however,
keep in mind that, for large values of $m$ and
low $f_{10}$'s, the constraint in
Eq.~\eqref{eq:sshiftPRvalue} cannot be
satisfied. Consequently, the curve in
Fig.~\ref{fig:sshiftns} applied to low values
of $f_{10}$ terminates on the right at a value
of $m$ which, of course, depends on $f_{10}$,
but is, in any case, higher than about
$2\ten{15}\units{GeV}$.

We have seen that, in minimal SUGRA, the model
develops a preference for values of $m$ near
$1.4\ten{15}\units{GeV}$. On the other hand,
for $f_{10}=0.10$, the prediction for the value
of $m$ which is derived from gauge coupling
constant unification is $m\simeq 2.085\ten{15}
\units{GeV}$, as the reader may find out in
Sec.~\ref{sec:SSHIFTgauge}. However, one can
see that, for this value of $m$, the predicted
spectral index is $\ns\simeq 1.0254$, which lies
inside the $1-\si$ range for $\ns$ given by the
fit in Ref.~\cite{bevis1} that we have been
using here.

\section{Gauge unification}
\label{sec:SSHIFTgauge}

We will now discuss the question of gauge
coupling constant unification in the model. As
already mentioned, the VEVs of the fields $H^c$,
$\Hb^c$ break the PS gauge group $G_{\rm PS}$
to $G_{\rm SM}$, whereas the VEV of the field
$\phi$ breaks it only to
$G_{\rm SM}\times{\rm U(1)_{B-L}}$. So, the gauge
boson $A^\perp$ corresponding to the linear
combination of $\rm U(1)_Y$ and
$\rm U(1)_{B-L}$ which is perpendicular to
$\rm U(1)_Y$, acquires its mass squared
$m^2_{A^\perp}=(5/2)g^2|\vev{H^c}|^2$ solely from
the VEVs of $H^c$, $\Hb^c$. On the other hand,
the masses squared $m_A^2$ and $m_{W_{\rm R}}^2$
of the color triplet, antitriplet ($A^\pm$) and
charged $\rm SU(2)_R$ ($W^\pm_{\rm R}$) gauge
bosons get contributions from $\vev{\phi}$ too.
Namely, $m_A^2=g^2(|\vev{H^c}|^2+(4/3)
|\vev{\phi}|^2)$ and $m_{W_{\rm R}}^2=g^2
(|\vev{H^c}|^2+2|\vev{\phi}|^2)$. Calculating
the full mass spectrum of the model in the
appropriate SUSY vacuum, one finds that there are
fields acquiring mass of order $m$ and others
that acquire mass of order $g|\vev{H^c}|$. The
presence of cosmic strings has forced the
magnitude of the VEV of the fields $H^c$,
$\Hb^c$ in the SUSY vacuum to be in the range
$(1.85-3.21)\ten{15}\units{GeV}$ (for
$f_{10}=0.02-0.18$), which is about an order of
magnitude below the SUSY GUT scale. Furthermore,
for all the values of the parameters encountered
here, the highest mass scale of the model in the
SUSY vacuum is $m_{A^\perp}=\sqrt{5/2}\,g|
\vev{H^c}|$. So, we set this scale equal to the
unification scale $M_x$. From all the above, it
is evident that the great desert hypothesis is
not satisfied in this model and the simple SUSY
unification of the gauge coupling constants is
spoiled.

\begin{figure}[tp]
\centering
\includegraphics[width=\figwidth]{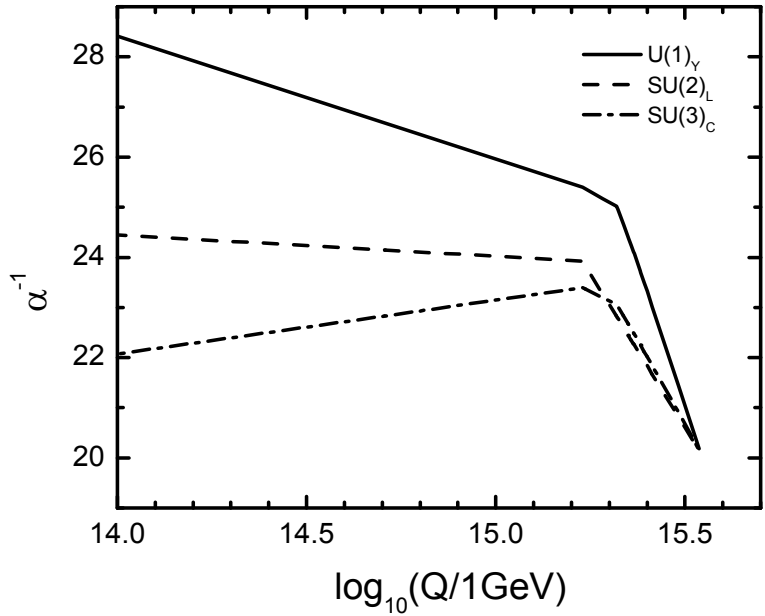}
\caption{Gauge coupling constant unification
for semi-shifted hybrid inflation in the case
of minimal SUGRA for $v=-\ga m/4\ka\la$,
$\ga/2\la=1$ and $f_{10}=0.10$. The parameter
$\alpha$ represents the three running SM fine
structure constants as indicated and $Q$ is
the running mass scale.}
\label{fig:gauge}
\end{figure}

One can easily see that, although there exist
many fields with $\rm SU(3)_c$ and $\rm U(1)_Y$
quantum numbers which can acquire heavy masses
below the unification scale and, thus, affect
the running of the corresponding gauge coupling
constants, the only heavy fields with
$\rm SU(2)_L$ quantum numbers are $h'$ and
$\bar{h}'$ belonging to the $({\bf 15,2,2})$
representation (see Chap.~\ref{sec:QUASI}).
However, these fields affect equally the running
of the $\rm U(1)_Y$ gauge coupling constant and,
consequently, cannot help us much in achieving
gauge unification. We, therefore, assume that
their masses are close to $M_x$ so that they do
not contribute to the renormalization group
running. As a consequence of these facts, the
$\rm SU(2)_L$ gauge coupling constant fails to
unify with the other gauge coupling constants.
One is, thus, forced to consider the inclusion
of some extra fields. There is a good choice
\cite{semishifted} using a single extra field,
namely a superfield $f$ belonging to the
$({\bf 15,3,1})$ representation. This field
affects mainly the running of the $\rm SU(2)_L$
gauge coupling constant. If we require that $f$
has charge $1/2$ under the global $\rm U(1)$ R
symmetry, then the only renormalizable
superpotential term in which this field is
allowed to participate is a mass term of the
form $\frac{1}{2}\,m_f f^2$. One can then tune the
new mass parameter $m_f$, along with the mass
$m$, so as to achieve unification of the gauge
coupling constants. In contrast to
Ref.~\cite{stsmooth} (see also
Chap.~\ref{sec:STSMOOTH}), we will not include
here the superpotential term $\phi^2\pb$ allowed
by the symmetries of the model since, as it turns
out, it is not so useful in the present case. So,
we assume that the corresponding coupling
constant is negligible.

\begin{figure}[tp]
\centering
\includegraphics[width=\figwidth]{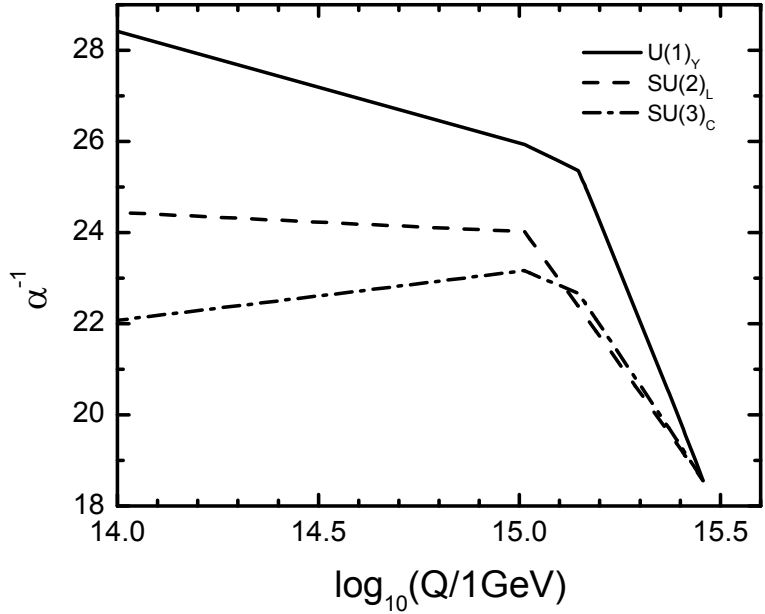}
\caption{Gauge coupling constant unification
for semi-shifted hybrid inflation in the case
of minimal SUGRA for $v=-\ga m/4\ka\la$,
$\ga/2\la=1$ and $\ns=1$. Same notation as
in Fig.~\ref{fig:gauge}.}
\label{fig:HZgauge}
\end{figure}

We have implemented \cite{semishifted} a code
that is built on top of the {\tt SOFTSUSY} code
of Ref.~\cite{allanach} and performs the running
of the gauge coupling constants at two loops. We
have incorporated six mass thresholds below the
unification scale $M_x$, namely $m_f$, $m$,
$[m^2+(4/3)\la^2|\vev{H^c}|^2]^{1/2}$,
$[m^2+2\la^2|\vev{H^c}|^2]^{1/2}$,
$g[|\vev{H^c}|^2+(4/3)|\vev{\phi}|^2]^{1/2}$
and $g[|\vev{H^c}|^2+2|\vev{\phi}|^2]^{1/2}$.
In Fig.~\ref{fig:gauge}, we present the
unification of the SM gauge coupling constants
in the $f_{10}=0.10$ case. We deduce that gauge
unification is achieved for
$m_f\simeq1.69\ten{15}\units{GeV}$ and
$m\simeq2.085\ten{15}\units{GeV}$ with
the values of the other parameters of the model
being $\ns\simeq1.0254$,
$M\simeq2.53\ten{15}\units{GeV}$,
$\ga\simeq0.515$, $\la\simeq0.2575$,
$\ka\simeq 0.713$,
$\si_Q\simeq 2.5\ten{17}\units{GeV}$,
$\si_f\simeq 4.5\ten{16}\units{GeV}$,
$N_Q\simeq54.2$,
$d\ns/d\ln k\simeq-0.8\ten{-6}$,
$r\simeq1.5\ten{-5}$ and the ratio
$\si_f/\si_c\simeq6.5$. The GUT gauge coupling
constant turns out to be $g\simeq0.789$ and the
unification scale
$M_x\simeq3.45\ten{15}\units{GeV}$.
In the HZ case (i.e. for $\ns=1$), gauge
unification is achieved for
$m_f\simeq1.025\ten{15}\units{GeV}$ and
$m\simeq1.40\ten{15}\units{GeV}$ (see
Fig.~\ref{fig:HZgauge}), which corresponds to
$f_{10}\simeq0.039$,
$M\simeq1.68\ten{15}\units{GeV}$,
$\ga\simeq0.367$, $\la\simeq0.1835$,
$\ka\simeq0.721$,
$\si_Q\simeq1.5\ten{17}\units{GeV}$,
$\si_f\simeq3.4\ten{16}\units{GeV}$,
$N_Q\simeq53.9$,
$d\ns/d\ln k\simeq-0.2\ten{-6}$,
$r\simeq0.3\ten{-5}$,
$\si_f/\si_c\simeq6.3$, $g\simeq0.823$ and
$M_x\simeq2.865\ten{15}\units{GeV}$.
Note that the unification scale $M_x$ turns out
to be somewhat small. This fact, however, does
not lead to unacceptably fast proton decay since
the relevant diagrams are suppressed by large
factors (for details, see Ref.~\cite{shifted}).

\newpage
\thispagestyle{empty}
\mbox{}

\chapter{New Smooth Hybrid Inflation}
\label{sec:NSMOOTH}

\section{Introduction}
\label{sec:NSMOOTHintro}

It has been shown in Chap.~\ref{sec:NSHIFT} that
shifted hybrid inflation can be realized within
the SUSY PS model even without invoking any
non-renormalizable superpotential terms,
provided that we supplement the model with
some extra Higgs superfields. Moreover, as we
saw in Chap.~\ref{sec:SSHIFT}, the same extended
SUSY PS model also incorporates an alternative,
``semi-shifted'' inflationary scenario, in which
the $\rm U(1)_{B-L}$ gauge symmetry remains
unbroken during inflation and it breaks
immediately after it, leading to a network of
local cosmic strings that can contribute a small
amount to the primordial curvature perturbations.
This extension of the SUSY PS model, described in
Chap.~\ref{sec:QUASI}, was actually introduced
\cite{quasi} for a very different reason. It is
well known \cite{copw} that in SUSY models with
exact Yukawa unification (or with large
$\tan\beta$ in general), such as the simplest
SUSY PS model, and universal boundary conditions,
the $b$-quark mass $m_b$ receives large SUSY
corrections, which, for $\mu>0$, lead to
unacceptably large values of $m_b$. Therefore,
Yukawa unification must be (moderately) violated
so that, for $\mu>0$, the predicted bottom quark
mass resides within the experimentally allowed
range even with universal boundary conditions.
This requirement has forced \cite{quasi} the
extension of the superfield content of this model
by including, among other superfields, an extra
pair of $\rm SU(4)_c$ non-singlet $\rm SU(2)_L$
doublets, which naturally develop \cite{wetterich}
subdominant vacuum expectation values and mix with
the main electroweak doublets of the model leading
to a moderate violation of Yukawa unification. It
is remarkable that the resulting extended SUSY PS
model automatically and naturally leads to the
aforementioned new versions of shifted hybrid
inflation based solely on renormalizable
superpotential terms.

In this chapter, we will show that the same
extension of the SUSY PS model can lead
\cite{newsmooth} to a new version of smooth
hybrid inflation based only on renormalizable
superpotential terms, provided that a particular
parameter of its superpotential is adequately
small. Indeed, the scalar potential of the model,
for a wide range of its other parameters,
possesses \cite{newsmooth} a valley of minima
which has an inclination already at the classical
level and can be used as inflationary path leading
to a novel realization of smooth hybrid inflation.
This scenario is referred to as ``new smooth''
hybrid inflation. The predictions of this
inflationary model can be easily made compatible
with CMBR measurements for natural values of the
parameters of the model. In particular, in global
SUSY, the spectral index turns out to be
adequately small so that it is consistent with
the fitting of the WMAP3 data \cite{WMAP3} by the
standard power-law cosmological model with cold
dark matter and a cosmological constant
($\Lambda$CDM). Finally, as in the
``conventional'' realization of smooth hybrid
inflation, $G_{\rm PS}$ is already broken to
$G_{\rm SM}$ during new smooth hybrid inflation
and, thus, no topological defects are formed at
the end of inflation.

The inclusion of SUGRA corrections with minimal
K\"{a}hler potential raises the spectral index
above the allowed range as in standard and
shifted hybrid inflation for relatively large
values of the relevant dimensionless coupling
constant and in smooth hybrid inflation for GUT
breaking scale close to its SUSY value (see
Ref.~\cite{SenoguzShafi}). However, the
introduction of a non-minimal term in the
K\"{a}hler potential with appropriately chosen
sign can help to reduce the spectral index so
that it becomes compatible with the data
(compare with Refs.~\cite{SUSYnonminimal,
smoothnonminimal,osamu}). This can be achieved
with the potential remaining a monotonically
increasing function of the inflaton field
everywhere on the inflationary path. So,
complications \cite{SUSYnonminimal,
smoothnonminimal} from the appearance of a
local maximum and minimum of the potential on
the inflationary path when such a non-minimal
K\"{a}hler potential is used, are avoided. One
possible complication is that the system gets
trapped near the minimum of the inflationary
potential and, consequently, no hybrid inflation
takes place. Another complication is that, even
if hybrid inflation of the so-called hilltop type
\cite{hilltop} occurs with the inflaton rolling
from the region of the maximum down to smaller
values, the spectral index can become compatible
with the data only at the cost of a mild tuning
of the initial conditions (see Ref.~\cite{gpp}).

\section{New smooth hybrid inflation in
global SUSY} \label{sec:NSMOOTHsusy}

We consider the extended SUSY PS model of
Chap.~\ref{sec:QUASI} as our starting point. As
already mentioned, this extended SUSY PS model
leads to two new versions of shifted hybrid
inflation, called ``new shifted'' and
``semi-shifted'' hybrid inflation, which are
based solely on renormalizable interactions.
The superpotential terms which are relevant for
these inflationary scenarios have been given in
Eqs.~\eqref{eq:Wnshift} and \eqref{eq:Wsshift}
respectively. They both represent the same
superpotential with different choices of the
phases of its coupling constants. Here we will
use the version of Eq.~\eqref{eq:Wsshift},
which reads
\beq\label{eq:Wnsmooth}
W=\ka S(M^2-\phi^2)-\ga S H^c\Hb^c+m\phi\pb
-\la\pb H^c\Hb^c,
\eeq
where $M$, $m>0$ are superheavy masses of the
order of $M_{\rm GUT}$ and $\ka$, $\ga$, $\la>0$
are dimensionless coupling constants. These
parameters are normalized so that they
correspond to the couplings between the SM
singlet components of the superfields. As we
have mentioned previously, in a general
superpotential of the type of
Eq.~\eqref{eq:Wnsmooth}, $M$, $m$ and any two
of the three dimensionless parameters $\ka$,
$\ga$, $\la$ can always be made real and
positive by appropriately redefining the
phases of the superfields. The third
dimensionless parameter, however, remains
generally complex. For definiteness, we have
chosen here this parameter to be real and
positive too.

In this chapter, we will show that the specific
superpotential of Eq.~\eqref{eq:Wnsmooth} leads
to a new version of smooth hybrid inflation (see
Sec.~\ref{sec:HYBRIDsmooth}) provided that the
parameter $\ga$ is taken to be adequately small.
We will first examine the case with $\ga$ set
equal to zero and then we will move on to allow
a small, but non-zero, value for this parameter.
Note that one could get rid of the $\ga$-term in
the superpotential entirely by introducing an
extra $Z_2$ symmetry under which $H^c$, $\phi$
and $\pb$ change sign. However, this would
disallow the solution of the $b$-quark mass
problem and, thus, invalidate the original
motivation for introducing this extended SUSY
PS model. This is due to the fact that the
superpotential term which generates the crucial
mixing between the $\rm SU(4)_c$ singlet and
non-singlet $\rm SU(2)_L$ doublets (see
Chap.~\ref{sec:QUASI}) is forbidden by this
discrete symmetry. Needless to say that, for
$\ga=0$, all the choices for the phases of the
parameters in Eq.~\eqref{eq:Wnsmooth} are
equivalent.

\subsection*{The $\ga=0$ case}

Setting $\ga=0$, the F-term scalar potential
obtained from $W$ is given by
\beq\label{eq:nsmVg0}
V=\ka^2|M^2-\phi^2|^2+|m\pb-2\ka S\phi|^2
+|m\phi-\la H^c\Hb^c|^2+\la^2|\pb|^2
\left(|H^c|^2+|\Hb^c|^2\right),
\eeq
where the complex scalar fields which belong to
the SM singlet components of the superfields are
denoted by the same symbol. We will ignore
throughout the soft SUSY breaking terms
\cite{sstad} in the scalar potential since their
effect on inflationary dynamics is negligible
in our case as in the case of the conventional
realization of smooth hybrid inflation
(see Ref.~\cite{smoothnonminimal}).

From the potential in Eq.~\eqref{eq:nsmVg0}, we
find that the SUSY vacua lie at
\beq
\pb=S=0,\quad \phi^2=M^2 ,\quad
H^c\Hb^c=\frac{m}{\la}\,\phi.
\eeq
The vanishing of the D-terms yields
$\Hb^{c*}=e^{i\theta}H^c$, which implies that we
have four distinct vacua:
\bea
\phi=M,\quad H^c=\Hb^c=\pm\sqrt{\frac{mM}{\la}}
\quad (\theta=0),\label{eq:nsmvac+}\\
\phi=-M,\quad H^c=-\Hb^c=\pm\sqrt{\frac{mM}{\la}}
\quad (\theta=\pi) \label{eq:nsmvac-}
\eea
with $\pb=S=0$. Here, for simplicity, $H^c$,
$\Hb^c$ have been rotated to the real axis by an
appropriate gauge transformation. However, we
should keep in mind that the fields $H^c$,
$\pm\Hb^{c*}$ (the plus or minus sign corresponds
to $\theta=0$ or $\pi$ respectively) can have an
arbitrary common phase in the vacuum and, thus,
the two distinct vacua in Eq.~\eqref{eq:nsmvac+}
or \eqref{eq:nsmvac-} are not, in reality,
discrete, but rather belong to a continuous $S^1$
vacuum submanifold. Note that the vacua in
Eq.~\eqref{eq:nsmvac+} are related to the ones
in Eq.~\eqref{eq:nsmvac-} by the $Z_2$ symmetry
mentioned above. As we will see later, the
specific point of the vacuum manifold towards
which the system is heading is already chosen
during inflation. So the model does not encounter
any topological defect problem. Actually, there
is no production of topological defects at all.

It is not very hard to show that, at any possible
minimum of the potential, $\epsilon=0$ or $\pi$
and $\epsilon=\bar{\epsilon}=-\theta$, where
$\epsilon$ and $\bar{\epsilon}$ are the phases of
$\phi$ and $\pb$ respectively ($S$ can be made
real by an appropriate global $\rm U(1)$ R
transformation). This remains true even at the
minima of $V$ with respect to $\phi$, $\pb$,
$H^c$ and $\Hb^c$ for fixed $S$. So, we will
restrict ourselves to these values of $\theta$
and phases of $\phi$ and $\pb$. The scalar
potential then takes the form
\beq\label{eq:nsmVmin}
V_{\rm min}=\ka^2\left(|\phi|^2-M^2\right)^2+
\left(2\ka|S||\phi|-m|\pb|\right)^2
+\left(m|\phi|-\la |H^c|^2\right)^2
+2\la^2|\pb|^2|H^c|^2.
\eeq
The derivatives of this potential with respect to
the norms of the fields are
\bea
\frac{\pd V_{\rm min}}{\pd |S|}
&=&4\ka\left(2\ka|S||\phi|-m|\pb|\right)|\phi|,
\label{eq:dVdSnsm}\\
\frac{\pd V_{\rm min}}{\pd |\phi|}
&=&4\ka^2\left(|\phi|^2-M^2\right)|\phi|
+4\ka\left(2\ka |S||\phi|-m|\pb|\right)|S|
+2m\left(m|\phi|-\la|H^c|^2\right),
\label{eq:dVdphinsm}\\
\frac{\pd V_{\rm min}}{\pd |\pb|}
&=&-2m\left(2\ka|S||\phi|-m|\pb|\right)
+4\la^2|\pb||H^c|^2,
\label{eq:dVdpbnsm}\\
\frac{\pd V_{\rm min}}{\pd |H^c|}
&=&-4\la\left(m|\phi|-\la|H^c|^2
-\la|\pb|^2\right)|H^c|.
\label{eq:dVdHnsm}
\eea

The potential $V_{\rm min}$ possesses two flat
directions. The first one is the trivial flat
direction at $|\phi|=|\pb|=|H^c|=0$ with
$V=V_{\rm tr}\equiv\ka^2M^4$. The second one
exists only if $\tilde{M}^2\equiv M^2-m^2/2
\ka^2>0$ and is the semi-shifted flat direction
(see Sec.~\ref{sec:SSHIFTsusy}), located at
\beq
|\phi|=\tilde{M},\quad
|\pb|=\frac{2\ka\tilde{M}}{m}\,|S|,\quad
|H^c|=0,
\eeq
where $\tilde{M}\equiv (M^2-m^2/2\ka^2)^{1/2}$,
with \mbox{$V=\ka^2(M^4-\tilde{M}^4)$}. The
mass-squared matrix of the variables $|S|$,
$|\phi|$, $|\pb|$ and $|H^c|$ on the trivial
flat direction is
\beq\label{eq:nsmmass2}
\left(\ba{cccc}
0 & 0 & 0 & 0 \\
0 & 4\ka^2(2|S|^2-\tilde{M}^2) & -4\ka m|S| & 0\\
0 & -4\ka m|S| & 2m^2 & 0 \\
0 & 0 & 0 & 0
\ea\right).
\eeq
If $M_{\phi\pb}$ denotes the $|\phi|$, $|\pb|$
sector of this matrix, then
\beq
\det\{M_{\phi\pb}\}=-8\ka^2m^2\tilde{M}^2,\quad
\tr\{M_{\phi\pb}\}=4\ka^2(2|S|^2-\tilde{M}^2)+2m^2.
\eeq
So, the matrix $M_{\phi\pb}$ has one positive and
one negative eigenvalue for $\tilde{M}^2>0$ and
two positive eigenvalues for $\tilde{M}^2<0$.
In the former case, the trivial flat direction is
a path of saddle points and the semi-shifted flat
direction is an honest candidate for the
inflationary path. However, in this chapter we
will concentrate on the latter case and set
$\tilde{\mu}^2\equiv -\tilde{M}^2>0$. Note that,
even in this case, the trivial flat direction may
not be a valley of local minima because of the
existence of the zero eigenvalue of the full
mass-squared matrix in Eq.~\eqref{eq:nsmmass2}
associated with the field $|H^c|$. It is
perfectly conceivable that, starting from any
point on the trivial flat direction, there exist
paths along which the potential decreases as we
move away from this flat direction (at least
initially). Actually, as we will show below,
this happens to be the case here.

To examine the stability of the trivial flat
direction, we consider a point on it and try to
see whether, starting from this point, one can
construct paths in the \mbox{$\left(|H^c|,|\phi|,
|\pb|\right)$} space along which the potential
in Eq.~\eqref{eq:nsmVmin} has a local maximum at
the point on the trivial flat direction. In
particular, we will try to find the path of
steepest descent. Throughout the analysis,
$|S|$ will be considered as a fixed parameter
characterizing the chosen point on the trivial
flat direction rather than as a dynamical
variable. Setting $|H^c|=\chi$, $|\phi|=\psi$
and $|\pb|=\omega$, we can parameterize any
path in the field space as \mbox{$\left(\chi,
\psi(\chi),\omega(\chi)\right)$}. We see, from
the form of the matrix in Eq.~\eqref{eq:nsmmass2},
that the required paths must be tangential to the
$|H^c|$ direction at their origin (because, for
$\tilde{\mu}^2>0$, displacement along the $|\phi|$
or $|\pb|$ direction enhances the potential
locally). Thus, the required initial conditions
for these paths are (the prime here denotes
derivation with respect to $\chi$)
\beq\label{eq:nsmincon1}
\chi=0,\quad \psi(0)=\omega(0)=0,\quad
\psi'(0)=\omega'(0)=0.
\eeq

The potential $V_{\rm min}$ on such a path can
be written as
\beq
F(\chi)=f\left(\chi,\psi(\chi),\omega(\chi)
\right),
\eeq
where $f(\chi,\psi,\omega)\equiv V_{\rm min}
(\chi,\psi,\omega)$. It is then obvious that
$F'(0)$ is zero by construction since
\beq\label{eq:nsmincon2}
\left(\bar\nabla V_{\rm min}\right)_0=0,
\eeq
where the subscript $0$ denotes the value at
$\chi=\psi=\omega=0$. Thus, the initial point
of the path is a critical point of $F(\chi)$
(as it should). Moreover, it is easily verified,
using Eqs.~\eqref{eq:nsmmass2},
\eqref{eq:nsmincon1} and \eqref{eq:nsmincon2},
that $F''(0)=0$, which means that we cannot
decide on the stability of the trivial flat
direction merely from the mass-squared matrix
in Eq.~\eqref{eq:nsmmass2}. Therefore, higher
derivatives of $F(\chi)$ must be considered. We
find that $F'''(0)=0$ and
\beq\label{eq:F4}
F''''(0)=\alpha+\zeta\psi''_0+\rho\omega''_0+
(\psi''_0,\omega''_0)
\left(\ba{cc}a & c \\c & b\ea\right)
\left(\ba{c}\psi''_0\\\omega''_0\ea\right)
\eeq
with $\psi''_0\equiv\psi''(0)$,
$\omega''_0\equiv\omega''(0)$ and
\bea
\alpha &\equiv&\left(\frac{\pd^4 f}
{\pd \chi^4}\right)_0=24\la^2,\nonumber\\
\zeta &\equiv&6\left(\frac{\pd^3 f}
{\pd\chi^2\pd\psi}\right)_0=-24\la m,\nonumber\\
\rho &\equiv& 6\left(\frac{\pd^3 f}
{\pd\chi^2\pd\omega}\right)_0=0,\nonumber\\
a &\equiv& 3\left(\frac{\pd^2 f}{\pd\psi^2}
\right)_0=12\ka^2(\tilde{\mu}^2+2|S|^2),\nonumber\\
b &\equiv&3\left(\frac{\pd^2 f}{\pd\omega^2}
\right)_0=6m^2,\nonumber\\
c &\equiv& 3\left(\frac{\pd^2 f}
{\pd\psi\pd\omega}\right)_0=-12\ka m|S|,\nonumber
\eea
where Eqs.~\eqref{eq:nsmVmin}, \eqref{eq:nsmincon1}
and \eqref{eq:nsmincon2} were used. Note that the
$2\times2$ matrix in the last term of the right
hand side of Eq.~\eqref{eq:F4} is just
$3M_{\phi\pb}$, which is positive definite for
$\tilde{\mu}^2>0$ (see the discussion following
Eq.~\eqref{eq:nsmmass2}).

By applying the transformation
\beq\label{eq:trans1}
\psi''_0=\hat{\psi}''_0+\delta\psi''_0,\quad
\omega''_0=\hat{\omega}''_0+\delta \omega''_0,
\eeq
one can show that Eq.~\eqref{eq:F4} can be
brought into the form
\beq\label{eq:F4trans}
F''''(0)=-\frac{24\la^2M^2}{\tilde{\mu}^2}+
(\delta\psi''_0,\delta \omega''_0)
\left(\ba{cc}a&c\\c&b\ea\right)
\left(\ba{c}\delta\psi''_0\\
\delta\omega''_0\ea\right),
\eeq
with
\beq
\hat{\psi}''_0=-\frac{\zeta b}{2(ab-c^2)}>0,
\quad
\hat{\omega}''_0=\frac{\zeta c}{2(ab-c^2)}
\geq 0.
\eeq
The last term in the right hand side of
Eq.~\eqref{eq:F4trans} is a positive
definite quadratic form in
\mbox{$\delta\psi''_0\geq-\hat{\psi}''_0$},
\mbox{$\delta\omega''_0\geq-\hat{\omega}''_0$}
(the non-positive lower bounds originate from the
fact that $\psi''_0$, $\omega''_0\geq 0$, which
in turn comes from Eq.~\eqref{eq:nsmincon1} and
the fact that $\psi$, $\omega\geq 0$ by their
definition). It is obvious then that there exist
choices of $\delta\psi''_0$, $\delta\omega''_0$
which render $F''''_0$ negative. Thus, on the
corresponding paths, $F(\chi)$ has a local
maximum at $\chi=0$. We conclude that the trivial
flat direction is a path of saddle points rather
than a valley of local minima. The path of
steepest decent corresponds to $\delta\psi''_0$,
$\delta\omega''_0=0$, which minimizes $F''''_0$.

We have just seen that, for any fixed value of
$|S|$, $V_{\rm min}$ has a local maximum on the
trivial flat direction at $|\phi|=|\pb|=|H^c|=0$.
Moreover, $V_{\rm min}\to\infty$ as
$|\phi|^2+|\pb|^2+|H^c|^2\to\infty$. This
means that, for each value of $|S|$,
$V_{\rm min}$ must have a non-trivial absolute
minimum (where at least one of the fields
$|\phi|$, $|\pb|$ and $|H^c|$ has a non-zero
value). These minima then form a valley, which
may be used as inflationary trajectory. Actually,
as we will show soon, this trajectory is not flat
and resembles the path described in
Sec.~\ref{sec:HYBRIDsmooth} for smooth hybrid
inflation. We can find the valley of minima of
$V_{\rm min}$ by minimizing this potential with
respect to $|\phi|$, $|\pb|$ and $|H^c|$,
regarding $|S|$ as a fixed parameter. This
amounts to solving the system of equations that
is formed by equating the partial derivatives in
Eqs.~\eqref{eq:dVdphinsm}-\eqref{eq:dVdHnsm} with
zero. We obtain three non-linear equations with
three unknowns, which cannot be solved
analytically. Though, as in the case of
conventional smooth hybrid inflation (see
Sec.~\ref{sec:HYBRIDsmooth}), we will try to find
a solution in the large $|S|$ limit. In particular,
we will try to find a power series solution with
respect to some parameter of the form
$\text{``mass''}/|S|$ which remains smaller than
unity throughout the entire range of $|S|$ which
is relevant for inflation. As it will become
clear below, a convenient quantity for the
``mass'' in the numerator is
$v_g\equiv\sqrt{mM/\la}$, which is just the
VEV $|\vev{H^c}|$ at the SUSY minima of the
potential. Re-expressing the system of equations
by using the dimensionless variables
$x\equiv|\phi|/M$, $y\equiv|\pb|/\sqrt{2}p\,v_g$,
$z\equiv|H^c|/v_g$ and $w\equiv v_g/|S|$, where
$p\equiv\sqrt{2}\ka M/m$ is a dimensionless
parameter, smaller than unity for
$\tilde{\mu}^2>0$, we obtain
\beq\label{eq:nsmsystem}
wx(x^2-1)+4yz^2+2wy^2=0, \quad
x-wy=\sqrt{2}\,\frac{\la}{\ka}\,pwyz^2,
\quad x=z^2+2p^2y^2.
\eeq
Writing the variables $x$, $y$ and $z^2$ as power
series in $w$ and equating the coefficients of
the corresponding powers of $w$ in the two sides
of Eqs.~\eqref{eq:nsmsystem}, we get
\beq\label{eq:VarExpans}
x=x_2w^2+x_4w^4+\dots,\quad
y=y_1w+y_3w^3+\dots,\quad
z^2=z_2w^2+z_4w^4+\dots,
\eeq
where the coefficients $x_i$, $y_i$ and $z_i$
depend only on the parameter $p$ and the ratio
$\la/\ka$ and are given by
\begin{gather}
x_2=y_1=\frac{3}{8p^2}\left(1-\sqrt{1-8p^2/9}
\right),\quad z_2=\frac{1}{4}\,(1-2x_2),
\label{eq:xyz1}\\[5pt]
x_4=\frac{\sqrt{2}}{8}\,\frac{\la}{\ka}\,p\;
\frac{x_2(1-2x_2)(3-10x_2)}{1-3x_2},\quad
y_3=\frac{1-4x_2}{3-10x_2}\;x_4,\quad
z_4=\frac{1+2(1-2p^2)x_2}{3-10x_2}\;x_4.
\label{eq:xyz2}
\end{gather}

A useful approximation to these coefficients can
be found by expanding them with respect to the
small parameter $p$ (see below). Thus, to first
non-trivial order in $p$, we find the following
simple expressions:
\beq
x_2=y_1=z_2=\frac{1}{6}, \quad
x_4=z_4=\frac{\sqrt{2}}{27}\,\frac{\la}{\ka}\,p\,,
\quad y_3=\frac{\sqrt{2}}{108}\,\frac{\la}{\ka}\,p
\eeq
and Eqs.~\eqref{eq:VarExpans} take the form
\bea
|\phi| &\simeq& \frac{Mv_g^2}{6|S|^2}
\left(1+\frac{2\sqrt{2}}{9}\,\frac{\la}{\ka}\,p\,
w^2+\dots\right),\nonumber\\
|\pb| &\simeq& \sqrt{2}p\,\frac{v_g^2}{6|S|}
\left(1+\frac{\sqrt{2}}{18}\,\frac{\la}{\ka}\,p\,
w^2+\dots\right), \label{eq:FieldExpans}\\
|H^c| &\simeq& \frac{v_g^2}{\sqrt{6}\,|S|}
\left(1+\frac{\sqrt{2}}{9}\,\frac{\la}{\ka}\,p\,
w^2+\dots\right). \nonumber
\eea
Taking into account the possible values of the
phases $\epsilon$, $\bar{\epsilon}$ and $\theta$
(and with $H^c$, $\Hb^c$ rotated to the real
axis), we see that the potential in
Eq.~\eqref{eq:nsmVg0} possesses four valleys
of absolute minima (for fixed $|S|$) which
presumably lead (for $|S|\to 0$) to the four
SUSY vacua in Eqs.~\eqref{eq:nsmvac+} and
\eqref{eq:nsmvac-}. We should keep in mind, though,
that the two valleys corresponding to the same
value of $\theta$ are not discrete, but
continuously connected since $H^c$, $\pm\Hb^{c*}$
can have an arbitrary common phase. The expansions
in Eq.~\eqref{eq:FieldExpans} hold as long as
$w<1$, that is $|S|>v_g$. In the following, we
will keep only the terms of leading order in
$w$ in the above equations. Although this might
seem somewhat arbitrary, we will justify it
later. Substituting the expansions in
Eq.~\eqref{eq:FieldExpans} into the potential
of Eq.~\eqref{eq:nsmVmin} and keeping only terms of
leading order in $w$, we get
\beq\label{eq:nsmVeff}
V_{\rm min}\simeq
\ka^2M^4\left(1-\frac{v_g^4}{54|S|^4}\right).
\eeq
This is exactly the form of the potential for
smooth hybrid inflation considered in
Sec.~\ref{sec:HYBRIDsmooth}. Thus, we have shown
that the present model possesses inflationary
paths leading to smooth hybrid inflation. We
call \cite{newsmooth} the resulting scenario
``new smooth'' hybrid inflation since, in
contrast to the conventional realization of
smooth hybrid inflation, it is achieved by
using only renormalizable interactions. It is
evident that, as the system follows the new
smooth inflationary path, the phases of the
various fields remain fixed. Moreover, the
particular point of the vacuum manifold
towards which the system is heading is
already chosen during inflation and we
encounter no cosmological defect problems.

Setting $S=\si/\sqrt{2}$, where $\si$ is the
canonically normalized real inflaton field
(recall that $S$ was made real by an R
transformation), we obtain the potential
along the new smooth path
\beq\label{eq:Vg0}
V\simeq v_0^4\left(1-\frac{2v_g^4}{27\si^4}
\right),
\eeq
where $v_0\equiv\sqrt{\ka}M$ is the inflationary
scale. The slow-roll parameters $\epsilon$, $\eta$
and the parameter $\xi^2$, which enters the
running of the spectral index, are (see
Sec.~\ref{sec:HYBRIDintro})
\bea
\epsilon &\equiv& \frac{\mP^2}{2}\,
\left(\frac{V'(\si)}{V(\si)}\right)^2
\simeq\frac{32\mP^2v_g^8}{729\si^{10}},\\
\eta &\equiv& \mP^2\,\left(
\frac{V''(\si)}{V(\si)}\right)
\simeq-\frac{40\mP^2v_g^4}{27\si^6},\\
\xi^2 &\equiv& \mP^4\,\left(
\frac{V'(\si)V'''(\si)}{V^2(\si)}\right)
\simeq\frac{640\mP^4v_g^8}{243\si^{12}},
\eea
where the prime here denotes (as is obvious)
derivation with respect to $\si$. Inflation ends
at $\si=\si_f$ (taken positive by an R
transformation) where $\eta=-1$, which gives
\beq\label{eq:nsmsigmaf}
\si_f^6\simeq\frac{40\mP^2v_g^4}{27}.
\eeq

The number of e-foldings from the time when the
pivot scale $k_0=0.002\units{Mpc^{-1}}$ crosses
outside the inflationary horizon until the end of
inflation is given by (see
Sec.~\ref{sec:HYBRIDintro})
\beq
N_Q\simeq\frac{1}{\mP^2}\,
\int_{\si_f}^{\si_Q}
\frac{V(\si)}{V'(\si)}\,d\si
\simeq\frac{9}{16\mP^2v_g^4}\,
\left(\si_Q^6-\si_f^6\right),
\eeq
where $\si_Q\equiv\sqrt{2}S_Q>0$ is the value
of the inflaton field at horizon crossing of the
pivot scale. Taking into account the fact that
$\si_f\ll\si_Q$, we can write
\beq\label{eq:nsmsigmaQ}
\si_Q^6\simeq\frac{16\mP^2v_g^4}{9}\,N_Q.
\eeq
The power spectrum $\PR$ of the primordial
curvature perturbation at the scale $k_0$ is
given by
\beq\label{eq:nsmPert}
\PR\simeq\frac{1}{2\pi\sqrt{3}}\,
\frac{V^{3/2}(\si_Q)}{\mP^3V'(\si_Q)}\simeq
\frac{3^{5/6}N_Q^{5/6}}{2^{2/3}\pi}\,
\left(\frac{v_0^3}{\mP^2v_g}\right)^{2/3}.
\eeq
The spectral index $\ns$, the tensor-to-scalar
ratio $r$ and the running of the spectral index
$d\ns/d\ln k$ are given by (see
Sec.~\ref{sec:HYBRIDintro})
\begin{gather}
\ns\simeq 1+2\eta-6\epsilon\simeq1
-\frac{5}{3N_Q},\nonumber\\[5pt]
r\simeq\,16\epsilon\simeq
\frac{2^{7/3}}{3^{8/3}N_Q^{5/3}}\,
\left(\frac{v_g}{\mP}\right)^{4/3},\\[5pt]
\frac{d\ns}{d\ln k}\simeq16\epsilon\eta
-24\epsilon^2-2\xi^2\simeq -\frac{5}{3N_Q^2},
\nonumber
\end{gather}
where $\epsilon$, $\eta$ and $\xi^2$ are evaluated
at $\si=\si_Q$. The number of e-foldings $N_Q$
required for solving the horizon and flatness
problems of standard HBB cosmology is given
approximately by (see e.g.~\cite{LazaridesReview})
\beq\label{eq:nsmNQvsVinf}
N_Q\simeq53.76\,+\frac{2}{3}\,\ln\left(\frac{v_0}
{10^{15}\units{GeV}}\right)+\frac{1}{3}\,\ln
\left(\frac{T_{\rm r}}{10^9\units{GeV}}\right),
\eeq
where $T_{\rm r}$ is the reheat temperature which
is expected not to exceed about $10^9\units{GeV}$,
which is the well-known gravitino bound
\cite{gravitino}.

Taking $v_g$ to have the SUSY GUT value, i.e.
$v_g\simeq 2.86\ten{16}\units{GeV}$ (see below),
$T_{\rm r}$ to saturate the gravitino bound, i.e.
$T_{\rm r}\simeq 10^9\units{GeV}$, and the WMAP3
\cite{WMAP3} normalization $\PR\simeq
4.85\ten{-5}$ at the comoving scale $k_0$, we
can solve Eqs.~\eqref{eq:nsmPert} and
\eqref{eq:nsmNQvsVinf} numerically. We obtain
\beq
N_Q\simeq53.78,\quad v_0
\simeq1.036\ten{15}\units{GeV}.
\eeq
The spectral index, the tensor-to-scalar ratio
and the running of the spectral index are then
\beq\label{eq:nsralpha}
\ns\simeq 0.969,\quad r\simeq 9.4\ten{-7},\quad
\frac{d\ns}{d\ln k}\simeq -5.8\ten{-4}.
\eeq
We see that the running of the spectral index
and the tensor-to-scalar ratio are negligible
and, thus, the standard power-law $\Lambda$CDM
cosmological model should hold to a very
good accuracy. Fitting the three-year results
from WMAP \cite{WMAP3} with this cosmological
model, one obtains that, at the pivot scale $k_0$,
\beq\label{eq:nsm_nswmap}
\ns=0.958\pm 0.016~\Rightarrow~0.926
\lesssim\ns\lesssim 0.99
\eeq
at 95$\%$ confidence level. So, the value of the
spectral index in Eq.~\eqref{eq:nsralpha} is
perfectly acceptable. It is, actually, the same
as in conventional smooth hybrid inflation
(see Sec.~\ref{sec:HYBRIDsmooth}) since the
inflationary potential for large $|S|$ is exactly
the same, as we already pointed out.

We have already fixed the values of the
parameters $v_0=\sqrt{\ka}M$ and
$v_g=\sqrt{mM/\la}$. So, we are free to
make two more choices in order to determine the
four parameters of the model $m$, $M$, $\ka$
and $\la$. A legitimate choice is to set
$\ka=\la$ and $m=M$ which leads to quite
natural values for the parameters, namely
\beq\label{eq:nsmpars}
m=M=\sqrt{v_0v_g}\simeq5.44\ten{15}\units{GeV},
\quad \ka=\la=\frac{v_0}{v_g}\simeq0.0362.
\eeq
For these values we find from
Eqs.~\eqref{eq:nsmsigmaf},
\eqref{eq:nsmsigmaQ}
that $\si_f\simeq 1.34\ten{17}\units{GeV}$
and $\si_Q\simeq 2.69\ten{17}\units{GeV}$.

\begin{figure}[tp]
\centering
\includegraphics[width=\figwidth]{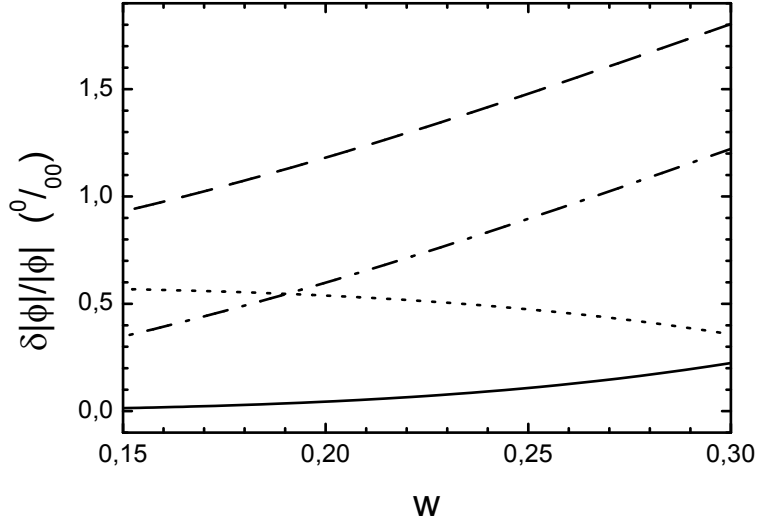}
\caption{Relative error in $|\phi|$ on the new
smooth inflationary path in global SUSY for the
values of the parameters in
Eq.~\eqref{eq:nsmpars} and $\ga=0$ when we
use the expansion of Eq.~\eqref{eq:VarExpans}
up to second order in $w$ with coefficient
evaluated to leading order in $p$ (dashed line) or
accurately (dot-dashed line) and up to fourth
order in $w$ with coefficients evaluated to
leading order in $p$ (dotted line) or accurately
(solid line).}
\label{fig:RelativeError}
\end{figure}

Let us now turn to the justification of the
expansions in Eqs.~\eqref{eq:VarExpans} and
\eqref{eq:FieldExpans}. The value of $|S|$
at the termination of inflation is approximately
\beq
S_f^6=\frac{\si_f^6}{2^3}\simeq
\frac{5\mP^2v_g^4}{27}.
\eeq
Therefore, the maximum value of $w$ during
inflation is
\beq
w_{\max}=\frac{v_g}{S_f}\simeq\frac{3^{1/2}}
{5^{1/6}}\left(\frac{v_g}{\mP}\right)^{1/3}
\simeq 0.3.
\eeq
Consequently, the condition $w<1$ is satisfied
during inflation and the expansions in
Eq.~\eqref{eq:VarExpans} are valid.
Moreover, $p\simeq 0.0512\ll 1$ for the values
in Eq.~\eqref{eq:nsmpars} and, thus, for
$\la\sim\ka$, the expansions in
Eq.~\eqref{eq:FieldExpans} are also
justified. We find numerically that these
expansions are actually justified in the entire
range $w\leq w_{\max}$ even for values of $p$
close to unity and $\la>\ka$. Rough estimates of
the maximum relative errors when only the leading
order term is kept in the expansions of
Eq.~\eqref{eq:FieldExpans} are given by the
second term in the parentheses in these equations
for $w=w_{\max}$. For the values in
Eq.~\eqref{eq:nsmpars} we get that the
maximum relative error in $|\phi|$, which seems
to be the largest of the errors in $|\phi|$,
$|\pb|$ and $|H^c|$, is given by the estimate
\beq\label{eq:ErrorInPhi}
\frac{\delta|\phi|}{|\phi|}
\simeq \frac{2\sqrt{2}}{9}\,\frac{\la}{\ka}\,p\;
w_{\max}^2\simeq 1.45\ten{-3}\sim
1\text{\textperthousand}.
\eeq
This is verified numerically as shown in
Fig.~\ref{fig:RelativeError}, where we plot the
relative error in $|\phi|$ during inflation when
we approximate the new smooth inflationary path
by the expansions in Eq.~\eqref{eq:VarExpans}.
Note that, in order to retain a precision better
than $1\%$ in $|\phi|$ keeping only the leading
order term in its expansion in
Eq.~\eqref{eq:FieldExpans}, the relation
$M\lesssim v_g/2$ has to hold, as can be seen from
Eq.~\eqref{eq:ErrorInPhi} for
$w_{\max}\simeq 0.3$.

The identification of $v_g$, which is the VEV
$|\vev{H^c}|$ or $|\vev{\Hb^c}|$, with the
SUSY GUT scale $M_{\rm GUT}$ can be easily
justified. As already mentioned, the VEVs of
$H^c$, $\Hb^c$ break the PS gauge group to
$G_{\rm SM}$, whereas the VEV of the field $\phi$
breaks it only to $G_{\rm SM}\times
{\rm U(1)_{B-L}}$. So, the gauge boson $A^\perp$
corresponding to the linear combination of
${\rm U(1)_{Y}}$ and ${\rm U(1)_{B-L}}$ which is
perpendicular to ${\rm U(1)_{Y}}$, acquires its
mass squared $m^2_{A^\perp}=(5/2)g^2|\vev{H^c}|^2$
solely from the VEVs of $H^c$, $\Hb^c$ ($g$ is
the SUSY GUT gauge coupling constant). On the
other hand, the masses squared $m_A^2$ and
$m_{W_{\rm R}}^2$ of the color triplet,
anti-triplet ($A^\pm$) and charged
${\rm SU(2)_R}$ ($W^\pm_{\rm R}$) gauge bosons
get contributions from $\vev{\phi}$ too. Namely,
$m_A^2=g^2(|\vev{H^c}|^2+(4/3)|\vev{\phi}|^2)$
and $m_{W_{\rm R}}^2=g^2(|\vev{H^c}|^2+
2|\vev{\phi}|^2)$. For the values in
Eq.~\eqref{eq:nsmpars}, however,
\beq
\frac{|\vev{\phi}|^2}{|\vev{H^c}|^2}=
\frac{\la M}{m}\simeq 0.0362\ll 1,
\eeq
which implies that $m_A\approx m_{W_{\rm R}}
\approx gv_g$ within a few per cent. So, $v_g$ is
approximately equal to the practically common
mass of the SM non-singlet superheavy gauge
bosons divided by $g\approx 0.7$, which is, in
turn, equal to $M_{\rm GUT}\simeq 2.86\ten{16}
\units{GeV}$ (the SM singlet gauge boson
$A^\perp$ does not affect the renormalization
group equations).

\subsection*{The $\ga\neq 0$ case}

We will now turn to the case of a non-vanishing,
but small value of the parameter $\ga$. The
scalar potential in this case takes the form
\beq\label{eq:Fpotg}
V=|\ka\,(M^2-\phi^2)-\ga H^c\Hb^c|^2
+|m\pb-2\ka S\phi|^2+|m\phi-\la H^c\Hb^c|^2
+|\ga S+\la\pb\,|^2\left(|H^c|^2+|\Hb^c|^2\right)
\eeq
and the SUSY vacua lie at
\beq
\phi=\frac{\ga m}{2\ka\la}\left(-1\pm\sqrt{1+
\frac{4\ka^2\la^2M^2}{\ga^2m^2}}\,\right)
\equiv\phi_{\pm},\quad \pb=S=0,\quad
H^c\Hb^c=\frac{m}{\la}\,\phi.
\eeq
Again, the vanishing of the D-terms yields
$\Hb^{c*}=e^{i\theta}H^c$, which implies that
we have four distinct SUSY vacua
(cf.~Eqs.~\eqref{eq:sshvac+},
\eqref{eq:sshvac-}):
\bea
\phi=\phi_{+},\quad H^c=\Hb^c=\pm
\sqrt{\frac{m\phi_{+}}{\la}}\quad(\theta=0),
\label{eq:nsmvacu1}\\
\phi=\phi_{-},\quad H^c=-\Hb^c=\pm
\sqrt{\frac{-m\phi_{-}}{\la}}\quad(\theta=\pi)
\label{eq:nsmvacu2}
\eea
with $\pb=S=0$. Here $H^c$, $\Hb^{c}$ are rotated
to the real axis, but we should again keep in
mind that the two vacua in Eq.~\eqref{eq:nsmvacu1}
or \eqref{eq:nsmvacu2} belong, in reality, to a
continuum of vacua. One can show that the
potential now generally possesses three flat
directions. The first one is the usual trivial
flat direction at $\phi=\pb=H^c=\Hb^c=0$ with
$V=V_{\rm tr}=\ka^2M^4$. The second one
exists only if $\tilde{M}^2>0$ and lies at
\beq\label{eq:sshpath}
\phi=\pm\,\tilde{M},\quad
\pb=\frac{2\ka\phi}{m}\,S,\quad
H^c=\Hb^c=0.
\eeq
It is the semi-shifted flat direction (see
Chap.~\ref{sec:SSHIFT}) with \mbox{$V_{\rm ssh}=
\ka^2(M^4-\tilde{M}^4)$} along which $G_{\rm PS}$
is broken to $G_{\rm SM}\times{\rm U(1)_{B-L}}$.
Note that the positions of the trivial and
semi-shifted flat directions remain the same as
in the $\ga=0$ case. The third flat direction,
which appears at
\begin{gather}
\phi=-\frac{\ga m}{2\ka\la},\quad
\pb=-\frac{\ga}{\la}\,S,\quad H^c\Hb^c=
\frac{\ka\ga(M^2-\phi^2)+\la m\phi}
{\ga^2+\la^2},\\[5pt]
V=V_{\rm nsh}\equiv\frac{\ka^2\la^2}{\ga^2+\la^2}
\left(M^2+\frac{\ga^2m^2}{4\ka^2\la^2}\right)^2,
\end{gather}
exists only for $\ga\neq 0$ and is analogous
to the trajectory for the new shifted hybrid
inflation of Chap.~\ref{sec:NSHIFT}. Along this
direction, $G_{\rm PS}$ is broken to $G_{\rm SM}$.
In our subsequent discussion, we will again
concentrate on the case where $\tilde{\mu}^2=
-\tilde{M}^2>0$. It is interesting to note that,
in this case, we always have $V_{\rm nsh}>
V_{\rm tr}$ and it is, thus, more likely that the
system will eventually settle down on the trivial
rather than the new shifted flat direction (the
semi-shifted flat direction in
Eq.~\eqref{eq:sshpath} does not
exist in this case).

If we expand the complex scalar fields $\phi$,
$\pb$, $H^c$, $\Hb^c$ in real and imaginary parts
according to the prescription $s=(s_1+is_2)/
\sqrt{2}$, we find that, on the trivial flat
direction, the mass-squared matrices
$M_{\phi1}^2$ of $\phi_1$, $\pb_1$ and
$M_{\phi2}^2$ of $\phi_2$, $\pb_2$ are
\beq
M_{\phi1(\phi2)}^2=\left(\ba{cc}
m^2+4\ka^2|S|^2\mp2\ka^2M^2 & -2\ka m S \\
-2\ka m S  & m^2 \ea\right)
\eeq
and the mass-squared matrices $M_{H1}^2$ of
$H^c_1$, $\Hb^c_1$ and $M_{H2}^2$ of $H^c_2$,
$\Hb^c_2$
are
\beq
M_{H1(H2)}^{2}=\left(\ba{cc}
\ga^2|S|^2  & \mp\ga\ka M^2  \\
\mp\ga\ka M^2 & \ga^2|S|^2 \ea\right).
\eeq
The matrices $M_{\phi1(\phi2)}^2$ are always
positive definite, while the matrices
$M_{H1(H2)}^2$ acquire one negative eigenvalue
for
\beq
|S|<S_c\equiv\sqrt{\frac{\ka}{\ga}}\;M.
\eeq
Thus, the trivial flat direction is now stable
for $|S|>S_c$ and unstable for $|S|<S_c$.
Yet, one can easily see that, for $\ga\to 0$,
$S_c\to\infty$ and we are led to the previous
($\ga=0$) case where the entire trivial flat
direction was a path of saddle points. So, one
can imagine that, for small enough values of the
parameter $\ga$, the trivial flat direction,
after its destabilization at the critical point,
forks into four valleys of local or global minima
(for fixed $|S|$) of the potential in
Eq.~\eqref{eq:Fpotg}, which resemble the valleys
for new smooth hybrid inflation described above
in the $\ga=0$ case.

Actually, the valleys for a small non-zero
$\ga$ are expected to differ from the ones for
$\ga=0$ by corrections involving the small
parameter $\ga$. The terms in the potential
of Eq.~\eqref{eq:Fpotg} which depend on $\ga$
and the phases $\epsilon$, $\bar{\epsilon}$
and $\theta$ are
\beq\label{eq:dVg}
\delta V=-2\ga|H^c|^2\left[\ka M^2\cos\theta
-2\la|S||\pb|\cos\bar{\epsilon}
-\ka|\phi|^2\cos(2\epsilon+\theta)\right].
\eeq
Estimating this expression on the valleys for
$\ga=0$ by using the leading terms in the
expansions of $|\phi|$ and $|\pb|$ in
Eq.~\eqref{eq:FieldExpans}, we
find that, for $v_g/|S|<1$,
\beq\label{eq:dVvalley}
\delta V\approx -2\ka\ga M^2|H^c|^2
\left[\cos\theta-\frac{2}{3}\cos\bar{\epsilon}
-\frac{1}{36}\left(\frac{v_g}{|S|}\right)^4
\cos(2\epsilon+\theta)\right].
\eeq
From this, we see that the $\ga$ dependent
corrections enhance the potential in the valleys
with $\epsilon=\bar{\epsilon}=\theta=\pi$ and reduce
it in the valleys with $\epsilon=\bar{\epsilon}=
\theta=0$. This fact can also be confirmed
numerically. So, as it turns out, the trivial
flat direction bifurcates at $|S|=S_c$ into two
valleys of {\em absolute} minima for fixed $|S|$
which correspond to $\theta\simeq 0$ and lead to
the two SUSY vacua in Eq.~\eqref{eq:nsmvacu1}.
They are the valleys for new smooth hybrid
inflation in the case with $\ga\neq 0$, but small.
Recall, however, that these two valleys are not
discrete, but belong to a continuum of valleys.

\begin{figure}[tp]
\centering
\includegraphics[width=\figwidth]{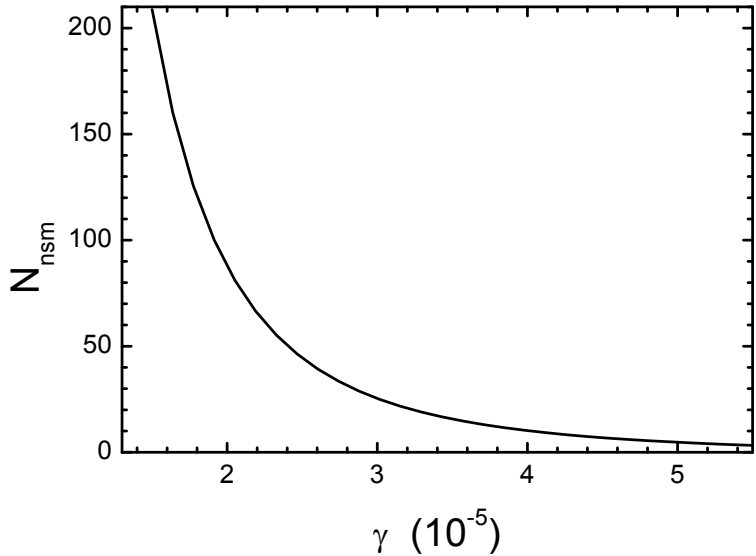}
\caption{Number of e-foldings $N_{\rm nsm}$ along
the new smooth inflationary path versus $\ga$ in
global SUSY when the system slowly rolls from
$\si=0.95\,\si_c$ down to $\si=\si_f$. The other
parameters of the model (except $\ga$) take on
the values in Eq.~\eqref{eq:nsmpars}.}
\label{fig:Nsmooth}
\end{figure}

Unfortunately, it is quite difficult to find a
reliable expansion for the fields on these
valleys, mainly because of the obstacle at
$|S|=S_c$, which prevents us from taking the
limit $v_g/|S|\to 0$. So, numerical computation
is our last resort. We have found numerically
that, when the system crosses the critical point
at $\si=\si_c$ ($\si_c\equiv\sqrt{2}S_c$)
after it has rolled down the trivial flat
direction, it does not immediately settle down
on the new smooth path. This takes place after
a while and at a value of $\si$ which is well
above $0.95\,\si_c$. Furthermore, quantum
fluctuations which could kick the system out
of the new smooth path are utterly suppressed
well before the system reaches this value of
$\si$. However, just to be on the safe side,
we will consider here the slow rolling of the
system along the new smooth path starting from
$\si=0.95\,\si_c$. In Fig.~\ref{fig:Nsmooth},
we plot the number of e-foldings $N_{\rm nsm}$
along the new smooth path as a function of the
parameter $\ga$ in global SUSY and with the
parameter values in Eq.~\eqref{eq:nsmpars},
when the system slowly rolls from $\si=0.95\,
\si_c$ down to $\si=\si_f$ where $\eta=-1$ and
the slow roll ends. We see that, for small
enough $\ga$, we can have an adequate number
of e-foldings for solving the horizon and
flatness problems of standard HBB cosmology.

\begin{figure}[tp]
\centering
\includegraphics[width=\figwidth]{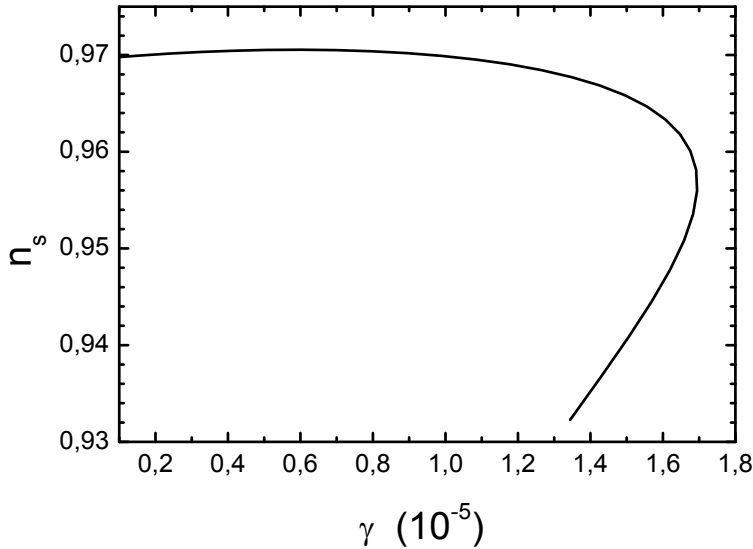}
\caption{Spectral index in new smooth hybrid
inflation as a function of $\ga$ in global
SUSY for $p\equiv\sqrt{2}\ka M/m=1/\sqrt{2}$
and $\ka=0.1$. The endpoint of the curve at
$\ns\simeq 0.932$ corresponds to the case
where our present horizon scale crosses
outside the inflationary horizon when
$\si=0.95\,\si_c$.}
\label{fig:nsNewSmooth}
\end{figure}

To pursue the investigation of the model further,
we set $p\equiv\sqrt{2}\ka M/m=1/\sqrt{2}$,
$\ka=0.1$ and fix the value of the power spectrum
$\PR$ of the primordial curvature perturbation to
the three-year WMAP \cite{WMAP3} result $\PR
\simeq 4.85\ten{-5}$. As already mentioned, on
the new smooth path the fields $H^c$ and
$\Hb^{c*}$ have practically the same phase
$(\theta\simeq 0)$. So, one of the vacua in
Eq.~\eqref{eq:nsmvacu1} is already selected
during inflation (i.e. the common phase of $H^c$
and $\Hb^{c*}$ is fixed during inflation). We set
the VEV $|\vev{H^c}|=\sqrt{m\phi_+/\la}$ equal to
the SUSY GUT scale (in practice, we just set
$v_g\equiv\sqrt{mM/\la}=2.86\ten{16}\units{GeV}$,
since the resulting error is very small). After
these choices, the only freedom left is the value
of $\ga$. In Fig.~\ref{fig:nsNewSmooth}, we plot
the predicted spectral index of the model as a
function of $\ga$. We terminate the curve when the
value of $\si$ at which our present horizon crosses
outside the inflationary horizon becomes as large
as $0.95\,\si_c$. We observe that there exists a
range of values for $\ga$ within which the system
admits two separate solutions, each corresponding
to a different value of $\la$. This new feature
of the model, which is not shared by conventional
smooth hybrid inflation, originates from the
presence of the critical point at $\si=\si_c$
blocking the extension of the new smooth path to
larger values of $\si$. The part of the curve
with $\ns<0.96$ corresponds to values of $\si_Q$
in the range $0.85<\si_Q/\si_c<0.946$, while its
branch with $\ns>0.96$ corresponds to $\si_Q/
\si_c<0.85$. We see that spectral indices
compatible with Eq.~\eqref{eq:nsm_nswmap} can
easily be obtained for $\ga$'s which are small
enough so that the number of e-foldings generated
is adequately large for solving the horizon and
flatness problems. It is important to point out
that, in global SUSY, the new smooth hybrid
inflation model is far ``superior'' to
conventional smooth hybrid inflation, which
predicts $\ns\simeq 0.97$, in that it can
easily accommodate much smaller values of $\ns$
and, thus, be more comfortably compatible with
the data. However, we should note that obtaining
values of $\ns$ which are very close to its lower
bound in Eq.~\eqref{eq:nsm_nswmap} would require
getting slightly above $\si=0.95\,\si_c$.
This is, though, not at all impossible, as we
already mentioned, since in many cases new
smooth hybrid inflation in global SUSY starts
well above that point.

For the values of $\ga$ which correspond to
the curve depicted in Fig.~\ref{fig:nsNewSmooth},
i.e. $\ga\simeq (0.3-1.7)\ten{-5}$, we find
that $\la\simeq (1.4-3.1)\ten{-3}$,
$M\simeq (2.4-3.6)\ten{16}\units{GeV}$,
$m\simeq (4.8-9.2)\ten{15}\units{GeV}$ and
$\si_c\simeq (3-9)\ten{17}\units{GeV}$. The number
of e-foldings from the time when the pivot scale
$k_0$ crosses outside the inflationary horizon
until the end of inflation is $N_Q\simeq
53.6-53.85$. The value $\si_f$ of $\si$
when inflation ends is about
$1.4\ten{17}\units{GeV}$ and $\si_Q$ lies in the
range $(2.85-3.025)\ten{17}\units{GeV}$. Finally,
$d\ns/d\ln k\simeq -(4.1-5.5)\ten{-4}$ and
$r\simeq (3-13)\ten{-7}$. Variations in the
values of $p$ and $\ka$ (which are the only
arbitrarily chosen parameters) have shown not to
have any significant effect on the results.
Contrary to the $\ga=0$ case, the numerical
results for $\ga\neq 0$ certainly depend on the
choice of the phases of the parameters in the
superpotential of Eq.~\eqref{eq:Wnsmooth}. As
already explained, only one of the dimensionless
parameters of this superpotential, say the
parameter $\ga$, is genuinely complex. Its phase
affects the position of the SUSY vacua in
Eq.~\eqref{eq:nsmvacu1} and presumably the
position of the new smooth paths which lead to
these vacua. However, the general qualitative
structure of the theory is not expected to be
affected.

\section{Supergravity corrections}
\label{sec:NSMOOTHsugra}

It has been shown in Ref.~\cite{newsmooth},
that when global SUSY is promoted to local,
some features of the model are sensitive to
non-minimal terms in the K\"{a}hler potential.
In particular, although SUGRA corrections with
a minimal K\"{a}hler potential raise the spectral
index above the allowed range, non-minimal terms
can help us reduce the spectral index so as to
become comfortably compatible with the data.
We will work again by first concentrating on
the $\ga=0$ case.

The F-term scalar potential in SUGRA is given by
(see Sec.~\ref{sec:MSSMsugra})
\beq\label{eq:nsmVSUGRA}
V=e^{K/\mP^2}\left[(K^{-1})_i^j\;
F^{i*}F_j-3|W|^2/\mP^2\right],
\eeq
where $K$ is the K\"{a}hler potential and
$F^{i*}=W^i+K^iW/\mP^2$. As usual, a superscript
(subscript) $i$ denotes derivation with respect
to the complex scalar field $s_i$ ($s^{i*}$) and
$(K^{-1})_i^j$ is the inverse of the K\"{a}hler
metric $K_i^j$. We will consider, at first, a
minimal K\"{a}hler potential and leave the
inclusion of non-minimal terms for later. The
minimal K\"{a}hler potential, in our case, has
the form
\beq\label{eq:nsmMinK}
K_0=|S|^2+|\phi|^2+|\pb|^2+|H^c|^2+|\Hb^c|^2
\eeq
and the scalar potential is given by
\beq\label{eq:nsmVSUGRAmin}
\tilde{V}_0\equiv\frac{V_0}{\ka^2M^4}=
e^{K_0/\mP^2}\;\left[\sum_{s}
\left|\tilde{W}_s+\frac{\tilde{W}s^*}
{\mP^2}\right|^2-3\,\frac{|\tilde{W}|^2}
{\mP^2}\right],
\eeq
where $\tilde{W}=W/\ka M^2$ and $s$ stands for
any of the five complex scalar fields appearing
in Eq.~\eqref{eq:nsmMinK}. It has been numerically
verified \cite{newsmooth} that, for the parameters
in Eq.~\eqref{eq:nsmpars} and $\ga=0$, the
potential is again minimized for fixed $|S|$ on
the new smooth path for $\phi=\pm|\phi|$,
$\pb=\pm|\pb|$ and $H^c\Hb^c=\pm|H^c|^2$, where
the signs are correlated. (Recall, also, that S
has been chosen real and positive). So, we will
restrict our attention again to these directions.
Furthermore, we have found that the relative
error in approximating the new smooth path by
Eq.~\eqref{eq:VarExpans} or
\eqref{eq:FieldExpans} is of the same order
of magnitude as that in the global SUSY limit
(see Fig.~\ref{fig:RelativeError}), namely $\sim
1\text{\textperthousand}$. So, we will use these
expansions for the new smooth path in the SUGRA
case as well.

Below, we give the expansions of the various
quantities entering the potential of
Eq.~\eqref{eq:nsmVSUGRAmin}, calculated on the new
smooth path for $\ga=0$. Note that, besides $w$,
we now have another small variable, namely
$|S|/\mP$, which is expected to be at least
one order of magnitude below unity during
inflation (e.g. $S_Q/\mP\sim 0.08$ for the
relevant value of $v_g$). In addition, the
constants $v_g/\mP$ and $M/\mP$ are also well
below unity. We will treat only $v_g/\mP$
as an independent small constant since
$M/\mP=M/v_g\cdot v_g/\mP$ with $M/v_g\sim 1$.
Using Eqs.~\eqref{eq:VarExpans} and
\eqref{eq:xyz1}-\eqref{eq:xyz2}, we can expand
the superpotential and its derivatives on the
new smooth path as follows:
\begin{gather}
\frac{\tilde{W}}{\mP}\simeq\frac{|S|}
{\mP}\Big[1-\frac{1}{2}\,x_2(1-4x_2)\,w^4+
\dots\Big],\label{eq:WExpan}\\
\tilde{W}_S\simeq\Big[1-x_2^2\,w^4+\dots\Big],
\quad\tilde{W}_{\phi}\simeq\pm
\Big[-4x_2z_2\frac{M}{v_g}\,w^3+\dots\Big],
\label{eq:WsphiExpan}\\
\tilde{W}_{\pb}\simeq\pm
\Big[2\sqrt{2}p\,x_2^2\,w^2+\dots\Big],\quad
\tilde{W}_{H^c} = \pm\tilde{W}_{\Hb^c} \simeq
\Big[-2x_2\sqrt{z_2}\,w^2+\dots\Big],
\label{eq:WpbhExpan}
\end{gather}
where the $\pm$ signs are again correlated, the
ellipses represent terms of higher order in $w$
and the last equation in
Eq.~\eqref{eq:WpbhExpan} has been written
in the case where $H^c>0$ (for $H^c<0$ we
should put an overall minus sign in front of
the bracket). Using Eq.~\eqref{eq:VarExpans},
we can write the expansions of the fields on the
new smooth path as
\begin{gather}
\frac{\phi}{\mP}\simeq\pm\frac{M}{\mP}
\Big[x_2w^2+x_4w^4+\dots\Big],\quad
\frac{\pb}{\mP}\simeq\pm\sqrt{2}p\frac{v_g}{\mP}
\Big[y_1w+y_3w^3+\dots\Big],\label{eq:phipbexp}\\
\frac{H^c}{\mP}=\pm\frac{\Hb^c}{\mP}\simeq
\frac{v_g}{\mP}\Big[\sqrt{z_2}w+\frac{z_4}
{2\sqrt{z_2}}w^3+\dots\Big],\label{eq:hexp}
\end{gather}
where the $\pm$ signs are correlated with the ones in
Eqs.~\eqref{eq:WsphiExpan}-\eqref{eq:WpbhExpan}
and we again take $H^c>0$.

We will seek for an expansion of the dimensionless
potential $\tilde{V}_0$ on the new smooth path
(for $\ga=0$) in powers of $|S|/\mP$ and
$w$. One can easily show, using
Eqs.~\eqref{eq:nsmVSUGRAmin}-\eqref{eq:hexp},
that only even powers of $|S|/\mP$ and $w$
enter this expansion. Thus, the dimensionless
potential expanded in these variables up to
fourth order takes the form
\beq\label{eq:VSUGRAguess}
\tilde{V}_0\simeq A_0+A_2\frac{|S|^2}{\mP^2}
+A_4\frac{|S|^4}{\mP^4}+B_2w^2+B_4w^4.
\eeq
To construct the expansion of the dimensionless
potential on the new smooth inflationary path, we
first classify the various possible types of
dimensionless quantities entering the calculation
of $\tilde{V}_0$ on this path. The dimensionless
parameters $p$, $x_i$, $y_i$, $z_i$, $\la/\ka$
and $M/v_g$ will be considered to be of order
unity and, as all the quantities of order unity,
will be called of type $\1$. Any quantity that is
proportional to some positive power of
$w=v_g/|S|$ with coefficient of order unity will
be called of type $\ti$. Note that all the terms
in the square brackets in
Eqs.~\eqref{eq:WExpan}-\eqref{eq:hexp}
are either of type $\1$ or $\ti$. Furthermore, any
quantity that is proportional to some positive
power of $|S|/\mP$ with coefficient of order
unity will be called of type $\tii$. Finally,
positive powers of the small constant $v_g/\mP$
with coefficients of order unity will be called
quantities of type $\m$. It is easy to see, using
Eqs.~\eqref{eq:nsmVSUGRAmin}-\eqref{eq:hexp}, that
only even powers of $v_g/\mP$ appear in the
expansion of $\tilde{V}_0$. Quantities of the
form $\ti\cdot\tii$ can only take one of the
forms $\m$, $\m\cdot\ti$ and $\m\cdot\tii$. So,
the final expansion of $\tilde{V}_0$ is expected
to contain only terms of the form $\1$, $\ti$,
$\tii$, $\m$, $\m\cdot\ti$ and $\m\cdot\tii$.

Now, we can split the relevant range $v_g\lesssim
|S|\lesssim \mP$ of $|S|$ into two intervals
according to which of the two fourth order
quantities, $v_g^4/|S|^4$ and $|S|^4/\mP^4$,
dominates. The former dominates in the interval
$v_g\lesssim |S|\lesssim (v_g \mP)^{1/2}$, while
the latter in the interval $(v_g \mP)^{1/2}
\lesssim |S|\lesssim \mP$. Comparing the quantity
$v_g^2/\mP^2$ with the two aforementioned
fourth order quantities, we find that, in each
of the two intervals, it is smaller than the
dominant fourth order quantity in this interval.
So, all the terms of type $\m$ can be neglected
in the final expression of the potential in
Eq.~\eqref{eq:VSUGRAguess} provided that $A_4$
and $B_4$ contain terms of type $\1$, which turns
out to be the case (see below). The same is true
for the terms of order $v_g^2/\mP^2\cdot
v_g^2/|S|^2$ and $v_g^2/\mP^2\cdot|S|^2/\mP^2$ as
well as all the higher order terms of the form
$\m\cdot\ti$ and $\m\cdot\tii$. According to the
above, the dimensionless potential, up to fourth
order in $|S|/\mP$ and $w$, should only contain
terms of type $\1$, $\ti$ and $\tii$, which is
equivalent to saying that the coefficients $A_i$
and $B_i$ in Eq.~\eqref{eq:VSUGRAguess} should
not contain terms of type $\m$.

Let us now find some rules which can help us
manipulate the expansion of $\tilde{V}_0$ on the
new smooth path. First of all, note that this
dimensionless potential consists of a sum of
products of $\tilde{W}/\mP$, $\tilde{W}_s$
and $|s|/\mP$, as seen from
Eq.~\eqref{eq:nsmVSUGRAmin}. The quantities
$|s|/\mP$ with $s\ne S$ in
Eqs.~\eqref{eq:phipbexp}-\eqref{eq:hexp}
consist of terms of the form $\m\cdot\ti$,
while $|S|/\mP$ and the quantities in
Eqs.~\eqref{eq:WExpan}-\eqref{eq:WpbhExpan}
contain terms of the form $\1$, $\ti$, $\tii$
and $\m\cdot\ti$. It is readily shown that
products of any of these quantities can only
contain terms of type $\1$, $\ti$, $\tii$, $\m$,
$\m\cdot\ti$ and $\m\cdot\tii$. Moreover, one
can easily see that, if a term of type $\m$,
$\m\cdot\ti$ or $\m\cdot\tii$ is encountered
at any intermediate stage of the calculation,
it is bound to yield terms of type $\m$,
$\m\cdot\ti$, or $\m\cdot\tii$ in the final
expansion of $\tilde{V}_0$. However, we have
already shown that such terms need not be kept
in the final form of the potential since they
give a negligible contribution. Thus, we
conclude that we can drop terms of the form $\m$,
$\m\cdot\ti$ and $\m\cdot\tii$ whenever we come
across them and maintain only terms of the form
$\1$, $\ti$ and $\tii$ in the various stages of
the calculation. A corollary to this is that we
can take $K_0$ in the exponential of
Eq.~\eqref{eq:nsmVSUGRAmin} to be simply
$|S|^2$ and $\tilde{W}/\mP$ in
Eq.~\eqref{eq:WExpan} to be simply $|S|/\mP$.

Taking all the above into account, we can now
quite easily find that the relevant terms in the
dimensionless potential of
Eq.~\eqref{eq:nsmVSUGRAmin} on the new smooth
path (for $\ga=0$), will be all contained in
\beq\label{eq:tV0}
\tilde{V}_0\simeq e^{|S|^2/\mP^2}\;\Bigg[
\tilde{V}_g+\frac{|\tilde{W}|^2|S|^2}
{\mP^4}+\left(\frac{\tilde{W}_S^*\tilde{W}S^*}
{\mP^2}+\text{c.c.}\right)
-3\,\frac{|\tilde{W}|^2}{\mP^2}\Bigg],
\eeq
where $\tilde{V}_g=\sum_s|\tilde{W}_s|^2$ is the
dimensionless scalar potential in the global SUSY
limit. Substituting
Eqs.~\eqref{eq:WExpan}-\eqref{eq:WpbhExpan}
into Eq.~\eqref{eq:tV0} and keeping only the
relevant terms, we obtain the potential
\beq\label{eq:Vminsugra}
V_0\simeq\ka^2M^4\left(
1+\frac{1}{2}\,\frac{|S|^4}{\mP^4}
-\frac{v_g^4}{54|S|^4}\right).
\eeq
Note that, in our case, the leading SUGRA
correction to the inflationary potential for
minimal K\"{a}hler potential, which corresponds
to the second term in the parenthesis of
Eq.~\eqref{eq:Vminsugra}, is the same as the one
found in the first of Ref.~\cite{HybridSUGRA},
in the case of standard hybrid inflation and in
Ref.~\cite{SenoguzShafi}, in the case of shifted
and smooth hybrid inflation. Actually, the
inflationary potential for conventional smooth
hybrid inflation in Ref.~\cite{SenoguzShafi}
coincides with the potential in
Eq.~\eqref{eq:Vminsugra}, which applies to new
smooth hybrid inflation for $\ga=0$.

Let us now turn to the consideration of a more
general K\"{a}hler potential containing
non-minimal terms. As we are interested in the
region of field space with $|s|\ll\mP$, we can
expand the K\"{a}hler potential as a power series
in the fields. The same rules that we have
extracted above for manipulating the expansion of
the potential on the new smooth path in the case
of minimal K\"{a}hler potential hold for this
case as well. In particular, in expanding the
potential up to fourth order in $|S|/\mP$ and $w$,
we can drop terms of the form $\m$, $\m\cdot\ti$
and $\m\cdot\tii$ whenever they appear at an
intermediate stage of the calculation. As a
consequence, we can take $K$ in the exponential
of Eq.~\eqref{eq:nsmVSUGRA} to consist only of
terms containing solely powers of the field $S$
and not the other fields (compare with the
similar argument above in the case of a minimal
K\"{a}hler potential). Since terms of the form
$|S|^n(S^m+S^{*m})$ with $n\geq 0$ and $m\geq 1$
are not allowed due to the R symmetry, the only
relevant non-minimal K\"{a}hler potential terms are
\beq\label{eq:KahlerTerms1}
|S|^4/\mP^2,\quad |S|^6/\mP^4
\eeq
up to order six in $|S|/\mP$. The same terms are
the only non-minimal K\"{a}hler potential terms
(up to sixth order) which can give a non-negligible
contribution to $K_i/\mP$. This is due to the fact
that, in $K$, we cannot have terms with a single
field $s\ne S$ multiplying powers of $S$ and $S^*$
since there exist no other gauge singlet fields in
the theory. So, all terms in $K$ other than the
ones of the form in Eq.~\eqref{eq:KahlerTerms1}
contain at least two fields $s\neq S$ and, thus,
give negligible contributions to $K_i/\mP$.
Finally, the inverse K\"{a}hler metric
$(K^{-1})_i^j$ can be expanded as a power series
of the higher order terms contained in the
K\"{a}hler metric $K_i^j$. Besides the terms of
the form in Eq.~\eqref{eq:KahlerTerms1}, other
K\"{a}hler potential terms that can contribute to
$(K^{-1})_i^j$ are certainly the ones of the form
\beq\label{eq:KahlerTerms2}
|S|^2|s|^2/\mP^2
\eeq
with $s$ being any of the fields $\phi$, $\pb$,
$H^c$ and $\Hb^c$. In general, any term
containing two of the four fields $\phi$, $\pb$,
$H^c$ and $\Hb^c$ multiplied by powers of $S$
and $S^*$ will contribute. The only possible
combinations of two fields $s\neq S$, other than
$|s|^2$, that respect gauge invariance are
$H^c\Hb^c$, $\phi^2$, $\phi\pb$, $\phi^*\pb$ and
$\pb^2$ along with their complex conjugates. The
first two can be multiplied by powers of $|S|^2$,
while the other three need some extra $S$ or
$S^*$ factors in order to become R symmetry
invariant. In summary, we can parameterize the
most general K\"{a}hler potential which is
relevant for our calculation here as follows:
\bea\label{eq:GeneralKahler}
K&=&K_0+\frac{k_S}{4}\,\frac{|S|^4}{\mP^2}+
\frac{k_{SS}}{6}\,\frac{|S|^6}{\mP^4}+
\sum_{s\ne S}k_{Ss}\,\frac{|S|^2|s|^2}{\mP^2}+
\Bigg(k_{\phi\pb S^*}\frac{\phi\pb S^*}{\mP}+
k_{\phi^*\pb S^*}\frac{\phi^*\pb S^*}{\mP}
\nonumber\\
& &+k_{\pb\pb S^*S^*}\frac{\pb^2 S^{*2}}{\mP^2}+
k_{\phi\phi SS^*}\frac{\phi^2 |S|^2}{\mP^2}+
k_{H\Hb SS^*}\frac{H^c\Hb^c |S|^2}{\mP^2}+
\text{c.c.}\Bigg),
\eea
where the various $k$ coefficients are considered
to be of order unity. From this, we get
\beq\label{eq:dKdS}
\frac{K_S}{\mP}\simeq\frac{S^*}{\mP}
\left(1+\frac{k_S}{2}\,\frac{|S|^2}{\mP^2}
+\frac{k_{SS}}{2}\,\frac{|S|^4}{\mP^4}
\right),
\eeq
while all the other first derivatives $K_s/\mP$
are of the form $\m$, $\m\cdot\ti$ or
$\m\cdot\tii$ and can be neglected. The relevant
contributions to the K\"{a}hler metric and its
inverse are
\beq\label{eq:InverseK}
\big[K_i^j\big]\simeq
\left(\ba{ccccc}
K_1^1 & 0 & 0 & 0 & 0 \\
0 & K_2^2 & K_2^3 & 0 & 0 \\
0 & K_3^2 & K_3^3 & 0 & 0 \\
0 & 0 & 0 & K_4^4 & 0 \\
0 & 0 & 0 & 0 & K_5^5
\ea\right),
\eeq
\beq
\big[(K^{-1})_i^j\big]\simeq
\left(\ba{ccccc}
\frac{1}{K_1^1} & 0 & 0 & 0 & 0 \\[5pt]
0 & \frac{K_3^3}{D} & -\frac{K_2^3}{D}
& 0 & 0 \\[5pt]
0 & -\frac{K_3^2}{D} & \frac{K_2^2}{D}
& 0 & 0 \\[5pt]
0 & 0 & 0 & \frac{1}{K_4^4} & 0 \\[5pt]
0 & 0 & 0 & 0 & \frac{1}{K_5^5}
\ea\right),
\eeq
where
\begin{gather}
K_1^1\simeq 1+k_S\frac{|S|^2}{\mP^2}+
\frac{3}{2}\,k_{SS}\frac{|S|^4}{\mP^4},\quad
K_2^2 = 1+k_{S\phi}\frac{|S|^2}{\mP^2},\quad
K_3^3 = 1+k_{S\pb}\frac{|S|^2}{\mP^2},
\label{eq:ddKdSdS}\\
K_4^4 = 1+k_{SH}\frac{|S|^2}{\mP^2},\quad
K_5^5 = 1+k_{S\Hb}\frac{|S|^2}{\mP^2},\quad
K_2^3=K_3^{2*}=k^*_{\phi^*\pb S^*}
\frac{S}{\mP},\quad D=K_2^2K_3^3-|K_2^3|^2
\end{gather}
and $i=1,2,3,4,5$ correspond to the fields
$S$, $\phi$, $\pb$, $H^c$, $\Hb^c$ respectively.

As can be seen from Eqs.~\eqref{eq:nsmVSUGRA} and
\eqref{eq:InverseK}, the only contribution to the
scalar potential on the new smooth path
originating from non-diagonal elements of the
inverse K\"{a}hler metric comes from the term
$(K^{-1})_2^3\,F^{2*}F_3+\text{c.c.}$, which, on
the new smooth path, is given to leading order by
\beq
\ka^2M^4\left(16\sqrt{2}p\,x_2^3z_2\,\Real\{
k_{\phi^*\pb S^*}\}\right)\frac{M}{\mP}\;w^4.
\eeq
It is, thus, of the form $\m\cdot\ti$ and can be
dropped. From the diagonal entries in
$(K^{-1})_i^j$, one finds that the relevant
contributions to the potential on the new smooth
path will come from
\beq\label{eq:nsmVDom}
V\simeq e^{K/\mP^2}\Bigg[\left|W_S+\frac{WK_S}
{\mP^2}\right|^2K^{S^*S}+\sum_{s\ne S}|W_s|^2-
3\,\frac{|W|^2}{\mP^2}\Bigg].
\eeq
Substituting
Eqs.~\eqref{eq:WExpan}-\eqref{eq:WpbhExpan},
\eqref{eq:dKdS}, \eqref{eq:InverseK} and
\eqref{eq:ddKdSdS} into Eq.~\eqref{eq:nsmVDom},
expanding in powers of $|S|/\mP$ and keeping
only terms of type $\1$, $\ti$ and $\tii$, we
finally obtain, for the potential on the new
smooth path for $\ga=0$ in SUGRA, the
approximation
\beq\label{eq:nsmVSUGRAeff}
V\simeq v_0^4\left(1-k_S\frac{|S|^2}{\mP^2}+
\frac{1}{2}\,\ga_S\frac{|S|^4}{\mP^4}-
\frac{v_g^4}{54|S|^4}\right),
\eeq
where $v_0=\sqrt{k}M$ and $\ga_S\equiv 1-
\frac{7}{2}\,k_S-3\,k_{SS}+2\,k_S^2$. We see
that, from
the variety of terms in the K\"{a}hler potential,
only those with coefficients $k_S$ and $k_{SS}$
contribute to the scalar potential on the new
smooth path expanded up to fourth order in
$|S|/\mP$ and $v_g/|S|$. Note that
Eq.~\eqref{eq:nsmVSUGRAeff} coincides
with the corresponding result found in
Ref.~\cite{smoothnonminimal} in the case of
conventional smooth hybrid inflation. Moreover,
the SUGRA correction to the inflationary
potential, which corresponds to the second and
third terms in the parenthesis in the right hand
side of Eq.~\eqref{eq:nsmVSUGRAeff}, coincides
with the SUGRA correction found in
Ref.~\cite{SUSYnonminimal} in the
case of standard hybrid inflation.

All the above results hold as long as
Eq.~\eqref{eq:FieldExpans} is a good
approximation of the new smooth path for $\ga=0$,
in the case of a non-minimal K\"ahler potential
too. We have checked \cite{newsmooth} numerically
that, at least for values of the parameters close
to the ones in Eq.~\eqref{eq:nsmpars}, the
relative error in the fields on the new smooth
path remains smaller than $2\%$ for a general
K\"ahler potential (which can include more terms
besides the ones in Eq.~\eqref{eq:GeneralKahler}),
even when the various $k$ coefficients are of
order unity. As in Ref.~\cite{SUSYnonminimal},
the new terms in the inflationary potential
that originate from the non-minimal terms in
the K\"{a}hler potential and are proportional
to $|S|^2$ and $|S|^4$, can give rise to a
local minimum at $|S|=|S|_{\min}$ and maximum
at $|S|=|S|_{\max}<|S|_{\min}$ on the
inflationary path. This means that, if the
system starts from a point with $|S|>|S|_{\max}$,
it can be trapped in the local minimum of the
potential. Nevertheless, as in
Ref.~\cite{smoothnonminimal} where conventional
smooth hybrid inflation was considered, in the
case of new smooth hybrid inflation too, there
exists a range of values for $k_S$ where the
minimum-maximum on the inflationary potential
does not appear and the system can start its
slow rolling from any point without the danger
of getting trapped.

Let us find the condition for the inflationary
potential in Eq.~\eqref{eq:nsmVSUGRAeff}, which
holds in the case $\ga=0$, not to have the
``minimum-maximum'' problem. Using the
dimensionless real inflaton field $\hat{\si}
\equiv\si/\mP$, this potential and its
derivative with respect to $\hat{\si}$
are given by
\bea
\tilde{V} &\equiv& \frac{V}{v_0^4}\simeq
1-\frac{1}{2}\,k_S\,\hat{\si}^2+
\frac{1}{8}\,\ga_S\,\hat{\si}^4-
\frac{2\hat{v}_g^4}{27\hat{\si}^4},\\[5pt]
\frac{d\tilde{V}}{d\hat{\si}} &\equiv&
\frac{1}{v_0^4}\,\frac{dV}{d\hat{\si}}
\simeq -k_S\,\hat{\si}+\frac{1}{2}\,\ga_S\,
\hat{\si}^3+\frac{8\hat{v}_g^4}{27\hat{\si}^5},
\eea
where $\hat{v}_g\equiv v_g/\mP$ and $\ga_S$ is
assumed positive. We can evade the local minimum
and maximum of the inflationary potential if we
require that $d\tilde{V}/d\hat{\si}$ remains
positive for any $\hat{\si}>0$ so that this
potential is a monotonically increasing function
of $\si$. This gives the condition
\beq
f(\hat{\si})\equiv\hat{\si}^8-\frac{2k_S}{\ga_S}\,
\hat{\si}^6+\frac{16\hat{v}_g^4}{27\ga_S}\geq 0.
\eeq
For $k_S>0$, which is the interesting case as we
will soon see, the minimum of the function
$f(\hat{\si})$ lies at $\hat{\si}_1=\left(
3k_S/2\ga_S\right)^{1/2}$, with $f(\hat{\si}_1)=
-27k_S^4/16\ga_S^4+16\hat{v}_g^4/27\ga_S$ and
the requirement $f(\hat{\si}_1)\geq 0$ yields
the restriction
\beq\label{eq:kmax}
k_S\leq k_S^{\max}\equiv\frac{4}{3\sqrt{3}}
\;\ga_S^{3/4}\frac{v_g}{\mP}.
\eeq
Note that, for $\ga_S\sim 1$, this inequality
implies that $k_S<1$ and thus $\hat{\si}_1<1$,
so, the minimum of $f(\hat{\si})$ lies in the
relevant region where $\si<\mP$.

For $k_S>k_S^{\max}$, on the other hand, the
inflationary potential has a local minimum and
maximum which approximately lie at
\beq
\si_{\min}\simeq \mP\left(\frac{2k_S}{\ga_S}
\right)^{1/2},\quad \si_{\max}\simeq \mP\left(
\frac{8v_g^4}{27k_S\mP^4}\right)^{1/6}.
\eeq
Even in this case, the system can always undergo
a stage of inflation with the required number of
e-foldings starting at a $\si<\si_{\max}$. This
is due to the vanishing of the derivative
$V'(\si)$ at $\si=\si_{\max}$. However, the more
the e-foldings we want to obtain, the closer we
must set the initial $\si$ to the maximum of the
potential, which leads to an initial condition
problem. Yet, as we will see, we can obtain a
spectral index as low as $0.95$ at $k_0=0.002
\units{Mpc^{-1}}$, in agreement with the WMAP
three-year value $0.958\pm0.016$ \cite{WMAP3},
maintaining the constraint $k_S\leq k_S^{\max}$.

Using the inflationary potential in
Eq.~\eqref{eq:nsmVSUGRAeff}, the
spectral index turns out to be
\beq
\ns\simeq 1+2\eta_Q\simeq
1-2k_S+3\,\ga_S\,\frac{\si^2_Q}{\mP^2}-
\frac{80v_g^4\mP^2}{27\si^6_Q},
\eeq
where $\eta_Q$ is the value of $\eta$ when the
pivot scale $k_0=0.002\units{Mpc^{-1}}$ crosses
outside the inflationary horizon. We can see
that the $k_S$ term in the K\"{a}hler potential
contributes to the lowering of the spectral
index if $k_S$ is positive. So, a $k_S$ with
this choice of sign can help us make the
spectral index comfortably compatible with the
three-year WMAP \cite{WMAP3} measurements.
However, since we cannot have any reliable and
convenient approximation for $\si_Q$, a
numerical investigation is required.

\begin{figure}[tp]
\centering
\includegraphics[width=\figwidth]{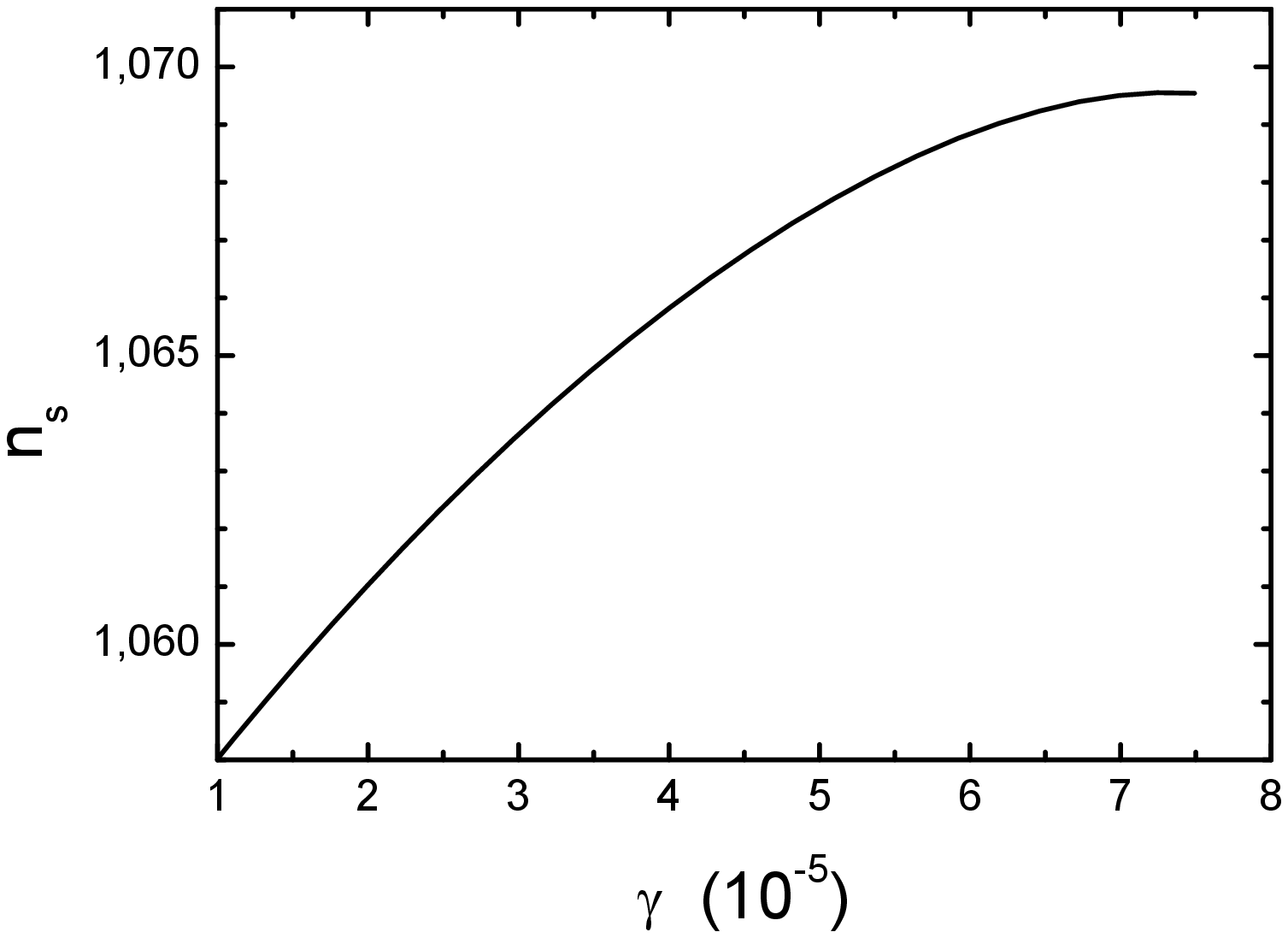}
\caption{Spectral index in new smooth hybrid
inflation as a function of $\ga$ in minimal
SUGRA for $p\equiv\sqrt{2}\ka M/m=1/\sqrt{2}$
and $\ka=0.1$. The endpoint of the curve at
$\ga\simeq 7.5\ten{-5}$ ($\ns\simeq 1.0695$)
corresponds to the case where our present horizon
scale crosses outside the inflationary horizon
when $\si=0.95\,\si_c$.}
\label{fig:nsMinSUGRA}
\end{figure}

Turning now to the case of small but non-zero
$\ga$, one can assert that, again, only the same
non-minimal terms of the K\"{a}hler potential
with coefficients $k_S$ and $k_{SS}$ will enter
the expansion of the potential on the new smooth
path, although the global SUSY potential for new
smooth hybrid inflation is not, in this case,
given by Eq.~\eqref{eq:Vg0} but has to be
calculated numerically. So, due to
the small value of $\ga$, we can assume that
the potential on the new smooth path in the case
of SUGRA with the non-minimal K\"{a}hler
potential of Eq.~\eqref{eq:GeneralKahler} and
$\ga\neq 0$ has the form
\beq
V\simeq v_0^4\left(\tilde{V}_\text{SUSY}
-\frac{1}{2}\,k_S\,\frac{\si^2}{\mP^2}
+\frac{1}{8}\,\ga_S\,\frac{\si^4}{\mP^4}
\right),
\eeq
where $\tilde{V}_\text{SUSY}\equiv V_\text{SUSY}/
v_0^4$ with $V_\text{SUSY}$ being the inflationary
potential in the case of global SUSY and
$\ga\neq0$. Note, also, that in the SUGRA case
with $\ga\neq0$, the critical value of $\si$
where the trivial flat direction becomes unstable
will be slightly different from the critical value
of $\si$ in the global SUSY case.

As in the global SUSY case with $\ga\neq 0$, we
take \cite{newsmooth} $p\equiv\sqrt{2}\ka M/m=
1/\sqrt{2}$, $\ka=0.1$ and fix numerically the
power spectrum $\PR$ of the primordial curvature
perturbation to the three-year WMAP \cite{WMAP3}
normalization. We also set the VEV $|\vev{H^c}|$
equal to the SUSY GUT scale, which, to a very
good approximation, means that we put $v_g\equiv
\sqrt{mM/\la}\simeq2.86\ten{16}\units{GeV}$. The
scalar spectral index in SUGRA with a minimal
K\"{a}hler potential (i.e. for $k_S=k_{SS}=0$)
as a function of the parameter $\ga$ is shown in
Fig.~\ref{fig:nsMinSUGRA}. We terminate the curve
when the value of $\si$ at which our present
horizon scale crosses outside the inflationary
horizon reaches $0.95\,\si_c$, as we did in
Fig.~\ref{fig:nsNewSmooth}. We see that minimal
SUGRA elevates the scalar spectral index above
the $95\%$ confidence level range obtained by
fitting the three-year WMAP data by the standard
power-law $\Lambda$CDM cosmological model ($\ns$
tends to approximately $1.055$ as $\ga\to0$).
This situation is readily rectified by the
inclusion of non-minimal terms in the K\"{a}hler
potential, as we will see below. For the range of
values of $\ga$ shown in Fig.~\ref{fig:nsMinSUGRA}
(i.e. for $\ga\simeq(1-7.5)\ten{-5}$), the ranges
of the other parameters of the model are as
follows \cite{newsmooth}:
$\la\simeq(1.33-1.68)\ten{-2}$,
$M\simeq(7.4-8.3)\ten{16}\units{GeV}$,
$m\simeq(1.48-1.66)\ten{16}\units{GeV}$,
$\si_c\simeq(4.2-9.8)\ten{17}\units{GeV}$,
$\si_Q\simeq(3.6-3.95)\ten{17}\units{GeV}$,
$\si_f\simeq(1.39-1.395)\ten{17}\units{GeV}$,
$N_Q\simeq54.3-54.4$, $d\ns/d\ln k\simeq-(2.1-2.6)
\ten{-3}$ and $r\simeq(2.4-3.8)\ten{-5}$.

\begin{figure}[tp]
\centering
\includegraphics[width=\figwidth]{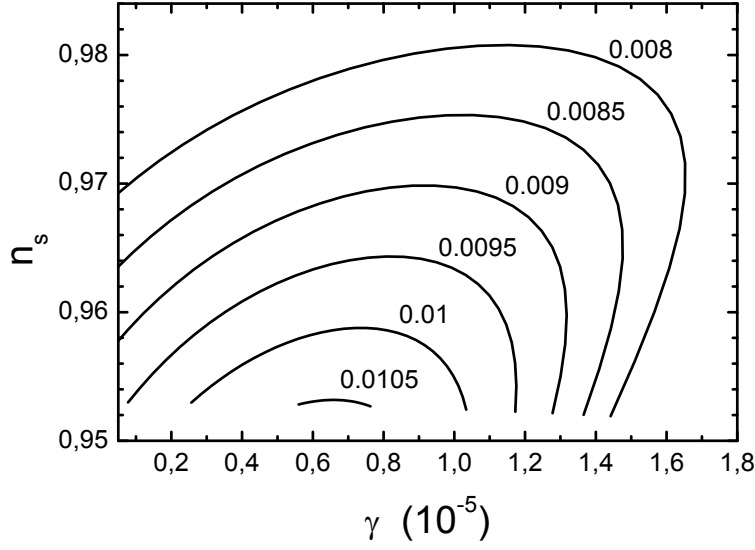}
\caption{Spectral index in new smooth hybrid
inflation in non-minimal SUGRA as a function of
$\ga$ for $p\equiv\sqrt{2}\ka M/m=1/\sqrt{2}$
and $\ka=0.1$. The values of $k_S$, which are
indicated on the curves, range from $0.008$ to
$0.0105$ and $k_{SS}=0$.}
\label{fig:SpectralIndex}
\end{figure}

Next, we consider the case where non-minimal
terms are present in the K\"{a}hler potential.
We will let $k_S$ have a non-zero positive value
but take $k_{SS}=0$ for simplicity. The spectral
index can again be numerically calculated
\cite{newsmooth} and plotted as a function
of the parameter $\ga$. This is shown in
Fig.~\ref{fig:SpectralIndex} where $\ns$ is
drawn for various values of $k_S$.
The limiting points on each curve correspond to
the situation where the potential on the new
smooth inflationary path starts developing a
local minimum and maximum. We observe that,
although all curves terminate on the right, only
curves that correspond to larger values of $k_S$
(and smaller values of $\ns$) have an endpoint
on the small $\ga$ side. It is instructive to
note that, for $\ga=0$, Eq.~\eqref{eq:kmax}
gives $k_S^{\max}\simeq 0.0088$, which is in
fairly good agreement with
Fig.~\ref{fig:SpectralIndex}. From this figure,
one can infer that the spectral index can be
readily set below unity in SUGRA with non-minimal
K\"{a}hler potential and that one can achieve a
value as low as $\ns\simeq 0.952$ without having
to put up with a local minimum and maximum of the
potential on the inflationary path. This minimum
value of $\ns$ corresponds to the endpoint of the
curve with $k_S=0.008$. The maximum allowed value
of $k_S$ is about 0.01054 corresponding to
$\ga\simeq 0.66\ten{-5}$ and $\ns\simeq 0.953$.
Finally, for the range of values of $\ga$ and
$k_S$ corresponding to the curves in
Fig.~\ref{fig:SpectralIndex}, the ranges of
variance of the other parameters of the model
are as follows \cite{newsmooth}:
$\la\simeq(1.5-2.6)\ten{-3}$,
$M\simeq(2.5-3.3)\ten{16}\units{GeV}$,
$m\simeq(0.5-0.66)\ten{16}\units{GeV}$,
$\si_c\simeq(0.45-1.7)\ten{18}\units{GeV}$,
$\si_Q\simeq(2.54-2.77)\ten{17}\units{GeV}$,
$\si_f\simeq(1.39-1.395)\ten{17}\units{GeV}$,
$N_Q\simeq53.6-53.8$, $d\ns/d\ln k\simeq-(7.2-9.2)
\ten{-4}$ and $r\simeq(3-9.6)\ten{-7}$. Variations
in the values of $p$ and $\ka$ have shown not to
have any significant effect on the results. In
particular, the spectral index cannot become
smaller than about 0.95 by varying these
parameters provided that the appearance of a
local minimum and maximum on the inflationary
potential is avoided. Note, however, that smaller
values of $\ns$ can be readily achieved but only
at the cost of the presence of the
``minimum-maximum'' problem.

\newpage
\thispagestyle{empty}
\mbox{}

\chapter{Standard-Smooth Hybrid Inflation}
\label{sec:STSMOOTH}

\section{Introduction}
\label{sec:STSMOOTHintro}

As we have mentioned before, it is well known
that the standard supersymmetric realization of
hybrid inflation in the context of grand unified
theories leads, at the end of inflation, to a
copious production of topological defects such as
cosmic strings \cite{string}, magnetic monopoles
\cite{monopole}, or domain walls \cite{domain},
if these defects are predicted by the underlying
symmetry breaking. In the case of magnetic
monopoles or domain walls, this causes a
cosmological catastrophe. The simplest GUT gauge
group whose breaking to the SM gauge group
predicts the existence of topologically stable
magnetic monopoles is the Pati-Salam group
$G_{\rm PS}=\rm SU(4)_c\times SU(2)_L\times
SU(2)_R$ \cite{PatiSalam}. (Note that the
PS monopoles carry \cite{magg} two units of
Dirac magnetic charge.) So, applying the
standard realization of hybrid inflation
within the SUSY PS GUT model, we encounter a
cosmologically disastrous overproduction of
magnetic monopoles at the end of inflation,
where the GUT gauge symmetry $G_{\rm PS}$
breaks spontaneously to $G_{\rm SM}$.

Possible ways out of this difficulty are provided
by the shifted or smooth variants of SUSY hybrid
inflation (see Chap.~\ref{sec:HYBRID}), which, in
their conventional realization, utilize
non-renormalizable superpotential terms. As we
have seen, in these inflationary scenarios the
GUT gauge symmetry $G_{\rm PS}$ is broken to
$G_{\rm SM}$ already during inflation and, thus,
no magnetic monopoles are produced at the
termination of inflation. It has also been shown
in Chaps.~\ref{sec:NSHIFT} and \ref{sec:NSMOOTH},
that hybrid inflation of both the shifted and
smooth type can be implemented within an extended
SUSY PS model without the need of non-renormalizable
superpotential interactions. It is very interesting
to point out that this extended SUSY PS model,
described in Chap.~\ref{sec:QUASI}, was initially
constructed \cite{quasi} for solving a very
different problem. In SUSY models with exact
Yukawa unification, such as the simplest SUSY PS
model, and universal boundary conditions, the
$b$-quark mass comes out unacceptably large for
$\mu>0$. Therefore, Yukawa unification must be
moderately violated so that, for $\mu>0$, the
predicted $b$-quark mass resides within the
experimentally allowed range even with universal
boundary conditions. This requirement has led
\cite{quasi} to the extension of the superfield
content of the SUSY PS model by including, among
other superfields, an extra pair of $\rm SU(4)_c$
non-singlet $\rm SU(2)_L$ doublets, which
naturally develop \cite{wetterich} subdominant
VEVs and mix with the main electroweak doublets
of the model leading to a moderate violation of
Yukawa unification (see Chap.~\ref{sec:QUASI} for
details). It is quite remarkable that the
resulting extended SUSY PS model can automatically
and naturally lead \cite{newshifted,newsmooth}
to a new version of shifted and smooth hybrid
inflation based solely on renormalizable
superpotential terms. As in the conventional
realization of shifted and smooth hybrid
inflation, the GUT gauge group $G_{\rm PS}$ is
broken to $G_{\rm SM}$ already during inflation
in these models too and monopole production at
the end of inflation is avoided.

Unfortunately, there is generally a tension
between the above mentioned well-motivated,
natural and otherwise successful hybrid
inflationary models and the recent three-year
results \cite{WMAP3} from the WMAP satellite.
Indeed, in global SUSY, these models, possibly
with the exception of the smooth \cite{smooth}
and new smooth \cite{newsmooth} hybrid inflation
models, predict that,  the spectral index $\ns$
is very close to unity and with no much running.
Moreover, inclusion of supergravity corrections
with canonical K\"{a}hler potential yields, in
all cases, $\ns$'s which are very close to unity
or even exceed it. On the other hand, fitting the
WMAP3 data with the $\Lambda$CDM cosmological
model, one obtains \cite{WMAP3} $\ns$'s clearly
lower than unity.

One possible resolution of this inconsistency is
\cite{SUSYnonminimal,smoothnonminimal,newsmooth}
to use a non-minimal K\"{a}hler potential with
a convenient choice of the sign of one of its
terms, as we did in Sec.~\ref{sec:NSMOOTHsugra}.
This generates a negative mass term for the
inflaton and the inflationary potential acquires,
in general, a local minimum and maximum. Then,
as the inflaton rolls from this maximum down to
smaller values, hybrid inflation of the hilltop
type \cite{hilltop} can occur. In this case,
$\ns$ can become consistent with the WMAP3
measurements, but only at the cost of a mild
tuning \cite{gpp} of the initial conditions. In
any case, we must make sure that the system is
not trapped in the local minimum of the
inflationary potential, which can easily happen
for general initial conditions. In such a case,
no hybrid inflation would take place. Note that,
in the cases of smooth and new smooth hybrid
inflation, acceptable $\ns$'s can be obtained
\cite{smoothnonminimal,newsmooth} even without
the appearance of this local minimum and maximum
and, thus, the related complications can be
avoided (see Sec.~\ref{sec:NSMOOTHsugra}).

Another possibility \cite{mhin} for reducing the
spectral index predicted by hybrid inflation
models is based on the observation that, in
such models, $\ns$ generally decreases with
the number of e-foldings suffered by our
present horizon scale during hybrid inflation.
So, restricting this number of e-foldings, we
can achieve values of $\ns$ which are compatible
with the WMAP3 data even with minimal K\"{a}hler
potential. The additional number of e-foldings
required for solving the horizon and flatness
problems of standard hot big bang cosmology can
be provided by a second stage of inflation which
follows hybrid inflation. In Ref.~\cite{mhin},
this complementary inflation was taken to be of
the modular type \cite{modular}, realized by a
string axion at an intermediate scale. Note, in
passing, that a restricted number of e-foldings
during hybrid inflation was previously used
\cite{yamaguchi} to achieve sufficient running
of the spectral index.

In this chapter, we will describe an alternative,
two-stage inflationary model \cite{stsmooth},
based on the same extended SUSY PS model of
Chap.~\ref{sec:QUASI} which, as we saw, can
lead to new shifted, semi-shifted or new smooth
hybrid inflation. We will restrict ourselves in
the range of parameters of this model that
corresponds to the last case. As shown in
Chap.~\ref{sec:NSMOOTH}, the relevant scalar
potential possesses, in this case, a trivial
classically flat direction which is stable for
large values of the inflaton field. Along this
direction the PS gauge group is unbroken. For
values of the inflaton field smaller than a
certain critical value, this flat direction is
destabilized giving its place to a classically
non-flat valley of minima along which new
smooth hybrid inflation can take place. The
GUT gauge group $G_{\rm PS}$ is broken to
$G_{\rm SM}$ in this valley.

In Chap.~\ref{sec:NSMOOTH}, we investigated the
possibility that all the cosmologically relevant
scales exit the horizon during new smooth hybrid
inflation, which is, thus, responsible for the
observed spectrum of primordial fluctuations.
Here, we will consider an alternative possibility.
As usual, the trivial flat direction acquires
a logarithmic slope from one-loop radiative
corrections which are due to the SUSY breaking
caused by the non-vanishing potential energy
density on this direction. So, a version of
standard hybrid inflation can easily take place
as the system slowly rolls down the trivial
flat direction. We will assume here that the
cosmologically relevant scales exit the horizon
during this inflationary stage. Then, as in
Ref.~\cite{mhin}, we can easily achieve, in
global SUSY, spectral indices which are
compatible with the data by restricting the
number of e-foldings suffered by our present
horizon scale during this inflationary period.
The additional number of e-foldings required
for solving the horizon and flatness problems is
naturally provided, in this case, by a second
stage of inflation consisting mainly of new
smooth hybrid inflation. So, the necessary
complementary inflation is automatically built
in the model itself and we do not have to invoke
an {\it ad hoc} second stage of inflation as in
Ref.~\cite{mhin}. Furthermore, the PS monopoles
which are formed at the end of the standard
hybrid stage of inflation can be adequately
diluted by the subsequent second stage of
inflation. The inclusion of SUGRA corrections
with minimal K\"{a}hler potential raises the
spectral index, which, however, remains
acceptable for a wide range of the model
parameters. So, in this model, there is no need
to include non-minimal terms in the K\"{a}hler
potential and, consequently, complications from
the possible appearance of a local minimum and
maximum on the inflationary potential are avoided.

\section{Standard-smooth hybrid inflation in
global SUSY} \label{sec:STSMOOTHsusy}

The superpotential terms which are relevant for
inflation are given in Eq.~\eqref{eq:Wsshift}
or Eq.~\eqref{eq:Wnsmooth}, which we repeat here
for convenience
\beq
\label{eq:Wstsm}
W=\ka S(M^2-\phi^2)-\ga S H^c\Hb^c+m\phi\pb
-\la\pb H^c\Hb^c,
\eeq
where $M$, $m$ are superheavy masses of the order
of the SUSY GUT scale $M_{\rm GUT}\simeq 2.86
\ten{16}\units{GeV}$ and $\ka$, $\ga$, $\la$
are dimensionless coupling constants. All these
parameters are normalized so that they correspond
to the couplings between the SM singlet
components of the superfields. As we said in
Sec.~\ref{sec:SSHIFTsusy} and repeated in
Sec.~\ref{sec:NSMOOTHsusy}, the mass parameters
$M$, $m$ and any two of the three dimensionless
parameters $\ka$, $\ga$, $\la$ can be made real
and positive by appropriately redefining the
phases of the superfields. The third
dimensionless parameter, however, remains in
general complex. For definiteness, we choose
this parameter to be real and positive too as
we did in Chaps.~\ref{sec:SSHIFT} and
\ref{sec:NSMOOTH}.

The F-term scalar potential obtained from the
superpotential $W$ in Eq.~\eqref{eq:Wstsm}
is given by
\beq\label{eq:Vstsm}
V=|\ka\,(M^2-\phi^2)-\ga H^c\Hb^c|^2
+|m\pb-2\ka S\phi|^2+|m\phi-\la H^c\Hb^c|^2
+|\ga S+\la\pb\,|^2\left(|H^c|^2+|\Hb^c|^2\right),
\eeq
where the complex scalar fields which belong to
the SM singlet components of the superfields are
denoted by the same symbol. In
Sec.~\ref{sec:NSMOOTHsusy}, it was shown that
this potential leads to a new version of smooth
hybrid inflation provided that
\beq\label{eq:stsmmu2}
\tilde{\mu}^2\equiv-M^2+\frac{m^2}{2\ka^2}>0
\eeq
and the parameter $\ga$ is adequately small.
It was argued that, under these circumstances,
there exists a trivial classically flat direction
at $\phi=\pb=H^c=\Hb^c=0$ with $V=V_{\rm tr}
\equiv\ka^2M^4$, which is a valley of local
minima for
\beq
|S|>S_c\equiv\sqrt{\frac{\ka}{\ga}}\;M
\eeq
and becomes unstable for $|S|<S_c$, giving its
place to a classically non-flat valley of minima
along which new smooth hybrid inflation can take
place.

We will now briefly summarize some of the main
results of Chap.~\ref{sec:NSMOOTH}, which are
relevant for our discussion here. The SUSY
vacua of the potential in Eq.~\eqref{eq:Vstsm}
lie at
\beq\label{eq:stsmSUSYvac}
\phi=\frac{\ga m}{2\ka\la}\left(-1\pm\sqrt{1+
\frac{4\ka^2\la^2M^2}{\ga^2m^2}}\,\right)\equiv
\phi_{\pm},\quad \pb=S=0,\quad
H^c\Hb^c=\frac{m}{\la}\,\phi.
\eeq
The vanishing of the D-terms yields $\Hb^{c*}=
e^{i\theta}H^c$, which implies that there exist
two distinct continua of SUSY vacua:
\bea
\phi=\phi_{+},\quad\Hb^{c*}=H^c,\quad |H^c|=
\sqrt{\frac{m\phi_{+}}{\la}} \quad (\theta=0),
\label{eq:stsmvac1}\\
\phi=\phi_{-},\quad\Hb^{c*}=-H^c,\quad |H^c|=
\sqrt{\frac{-m\phi_{-}}{\la}} \quad
(\theta=\pi),
\label{eq:stsmvac2}
\eea
with $\pb=S=0$. One can show that the potential,
besides the trivial flat direction, possesses
generally two non-trivial flat directions too.
One of them exists only if $\tilde{M}^2\equiv
-\tilde{\mu}^2>0$ and lies at
\beq\label{eq:sshpathstsm}
\phi=\pm\,\tilde{M},\quad
\pb=\frac{2\ka\phi}{m}\,S,\quad H^c=\Hb^c=0.
\eeq
It is the semi-shifted flat direction discussed
in Chap.~\ref{sec:SSHIFT}, with $V=V_{\rm ssh}
\equiv\ka^2(M^4-\tilde{M}^4)$, along which
$G_{\rm PS}$ is broken to $G_{\rm SM}\times
{\rm U(1)_{B-L}}$. The second non-trivial flat
direction, which appears at
\begin{gather}
\phi=-\frac{\ga m}{2\ka\la},\quad
\pb=-\frac{\ga}{\la}\,S,\quad H^c\Hb^c=
\frac{\ka\ga(M^2-\phi^2)+\la m\phi}{\ga^2+\la^2},\\
V=V_{\rm nsh}\equiv\frac{\ka^2\la^2}{\ga^2+\la^2}
\left(M^2+\frac{\ga^2m^2}{4\ka^2\la^2}\right)^2,
\end{gather}
exists only for $\ga\neq 0$ and is analogous
to the trajectory for the new shifted hybrid
inflation of Chap.~\ref{sec:NSHIFT}. Along
this direction, $G_{\rm PS}$ is broken to
$G_{\rm SM}$. In our subsequent discussion, we
will concentrate on the case $\tilde{\mu}^2>0$,
where the shifted flat direction in
Eq.~\eqref{eq:sshpathstsm} does not exist. It is
interesting to point out that, in this case, we
always have $V_{\rm nsh}>V_{\rm tr}$ and it
is, thus, more likely that the system will
eventually settle down on the trivial rather
than the new shifted flat direction.

If we expand the complex scalar fields $\phi$,
$\pb$, $H^c$ and $\Hb^c$ in real and imaginary
parts according to the scheme $s=(s_1+is_2)/
\sqrt{2}$, we find that, on the trivial flat
direction, the mass-squared matrices
$M_{\phi1}^2$ of $\phi_1$, $\pb_1$ and
$M_{\phi2}^2$ of $\phi_2$, $\pb_2$ are
\beq\label{eq:stsmM2phi}
M_{\phi1(\phi2)}^2=\left(\ba{cc}
m^2+4\ka^2|S|^2\mp2\ka^2M^2 & -2\ka m S \\
-2\ka m S  & m^2 \ea\right)
\eeq
and the mass-squared matrices $M_{H1}^2$ of
$H^c_1$, $\Hb^c_1$ and $M_{H2}^2$ of $H^c_2$,
$\Hb^c_2$ are
\beq\label{eq:stsmM2H}
M_{H1(H2)}^{2}=\left(\ba{cc}
\ga^2|S|^2  & \mp\ga\ka M^2  \\
\mp\ga\ka M^2 & \ga^2|S|^2 \ea\right).
\eeq
The matrices $M_{\phi1(\phi2)}^2$ are always
positive definite, while the $M_{H1(H2)}^2$
acquire one negative eigenvalue for $|S|<S_c$.
Thus, the trivial flat direction is stable for
$|S|>S_c$ and unstable for $|S|<S_c$.

It has been shown in Sec.~\ref{sec:NSMOOTHsusy}
that, for small enough values of the parameter
$\ga$, the trivial flat direction, after
its destabilization at the critical point,
gives its place to a valley of {\em absolute}
minima for fixed $|S|$, which correspond to
$\theta\simeq 0$ and lead to the SUSY vacua in
Eq.~\eqref{eq:stsmvac1}. This valley possesses
an inclination already at the classical level
and can accommodate a stage of inflation with
the properties of smooth hybrid inflation. The
name ``new smooth'' hybrid inflation has been
coined \cite{newsmooth} for the inflationary
scenario obtained when all the e-foldings
required for solving the horizon and flatness
problems of standard hot big bang cosmology are
obtained when the system follows this valley.
In this chapter, as we have already mentioned,
we will study the case when the total required
number of e-foldings splits between two stages
of inflation, the standard hybrid inflation
stage for $|S|>S_c$ and the new smooth hybrid
inflation stage, including an intermediate
short inflationary period, for $|S|<S_c$.

The general outline of this scenario, which has
been named \cite{stsmooth} ``standard-smooth''
hybrid inflation, goes as follows. We assume
that the system, possibly after a period of
pre-inflation at the Planck scale, settles down
at a point on the trivial flat direction with
$|S|>S_c$ (see e.g.~\cite{init}). The constant
classical potential energy density on this
direction breaks SUSY explicitly and implies
the existence of one-loop radiative corrections,
which lift the flatness of the potential producing
the necessary inclination for driving the inflaton
towards the critical point at $|S|=S_c$. So the
standard hybrid inflation stage of the scenario
can be realized along this path. As the system
moves below the critical point, some of the
masses squared of the fields become negative,
resulting to a phase of spinodal decomposition.
This phase is relatively fast, causes the
spontaneous breaking of $G_{\rm PS}$ to
$G_{\rm SM}$ and generates a limited number of
e-foldings. After this intermediate inflationary
phase, the system settles down on the new smooth
hybrid inflationary path and, thus, new smooth
hybrid inflation takes place. The second stage of
inflation, consisting of the intermediate phase
and the subsequent new smooth hybrid inflation,
yields the additional number of e-foldings
required for solving the horizon and flatness
problems of standard hot big bang cosmology. At
the end of this stage, the system falls rapidly
into the appropriate SUSY vacuum of the theory
leading, though, to no topological defect
production, since the GUT gauge group is already
broken to the SM gauge group during this
inflationary stage. Two more requirements need
to be fulfilled in order for this scenario to
be viable. First, one has to make sure that
the number of e-foldings generated during the
second stage of inflation is adequate for
diluting any monopoles generated during the
phase transition at the end of the first stage of
inflation. Secondly, one must ensure that all the
cosmologically relevant scales receive
inflationary perturbations only from the first
stage of inflation, so that the existence of
measurable perturbations originating from the
phase of spinodal decomposition, which are of a
rather obscure nature, is avoided. Both of these
requirements are very easily satisfied in this
model, as we will see in the course of the
subsequent discussion.

The one-loop radiative correction to the
potential due to the SUSY breaking on the
trivial inflationary path is calculated,
as usual, by the Coleman-Weinberg formula:
\beq
\Delta V=\frac{1}{64\pi^2}\,\sum_i(-1)^{F_i}
M_i^4\ln\frac{M_i^2}{\Lambda^2},
\eeq
where the sum extends over all helicity states
$i$, $F_i$ and $M_i^2$ are the fermion number
and mass squared of the $i$th state and $\Lambda$
is a renormalization mass scale. In order to use
this formula for creating a logarithmic slope in
the inflationary potential, we have first to
derive the mass spectrum of the model on the
trivial inflationary path. It is easy to see
that, in the bosonic sector, we obtain two
groups of 45 pairs of real scalars with the
mass-squared matrices
\beq
M_{-(+)}^2=\left(\ba{cc}
m^2+4\ka^2|S|^2\mp2\ka^2M^2 & -2\ka m S \\
-2\ka m S  & m^2 \ea\right)
\eeq
and two more groups of 8 pairs of real scalars
with mass-squared matrices
\beq
M_{1(2)}^{2}=\left(\ba{cc}
\ga^2|S|^2  & \mp\ga\ka M^2  \\
\mp\ga\ka M^2 & \ga^2|S|^2 \ea\right).
\eeq
Note that $M_{-(+)}^2$ equals $M_{\phi1(\phi2)}^2$
of Eq.~\eqref{eq:stsmM2phi} and $M_{1(2)}^2$
equals $M_{H1(H2)}^{2}$ of Eq.~\eqref{eq:stsmM2H}.
In the fermionic sector of the theory, we obtain
45 pairs of Weyl fermions with mass-squared matrix
\beq
M_0^2=\left(\ba{cc}
m^2+4\ka^2|S|^2 & -2\ka m S \\
-2\ka m S  & m^2 \ea\right)
\eeq
and 8 more pairs of Weyl fermions with
mass-squared matrix
\beq
\bar{M}_0^2=\left(\ba{cc}
\ga^2|S|^2  & 0 \\
0 & \ga^2|S|^2 \ea\right).
\eeq
Note that the matrices $M_0^2$, $\bar{M}_0^2$
equal $M_{-(+)}^2$, $M_{1(2)}^{2}$ respectively,
without the $\mp$ terms in those. The one-loop
radiative correction to the inflationary
potential then takes the form
\bea\label{eq:stsmRadCorr}
\Delta V &=& \frac{45}{64\pi^2}\;\tr\Big\{
M_{+}^4\ln\frac{M_{+}^2}{\Lambda^2}+
M_{-}^4\ln\frac{M_{-}^2}{\Lambda^2}
-2M_{0}^4\ln\frac{M_{0}^2}{\Lambda^2}\Big\}
\nonumber\\
& &+\frac{8}{64\pi^2}\;\tr\Big\{M_{1}^4\ln
\frac{M_{1}^2}{\Lambda^2}+M_{2}^4\ln
\frac{M_{2}^2}{\Lambda^2}-2\bar{M}_{0}^4
\ln\frac{\bar{M}_{0}^2}{\Lambda^2}\Big\}.
\eea
The total effective potential on the trivial
inflationary path will be given by
$V_{\rm tr}^{\rm eff}=v_0^4+\Delta V$, where
$v_0\equiv\sqrt{\ka}M$ is the inflationary
scale. As already mentioned, the one-loop
radiative correction to the inflationary
potential lifts its classical flatness and
generates a logarithmic slope which is necessary
for driving the system towards the critical point
at $|S|=S_c$. It is important to note that the
$\sum_i(-1)^{F_i}M_i^4=8v_0^4\,(45\ka^2+4\ga^2)$
is $S$-independent, which implies that the slope
is $\Lambda$-independent and the scale $\Lambda$,
which remains undetermined, does not enter the
inflationary observables.

Making the complex scalar field $S$ real by an
appropriate global $\rm U(1)$ R transformation
and defining the canonically normalized real
inflaton field $\si\equiv\sqrt{2}S$, the
slow-roll parameters $\epsilon$, $\eta$ and the
parameter $\xi^2$, which enters the running of
the spectral index, are (see
Sec.~\ref{sec:HYBRIDintro})
\beq
\epsilon \equiv \frac{\mP^2}{2}\,\left(
\frac{V'(\si)}{V(\si)}\right)^2,\quad
\eta \equiv \mP^2\,\left(\frac{V''(\si)}
{V(\si)}\right),\quad \xi^2 \equiv \mP^4
\left(\frac{V'(\si)V'''(\si)}{V^2(\si)}\right),
\eeq
where the prime denotes derivation with respect
to the inflaton $\si$ and $\mP\simeq 2.44\ten{18}
\units{GeV}$ is the reduced Planck mass. In these
equations, $V$ is either the effective potential
$V_{\rm tr}^{\rm eff}$ on the trivial inflationary
path defined above, if we are referring to the
standard hybrid stage of inflation, or the
effective potential $V_{\rm nsm}^{\rm eff}$ for
new smooth hybrid inflation, which has to be
calculated numerically (see
Chap.~\ref{sec:NSMOOTH}), if we are referring
to the new smooth hybrid inflationary phase.

Numerical simulations have shown \cite{stsmooth}
that, after crossing the critical point at $\si=
\si_c\equiv\sqrt{2}S_c$, the system continues
evolving, for a while, with the Hubble parameter
$H$ remaining approximately constant and equal to
$H_0\equiv v_0^2/\sqrt{3}\mP$, until it settles
down on the new smooth hybrid inflationary path
at $\si\approx0.99\,\si_c$. The scale factor of
the universe increases by about 8 e-foldings
during this intermediate period. The fields $H^c$
and $\Hb^c$ are effectively massless at $\si=\si_c$
and, thus, acquire inflationary perturbations
$\delta H^c=\delta\Hb^c\approx H_0/2\pi$. So, for
the purpose of numerical simulation, their initial
values at the critical point have been taken equal
to these perturbations. The inflaton $\si$ is
assumed to have an initial velocity given by the
slow-roll equation
\beq
\dot{\si}=-\frac{V_{\rm tr}^{\rm eff\prime}
(\si_c)}{3H_0},
\eeq
where the overdot denotes derivation with respect
to the cosmic time $t$ and the inclination
$V_{\rm tr}^{\rm eff\prime}(\si_c)$ is provided
by the radiative corrections on the trivial flat
direction (for the parameter values that are of
interest, the slow-roll conditions $\epsilon
\leq 1$, $|\eta|\leq 1$ for the first stage of
inflation are violated only very close to the
critical point). Although the above results
are not independent from the values of the
model parameters, they represent legitimate
mean values. Moreover, inflationary observables
like the spectral index have shown
\cite{stsmooth} not to depend significantly
on the properties of this intermediate phase.

From the above we see that the number of
e-foldings from the time when the pivot scale
$k_0=0.002\units{Mpc^{-1}}$ crosses outside the
inflationary horizon until the end of inflation
is (see Sec.~\ref{sec:HYBRIDintro})
\beq\label{eq:stsmNQ}
N_Q\approx\frac{1}{\mP^2}\,
\int_{\si_f}^{0.99\,\si_c}
\frac{V_{\rm nsm}^{\rm eff}(\si)}
{V_{\rm nsm}^{\rm eff\prime}(\si)}\,d\si+
8+\frac{1}{\mP^2}\,\int_{\si_c}^{\si_Q}
\frac{V_{\rm tr}^{\rm eff}(\si)}
{V_{\rm tr}^{\rm eff\prime}(\si)}\,d\si,
\eeq
where $\si_Q\equiv\sqrt{2} S_Q>0$ is the value
of the inflaton field at horizon crossing of the
pivot scale and $\si_f$ refers to the value of
$\si$ at the end of the second stage of
inflation, which can be found from the
corresponding slow-roll conditions. The power
spectrum $\PR$ of the primordial curvature
perturbation at the scale $k_0$ is given by
\beq\label{eq:stsmPert}
\PR\simeq\frac{1}{2\pi\sqrt{3}}\,
\frac{[V_{\rm tr}^{\rm eff}(\si_Q)]^{3/2}}
{\mP^3V_{\rm tr}^{\rm eff\prime}(\si_Q)}.
\eeq
The spectral index $\ns$, the tensor-to-scalar
ratio $r$ and the running of the spectral index
$d\ns/d\ln k$ can be written as
\beq
\ns\simeq 1+2\eta-6\epsilon,\quad
r\simeq\,16\epsilon,\quad\frac{d\ns}{d\ln k}
\simeq16\epsilon\eta-24\epsilon^2-2\xi^2,
\eeq
where $\epsilon$, $\eta$ and $\xi^2$ are evaluated
at $\si=\si_Q$ (see Sec.~\ref{sec:HYBRIDintro}).
The number of e-foldings $N_Q$ that is required
for solving the horizon and flatness problems
of standard hot big bang cosmology is given
approximately by (see e.g.~\cite{LazaridesReview})
\beq\label{eq:stsmNQvsVinf}
N_Q\simeq53.76+\frac{2}{3}\,\ln\left(\frac{v_0}
{10^{15}\units{GeV}}\right)+\frac{1}{3}\,\ln
\left(\frac{T_{\rm r}}{10^9\units{GeV}}\right),
\eeq
where $T_{\rm r}$ is the reheat temperature
that is expected not to exceed about $10^9
\units{GeV}$, which is the well-known gravitino
bound \cite{gravitino}.

As already explained, magnetic monopoles are
produced at the end of the standard hybrid stage
of inflation, where $G_{\rm PS}$ breaks down
to $G_{\rm SM}$. We will now discuss, in some
detail, this production of magnetic monopoles and
their dilution by the subsequent second stage of
inflation. The masses of the fields $H^c$ and
$\Hb^c$, which vanish at $\si=\si_c$, grow very
fast as the system moves to smaller values of
$\si$. Actually, as one can show numerically
\cite{stsmooth}, they become of order $H_0$ when
the system is still very close to the critical
point. At that time the inflationary
perturbations of $H^c$ and $\Hb^c$ become
suppressed. After this, the system evolves
essentially classically. It remains, for a while,
close to the trivial flat direction (which, for
$\si<\si_c$, is unstable as it consists of saddle
points) yielding about 8 e-foldings, as mentioned
above. It finally settles down on the new smooth
hybrid inflationary path at $\si\approx 0.99\,
\si_c$. To be more precise, it ends up at a point
of the manifold which consists of the
{\em absolute} minima of the potential for fixed
$\si\approx 0.99\,\si_c$. The particular choice
of this point is made by the inflationary
perturbations of $H^c$ and $\Hb^c$, which cease
to operate when the masses of these fields reach
the value $H_0$. This happens after crossing the
critical point, but ``infinitesimally'' close to
it, as we already mentioned. So, the correlation
length that is relevant for magnetic monopole
production by the Kibble mechanism \cite{Kibble}
is approximately $H_0^{-1}$.

The initial monopole number density can then be
estimated \cite{Kibble} as
\beq\label{eq:inmondens}
n_{\rm M}^{\rm init}\approx
\frac{3{\sf p}}{4\pi}H_0^3,
\eeq
where ${\sf p}\sim 1/10$ is a geometric factor.
After inflation, the monopole number
density becomes
\beq\label{eq:finmondens}
n_{\rm M}^{\rm fin}\approx
\frac{3{\sf p}}{4\pi}H_0^3e^{-3\delta N},
\eeq
where $\delta N$ is the total number of e-foldings
during the intermediate period and the subsequent
new smooth hybrid inflation phase. Dividing
$n_{\rm M}^{\rm fin}$ by the number density
$n_{\rm infl}\approx V_{\rm tr}/m_{\rm infl}$ of
the inflatons that are produced at the termination
of inflation ($m_{\rm infl}$ is the inflaton mass
and $V_{\rm tr}\equiv v_0^4$), we obtain that, at
the end of inflation, the number density of
monopoles $n_{\rm M}$ is given by
\beq\label{eq:monoverinfl}
\frac{n_{\rm M}}{n_{\rm infl}}\approx
\frac{3{\sf p}}{4\pi}H_0^3e^{-3\delta N}
\frac{m_{\rm infl}}{V_{\rm tr}}.
\eeq
This ratio remains practically constant until
reheating, where the relative number density
of mono\-poles can be estimated as
(c.f.~\cite{thermal})
\beq\label{eq:relmon}
\frac{n_{\rm M}}{\sf s}=\frac{n_{\rm M}}
{n_{\rm infl}}\frac{n_{\rm infl}}{\sf s}
\approx\frac{3{\sf p}}{16\pi}
\frac{H_0T_{\rm r}}{\mP^2}e^{-3\delta N},
\eeq
where ${\sf s}$ is the entropy density and the
relations $n_{\rm infl}/{\sf s}=3T_{\rm r}/4
m_{\rm infl}$ (in the instantaneous inflaton
decay approximation) and $3H_0^2=V_{\rm tr}/
\mP^2$ were used. After reheating, the relative
number density of monopoles remains essentially
unaltered provided that there is no entropy
production at subsequent times. Taking
$n_{\rm M}/{\sf s}\lesssim 10^{-30}$, which
corresponds \cite{monreldens} to the Parker
bound \cite{Parker} on the present magnetic
monopole flux in our galaxy derived from
galactic magnetic field considerations,
$T_{\rm r}\sim 10^9\units{GeV}$ and $H_0
\sim 10^{12}\units{GeV}$, we obtain from
Eq.~\eqref{eq:relmon} that $\delta N\gtrsim 9.2$.
Using Eq.~\eqref{eq:stsmNQvsVinf}, this implies
that $N_{\rm st}\lesssim 45$, where $N_{\rm st}$
is the number of e-foldings of the pivot scale
$k_0$ during the standard hybrid stage of
inflation. Saturating this bound, we obtain a
monopole flux which may be measurable. However,
the interesting values of $N_{\rm st}$
encountered here, in the global SUSY case,
are much smaller (see below) and, thus, the
predicted magnetic monopole flux is unlikely
to be measurable. In the minimal SUGRA case,
$N_{\rm st}$ is restricted to quite small values
(see Sec.~\ref{sec:STSMOOTHsugra}) and the
monopole flux is predicted utterly negligible.

The model contains five free parameters, namely
$M$, $m$, $\ka$, $\ga$ and $\la$. As already
mentioned, the VEVs of $H^c$, $\Hb^c$ break
the PS gauge group to $G_{\rm SM}$, whereas the
VEV of the field $\phi$ breaks it only to
$G_{\rm SM}\times {\rm U(1)_{B-L}}$. So, the
gauge boson $A^\perp$ corresponding to the linear
combination of $\rm U(1)_Y$ and
${\rm U(1)_{B-L}}$ which is perpendicular to
$\rm U(1)_{Y}$, acquires its mass squared
$m^2_{A^\perp}=(5/2)g^2|\vev{H^c}|^2$ solely
from the VEVs $\vev{H^c}$, $\vev{\Hb^c}$ ($g$
is the SUSY GUT gauge coupling constant). On
the other hand, the masses squared $m_A^2$
and $m_{W_{\rm R}}^2$ of the color triplet,
anti-triplet ($A^\pm$) and charged
${\rm SU}(2)_{\rm R}$ ($W^\pm_{\rm R}$)
gauge bosons get contributions from $\vev{\phi}$
too. Namely, $m_A^2=g^2(|\vev{H^c}|^2+(4/3)
|\vev{\phi}|^2)$ and $m_{W_{\rm R}}^2=g^2
(|\vev{H^c}|^2+2|\vev{\phi}|^2)$. As we will see
below, the VEVs of $H^c$ and $\phi$ in the SUSY
vacua of the model turn out to be of the same
order of magnitude. Since the $A^\pm$ gauge
bosons are expected to affect the renormalization
group equations to a greater extent than the
$W^\pm_{\rm R}$ ones (the SM singlet gauge boson
$A^\perp$ does not affect them at all), we set
the mass $m_A$ divided by $g\approx 0.7$ equal to
the SUSY GUT scale $M_{\rm GUT}$. We also set the
value of the parameter $p\equiv \sqrt{2}\ka M/m$
equal to $1/\sqrt{2}$. Note that, for
$\tilde{\mu}^2>0$, this parameter is smaller
than unity, as seen from Eq.~\eqref{eq:stsmmu2}.
Finally, we take $T_{\rm r}$ to saturate the
gravitino bound \cite{gravitino}, i.e. $T_{\rm r}
\simeq 10^9\units{GeV}$, and fix the power
spectrum of the primordial curvature perturbation
to the WMAP3 \cite{WMAP3} normalization $\PR
\simeq 4.85\ten{-5}$ at the pivot scale $k_0$.
These choices fix three of the five parameters
of the model. So, we are left with two free
parameters. We take the ratio $\alpha\equiv
|\vev{H^c}|/|\vev{\phi}|$, which, for $\ga$
adequately small, approximately equals
$\sqrt{m/\la M}$ (see Eq.~\eqref{eq:stsmSUSYvac}),
to be one of them. The second free parameter can
be chosen to be the number of e-foldings
$N_{\rm st}$ of the pivot scale $k_0$ during the
standard hybrid stage of inflation ($N_{\rm st}$
can be fixed by adjusting e.g. the parameter
$\ga$). We plot our results \cite{stsmooth}
as functions of these two free parameters.

\begin{figure}[tp]
\centering
\includegraphics[width=\figwidth]{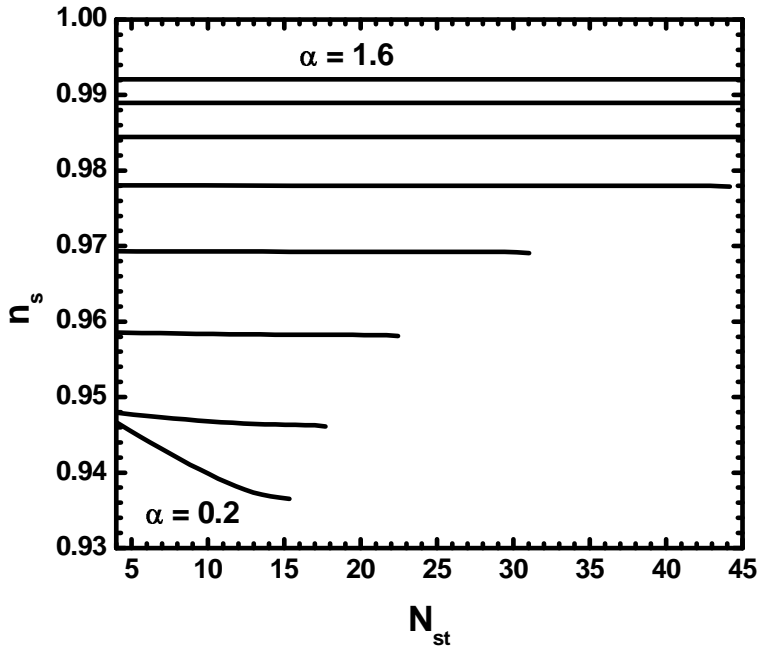}
\caption{Spectral index in standard-smooth
hybrid inflation versus $N_{\rm st}$ in global
SUSY for $p\equiv\sqrt{2}\ka M/m=1/\sqrt{2}$. The
parameter $\alpha\equiv|\vev{H^c}|/|\vev{\phi}|$
ranges from $0.2$ to $1.6$ with steps of $0.2$.}
\label{fig:stsm_nsSUSY}
\end{figure}

In Fig.~\ref{fig:stsm_nsSUSY}, we plot the
predicted spectral index of the model versus the
number of e-foldings $N_{\rm st}$ suffered by the
pivot scale $k_0$ during the standard hybrid stage
of inflation, for various values of the parameter
$\alpha$. Note that $N_{\rm st}$ is given by the
last term in the right-hand side of
Eq.~\eqref{eq:stsmNQ}. We have restricted ourselves
to $N_{\rm st}$'s between 4 and 45. The lower
limit guarantees the validity of our requirement
that all the cosmological scales receive
perturbations from the first stage of inflation.
Indeed, the number of e-foldings that elapse
between the horizon crossing of the pivot scale
$k_0$ and the largest cosmological scale
$0.1\units{Mpc^{-1}}$ is about 4. The upper
limit on $N_{\rm st}$ ensures that the present
flux of magnetic monopoles in our galaxy does not
exceed the Parker bound, as we showed above. The
parameter $\alpha$ is limited between $0.2$ and
$1.6$. Values of $\alpha$ lower than about $0.2$
require non-perturbative values of $\la$, whereas
$\alpha=1.6$ or higher is of no much interest
since it leads to unacceptably large $\ns$'s.
Whenever a curve in Fig.~\ref{fig:stsm_nsSUSY}
terminates on the right, this means that the
constraint on $\PR$ cannot be satisfied beyond
this endpoint. The WMAP3 data fitted by the
standard power-law $\Lambda$CDM cosmological
model predict \cite{WMAP3} that, at the pivot
scale $k_0$,
\beq
\label{eq:stsm_nswmap}
\ns=0.958\pm 0.016~\Rightarrow~0.926
\lesssim \ns \lesssim 0.99
\eeq
at $95\%$ confidence level. We see, from
Fig.~\ref{fig:stsm_nsSUSY}, that one can readily
obtain spectral indices that lie within this
2-$\si$ allowed range. Moreover, the 1-$\si$
range is fully covered by the predicted values
of the spectral index. Note, however, that one
cannot obtain spectral indices lower than about
$0.936$. It is obvious that large values of
$N_{\rm st}$, close to the Parker bound, are
of no much interest in our case, since they
yield large values for the spectral index. So,
a possibly measurable flux of monopoles at the
level of the Parker bound is rather unlikely
in this model.

For the curves depicted in
Fig.~\ref{fig:stsm_nsSUSY}, $\ga$ varies
\cite{stsmooth} in the range
$\ga\simeq(0.04-6)\ten{-3}$. It increases as
$\alpha$ decreases or $N_{\rm st}$ increases,
with its dependence on $N_{\rm st}$ being much
milder. The ranges of the other parameters of
the model are \cite{stsmooth}:
$\ka\simeq(0.46-3.62)\ten{-2}$,
$\la\simeq 0.004-1.56$,
$M\simeq(1.45-2.44)\ten{16}\units{GeV}$,
$m\simeq(0.13-1.56)\ten{15}\units{GeV}$,
$\si_Q\simeq(0.9-8.8)\ten{17}\units{GeV}$,
$\si_c\simeq(0.8-2.3)\ten{17}\units{GeV}$ and
$\si_f\simeq(0.5-1.5)\ten{17}\units{GeV}$.
The total number of e-foldings
from the time when the pivot scale $k_0$
crosses outside the inflationary horizon until
the end of the second stage of inflation is
$N_Q\simeq 53.7-54.7$. Finally,
$d\ns/d\ln k\simeq-(0.06-4)\ten{-3}$ and
the tensor-to-scalar ratio
$r\simeq(0.008-2.8)\ten{-4}$. A decrease in the
value of $p$, which is the only arbitrarily
chosen parameter, generally leads to an increase
of the spectral index. Thus, smaller values of
$p$ are expected to shift the curves in
Fig.~\ref{fig:stsm_nsSUSY} upwards, but
otherwise do not change the qualitative
features of the model.

\section{Supergravity corrections}
\label{sec:STSMOOTHsugra}

We now turn to the discussion of the SUGRA
corrections to the inflationary potentials of
the model. The F-term scalar potential in SUGRA
is given, as usual, by
\beq\label{eq:stsmVSUGRA}
V=e^{K/\mP^2}\left[(K^{-1})_i^j\;
F^{i*}F_j-3|W|^2/\mP^2\right],
\eeq
with $K$ being the K\"{a}hler potential and
$F^{i*}=W^i+K^iW/\mP^2$. A superscript (subscript)
$i$ denotes derivation with respect to the complex
scalar field $s_i$ ($s^{i*}$) and $(K^{-1})_i^j$
is the inverse K\"{a}hler metric. We will only
consider supergravity with minimal K\"{a}hler
potential and show that the WMAP3 results can
be met for a wide range of values of the
parameters of the model.

The minimal K\"{a}hler potential in the model
under consideration has, again, the form
\beq\label{eq:stsmMinKahler}
K^{\rm min}=|S|^2+|\phi|^2+|\pb|^2+|H^c|^2+|\Hb^c|^2
\eeq
and the corresponding F-term scalar potential is
\beq
\label{eq:stsmVSUGRAmin}
V^{\rm min}=e^{K^{\rm min}/\mP^2}\;\left[
\sum_{s}\left|W_s+\frac{Ws^*}{\mP^2}
\right|^2-3\,\frac{|W|^2}{\mP^2}\right],
\eeq
where $s$ stands for any of the five complex
scalar fields appearing in
Eq.~\eqref{eq:stsmMinKahler}. It is very easily
verified that, on the trivial flat direction,
this scalar potential expanded up to fourth
order in $|S|$ takes the form
\beq
V^{\rm min}_{\rm tr}\simeq v_0^4\left
(1+\frac{1}{2}\,\frac{|S|^4}{\mP^4}\right).
\eeq
Thus, after including the SUGRA corrections with
minimal K\"{a}hler potential, the effective
potential during the standard hybrid stage of
inflation becomes
\beq
\label{stsmVtrSUGRA}
V^{\rm SUGRA}_{\rm tr}\simeq
V^{\rm min}_{\rm tr}+\Delta V,
\eeq
with $\Delta V$ representing the one-loop
radiative correction given in
Eq.~\eqref{eq:stsmRadCorr}. Furthermore, it has
been shown in Sec.~\ref{sec:NSMOOTHsugra} that
the effective potential on the new smooth hybrid
inflationary path in the presence of minimal
SUGRA takes the form
\beq
V^{\rm SUGRA}_{\rm nsm}\simeq v_0^4
\left(\tilde{V}_{\rm nsm}+\frac{1}{2}\,
\frac{|S|^4}{\mP^4}\right),
\eeq
where $\tilde{V}_{\rm nsm}\equiv V_{\rm nsm}/
v_0^4$ and $V_{\rm nsm}$ represents the effective
potential on the new smooth hybrid inflationary
path in the case of global SUSY. Note that, in
the minimal SUGRA case, the critical value of
$\si$ where the trivial flat direction becomes
unstable, will be slightly different from the
critical value of $\sigma$ in the global SUSY
case.

The cosmology of the model after including the
minimal SUGRA corrections follows
straightforwardly from that of the global SUSY
case, if one replaces the inflationary effective
potentials of the latter by the ones derived
above and take into account some changes in
the intermediate phase between the two main
inflationary periods. Actually, one finds
\cite{stsmooth} numerically that, due to the
larger inclination of the inflationary path
provided by the minimal SUGRA corrections, the
number of e-foldings during the intermediate
period of inflation is reduced to about 2 or 3.
Also, the value of $\si$ at which the system
settles down on the new smooth hybrid inflationary
path decreases to about $\si\approx 0.95\,\si_c$.
Moreover, as it turns out, the evolution of the
system can be very well approximated by the
simplifying assumption that, during the
intermediate phase, the system also follows the
new smooth hybrid inflationary path. Therefore,
we remove the term 8 from the right-hand side
of Eq.~\eqref{eq:stsmNQ} and replace the upper
limit in the first integral by $\si_c$.

We again set the mass $m_A$ of the color
triplet, anti-triplet gauge bosons divided by
$g\approx 0.7$ equal to the SUSY GUT scale
$M_{\rm GUT}$ and the value of the parameter
$p\equiv\sqrt{2}\ka M/m$ equal to $1/\sqrt{2}$.
We also take $T_{\rm r}$ to saturate the
gravitino bound \cite{gravitino}, i.e. $T_{\rm r}
\simeq 10^9\units{GeV}$, and fix the power
spectrum of the primordial curvature perturbation
to the WMAP3 \cite{WMAP3} normalization
$\PR\simeq 4.85 \ten{-5}$ at the pivot scale
$k_0$. Finally, we will again plot our results
against the parameter $\alpha\equiv|\vev{H^c}|/
|\vev{\phi}|$ and the number of e-foldings
$N_{\rm st}$ of the pivot scale $k_0$ during the
standard hybrid stage of inflation.

\begin{figure}[tp]
\centering
\includegraphics[width=\figwidth]{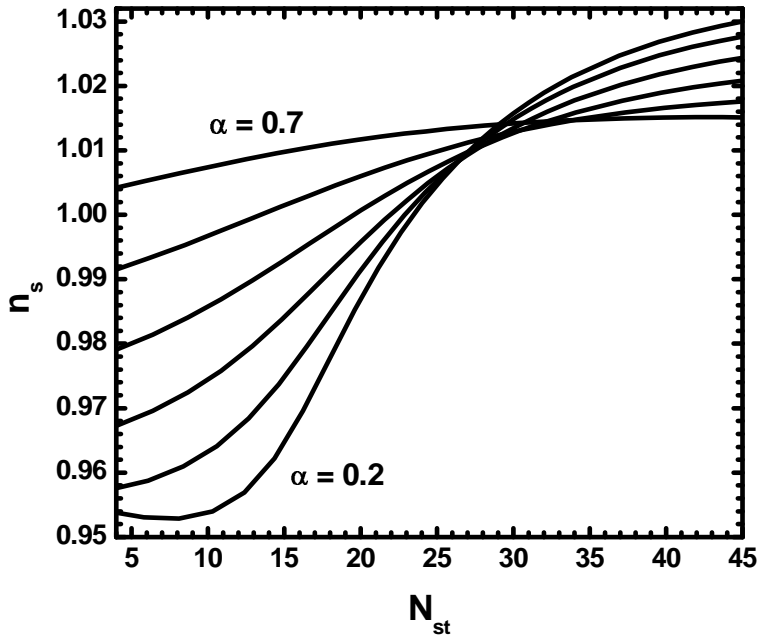}
\caption{Spectral index in standard-smooth hybrid
inflation versus $N_{\rm st}$ in minimal SUGRA
for $p\equiv\sqrt{2}\ka M/m=1/\sqrt{2}$. The
values of the parameter $\alpha$ range from
$0.2$ to $0.7$ with steps of $0.1$.}
\label{fig:stsm_nsSUGRA}
\end{figure}

In Fig.~\ref{fig:stsm_nsSUGRA}, we plot the
predicted spectral index of the model in minimal
SUGRA versus $N_{\rm st}$ for various values of
the parameter $\alpha$. We have allowed
$N_{\rm st}$ to vary only between 4 and 45 for
the same reasons mentioned in the global SUSY
case. For $\alpha$ smaller than about 0.2, the
required values of $\la$ turn out again to be
non-perturbative, whereas, for $\alpha$ greater
than about 0.7, the constraint on $\PR$ can not
be satisfied. We see that spectral indices below
unity are readily obtainable and that the central
value $\ns=0.958$ from the WMAP3 results is
achievable. Though, the spectral index cannot be
reduced below $\ns\simeq 0.953$, as is evident
from the curve with $\alpha=0.2$. Note that
values of $\ns$ in the $95\%$ confidence level
range of Eq.~\eqref{eq:stsm_nswmap} can be
obtained only if $N_{\rm st}$ is lower than
about 21. So, the predicted magnetic monopole
flux in our galaxy is utterly negligible.

The range of values of the parameter $\ga$
on the curves of Fig.~\ref{fig:stsm_nsSUGRA} is
\cite{stsmooth} $\ga\simeq(0.17-3.43)\ten{-3}$
with $\ga$ increasing with decreasing $\alpha$
and slightly increasing with increasing
$N_{\rm st}$. The ranges of the other parameters
of the model on these curves are \cite{stsmooth}:
$\ka\simeq(0.66-1.35)\ten{-2}$,
$\la\simeq 0.027-0.68$,
$M\simeq(2.12-2.44)\ten{16}\units{GeV}$,
$m\simeq(2.8-6.6)\ten{14}\units{GeV}$,
$\si_Q\simeq(0.95-3.05)\ten{17}\units{GeV}$,
$\si_c\simeq(0.6-2)\ten{17}\units{GeV}$
and $\si_f\simeq(4.9-9.9)\ten{16}\units{GeV}$.
The total number of e-foldings from the time
when the pivot scale $k_0$ crosses outside the
inflationary horizon until the end of the second
stage of inflation is $N_Q\simeq 54.1-54.5$.
Finally, $d\ns/d\ln k\simeq-(0.77-3.76)\ten{-3}$
and $r\simeq(0.7-5.3)\ten{-5}$. Again, a decrease
in the value of $p$ generally leads to a shift of
the curves in Fig.~\ref{fig:stsm_nsSUGRA} upwards,
without though affecting the other qualitative
features of the model.

\section{Gauge unification}
\label{sec:STSMOOTHgauge}

We will now briefly address the question of gauge
unification in the model. As the careful reader
may have noticed, cosmological considerations
have constrained the mass parameter $m$ to be
significantly lower than $M_{\rm GUT}$,
especially in the case of minimal SUGRA. This
could easily jeopardize the unification of gauge
coupling constants, and indeed it does, as it
turns out, since some of the fields that
contribute significantly to the gauge running
acquire masses of order $m$. Actually, there
are two different scales below $M_{\rm GUT}$
that give masses to fields contributing to the
renormalization group equations for the gauge
coupling constants. One of them is, as already
mentioned, around $m$ and the other is around
$|\vev{H^c}|=\sqrt{m|\vev{\phi}|/\la}$. This
holds in the minimal SUGRA case and, for not
too large $\ns$'s, in the global SUSY case too.
Gauge unification is destroyed for two reasons.
First of all, the fields which acquire masses
below $M_{\rm GUT}$ are too many and this causes
the appearance of Landau poles in the running of
the gauge coupling constants. Secondly, none of
these fields has $\rm SU(2)_L$ quantum numbers
and thus, even if divergences were not present,
the $\rm SU(2)_L$ gauge coupling constant would
fail to unify with the others .

The first problem can be avoided by considering
\cite{stsmooth} the superpotential term $\xi\phi^2
\pb$, which is allowed by all symmetries of the
theory (see Chap.~\ref{sec:QUASI}). The reason
for not including
this term in our discussion from the beginning is
that it does not contain a coupling between the
SM singlet components of $\phi$, $\pb$ and so it
does not affect the inflationary dynamics. This
is because $\phi^2\pb$ is the mixed product of
the three vectors $\phi$, $\phi$ and $\pb$ in
the 3-dimensional space in which the $\rm SO(3)$
group that is locally equivalent to $\rm SU(2)_R$
operates. Nevertheless, this term generates extra
contributions of order $|\xi\vev{\phi}|^2$ to the
mass squared of some fields and can, thus, help
us avoid the Landau poles.

The second problem can be solved only by
including extra fields in the model which affect
the running of the $\rm SU(2)_L$ gauge coupling
constant (c.f.~Sec.~\ref{sec:SSHIFTgauge}).
Note that, although the extended PS model under
consideration already contains fields with
$\rm SU(2)_L$ quantum numbers which are not
present in the minimal SUSY PS model, namely
the fields $h'$ and $\bar{h}'$ belonging to
the $({\bf 15,2,2})$ representation (see
Chap.~\ref{sec:QUASI}), these fields are not
sufficient for achieving the desired gauge
unification since they do not affect the running
of the $\rm SU(2)_L$ gauge coupling constant as
much as it is required. Consequently, one has
to consider the inclusion of some extra fields.
There is a good choice \cite{semishifted,stsmooth}
which utilizes a single extra field, namely a
superfield $\chi$ belonging to the $({\bf 15,3,1})$
representation. If we require that this field has
charge $1/2$ under the global $\rm U(1)$ R
symmetry, then the only superpotential term in
which this field is allowed to participate is a
mass term of the form $\frac{1}{2}m_{\chi}\chi^2$.
One can then tune the new mass parameter $m_{\chi}$
so as to achieve unification of the gauge coupling
constants. We find \cite{stsmooth} that this mass
should be $\approx 8\ten{14}\units{GeV}$.

It turns out that one can achieve gauge unification
at the appropriate scale ($\approx 2\ten{16}
\units{GeV}$) as long as the mass parameter $m$
is constrained to lie above $3\ten{14}\units{GeV}$.
This condition is fulfilled for almost all curves of
Figs.~\ref{fig:stsm_nsSUSY} and \ref{fig:stsm_nsSUGRA}
except for the curves with $\alpha=1.2$, $1.4$ and
$1.6$ in Fig.~\ref{fig:stsm_nsSUSY}. Note that this
constraint is equivalent to the statement that the
spectral index in the global SUSY case is less than
about $0.98$. So, the low spectral index regime is
not affected. Furthermore, if one wants to be on the
safe side, avoiding marginal gauge unification (the
value $m\approx 3\ten{14}\units{GeV}$ leads to gauge
unification with a rather large GUT gauge coupling
constant, which is of order unity or larger), then
one can impose the restriction $m\gtrsim 4\ten{14}
\units{GeV}$, which leads to the constraints $\alpha
\lesssim 0.8$ for Fig.~\ref{fig:stsm_nsSUSY} and
$\alpha\lesssim 0.5$ for Fig.~\ref{fig:stsm_nsSUGRA}.

\newpage
\thispagestyle{empty}
\mbox{}

\chapter{Conclusions}
\label{sec:CONC}

In this thesis we embarked upon the survey of the
diverse inflationary cosmology coming from a
specific particle physics model, namely the
extended SUSY Pati-Salam model with Yukawa
quasi-unification described in
Chap.~\ref{sec:QUASI}. We found that this model,
with the specific supermultiplet content,
symmetries and superpotential terms, can lead
to four distinct hybrid inflation scenarios of
different types and provides a very flexible
framework for inflationary phenomenology.

Despite this fact, this model was not first
constructed for cosmological purposes. It was
designed (see Chap.~\ref{sec:QUASI}) to cure
the problem that, in SUSY models with exact
Yukawa unification (such as the simplest SUSY
PS model) and universal boundary conditions,
the $b$-quark mass receives unacceptably large
values, for $\mu>0$. One way to deal with this
problem is to allow for a moderate violation of
Yukawa unification. This requirement has led to
the extension of the superfield content of the
SUSY PS model by including, among other things,
an extra pair of $\rm SU(4)_c$ non-singlet
$\rm SU(2)_L$ doublets, which naturally develop
subdominant VEVs and mix with the main electroweak
doublets of the model, leading to a moderate
violation of Yukawa unification. Also, the
presence of two extra superfields $\phi$, $\pb$
in the $({\bf 15,1,3})$ representation of
$G_{\rm PS}$ is necessitated by the requirement
that the violation of Yukawa unification is of
adequate magnitude. (Note, in passing, that this
mechanism applied to the $\mu<0$ case does not
lead \cite{cd2} to a viable scheme.) It is quite
remarkable that the resulting extended SUSY PS
model automatically and naturally incorporates
such a variety of inflationary models.

First, we reviewed the ``new shifted'' hybrid
inflation scenario, which was historically the
first to arise from the extended SUSY PS model
under consideration. In this model, the
inflationary superpotential contains only
renormalizable terms. In particular, the fields
$\phi$, $\pb$ lead to three new renormalizable
terms which are added to the standard
superpotential for SUSY hybrid inflation. We
showed that the resulting potential possesses
a ``shifted'' classically flat direction which
can serve as inflationary path. We analyzed
the mass spectrum of the model on this path and
constructed the one-loop radiative corrections
to the potential. These corrections generate a
slope along this path which can drive the system
towards the SUSY vacuum. The observational
constraint on the power spectrum amplitude of
the primordial curvature perturbation can be
easily satisfied with natural values of the
relevant parameters of the model. The slow roll
conditions are violated well before the
instability point of the new shifted path and,
thus, inflation terminates smoothly. The system
then quickly approaches the critical point and,
after reaching it, enters into a waterfall
regime during which it falls towards the SUSY
vacuum and oscillates about it. However, there
is no monopole production at the waterfall since
$G_{\rm PS}$ is broken to $G_{\rm SM}$ already
on the new shifted path.

As it turns out, the relevant part of inflation
occurs at values of the inflaton field which are
quite close to the reduced Planck scale. We
cannot, thus, ignore the SUGRA corrections which
can easily invalidate inflation by generating an
inflaton mass of the order of the Hubble constant.
In order to avoid this disaster, we described how
a particular mechanism \cite{panagiotak} can be
employed, leading to an exact cancellation of the
inflaton mass on the inflationary path. This
mechanism relies on a specific K\"{a}hler
potential and an extra gauge singlet with a
superheavy VEV via D-terms. The observational
constraint on $\PR$ can again be met by
readjusting the input values of the free
parameters which were obtained with global SUSY.

When, in Sec.~\ref{sec:NSHIFTsusy}, we searched
for flat directions in the potential, we pointed
out the existence of an extra flat direction,
apart from the new shifted and the trivial ones.
We discussed the properties of this direction and
the resulting inflationary scenario in
Chap.~\ref{sec:SSHIFT}. Since the fields $H^c$
and $\Hb^c$ do not have VEVs on this direction,
in contrast to the fields $\phi$ and $\pb$,
$G_{\rm PS}$ is not broken to $G_{\rm SM}$ but
to $G_{\rm SM}\times\rm U(1)_{B-L}$. Thus, we
have coined the name ``semi-shifted'' hybrid
inflation for this inflationary scenario. This
direction acquires a slope from one-loop
radiative corrections originating from the SUSY
breaking caused by the non-zero potential energy
density on this trajectory. As it turns out,
inflation terminates by violating the slow-roll
conditions well before the system reaches the
critical point of the semi-shifted path. The
subsequent breaking of the $\rm U(1)_{B-L}$
symmetry following the end of inflation leads
to the formation of local cosmic strings, which
contribute a small amount to the primordial
curvature perturbations.

It is known that, in the presence of a network of
cosmic strings, the present CMBR data can easily
become compatible with values of the spectral
index that are close to unity or even exceed it.
We have used a recent fit \cite{bevis1} to CMBR
and SDSS data which is based on field-theory
simulations of a dynamical network of local
cosmic strings. For the power-law $\Lambda$CDM
cosmological model this fit implies that, at
$95\%$ c.l., the spectral index is $\ns=0.94-1.06$
and the fractional contribution of cosmic strings
to the temperature power spectrum at $\ell=10$ is
$f_{10}=0.02-0.18$. Our numerical results show
that semi-shifted hybrid inflation with inclusion
of SUGRA corrections can easily become compatible
with this fit even without the need of non-minimal
terms in the K\"{a}hler potential or a subsequent
second stage of inflation. Taking into account the
constraints from the unification of the gauge
coupling constants, we have found that, for a
certain choice of parameters, the model yields
$f_{10}\simeq0.039$ in the HZ case (i.e.~for
$\ns=1$) and $\ns\simeq1.0254$ for the best-fit
value of $f_{10}$ ($=0.10$). Spectral indices
which are lower than about $0.98$ cannot be
obtained. So, the model shows a slight preference
to blue spectra. The cosmological disaster from
the possible overproduction of PS magnetic
monopoles is avoided since there is no production
of such monopoles at the end of inflation.

A very different scenario can arise from the same
SUSY PS model, for a wide range of the parameter
space, in the limit where one of the dimensionless
couplings of the theory, namely the parameter
$\ga$, becomes small. This is a new version of
smooth hybrid inflation, which, in contrast to
the conventional realization, is based only on
renormalizable interactions. An important
prerequisite for the viability of this model is,
as we pointed out, that a particular parameter of
the superpotential is adequately small. Then the
scalar potential of the model possesses, for a
wide range of its other parameters, valleys of
minima with classical inclination which can be
used as inflationary paths. This scenario, in
global SUSY, is naturally consistent with the
fitting of the three-year WMAP data by the
standard power-law $\Lambda$CDM cosmological
model. In particular, the spectral index turns
out to be adequately small so that it is
compatible with the data. Moreover, as in the
conventional realization of smooth hybrid
inflation, the PS gauge group is already
broken to the SM gauge group during inflation
and, thus, no topological defects are formed
at the end of inflation. Therefore, the
problem of possible overproduction of PS
magnetic monopoles is avoided.

Embedding the model in SUGRA with a minimal
K\"{a}hler potential raises the scalar spectral
index to values which are too high to be
compatible with the recent data. However,
inclusion of the leading non-minimal term in the
K\"{a}hler potential with appropriately chosen
sign can help to reduce the spectral index, so
that it resides comfortably within the allowed
range. Furthermore, the potential along the new
smooth inflationary path can remain everywhere
a monotonically increasing function of the
inflaton field. So, unnatural restrictions on
the initial conditions for inflation due to the
appearance of a maximum and a minimum on the
inflationary potential, which is common when
such a non-minimal K\"{a}hler term is used,
are avoided.

As we have seen, the extended SUSY PS model
incorporating Yukawa quasi-unification, can
automatically lead to new versions of the
shifted and smooth hybrid inflationary scenarios
based solely on renormalizable superpotential
interactions. In both of these cases, the PS
GUT gauge group is broken to the SM gauge group
already during inflation and, thus, no PS
magnetic monopole production takes place at the
end of inflation. In contrast to new smooth
hybrid inflation, the new shifted one yields,
in global SUSY, spectral indices which are too
close to unity and without much running, in
conflict with the recent WMAP data. Moreover,
inclusion of minimal SUGRA raises $\ns$ to
unacceptably large values in both of these
inflationary scenarios. It turns out that
this drawback can also be worked out within
the same extended SUSY PS model. In
Chap.~\ref{sec:STSMOOTH} we saw that this model
can also give rise a two-stage inflationary
scenario which can give acceptable $\ns$'s even
in minimal SUGRA. This scenario is naturally
realized for the range of values of the
parameters that lead to new smooth hybrid
inflation. The first stage of inflation is of
the standard hybrid type and takes place along
the trivial classically flat direction of the
scalar potential, which is stable for values of
the inflaton field larger than a certain critical
point. The inflaton is driven by the logarithmic
slope acquired by this direction from one-loop
radiative corrections, which are due to the SUSY
breaking caused by the non-vanishing potential
energy density on this direction. Note that, on
the trivial flat direction, the PS gauge group
is unbroken. Assuming that the cosmological
scales exit the horizon during the first stage
of inflation, we can achieve, in global SUSY,
spectral indices compatible with the WMAP3 data
by restricting the number of e-foldings suffered
by our present horizon scale during this
inflationary stage.

The system, after crossing the critical point of
the trivial flat direction, undergoes a
relatively short intermediate inflationary phase
and then falls rapidly into the new smooth hybrid
inflationary path along which it continues
inflating as it slowly rolls towards the vacua.
Note that this path appears right after the
destabilization of the trivial flat direction at
its critical point. During this second stage of
(intermediate plus new smooth hybrid) inflation,
the additional number of e-foldings needed for
solving the horizon and flatness problems is
naturally generated and $G_{\rm PS}$ is broken to
$G_{\rm SM}$. So, we see that the necessary
complementary inflation is automatically built in
the model itself and we do not have to invoke an
{\it ad hoc} second stage of inflation as in
other scenarios. Moreover, large reheat
temperatures can be achieved after the second
stage of inflation since this stage is realized
at a superheavy scale. Therefore, baryogenesis
via (non-thermal) leptogenesis may work in
this case in contrast to other models where the
reheat temperature is too low for sphalerons to
operate. Finally, the PS monopoles that are
formed at the end of the standard hybrid stage
of inflation can be adequately diluted by the
second stage of inflation. The monopole flux
in our galaxy in the case of global SUSY is
expected to be utterly negligible for values
of the spectral index that are of importance.

Including SUGRA corrections with minimal
K\"{a}hler potential enhances the predicted
values of the spectral index, which, however,
remain within the allowed interval for a wide
range of the model parameters. So, in this model,
there is no need to include non-minimal terms in
the K\"{a}hler potential and, thus, complications
from the possible appearance of a local maximum
and minimum on the inflationary potential are
avoided. The monopole flux in the SUGRA case
turns out not to be measurable for all the
allowed values of the model parameters.

\newpage
\thispagestyle{empty}
\mbox{}

\def\ijmp#1#2#3{{Int. Jour. Mod. Phys. }{\bf #1},~#3~(#2)}
\def\ijmpa#1#2#3{{Int. J. Mod. Phys.  A }{\bf #1},~#3~(#2)}
\def\plb#1#2#3{{Phys. Lett. B }{\bf #1},~#3~(#2)}
\def\zpc#1#2#3{{Z. Phys. C }{\bf #1},~#3~(#2)}
\def\prl#1#2#3{{Phys. Rev. Lett. }{\bf #1},~#3~(#2)}
\def\rmp#1#2#3{{Rev. Mod. Phys. }{\bf #1},~#3~(#2)}
\def\prep#1#2#3{{Phys. Rept. }{\bf #1},~#3~(#2)}
\def\prd#1#2#3{{Phys. Rev. D }{\bf #1},~#3~(#2)}
\def\prev#1#2#3{{Phys. Rev. }{\bf #1},~#3~(#2)}
\def\npb#1#2#3{{Nucl. Phys. B }{\bf #1},~#3~(#2)}
\def\npps#1#2#3{{Nucl. Phys. B (Proc. Sup.) }{\bf #1},~#3~(#2)}
\def\mpl#1#2#3{{Mod. Phys. Lett. }{\bf #1},~#3~(#2)}
\def\arnps#1#2#3{{Annu. Rev. Nucl. Part. Sci. }{\bf #1},~#3~(#2)}
\def\sjnp#1#2#3{{Sov. J. Nucl. Phys. }{\bf #1},~#3~(#2)}
\def\jetpl#1#2#3{{JETP Lett. }{\bf #1},~#3~(#2)}
\def\app#1#2#3{{Acta Phys. Polon. }{\bf #1},~#3~(#2)}
\def\rnc#1#2#3{{Riv. Nuovo Cim. }{\bf #1},~#3~(#2)}
\def\ap#1#2#3{{Ann. Phys. }{\bf #1},~#3~(#2)}
\def\ptp#1#2#3{{Prog. Theor. Phys. }{\bf #1},~#3~(#2)}
\def\rpp#1#2#3{{Rep. Prog. Phys. }{\bf #1},~#3~(#2)}
\def\n#1#2#3{{Nature }{\bf #1},~#3~(#2)}
\def\mnras#1#2#3{{MNRAS }{\bf #1},~#3~(#2)}
\def\s#1#2#3{{Science }{\bf #1},~#3~(#2)}
\def\cpc#1#2#3{{Comput. Phys. Commun. }{\bf #1},~#3~(#2)}
\def\epjc#1#2#3{{Eur. Phys. J. C }{\bf #1},~#3~(#2)}
\def\nima#1#2#3{{Nucl. Instrum. Meth. A }{\bf #1},~#3~(#2)}
\def\jhep#1#2#3{{JHEP }{\bf #1},~#3~(#2)}
\def\lnp#1#2#3{{Lect. Notes Phys. }{\bf #1},~#3~(#2)}
\def\appb#1#2#3{{Acta Phys. Polon. B }{\bf #1},~#3~(#2)}
\def\njp#1#2#3{{New J. Phys. }{\bf #1},~#3~(#2)}
\def\pl#1#2#3{{Phys. Lett. }{\bf #1B},~#3~(#2)}
\def\mpla#1#2#3{{Mod. Phys. Lett. A }{\bf #1},~#3~(#2)}
\def\jpcs#1#2#3{{J. Phys. Conf. Ser. }{\bf #1},~#3~(#2)}
\def\grg#1#2#3{{Gen. Rel. Grav. }{\bf #1},~#3~(#2)}
\def\baas#1#2#3{{Bull. Am. Astron. Soc. }{\bf #1},~#3~(#2)}
\def\jcap#1#2#3{{J. Cosmol. Astropart. Phys. }{\bf #1},~#3~(#2)}
\def\astp#1#2#3{{Astropart. Phys. }{\bf #1},~#3~(#2)}
\def\anj#1#2#3{{Astron. J. }{\bf #1},~#3~(#2)}
\def\apjss#1#2#3{{Astrophys. J. Suppl. Ser. }{\bf #1},~#3~(#2)}
\def\apjs#1#2#3{{Astrophys. J. Suppl. }{\bf #1},~#3~(#2)}
\def\apj#1#2#3{{Astrophys. J. }{\bf #1},~#3~(#2)}
\def\apjl#1#2#3{{Astrophys. J. Lett. }{\bf #1},~#3~(#2)}
\def\ibid#1#2#3{{\it ibid. }{\bf #1},~#3~(#2)}

\newpage
\thispagestyle{empty}
\mbox{}

\end{document}